\documentclass[onecolumn,11pt]{article} 

\usepackage[margin=1.1in]{geometry}
  
 \usepackage{setspace}
 \usepackage{float} 
\usepackage{graphicx}
\usepackage{amsmath}
\usepackage{amssymb}
\usepackage{dsfont}
\usepackage[english]{babel}
\usepackage[latin1]{inputenc}
\usepackage[T1]{fontenc}

\usepackage{algorithm, algorithmic}

\usepackage[authoryear, round]{natbib}

 \usepackage[
                  breaklinks = true,
                 colorlinks = true,
                 linkcolor = red,
                 urlcolor  = black, 
                 citecolor = blue,
                 anchorcolor = green,
                 ]{hyperref}

\graphicspath{
{Figures/}
{Figures/toy-nonlin/}
{Figures/toy-nonlin/PRM/}
{Figures/toy-nonlin/SRM/}
{Figures/toy-nonlin/bSRM/}
{Figures/waveform/}
{Figures/waveform/PRM/}
{Figures/waveform/SRM/}
{Figures/waveform/bSRM/}
{Figures/phonemes/}
{Figures/phonemes/PRM/}
{Figures/phonemes/PSRM/}
{Figures/phonemes/PbSRM/}
{Figures/yeast-cellcycle/}
{Figures/yeast-cellcycle/PRM/}
{Figures/yeast-cellcycle/SRM/}
{Figures/yeast-cellcycle/bSRM/}
{Figures/satellite/}
{Figures/satellite/PRM/}
{Figures/satellite/PSRM/}
{Figures/satellite/PbSRM/}
}

\newcommand{\bB}{\mathbf{B}}
\newcommand{\bO}{\mathbf{0}}

\newcommand{\bb}{\mathbf{b}}

\newcommand{\bs}{\mathbf{s}}

\newcommand{\bS}{\mathbf{S}}

\newcommand{\bx}{\mathbf{x}}
\newcommand{\bX}{\mathbf{X}}
\newcommand{\by}{\mathbf{y}}

\newcommand{\bz}{\mathbf{z}}

\newcommand{\bsx}{\boldsymbol{x}}

\newcommand{\cL}{\mathcal{L}}

\newcommand{\cD}{\mathcal{D}}

\newcommand{\cJ}{\mathcal{J}}

\newcommand{\bsmu}{\boldsymbol{\mu}}

\newcommand{\bsbeta}{\boldsymbol{\beta}}

\newcommand{\bsSigma}{\boldsymbol{\Sigma}}

\newcommand{\bstheta}{\boldsymbol{\theta}}

\newcommand{\bsepsilon}{\boldsymbol{\epsilon}}

\newcommand{\bsxi}{\boldsymbol{\xi}}

\newcommand{\Indicatrice}{\mathds{1}}

\newcommand{\Identity}{\textbf{I}}

\newcommand{\E}{\mathbb{E}}

\newcommand{\R}{\mathbb{R}}
	
\newcommand{\N}{\mathcal{N}}

\newcommand{\iid}{i.i.d.}

\date{}
\begin{document}

\title{{\bfseries Unsupervised learning of regression mixture models with unknown number of components}} 
\author{Faicel Chamroukhi}
 \maketitle
\begin{center}
Aix Marseille Universit\'e, CNRS, ENSAM, LSIS, UMR 7296, 13397 Marseille, France\\
Universit\'e de Toulon, CNRS, LSIS, UMR 7296, 83957 La Garde, France\\
\href{mailto:chamroukhi@univ-tln.fr}{chamroukhi@univ-tln.fr}
\end{center}

\begin{abstract}
Regression mixture models are widely studied in statistics, machine learning and data analysis. Fitting regression mixtures is challenging and is usually performed by maximum likelihood by using the expectation-maximization (EM) algorithm. However, it is well-known that the initialization is crucial for EM. If the initialization is inappropriately performed, the EM algorithm may lead to unsatisfactory results. The EM algorithm also requires the number of clusters to be given a priori; the problem of selecting the number of mixture components requires using model selection criteria to choose one from a set of pre-estimated candidate models. 
We propose a new fully unsupervised algorithm to learn regression mixture models with unknown number of components. 
The developed unsupervised learning approach consists in a penalized maximum likelihood estimation carried out by a robust expectation-maximization (EM) algorithm for fitting polynomial,  spline and B-spline regressions mixtures.
The proposed learning approach is fully unsupervised: 1) it simultaneously infers the model parameters and the optimal number of the regression mixture components from the data as the learning proceeds, rather than in a two-fold scheme as in standard model-based clustering using afterward model selection criteria, and 2) it does not require accurate initialization unlike the standard EM for regression mixtures.
The developed approach is applied to curve clustering problems. Numerical experiments on simulated data show that the proposed robust EM algorithm performs well and provides accurate results in terms of robustness with regard initialization and retrieving the optimal partition with the actual number of clusters. An application to real data in the framework of functional data clustering, confirms the benefit of the proposed approach for practical applications. 
\end{abstract}

%
 
\section{Introduction}

Mixture modeling \citep{mclachlanFiniteMixtureModels} is one of the most popular and successful approaches in  density estimation and cluster and discriminant analysis \citep{mclachlanFiniteMixtureModels, titteringtonBookMixtures, mclachlan_basford88, banfield_and_raftery_93, Fraley1998, Fraley2002_model-basedclustering, fraley_and_raftery_2007}.
In this paper we focus on clustering using the finite mixture model \citep{mclachlanFiniteMixtureModels}, that is model-based clustering \citep{banfield_and_raftery_93, mclachlan_basford88, Fraley2002_model-basedclustering} which is one of the most popular and successful approaches in cluster analysis. In the finite mixture approach for cluster analysis, the data probability density function is assumed to be a mixture density, each component density being associated with a cluster. The problem of clustering, therefore, becomes the one of estimating the parameters of the assumed mixture model (e.g, estimating the mean vector and the covariance matrix for each component density in the case of Gaussian mixture models). 
The mixture density estimation in this case is generally performed by maximizing the observed data log-likelihood. This can be achieved by the well-known expectation-maximization (EM) algorithm \citep{dlr, mclachlanEM} 
thanks to its good desirable properties of stability and reliable convergence.
One of the main model-based clustering approaches is the one based on Gaussian mixture models and the EM algorithm or its extensions \citep{mclachlanFiniteMixtureModels, mclachlanEM, Robust_EM_GMM}. It is in general concerned with multivariate data vectors. However, in many areas of application, 
the data are curves or functions. The analysis approaches in this case relate the  Functional Data Analysis (FDA)  framework  \citep{ramsayandsilvermanFDA2005} which concerns the paradigm of data analysis for which the individuals are entire functions or curves rather than vectors of reduced dimensions.
When the observations are curves or time series, the clustering can therefore be performed by using model-based curve clustering approaches, namely the regression mixture model  including polynomial regression mixtures, spline and B-spline regression mixtures \citep{Gaffney99trajectoryclustering, GaffneyThesis, chamroukhi_PhD_2010}. 
%
Modeling with regression mixtures is an important topic within the general family of mixture models. The regression mixture model \citep{Quandt1972, QuandtANDRamsey1978, DeVeaux1989, JonesANDMcLachlan1992, Gaffney99trajectoryclustering, VieleANDTong2002, FariaANDSoromenho2010, chamroukhi_PhD_2010, YoungANDHunter, HunterANDYoung} arises when we assume that the observed response $\by_i$ for the predictor variable $\bx_i$ is generated from one of $K$ possibly parametric regression functions $g(\by_i|\bx_i;\bstheta_k)$ of parameters  $\bstheta_k$ with prior probability $\pi_k$.
This includes polynomial regression mixtures, spline regression mixtures and B-spline regression mixtures. 
The problem of fitting regression mixture models is a widely studied problem in statistics, machine learning and data analysis, in particular for cluster analysis. 
It is usually performed by maximum likelihood using the expectation-maximization (EM) algorithm \citep{dlr, Gaffney99trajectoryclustering, JonesANDMcLachlan1992, mclachlanEM}. It is well-known that the initialization is crucial for EM. 
The EM algorithm also requires the number of clusters to be given a priori; the problem of selecting the number of components in this case requires using model selection criteria to choose one from a set of pre-estimated candidate models.
Some solutions have been provided in \cite{Robust_EM_GMM} and \cite{FigueiredoUnsupervisedlearningMixtures} for model-based clustering with Gaussian mixture models for multivariate data. 
In this paper we focus on model-based clustering approaches using regression mixtures for curves or functional data, rather than vectorial data.
We propose a new fully  unsupervised learning algorithm  to learn regression mixture models with unknown number of components. The developed unsupervised learning approach consists in a penalized maximum likelihood estimation carried out by a robust expectation-maximization (EM) algorithm for fitting polynomial,  spline and B-spline regressions mixtures. The proposed learning approach is fully unsupervised. It simultaneously infers the model parameters and the optimal number of the regression mixture components from the data as the learning proceeds, rather than in a two-fold scheme as in standard model-based clustering using afterward model selection criteria. Furthermore, it does not require accurate initialization unlike the standard EM for regression mixtures.

This paper is organized as follows. In section \ref{sec: related work MBC} we give a brief background on  model-based clustering using regression mixtures. In section \ref{sec: proposed robust EM-Mix-Reg} we present the proposed approach and the robust EM algorithm which maximizes a penalized log-likelihood criterion for regression mixtures in model-based  (curve) clustering, and derive the corresponding parameter updating formulas. Then, in section \ref{sec: Experiments} we present experimental results on both simulated data and real data to evaluate the proposed approach. Finally, section \ref{sec: conclusion and discussions} we discuss the proposal and provide concluding remarks and future directions.

\section{Model-based clustering with regression mixtures}
\label{sec: related work MBC}

\subsection{Model-based clustering} 
\label{ssec: related work MBC}

Model-based clustering  is a probabilistic generative approach for clustering that is based on the finite mixture model formulation \citep{banfield_and_raftery_93, mclachlan_basford88, Fraley2002_model-basedclustering, mclachlanFiniteMixtureModels}. It assumes that an observed data point $\bx_i$ from a sample $(\bx_1,\ldots,\bx_n)$, which is in general a multidimensional vector in $\R^d$, is (independently) generated from one of $K$ clusters with prior probability $\pi_k$. Each cluster being modeled by its specific density $f_k$ with possibly a set of parameters $\bstheta_k$.
The data are thus modeled by a finite mixture model of $K$ components $f_k$, which can be written as:
\begin{equation}
f(\bx_i;\bstheta) = \sum_{k=1}^K \pi_{k}f_k \big(\bx_i;\bstheta_k \big).
\label{eq: mixture model}
\end{equation}For example, in the case of a Gaussian mixture model, each of the density components $f_k \big(\bx_i;\bstheta_k \big)$ is a Gaussian density $\N(.;\bsmu_k,\bsSigma_k)$  with parameters $\bstheta_k=(\bsmu_k,\bsSigma_k)$ where $\bsmu_k$ is the mean vector and $\bsSigma_k$ is the covariance matrix.
%
The problem of clustering therefore becomes the one of estimating the mixture parameters $\bstheta$. The parameter estimation can be performed by maximizing the observed-data log-likelihood: 
\begin{equation}
\cL(\bstheta) = \sum_{i=1}^{n}\log\sum_{k=1}^K \pi_{k}f_k \big(\bx_i;\bstheta_k \big)
\label{eq: loglik normal mixture}
\end{equation}by using the EM algorithm \citep{dlr, mclachlanEM}. 

The standard model-based clustering technique based on multivariate unconditional mixture (\ref{eq: mixture model}) has been extended to the problem of clustering curves or functions, where the observed data are composed of both responses $\by_i$ and corresponding covariates $\bx_i$. In this case, the mixture component densities $f_k \big(\bx_i;\bstheta_k \big)$ in (\ref{eq: mixture model}) are replaced by conditional densities $f_k \big(\by_i|\bx_i;\bstheta_k \big)$, and  
a common way to model the conditional dependence in the observed data is to use regression functions. The resulting model is thus a regression mixture model, including polynomial regression mixture and (B-)spline regression mixtures, 
 see for example \citep{DeSarboAndCron1988, JonesANDMcLachlan1992, GaffneyThesis}. 
These approaches use the maximum likelihood estimation approach with the EM algorithm to estimate the model parameters. 
%
In the following section, we give an overview on the related model-based curve clustering approaches based on regression mixture models and the EM algorithm. 
 
\subsection{Regression mixtures for model-based clustering}
\label{ssec: polynomial and spline regression mixture}


The regression mixtures approaches assume that each observation $(\bx_i,\by_i)$ is drawn from one of $K$ clusters whose proportions (prior probabilities) are $(\pi_1,\ldots,\pi_K)$ and each cluster $k$ is supposed to be a set of realizations from a  polynomial 
regression model with coefficients $\bsbeta_k$ corrupted by a standard zero-mean Gaussian noise with a covariance-matrix $\sigma_k^2\Identity_{m_i}$:
\begin{equation}
\by_i = \bX_i \bsbeta_k + \bsepsilon_{ik}
\label{eq: polynomial regression model}
\end{equation}where $\by_i = (y_{i1},\ldots,y_{im_i})^T$ is an $m_i \times 1$ vector, $\bX_i$ 
is the $m_i \times (p+1)$ regression matrix (Vandermonde matrix) with rows 
\begin{equation}
\bsx_j = (1, x_{ij},x_{ij}^{2} \ldots, x_{ij}^{p}),
\end{equation}$p$ being the polynomial degree, $\bsbeta_k=(\beta_{k0},\ldots,\beta_{kp})^T$ is the  $(p+1) \times 1$ vector of regression coefficients for class $k$, 
 $\bsepsilon_{ik} \sim \N(\bO,\sigma_k^2\Identity_{m_i})$ is a multivariate standard zero-mean Gaussian vector with a covariance matrix $\sigma_k^2\Identity_{m_i}$,
 representing the corresponding additive Gaussian noise and $\Identity_{m_i}$ denotes the $m_i \times m_i$ identity matrix.  
Thus, the conditional density of curves for cluster $k$ is therefore given by the following functional normal density
\begin{equation}
f_k(\by_i|\bx_i, z_i = k;\bstheta_k) = \N (\by_{i};\bX_i \bsbeta_k,\sigma_k^2\Identity_{m_i}), 
\label{eq: polynomial regression density model}
\end{equation}
and the polynomial regression mixture is given by the following conditional mixture density:
\begin{eqnarray}
f(\by_i|\bx_i;\bstheta) &= &\sum_{k=1}^K \pi_k \  \N (\by_{i};\bX_i \bsbeta_k,\sigma_k^2\Identity_{m_i}). 
\label{eq: polynomial regression mixture}
\end{eqnarray}The model parameters are given by the parameter vector $\bstheta = (\pi_1,\ldots,\pi_K,\bstheta_1,\ldots,\bstheta_K)$ where the $\pi_k$'s are the non-negative mixing proportions that sum to 1,
 $\bstheta_k=(\bsbeta_{k},\sigma_{k}^2)$, $\sigma_{k}^2$ represents the regression parameters and the noise variance for cluster $k$. The unknown parameter vector $\bstheta$ is generally estimated by maximizing the observed-data log-likelihood given an $\iid$ dataset: 
\begin{equation}
\cL(\bstheta) = 
\sum_{i=1}^n  \log \sum_{k=1}^K \pi_k \ \N (\by_{i};\bX_i \bsbeta_k,\sigma_k^2\Identity_{m_i}).
\label{eq: log-lik PRM}
\end{equation}This is usually performed iteratively via the EM algorithm \citep{dlr, Gaffney99trajectoryclustering, GaffneyThesis}. 

  \subsection{Standard EM for fitting regression mixtures}
  \label{ssec: EM_MixReg}
The EM algorithm is the standard tool to maximize  the observed-data log-likelihood (\ref{eq: log-lik PRM}). This
is performed iteratively by maximizing, at each EM step, the conditional expectation of the log-likelihood of $\bstheta$ given the complete data, that is the complete-data log-likelihood, and a current parameter estimation of $\bstheta$. 
The complete data in this case are composed of the observed data $\cD = ((\bx_1,\by_1),\ldots,(\bx_n,\by_n))$ and 
the corresponding unknown cluster labels $\bz=(z_1,\ldots,z_n)$ with $z_i \in \{1,\ldots,K\}$.
Thus, the complete-data log-likelihood of $\bstheta$ is given by:
\begin{eqnarray}
\cL_c(\bstheta) & = & \sum_{i=1}^{n}\sum_{k=1}^{K} z_{ik} \log \big[\pi_k \N  (\by_{i};\bX_i\bsbeta_{k},\sigma^2_{k}\Identity_{m_i})\big] 
 \label{eq: complete log-lik PRM}
\end{eqnarray}where $z_{ik}$ is an indicator binary-valued variable such that $z_{ik}=1$ if $z_i=k$ (i.e., if $\by_i$ is generated by the polynomial regression component $k$) and $z_{ik}=0$ otherwise.

The EM algorithm for regression mixtures starts with an initial parameter vector $\bstheta^{(0)}$ and alternates between the two following steps until convergence:
\paragraph{E-step:}
\label{ssec: E-step EM-MixReg}
This step computes the expected complete-data log-likelihood given the observed data $\cD$ and a current estimation $\bstheta^{(q)}$ of the parameter vector $\bstheta$, $q$ being the current iteration number:  
{\begin{eqnarray}
Q(\bstheta,\bstheta^{(q)})&=& \E\big[\cL_c(\bstheta)|\cD;\bstheta^{(q)}\big]
\nonumber \\
&=& \sum_{i=1}^{n}\sum_{k=1}^{K}\tau_{ik}^{(q)} \log \big[\pi_k  \N  (\by_i;\bX \bsbeta_{k},\sigma^2_{k}\Identity_{m_i})\big]\cdot
\label{eq: Q-function PRM}
\end{eqnarray}}where
\begin{eqnarray}
\tau_{ik}^{(q)}& = & p(z_i=k|\by_{i},\bx_i;\bstheta^{(q)}) \nonumber \\
& = & \frac{\pi_k^{(q)} \N\big(\by_i;\bX_i \bsbeta^{(q)}_{k},\sigma^{2(q)}_{k}\Identity_{m_i}\big)}{\sum_{h=1}^K \pi_{h}^{(q)} \N(\by_i;\bX_i \bsbeta^{(q)}_{h},\sigma^{2(q)}_{h}\Identity_{m_i})} 
\label{eq: post prob tauik PRM}
\end{eqnarray}is the posterior probability that the $i$th curve is generated from cluster $k$.
This step therefore simply consists in computing the posterior cluster probabilities. 

\paragraph{M-step:}
\label{ssec: M-step EM-MixReg}This step updates the model parameters and provides the parameter vector $\bstheta^{(q+1)}$ by maximizing the $Q$-function (\ref{eq: Q-function PRM})  with respect to $\bstheta$.  
The mixing proportions are updated by maximizing $\sum_{i=1}^{n}\sum_{k=1}^{K} \tau^{(q)}_{ik} \log \pi_k$ with respect to $(\pi_1,\ldots,\pi_K)$ subject to the constraint $\sum_{k=1}^K \pi_k = 1$ using Lagrange multipliers, which gives the following updates \citep{mclachlanEM,mclachlanFiniteMixtureModels}:
\begin{eqnarray}
\pi_k^{(q+1)} &=& \frac{1}{n}\sum_{i=1}^n \tau^{(q)}_{ik} \quad (k=1,\ldots,K).
\label{eq: EM-MixReg pik update}
\end{eqnarray}
The maximization of each of the $K$ functions $\sum_{i=1}^{n} \tau_{ik}^{(q)} \log  \N  (\by_i;\bX \bsbeta_{k},\sigma^2_{k}\Identity_{m_i})$ in (\ref{eq: Q-function PRM}) consists in solving a weighted least-squares problem. The solution of this problem is straightforward and the updating equations are equivalent to the well-known weighted least-squares solutions (see for example \citep{GaffneyThesis, chamroukhi_PhD_2010}):
\begin{eqnarray}
\bsbeta_k^{(q+1)}  &=& \Big[\sum_{i=1}^{n}\tau^{(q)}_{ik} \bX^T_i\bX_i \Big]^{-1} \sum_{i=1}^{n}\tau^{(q)}_{ik} \bX_i^T \by_i \\
\label{eq: EM-MixReg beta_k update PRM}
\sigma_k^{2(q+1)} &=& \frac{1}{\sum_{i=1}^{n}\tau^{(q)}_{ik}} \sum_{i=1}^{n}\tau^{(q)}_{ik} \parallel \by_i - \bX_i\bsbeta_k\parallel^2.
\label{eq: EM-MixReg sigma_k update PRM}
\end{eqnarray}

From a clustering prospective, the estimated mixture components, are interpreted as $K$ clusters, where each cluster is associated to a mixture component.
Thus, once the model parameters have been estimated, the posterior probability that an individual was generated by the
$k$th cluster (or component) given by (\ref{eq: post prob tauik PRM}) represent a fuzzy partition of the data into $K$ clusters. The posterior cluster probabilities indeed represent a fuzzy cluster memberships. The represent the a posteriori uncertainty given the observed data and the estimated model about which cluster $k$ each observed data $(\bx_i,\by_i)$ is originated from. 
A hard partition can then be obtained by assigning each observed curve to the cluster with the highest posterior probability:
\begin{eqnarray}
\hat{z}_i = \arg \max_{k=1}^K \hat \tau_{ik}.
\end{eqnarray}
The pseudo code \ref{algo: EM-MixReg} summarizes the standard EM algorithm for regression mixtures. 
\begin{algorithm}[htbp]
\caption{\label{algo: EM-MixReg} Pseudo code of the standard EM algorithm for regression mixtures.}
{\bf Inputs:} Set of curves $\cD = ((\bx_1,\by_1),\ldots,(\bx_n,\by_n))$, number of clusters $K$ and polynomial degree $p$\\
\begin{algorithmic}[1]
\STATE fix a threshold $\epsilon>0$
;  set $q \leftarrow 0$ (iteration)

\textbf{Initialization:} $\bstheta^{(0)}= (\pi_1^{(0)},\ldots,\pi_K^{(0)}, \bstheta_1^{(0)},\ldots,\bstheta_K^{(0)})$: 
\STATE Initialize randomly the partition or by running $K$-means and initialize the $\pi_k$'s
\STATE Fit a regression model with parameters $\bstheta_k^{(0)} = (\bsbeta_k^{(0)}, \sigma_k^{2(0)})$ to each cluster $k$ 
\WHILE{increment in log-likelihood $> \epsilon$}
	\STATE \verb|//E-step:| 
		\FOR{$k=1,\ldots,K$}
		\STATE Compute $\tau_{ik}^{(q)}$ for $i=1,\ldots,n$  using Equation (\ref{eq: post prob tauik PRM})  
	\ENDFOR
	\STATE \verb|//M-step:| 
	\FOR{$k=1,\ldots,K$} 
		\STATE Compute $\pi_k^{(q+1)}$ using Equation (\ref{eq: EM-MixReg pik update})
		\STATE Compute $\bsbeta_k^{(q+1)}$ using Equation (\ref{eq: EM-MixReg beta_k update PRM})
		\STATE Compute $\sigma_k^{2(q+1)}$ using Equation (\ref{eq: EM-MixReg sigma_k update PRM})
	\ENDFOR
	\STATE $q \leftarrow q+1$ 
\ENDWHILE
\end{algorithmic}
{\bf Outputs:} $\hat{\bstheta} = \bstheta^{(q)}$, \quad $\hat{\tau}_{ik}=  \tau_{ik}^{(q)}$ 
\end{algorithm}

\subsection{Spline and B-spline regression mixtures}
The (B-)spline regression mixture is an extension of the previously described polynomial regression mixture to a semiparametric modeling by relying on splines or B-splines. 
\subsubsection{Spline and B-spline regression}

Splines \citep{deboor1978} are widely used for function approximation based on constrained piecewise polynomials. Let $\bsxi = \xi_0 < \xi_1, . . . , < \xi_L < \xi_{L+1}$
be ordered knots, including $L$ internal knots, $\xi_0$ and $\xi_{L+1}$ being the two boundary knots.
An order-$M$ spline with knots  $\bsxi$ is a piecewise-polynomial of degree $p = M-1$ with continuous derivatives at the interior knots up to order $M-2$. For example an order-2 spline is a continuous piecewise linear function. The spline regression function can be written as:
\begin{equation}
y_{ij} =\sum_{\ell=0}^{p} \beta_{\ell} x_{ij}^{\ell} + \sum_{\ell=1}^{L} \beta_{\ell+p} (x_{ij} - \xi_{\ell})^{p}_{+} + \epsilon_{ij}
\label{eq: PSR model}
\end{equation}
where $(x_{ij} - \xi_{\ell})_{+} = x_{ij} - \xi_{\ell} $ if $x_{ij} \geq \xi_{k}$ and $(x_{ij} - \xi_{\ell})_+ = 0$ otherwise,
$\bsbeta=(\beta_0,\ldots,\beta_{L+p})^T$ is the vector of spline coefficients in $\R^{L+M}$ and $\epsilon$ is and additive Gaussian noise. This spline regression model can be written in a vectorial form as
\begin{equation}
\by_{i} =\bS_{i} \bsbeta + \bsepsilon_i
\label{eq: vectorial PSR model}
\end{equation} where $\bS_i$ is the $m_i\times(L+M)$ spline regression matrix with rows 
\begin{equation}
\bs_{j}=(1, x_{ij},x_{ij}^{2} \ldots, x_{ij}^{p},(x_{ij} - \xi_{1})^{p}_{+},\ldots,(x_{ij} - \xi_{L})^{p}_{+}).
\label{eq: rows spline regression matrix}
\end{equation}
For splines, the columns of the regression matrix $\bX$ tend to be highly correlated since each column is a transformed version of $x$. This collinearity may result in a nearly singular matrix and imprecision in the spline fit \citep{ruppert_etal_semiparametricregression}. B-splines allows for efficient computations thanks to the block matrix form of the regression matrix.  An order-$M$ B-spline function is defined as a sum of linear combination of specific basis functions  $B_{\ell,M}$ as:
\begin{equation}
y_{ij} = \sum_{\ell=1}^{L+M} \beta_{\ell} B_{\ell,M} (x_{ij}), \; x_{ij} \in[\zeta_{\ell} , \zeta_{\ell+M}]
\end{equation}
where each $M$th order B-spline $B_{\ell,M}$ is  a piecewise polynomial of degree $p=M-1$ that has finite support over $[\zeta_{l} , \zeta_{\ell+M}]$ and is zero everywhere else.  Each of  the basis functions $B_{\ell,M} (x_{ij})$ can be computed as in Appendix  \ref{apxeq: construction of B-spline basis functions}  \citep{hastieTibshiraniFreidman_book_2009}. For the B-spline regression model, each row of the $m_i \times (L+M)$ B-spline regression matrix $\bB_i$ for the $i$th curve is thus given by:
\begin{equation}
\bb_{j}= (B_{1,M}(x_{ij}), \; B_{2,M}(x_{ij}),\ldots,\;B_{L+M,M}(x_{ij})).
 \label{eq: rows B-spline regression matrix}
\end{equation}

\subsubsection{Spline and B-spline regression mixtures and the EM algorithm}

The spline regression mixture (respectively the B-spline regression mixture) is similarly derived as the polynomial regression mixture described previously. The mixture density in this case is given by Eq. (\ref{eq: polynomial regression mixture}) where the regression matrix $\bX_i$ is replaced by the spline regression matrix $\bS_i$ (respectively the B-slipne regression matrix $\bB_i$).

The parameter estimation procedure for the (B-)spline regression mixture is the same as the one used for polynomial regression mixture, that is maximum likelihood estimation via the EM algorithm. The E- and M-steps are still the same, as well as as the initialization procedure and the convergence conditions.

\bigskip

However, it can be noticed that, the standard EM algorithm for all theses regression mixture models is sensitive to initialization.
It might yield poor estimations if the mixture parameters are not initialized properly.
The EM initialization in general can be performed from a randomly chosen partition of the data or by computing a  partition from another clustering algorithm such as $K$-means, Classification EM \citep{celeux_et_diebolt_SEM_85}, Stochastic EM \citep{celeuxetgovaert92_CEM}, etc or by initializing EM with a few number of steps of EM itself.
Several works have been proposed in the literature in order to overcome this drawback and  making the EM algorithm for Gaussian mixtures robust with regard initialization \citep{biernacki_etal_startingEM_CSDA03, Reddy:2008, Robust_EM_GMM}. 
Further details about choosing starting values for the EM algorithm for Gaussian mixtures can be found for example in \citep{biernacki_etal_startingEM_CSDA03}. 
 In addition, the EM algorithm requires the number of mixture components (clusters) to be known. While the number of cluster can be chosen by some model selection criteria such as BIC \citep{BIC}, AIC \citep{AIC} or ICL \citep{ICL}, this requires performing an afterward model selection procedure.
 Some authors have considered this issue in order to estimate the unknown number of mixture components in Gaussian mixture models, for example as in \citep{Richardson_BayesianMixtures, Robust_EM_GMM}.

In general, theses two issues have been considered each separately. Among the approaches considering both the problem of robustness with regard to initial values and automatically estimating the number of mixture components in the same algorithm, one can cite the EM algorithm proposed in \cite{FigueiredoUnsupervisedlearningMixtures}.
The authors proposed an EM algorithm that is capable of selecting the number of components and it attempts to be not sensitive with regard to initial values.
Indeed, the algorithm developed by \cite{FigueiredoUnsupervisedlearningMixtures} optimizes a particular criterion called the minimum message length (MML), which is a penalized negative log-likelihood rather than the observed data log-likelihood. The penalization term allows to control the model complexity. It starts by fitting a mixture model with large number of clusters and discards illegitimate clusters as the learning proceeds.
The degree of legitimate of each cluster is measured through the penalization term which includes  the mixing proportions to know if the cluster is small or not to be discarded, and therefore to reduce the number of clusters.
More recently, in \cite{Robust_EM_GMM}, the authors developed a robust EM algorithm for model-based clustering of multivariate data using Gaussian mixture models that simultaneously addresses the problem of initialization and estimating the number of mixture components. This algorithm overcome some initialization drawback of the EM algorithm proposed in \cite{FigueiredoUnsupervisedlearningMixtures}, which is still having an initialization problem. As shown in  \cite{Robust_EM_GMM}, this problem can become more serious especially for a dataset with a large number of clusters. 
%
%
However, these presented model-based clustering approaches, namely in \cite{Robust_EM_GMM} and \cite{FigueiredoUnsupervisedlearningMixtures}, are concerned with vectorial data where the observations are assumed to be vectors of reduced dimension.
When the data are rather curves or functions, they are not adapted. Indeed, when the data are functional where the individuals are presented as curves or surfaces rather than vectors, they are in general very structured. Relying on standard multivariate mixture analysis may therefore lead to unsatisfactory results in terms of modeling and classification accuracy \citep{chamroukhi_et_al_NN2009, chamroukhi_et_al_neurocomputing2010, chamroukhi_ijcnn_2011}. 
However, addressing the problem from a functional data analysis prospective, that is formulating
"functional" mixture models, allows to overcome these limitations, e.g.,
\citep{chamroukhi_et_al_NN2009, chamroukhi_et_al_neurocomputing2010, chamroukhi_ijcnn_2011, GaffneyThesis}.  
In this case of model-based functional data clustering, one can rely on the regression mixture approaches \citep{Gaffney99trajectoryclustering, GaffneyThesis, chamroukhi_PhD_2010} or generative hidden process regression \citep{chamroukhi_et_al_neurocomputing2010, chamroukhi_PhD_2010} which are adapted for curves with regime changes.
In this paper, we attempt to overcome the limitations of the EM algorithm in the case of  regression mixtures and model-based curve clustering by proposing an EM algorithm which is robust with regard initialization and automatically estimates the optimal number of clusters as the learning proceeds. We consider the polynomial, spline and B-spline regression mixtures. 

In the next section we derive our robust EM algorithm  for fitting regression mixtures including polynomial, spline and B-spline regression mixtures. 

\section{Penalized maximum likelihood via a robust EM algorithm for fitting regression mixtures}
\label{sec: proposed robust EM-Mix-Reg}
In this section we present the proposed EM algorithm for model-based curve clustering using regression mixtures.
The present work is in the same spirit of the EM algorithm presented in  \cite{Robust_EM_GMM}
 but by extending the idea to the case of functional data (curve) clustering rather than multivariate data clustering. It is therefore  concernned with regression mixture models rather than multivariate GMMs.
 Indeed, the data here are assumed to be curves rather than vectors of reduced dimensions.
This leads to fitting regression mixture models (including splines or B-splines), rather than fitting standard Gaussian mixtures. %
We start by describing the maximized objective function and then we derive the proposed EM algorithm to estimate the regression mixture model parameters.

\subsection{Penalized maximum likelihood estimation}
For estimating the regression mixture model (\ref{eq: polynomial regression mixture}), we attempt to maximize a penalized log-likelihood function rather than the standard observed-data log-likelihood (\ref{eq: log-lik PRM}). 
%
%
This criterion consists in penalizing the observed-data log-likelihood (\ref{eq: log-lik PRM}) by a term accounting for the model complexity. As the model complexity is related to in particular the number of clusters and therefore the structure of the hidden variables $z_i$, we chose to use the entropy of the hidden variable $z_i$ as penalty, $z_i$being the class label of the $i$th curve.
The penalized log-likelihood criterion is therefore derived as follows.
The discrete-valued variable $z_i \in \{1,\ldots,K\}$ with probability distribution $p(z_i)$ represents the hidden class label of the $i$th observed curve. The (differential) entropy of this variable is defined by
\begin{eqnarray}
H(z_i) &=& - \E[\log p(z_i)]= - \sum_{k=1}^K  p(z_i = k) \log p(z_i = k) \nonumber \\
 &=& - \sum_{k=1}^K \pi_k \log \pi_k\cdot
\end{eqnarray} 
By assuming that the variables $\bz = (z_1,\ldots,z_n)$ are independent and identically distributed (i.i.d),
which is in general the assumption in clustering using mixtures where the cluster labels are assumed to be distributed according to a Multinomial distribution, the whole entropy for $\bz$ is therefore additive and we have
\begin{eqnarray}
H(\bz) &=& - \sum_{i=1}^n\sum_{k=1}^K \pi_k \log \pi_k\cdot
\label{eq: entropy of z}
\end{eqnarray}
The penalized log-likelihood function we propose to maximize is thus constructed by penalizing the observed data log-likelihood (\ref{eq: log-lik PRM}) by the entropy term (\ref{eq: entropy of z}), that is
\begin{equation}
\cJ(\lambda,\bstheta) = \cL(\bstheta) - \lambda H(\bz), \quad \lambda \geq 0
\end{equation}
which leads to the following penalized log-likelihood criterion: 
 {\small \begin{eqnarray} 
\cJ(\lambda,\bstheta) &=& \sum_{i=1}^n  \log \sum_{k=1}^K \pi_k \N (\by_{i};\bX_i \bsbeta_k,\sigma_k^2\Identity_{m_i}) +  \lambda n \sum_{k=1}^K  \pi_k \log \pi_k 
\label{eq: penalized log-lik for PRM}
\end{eqnarray}}where $\cL(\bstheta)$ is the observed-data log-likelihood maximized by the standard EM algorithm for regression mixtures (see Equation (\ref{eq: log-lik PRM})) and $\lambda \geq 0$ is a parameter of control that controls the complexity of the fitted model.
This penalized log-likelihood function (\ref{eq: penalized log-lik for PRM}) we attempt to optimize allows to control the complexity of the model fit through the roughness penalty $\lambda  n \sum_{k=1}^K  \pi_k \log \pi_k$. The entropy term  $ -\sum_{i=1}^n \sum_{k=1}^K  \pi_k \log \pi_k$ in the penalty measures the complexity of a fitted model for $K$ clusters. When
the entropy
 is large, the fitted model is rougher, and when it is small, the fitted model is smoother. The non-negative smoothing parameter $\lambda$ is for establishing a trade-off between closeness of fit to the data and a smooth fit. As $\lambda$ decreases, the fitted model tends to be less complex, and we get a
 smooth fit. However, when $\lambda$ increases, the result is a rough fit.
We discuss in the next section how to set this regularization coefficient in an adaptive way.
%
 %
The next section shows how the penalized observed-data log-likelihood $\cJ(\lambda, \bstheta)$ is maximized w.r.t the model parameters $\bstheta$ by a robust EM algorithm for regression mixtures and model-based curve clustering.

\subsection{Robust EM algorithm for model-based curve clustering suing regression mixtures}
\label{ssec: Robust EM-MixReg}
Given an i.i.d training dataset of $n$ curves $\cD = ((\bx_1,\by_1),\ldots,(\bx_n, \by_n))$, 
the penalized log-likelihood (\ref{eq: penalized log-lik for PRM}) is iteratively maximized by using the following robust EM algorithm for model-based curve clustering.
Before giving the EM steps, we give the penalized complete-data log-likelihood, on which the EM formulation is based. The complete-data log-likelihood, in this penalized case, is given by
{\small \begin{eqnarray}
\cJ_c(\lambda, \bstheta) & = & \sum_{i=1}^{n}\sum_{k=1}^{K} z_{ik} \log \big[\pi_k \N  (\by_{i};\bX_i\bsbeta_k ,\sigma^2_{k}\Identity_{m_i})\big] 
+ \lambda  n \sum_{k=1}^K  \pi_k \log \pi_k \cdot 
 \label{eq: penalized complete log-lik for regression mixtures}
\end{eqnarray}}
%
After starting with an initial solution (see section \ref{ssec: initialization and stopping for Robust EM Mix-Reg} for the initialization strategy and stopping rule), the proposed algorithm  alternates between the two following steps until convergence. 
\subsubsection{E-step}
\label{ssec: M-step Robust EM-MixReg} 
This step computes the expectation of complete-data log-likelihood (\ref{eq: penalized complete log-lik for regression mixtures}) over the hidden variables $z_i$, given the observations $\cD$ and the current parameter estimation  $\bstheta^{(q)}$, $q$ being the current iteration number: 
{\small
\begin{eqnarray}
Q(\lambda, \bstheta;\bstheta^{(q)}) &=&  \E\big[\cJ_c(\lambda, \bstheta)|\cD;\bstheta^{(q)}\big] \nonumber \\
& = & \sum_{i=1}^{n}\sum_{k=1}^{K}  \E \big[z_{ik}|\cD;\bstheta^{(q)}\big] \log \big[\pi_k \N  (\by_{i};\bX_i\bsbeta_k ,\sigma^2_{k}\Identity_{m_i})\big] 
+ \lambda  n \sum_{k=1}^K  \pi_k \log \pi_k
\nonumber \\
& = & \sum_{i=1}^{n}\sum_{k=1}^{K}\tau_{ik}^{(q)} \log \N  (\by_i;\bX_i \bsbeta_{k},\sigma^2_{k}\Identity_{m_i})+ \sum_{i=1}^{n}\sum_{k=1}^{K}\tau_{ik}^{(q)} \log \pi_k + \lambda  n \sum_{k=1}^K  \pi_k \log \pi_k  
\label{eq: Q-function for the regression mixtures}
\end{eqnarray}}where
\begin{eqnarray}
\tau_{ik}^{(q)}= p(z_i=k|\by_{i},\bx_i;\bstheta^{(q)}) = \frac{\pi_k^{(q)} \N\big(\by_i;\bX_i \bsbeta^{T(q)}_{k},\sigma^{2(q)}_{k}\Identity_{m_i}\big)}{\sum_{h=1}^K \pi_{h}^{(q)} \N(\by_i;\bX_i \bsbeta^{(q)}_{h},\sigma^{2(q)}_{h}\Identity_{m_i})} 
\label{eq: Robust-MixReg post prob PRM}
\end{eqnarray}
is the posterior probability that the curve $(\bx_i,\by_i)$ is generated by  the cluster $k$. This step therefore only requires  the computation of the posterior cluster probabilities $\tau^{(q)}_{ik}$ $(i=1,\ldots,n)$ for each of the $K$ clusters.

\subsubsection{M-step}
\label{ssec: M-step Robust EM-MixReg} 
This step updates the value of the parameter vector $\bstheta$ by maximizing the $Q$-function (\ref{eq: Q-function for the regression mixtures}) with respect to $\bstheta$, that is: 
$\bstheta^{(q+1)} = \arg \max_{\bstheta} Q(\lambda, \bstheta;\bstheta^{(q)}).$  
It can be shown that this maximization can be performed by separate  maximizations w.r.t the mixing proportions $(\pi_{1},\ldots,\pi_{K})$ subject to the constraint $\sum_{k=1}^{K} \pi_{k} = 1$, 
and w.r.t the regression parameters $\{\bsbeta_{k},\sigma^2_{k}\}$.
The mixing proportions updates are obtained by maximizing the function
\begin{equation}
Q_{\pi}(\lambda; \bstheta^{(q)}) = \sum_{i=1}^{n}\sum_{k=1}^{K}\tau_{ik}^{(q)} \log \pi_k + \lambda  n \sum_{k=1}^K  \pi_k \log \pi_k \nonumber
\label{eq: J(pik) pinalized log-lik PRM} 
\end{equation}w.r.t the mixing proportions $(\pi_{1},\ldots,\pi_{K})$ subject to the constraint $\sum_{k=1}^{K} \pi_{k} = 1$. This can be solved using Lagrange multipliers (see Appendix \ref{apx. Estimation of the mixing proportion}) 
and the obtained updating formula is  given by:
\begin{equation}
\pi_{k}^{(q+1)} = \frac{1}{n}\sum_{i=1}^n \tau_{ik}^{(q)} + \lambda \pi_{k}^{(q)}\left(\log \pi_{k}^{(q)} - \sum_{h=1}^K\pi_{h}^{(q)}\log \pi_{h}^{(q)}\right)\cdot
\label{eq: Robust EM-MixReg pi_k update}
\end{equation}
%
We notice here that in the update of the mixing proportions (\ref{eq: Robust EM-MixReg pi_k update}) the update is close to the standard EM update $\big(\frac{1}{n}\sum_{i=1}^n \tau_{ik}^{(q)}$ see Eq. (\ref{eq: EM-MixReg pik update})$\big)$ for very small value of $\lambda$. However, for a large value of $\lambda$, the penalization term will play its role in order to make clusters competitive and thus allows for discarding illegitimate clusters and enhancing actual clusters.
Indeed, in the updating formula (\ref{eq: Robust EM-MixReg pi_k update}), we can see that  for cluster $k$ if
\begin{equation}
\left(\log \pi_{k}^{(q)} - \sum_{h=1}^K\pi_{h}^{(q)}\log \pi_{h}^{(q)}\right) > 0
\label{eq: step of pik update Robust EM-MixReg}
\end{equation}
that is, for the (logarithm of the) current proportion $\log \pi_{k}^{(q)}$, the entropy of the hidden variables is decreasing, and therefore the model complexity tends to be stable, 
the cluster $k$ has therefore to be enhanced. This explicitly results in the fact that the update of the $k$th mixing proportion $\pi_{k}^{(q+1)}$ in (\ref{eq: Robust EM-MixReg pi_k update}) will increase. 
%
On the other hand, if (\ref{eq: step of pik update Robust EM-MixReg}) is less than $0$, the cluster proportion will therefore decrease as is not very informative in the sense of the entropy.

Finally, the penalization coefficient $\lambda$ has to be set as described previously in such a way to be large for enhancing competition when the proportions are not increasing enough 
from one iteration to another. In this case, the robust algorithm plays its role for estimating the number of clusters (which is decreasing in this case by discarding small illegitimate clusters). We note that here a cluster $k$ can be discarded if its proportion is less than $\frac{1}{n}$, that is $\pi_{k}^{(q)}<\frac{1}{n}$.
On the other hand,  $\lambda$ has to become small when the proportions are sufficiently increasing as the learning proceeds and the partition can therefore be considered as stable. In this case, the robust EM algorithm tends to have the same behavior as the stand EM described in section \ref{ssec: EM_MixReg}.
The regularization coefficient $\lambda$ can be set in $[0,1]$ to prevent very large values. Furthermore, 
it can be set in an adaptive way similarly as described in \cite{Robust_EM_GMM} for multivariate Gaussian mixtures according to the following adaptive formula to adapt it as the learning proceeds:  
{\small \begin{eqnarray}
\lambda^{(q+1)}&=& \min \left\{\frac{\sum_{k=1}^K \exp\left(\eta n |\pi_{k}^{(q+1)} - \!  \pi_{k}^{(q)}|\right)}{K}, \frac{1- \max_{k=1}^K\left(\frac{\sum_{i=1}^n\tau^{(q)}_{ik}}{n}\right)}{-\pi_{k}^{(q)} \sum_{k=1}^K \pi_{k}^{(q)} \log \pi_{k}^{(q)}}\right\} 
\label{eq: lambda update Robust EM-MixReg}
\end{eqnarray}}%
where $\eta$ can be set as $\min(1, 0.5^{\lfloor\frac{m}{2}-1\rfloor})$, $m$ being the number of observations per curve and $\lfloor x \rfloor$ denotes the largest integer that is no more than $x$.
%
%

\bigskip 
The maximization w.r.t the regression parameters however consists in separately maximizing for each class $k$ the function
{\small \begin{eqnarray}
Q_{\bstheta_k}(\lambda,\bsbeta_k,\sigma^2_k; \bstheta^{(q)}) &=& \sum_{i=1}^{n} \tau_{ik}^{(q)} \log  \N  (\by_i;\bX_i \bsbeta_{k},\sigma^2_{k}\Identity_{m_i}) \nonumber \\
&=&\sum_{i=1}^{n}\tau^{(q)}_{ik}\left[-\frac{m}{2}\log 2 \pi - \frac{m}{2}\log \sigma^2_{k} - \frac{1}{2 \sigma^2_{k}} \parallel \by_i - \bX_i\bsbeta_k\parallel^2\right] \nonumber 
\end{eqnarray}}w.r.t $(\bsbeta_{k},\sigma^2_{k})$.
This maximization w.r.t the regression parameters yields to analytic solutions of  weighted least-squares problems where the weights are the posterior cluster probabilities $\tau_{ik}^{(q)}$. The updating formula are given by:
\begin{eqnarray}
\bsbeta_k^{(q+1)}  &=& \Big[\sum_{i=1}^{n}\tau^{(q)}_{ik} \bX^T_i\bX_i \Big]^{-1} \sum_{i=1}^{n}\tau^{(q)}_{ik} \bX_i^T \by_i
\label{eq: Robust EM-MixReg beta_k update}
\end{eqnarray} 
\vspace{-.4cm}
\begin{eqnarray}
\sigma_k^{2(q+1)} &=& \frac{1}{m\sum_{i=1}^{n}\tau^{(q)}_{ik}} \sum_{i=1}^{n}\tau^{(q)}_{ik} \parallel \by_i - \bX_i\bsbeta_k\parallel^2 
\label{eq: Robust EM-MixReg sigma_k update}
\end{eqnarray}where the posterior cluster probabilities $\tau^{(q)}_{ik}$ given by (\ref{eq: Robust-MixReg post prob PRM})
are computed using the mixing proportions derived in (\ref{eq: Robust EM-MixReg pi_k update}).
Then, once the model parameters have been estimated, a fuzzy partition of the data into $K$ clusters is then represented by the posterior cluster probabilities $\tau_{ik}$. A hard partition can then be computed according to the MAP principle by maximizing the posterior cluster probabilities, that is
\begin{eqnarray} 
\hat{z}_i = \arg \max_{k} \tau_{ik} 
\end{eqnarray}
where $\hat{z}_i$ denotes the estimated cluster label of the $i$th observation.

\subsection{Initialization strategy and stopping rule}
\label{ssec: initialization and stopping for Robust EM Mix-Reg} 
%
The initial number of clusters is $K^{(0)}= n$, $n$ being the total number of curves and the initial mixing proportions are $\pi_k^{(0)}= \frac{1}{K^{(0)}}$, ($k=1,\ldots, K^{(0)}$). Then, to initialize the regression parameters $\bsbeta_k$ and the noise variances $\sigma_k^{2(0)}$ ($k=1,\ldots, K^{(0)}$), we fitted a polynomial regression models on each curve $k$, ($k=1,\ldots, K^{(0)}$) and the initial values of the regression parameters are therefore given by
$\bsbeta_k^{(0)}  = \Big(\bX_k^T\bX_k \Big)^{-1}\bX_k \by_k$
and the noise variance can be deduced as
$\sigma_k^{2(0)} = \frac{1}{m} \parallel\by_k - \bX_k\bsbeta_k^{(0)}\parallel^2$. However, to avoid singularities at the starting point, we set $\sigma_k^{2(0)}$ as a middle value in the following sorted range $\parallel\by_i- \bX_i\bsbeta_k^{(0)}\parallel^2$ for $i=1,\ldots,n$.
 
The proposed EM algorithm is stopped when the maximum variation of the estimated regression parameters between two iterations $\max_{1\leq k \leq K^{(q)}} \parallel \bsbeta_k^{(q+1)} - \bsbeta_k^{(q)}\parallel$ was less than a fixed threshold $\epsilon$ (e.g., $10^{-6}$). 
The pseudo code \ref{algo: proposed Robust EM-MixReg} summarizes the proposed robust EM algorithm for model-based curve clustering using regression mixtures. 
\begin{algorithm}[H]
\caption{\label{algo: proposed Robust EM-MixReg} Pseudo code of the proposed Robust EM algorithm for polynomial regression mixtures.}
{\bf Inputs:} Set of curves $\cD = ((\bx_1,\by_1),\ldots,(\bx_n,\by_n))$ and polynomial degree $p$\\
\begin{algorithmic}[1]
\STATE $\epsilon \leftarrow 10^{-6}$; $q \leftarrow 0$;  $converge \leftarrow 0$ 

\verb|//Initialization:| 
\STATE $\bstheta^{(0)}= (\pi_1^{(0)},\ldots,\pi_K^{(0)}, \bstheta_1^{(0)},\ldots,\bstheta_K^{(0)})$; $K^{(0)} = n$

\FOR{$k=1,\ldots,K^{(q)}$}
 \STATE Compute $\tau_{ik}^{(q)}$ for $i=1,\ldots,n$  using Equation (\ref{eq: Robust-MixReg post prob PRM})  
 \STATE Compute $\pi_k^{(q)}$ using Equation (\ref{eq: Robust EM-MixReg pi_k update})
\ENDFOR

\verb|//EM loop:| 
\WHILE{{\small $(! \ converge)$}}
		\FOR{$k=1,\ldots,K^{(q)}$}
		\STATE Compute $\tau_{ik}^{(q)}$ for $i=1,\ldots,n$  using Equation (\ref{eq: Robust-MixReg post prob PRM})  
	\ENDFOR
	\FOR{$k=1,\ldots,K^{(q)}$} 
		\STATE Compute $\pi_k^{(q+1)}$ using Equation (\ref{eq: Robust EM-MixReg pi_k update})
	\ENDFOR		
	 \STATE Compute $\lambda^{(q+1)}$ using Equation (\ref{eq: lambda update Robust EM-MixReg})
	  ; Discard illegitimate clusters with small proportions  $\pi_k^{(q)}< \frac{1}{n}$
	  ; Set $K^{(q+1)} = K^{(q)} - \#\{\pi_k^{(q)}|\pi_k^{(q)}< \frac{1}{n}\}$
	  ; normalize $\tau_{ik}^{(q+1)}$ and $\pi_k^{(q+1)}$ so that their columns sum to one 
	\FOR{$k=1,\ldots,K^{(q)}$} 
		\STATE Compute $\bsbeta_k^{(q+1)}$ using Equation (\ref{eq: Robust EM-MixReg beta_k update})
		\STATE Compute $\sigma_k^{2(q+1)}$ using Equation (\ref{eq: Robust EM-MixReg sigma_k update})
	\ENDFOR
	\IF{$\max_{1\leq k \leq K^{(q)}} \parallel \bsbeta_k^{(q+1)} - \bsbeta_k^{(q)}\parallel <\epsilon$}
	 \STATE $converge =1$
	\ENDIF 
	\STATE $q \leftarrow q+1$ 
\ENDWHILE
\end{algorithmic}
{\bf Outputs:} $K= K^{(q)}, \quad \hat{\bstheta} = \bstheta^{(q)}$, \quad $\hat{\tau}_{ik}=  \tau_{ik}^{(q)}$ 
\end{algorithm}

 \subsection{Choosing the order of regression and spline knots number and locations}
For a general use of the proposed algorithm for the polynomial regression mixture, the order of regression can be chosen by cross-validation techniques as in \citep{GaffneyThesis}. 
In our experiments, we report the results corresponding to the for which th polynomial regression mixture provides the best fit.
However, in some situations, the PRM model may be too simple to capture the full structure of the data, in particular for curves with high non-linearity of with regime changes, even if they can be seen as providing a useful first-order approximation of the data structure. The (B)-spline regression models in this case are more adapted. 
For these models, on may need to choose the spline order and the number of knots and their locations.  
For the order of regression in (B)-splines, we notice that, in practice, the most widely used orders are $M$ = 1,2 and 4 \citep{hastieTibshiraniFreidman_book_2009}. For smooth functions approximation,  cubic (B)-splines (of order 4), which correspond to a (B)-spline of order 4 and thus with twice continuous derivatives, are sufficient to approximate smooth functions.
When the data present  irregularity, such as a kind of piecewise non continuous functions, a linear spline (of order 2) should be more adapted. This was namely used for the satellite dataset. The order 1 can be chosen for piecewise constant data.
Concerning the choice of the number of knots and their locations,  a common choice is to place a number of knots uniformly spaced across the range of $x$.   In general  more knots are needed for functions with regime changes. 
One can also use automatic techniques for the selection of the number of knots and their locations as reported in \cite{GaffneyThesis}.   For example, this can be performed by using cross validation as in \cite{ruppert_etal_semiparametricregression}. In \cite{KooperbergANDStone1991}, the knots are placed at selected order statistics of the sample data and the number of knots is determined either by a simple rule or by minimizing a variant of AIC.
The general goal is to use a sufficient number of knots to fit the data while and at the same time to avoid over-fitting and to not make the computation excessive. 
The current algorithm can be easily extended  to handle this type of automatic selection of spline knots placement, but as the unsupervised clustering problem itself requires much attention and is difficult, it is wise to fix the number and location of knots. In this paper, we will use knot sequences which are uniformly spaced across the range of $x$. The studied problems are not not very sensitive to the number and location of knots. Few number of equispaced knots (less than ten for the data studied in this paper) are sufficient to fit the data

\section{Experimental study}
\label{sec: Experiments}
This section is dedicated to the evaluation of the proposed approach on simulated data and real-world data. 
The algorithms have been implemented on Matlab and the esperiments were performed on a personal laptop. The developed Matlab codes are available upon request from the author.
We evaluate the proposed EM algorithm for the three regression mixtures models: the polynomial regression mixture, spline regression mixture and B-spline regression mixture, respectively abbreviated as PRM, SRM and bSRM. The evaluation is performed in terms of estimating the actual partition by considering the estimated number of clusters and the clustering accuracy (misclassification error). We first consider simulated data and the waveforms benchmark. Then, we consider three real-world data sets issued covering three different application area: phoneme recognition, clustering gene time course data for bio-informatics and clustering satellite data.
\subsection{Simulation study}  

The simulated data are simulated arbitrary curves and a the well-known benchmark of waveform data of \cite{Breiman1984}. 
%
%
%

\subsubsection{Experiments on simulated curves} 
\begin{figure}[H]
 \centering  
\includegraphics[width=6cm]{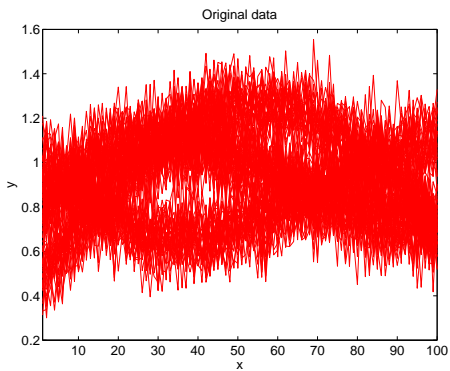} 
 \caption{\label{fig: toy-nonlin-data}Simulated arbitrary nonlinear curves from a three-class problem.}
\end{figure}

\begin{figure}[H]
\centering  
\includegraphics[width=5.2cm]{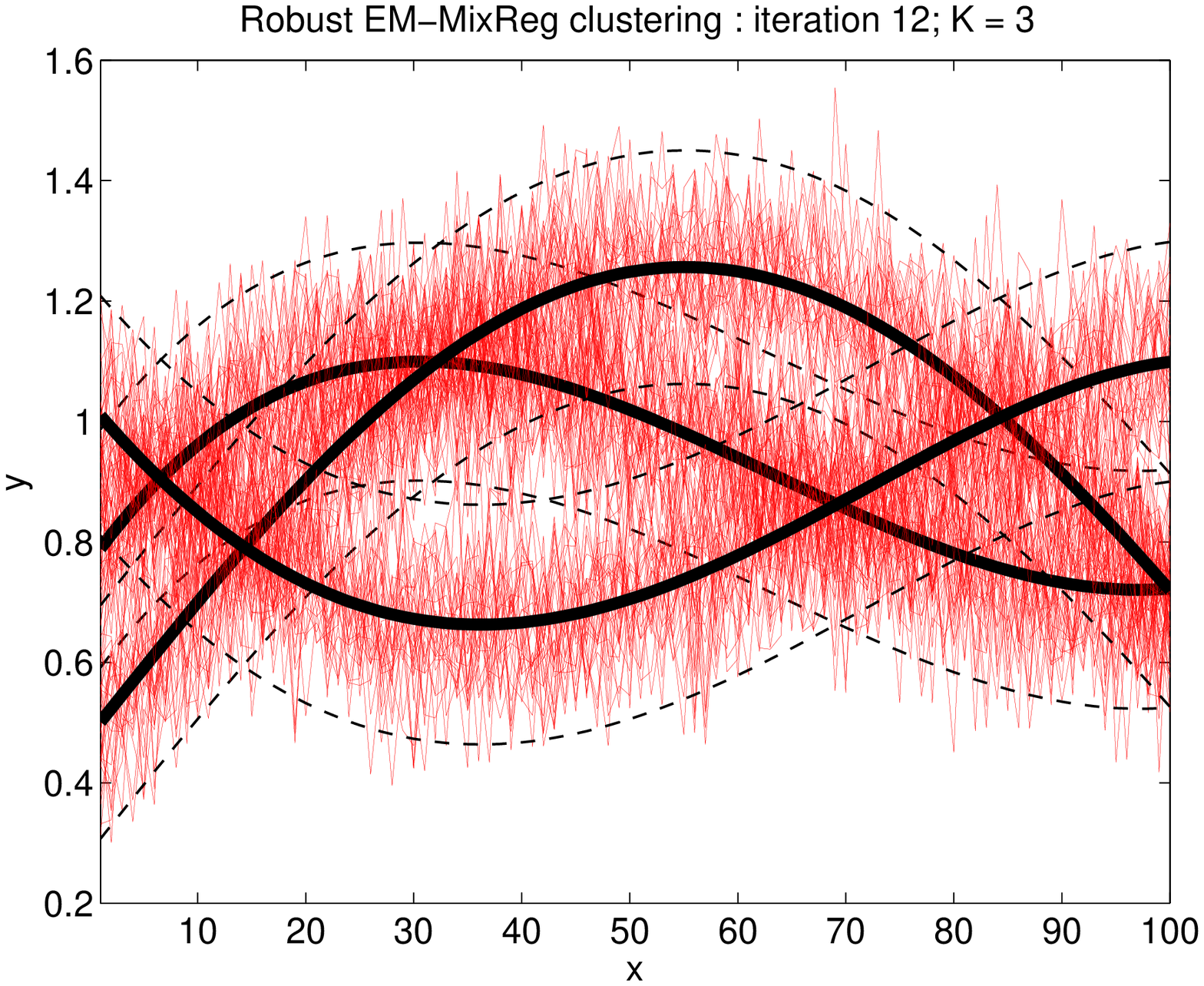} 
\includegraphics[width=5.2cm]{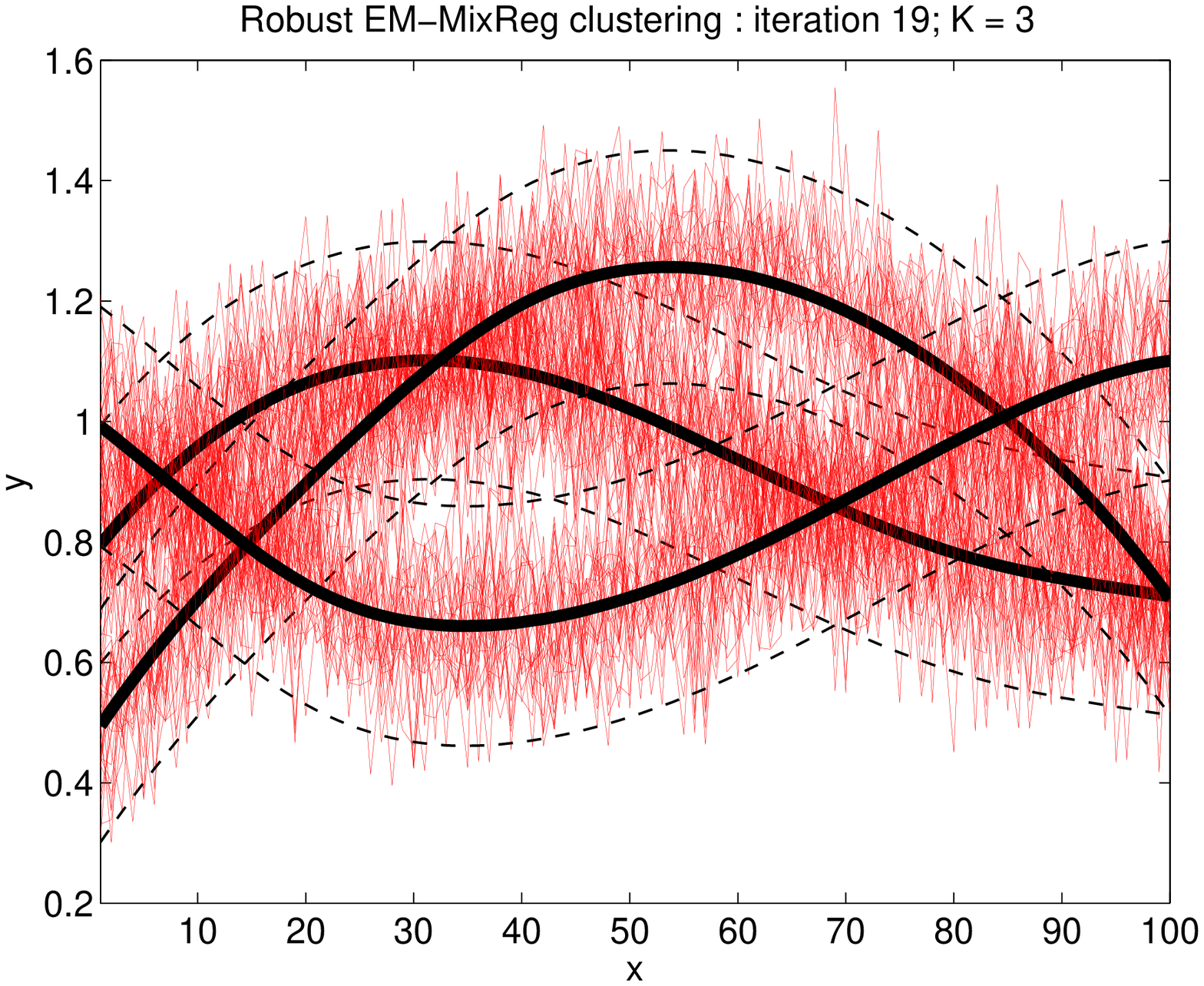} 
\includegraphics[width=5.2cm]{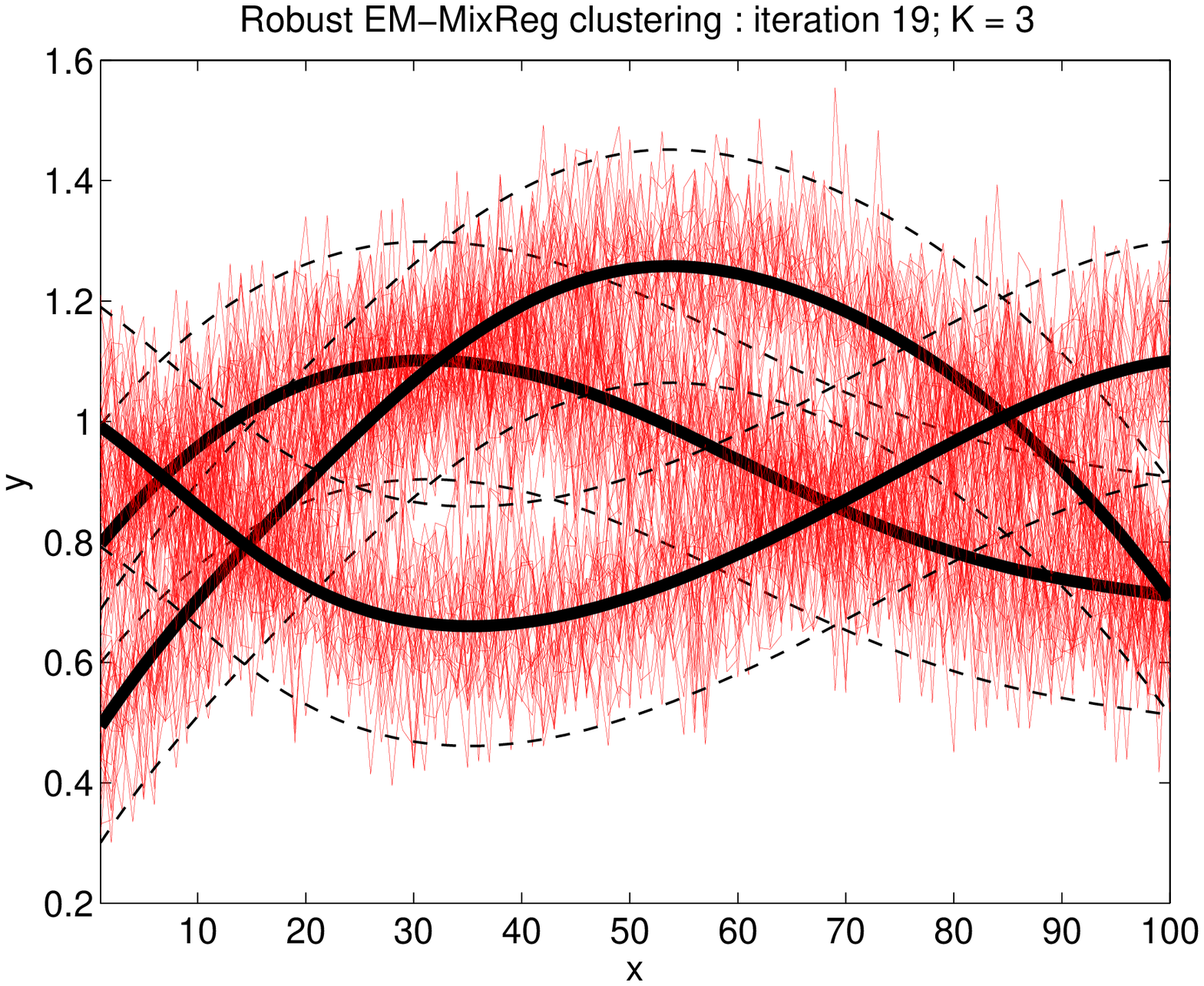}
 \caption{\label{fig: xRM partitions toy nonlin}Clustering results obtained with a polynomial regression mixture (PRM) (left) with $p=4$, a cubic spline regression mixture (SRM) (middle) and a cubic B-spline regression mixture (bSRM) (right) with 4 knots.} 
\end{figure}

\begin{figure}[H]
\centering  
\includegraphics[width=5.2cm]{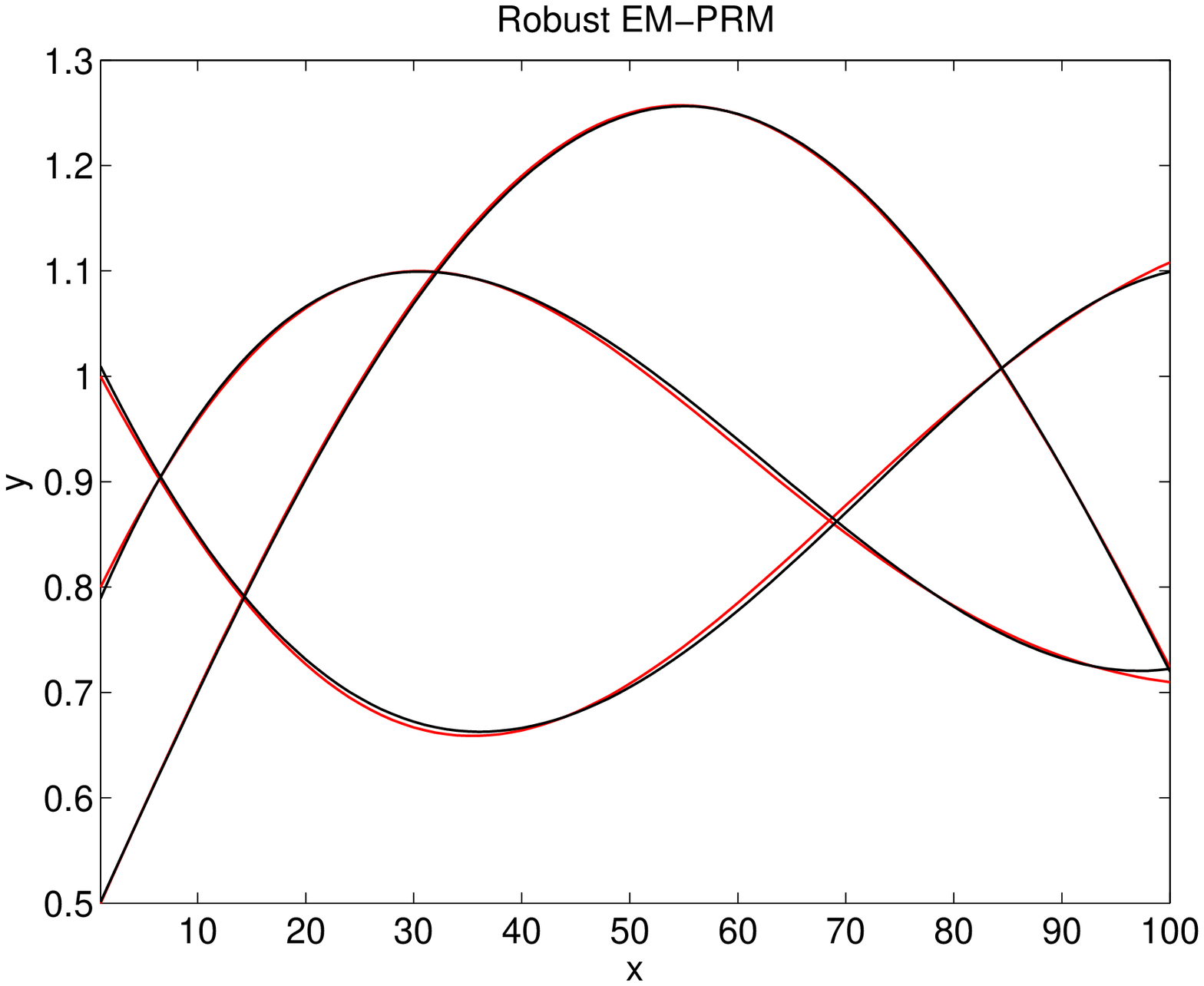}
\includegraphics[width=5.2cm]{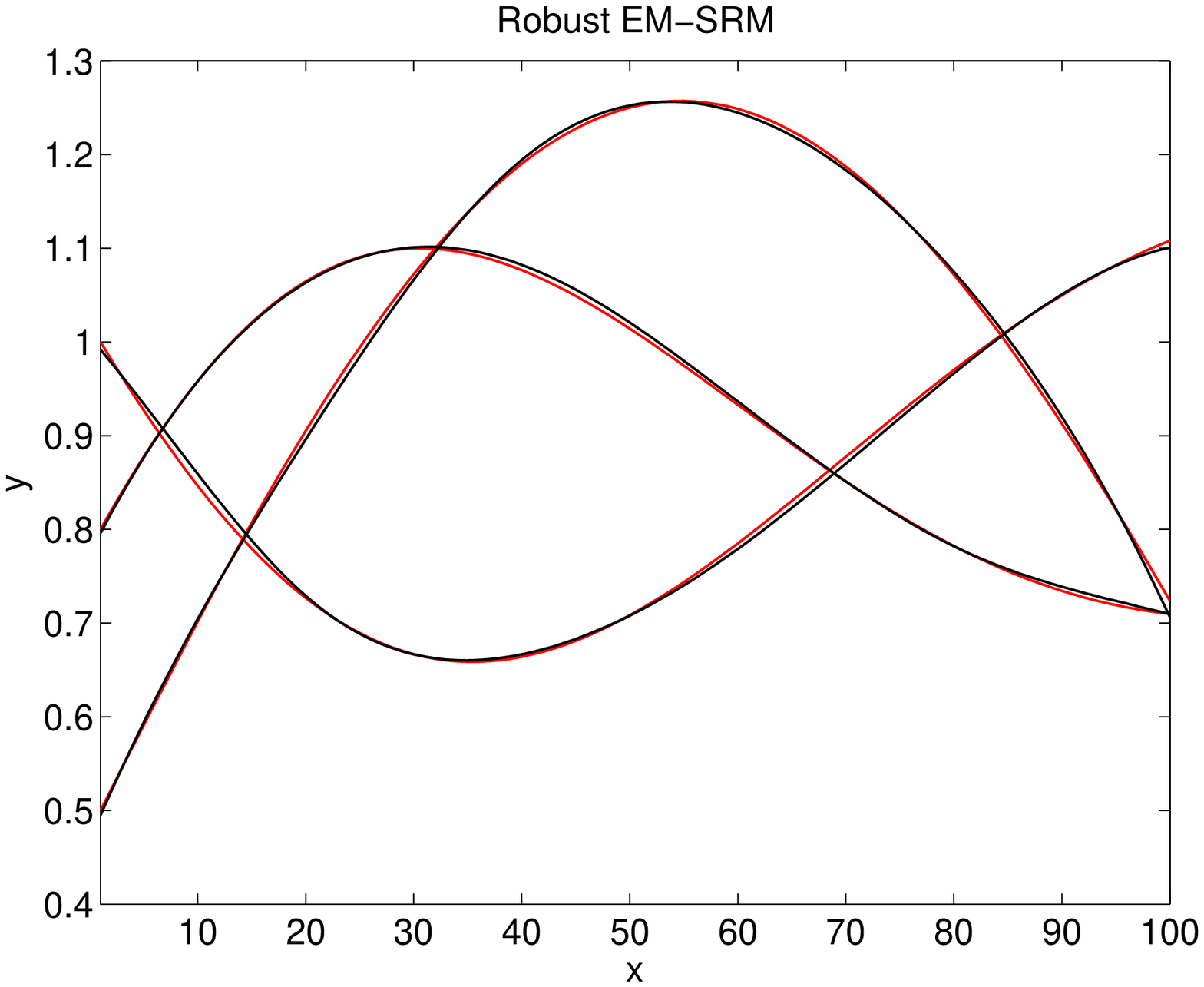}
\includegraphics[width=5.2cm]{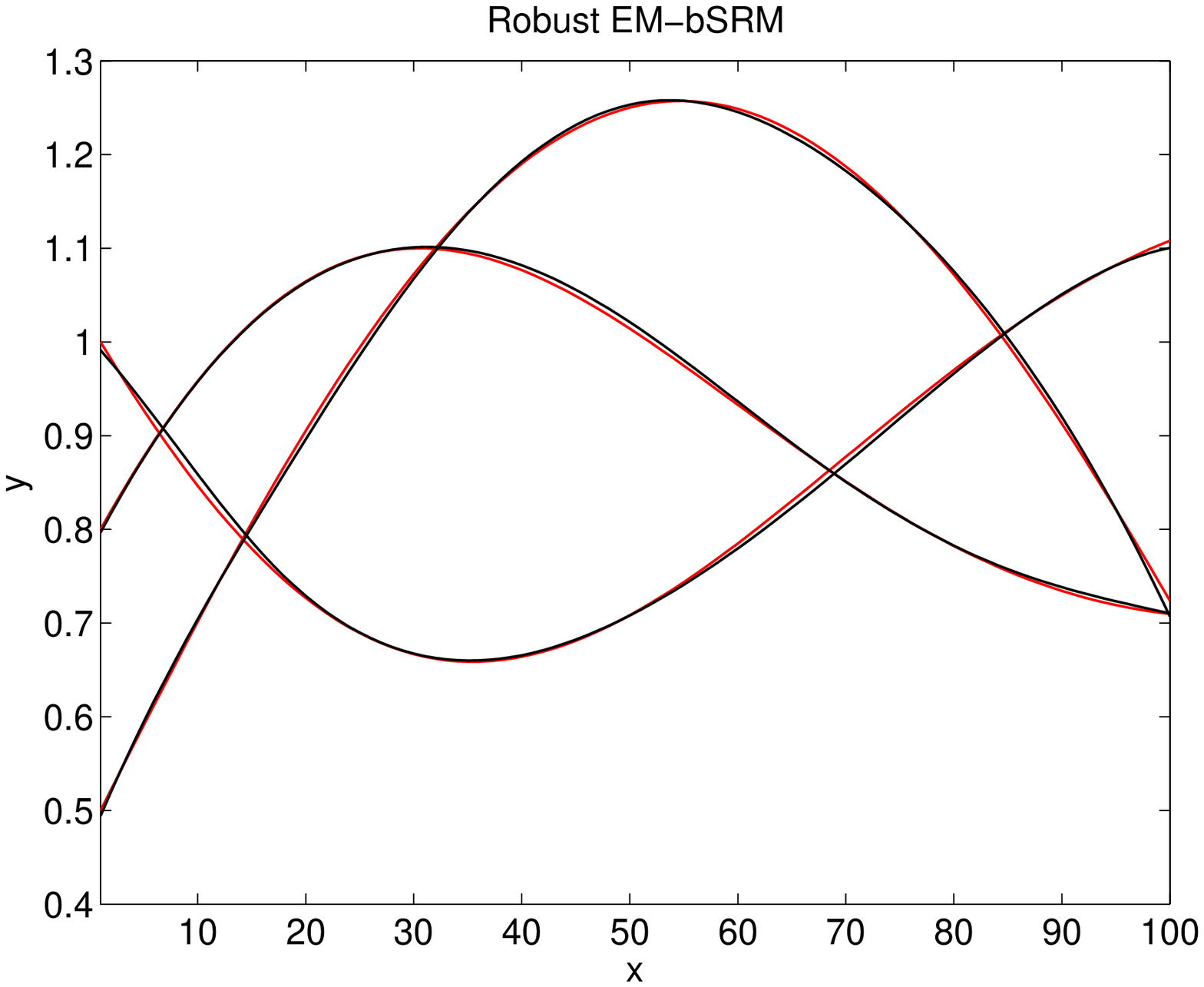}
 \caption{\label{fig: xRM mean-curves toy nonlin}Actual (red) and estimated cluster mean functions (black) for  the polynomial regression PRM (left), the SRM (middle) and the bSRM (right).} 
\end{figure}

 In this experiment we consider non-linear arbitrary curves simulated as follows. The curves are simulated from situation representing a three-class problem and consisted of $n=100$ arbitrary non-linear curves, each curve is composed of $m=50$ observations. For the $i$th curve ($i=1,\ldots,n$), the $j$th observation ($j=1,\ldots,m$) is generated as follows: 
\begin{itemize}
\item $y_{ij}=  0.8 + 0.5\,\exp(-1.5\, x)\,\sin(1.3 \pi\, x) + \sigma_1\,\epsilon_{ij}$;
\item $y_{ij}=  0.5 + 0.8\,\exp(-x)\,\sin(0.9 \pi\, x) + \sigma_2\,\epsilon_{ij}$;
\item $y_{ij}=  1 + 0.5\,\exp(-x)\,\sin(1.2 \pi\, x) + \sigma_3\,\epsilon_{ij}$;
\end{itemize} with $\sigma_1 =  \sigma_2 =  \sigma_3 = 0.1$.
The classes have respectively proportions $\pi_1 = 0.4, \pi_2 = 0.3, \pi_3 = 0.3$. 
Figure \ref{fig: toy-nonlin-data} shows a sample of the simulated data.
Figure \ref{fig: xRM partitions toy nonlin}  shows the clustering results obtained by a polynomial regression mixture model, a cubic spline and B-spline regression mixtures. 
Each cluster is shown with its estimated mean function and the confidence interval of $\pm 2\hat \sigma$. 
It can be observed that the actual number of clusters (three) is correctly estimated by the proposed algorithm for the three models.  for this simulated sample,  the actual partition is also correctly recovered for the three models and all the observations are correctly classified.
%
%
Figure \ref{fig: xRM mean-curves toy nonlin} shows the actual and estimated cluster mean functions for each regression mixture model. We ccan observed that the results are quasi-identical and the real mean functions are estimated accurately. 
Table \ref{tab. results toy nonlin} gives numerical results for this data set. We include the approximation error which is computed as the normalized squared error between the true cluster mean functions and the estimated ones. The data are correctly classified by the three models and the actual number of clusters is correctly estimated. 
We can also see that the errors in terms of approximation of the true cluster mean functions are very low, as well as the ones of the true cluster variance estimation.
\begin{table}[H]
\centering
\small
\begin{tabular}{|c  c c c   c c c   c c c|}
\hline
  & $K$ & misc. error rate& $\sigma_1$&$\sigma_2$&$\sigma_3$& $\pi_1$&$\pi_2$&$\pi_3$& Approx. Error \\
actual& 3 & $-$ & 1 & 1& 1& 0.4 & 0.3 & 0.3 & $-$ \\
 \hline
EM-PRM &  3 & 0\%  & 0.0988 & 0.0969& 0.0995& 0.4 & 0.3 & 0.3 & $4.4979e-05$ \\
EM-SRM &  3 & 0\% & 0.0987 & 0.0968 & 0.0994& 0.4 & 0.3 & 0.3 & $6.8076e-05$ \\
EM-bSRM &  3 & 0\%  & 0.0987 & 0.0968 & 0.0994& 0.4 & 0.3 & 0.3 & $6.4905e-05$ \\
 \hline
\end{tabular}
\caption{\label{tab. results toy nonlin}Clustering results for the simulated curves.}
\end{table}

\begin{figure}[H]
   \centering  
   \includegraphics[width=5.2cm]{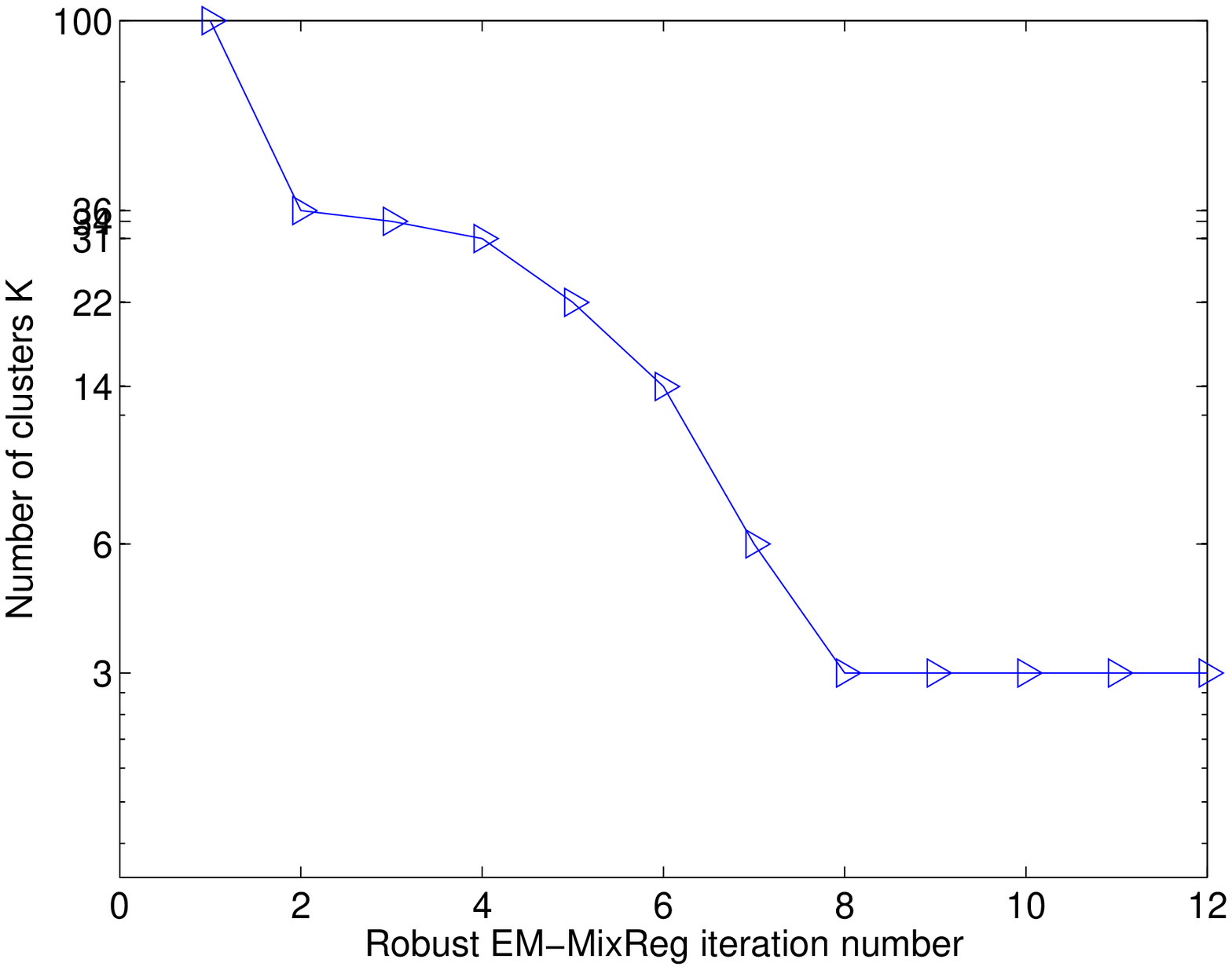}
   \includegraphics[width=5.2cm]{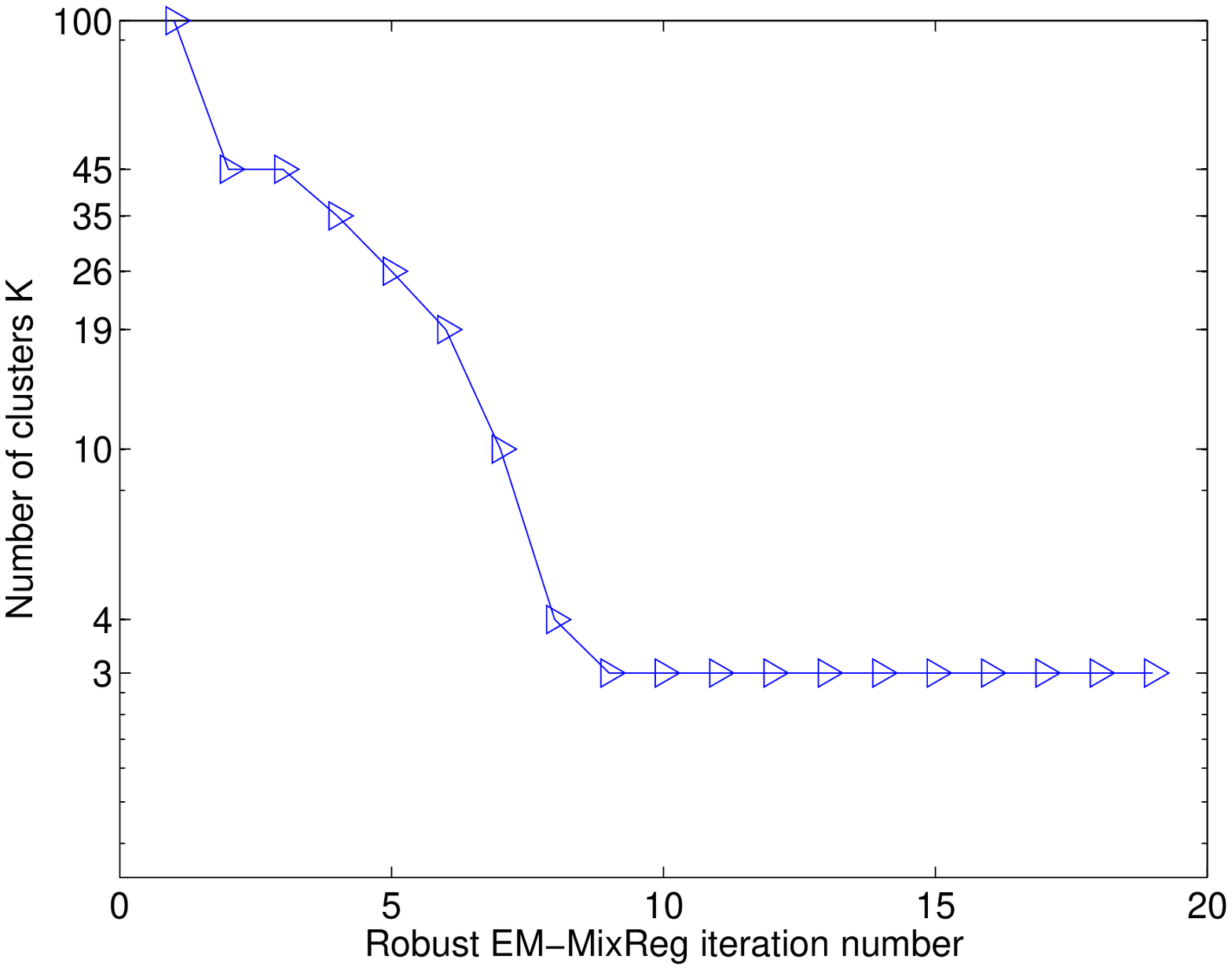}
   \includegraphics[width=5.2cm]{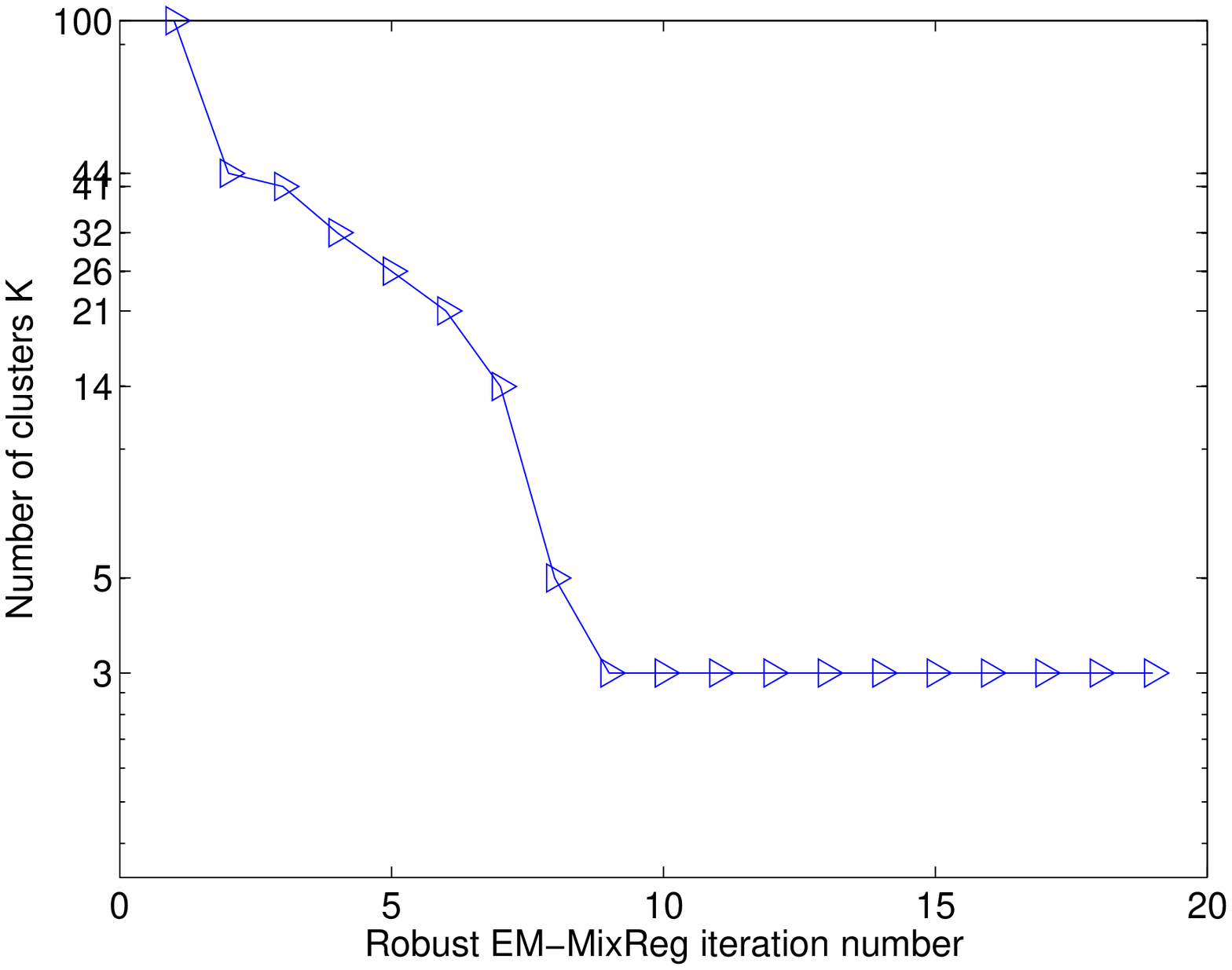}\\
   \includegraphics[width=5.2cm]{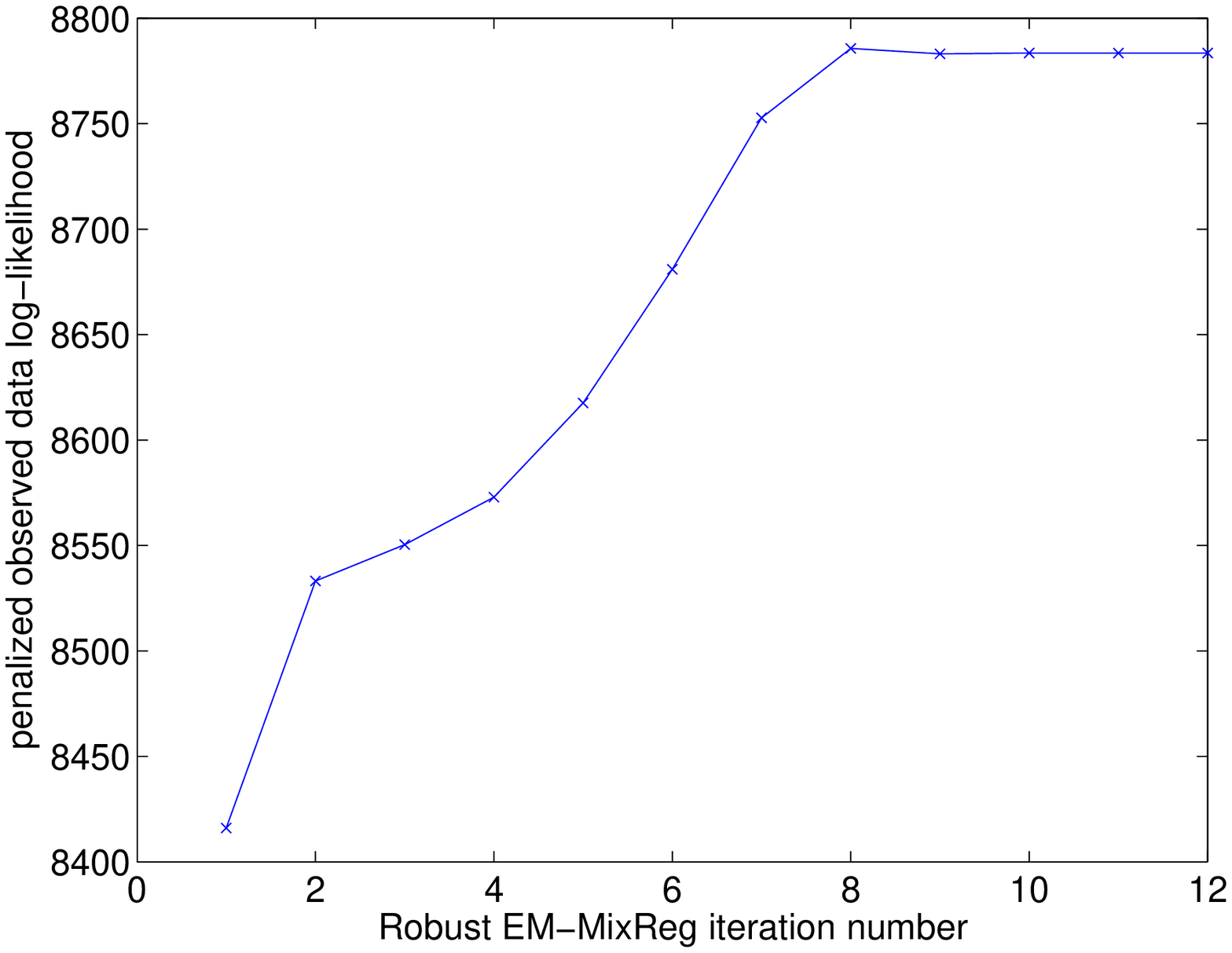}
   \includegraphics[width=5.2cm]{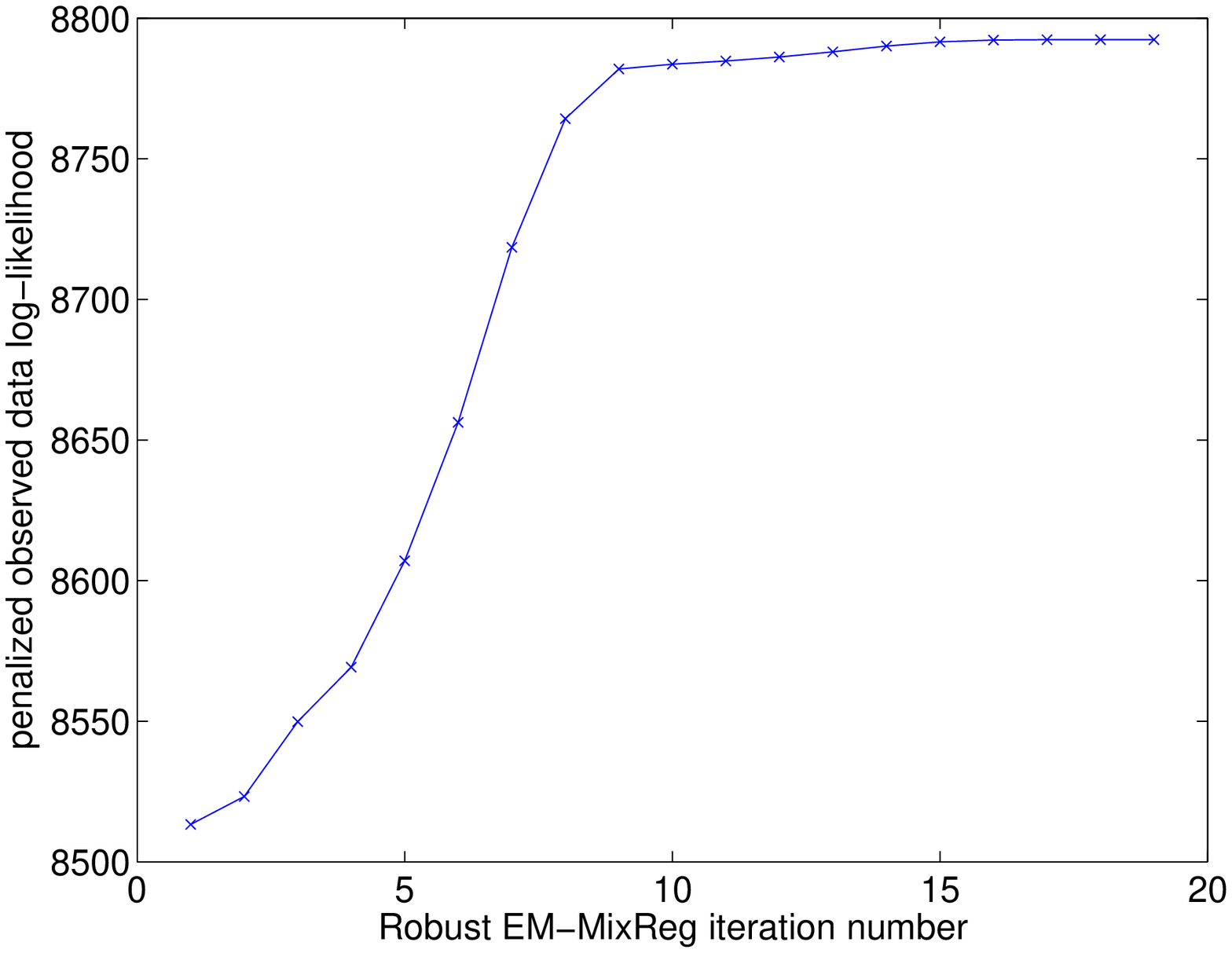}
   \includegraphics[width=5.2cm]{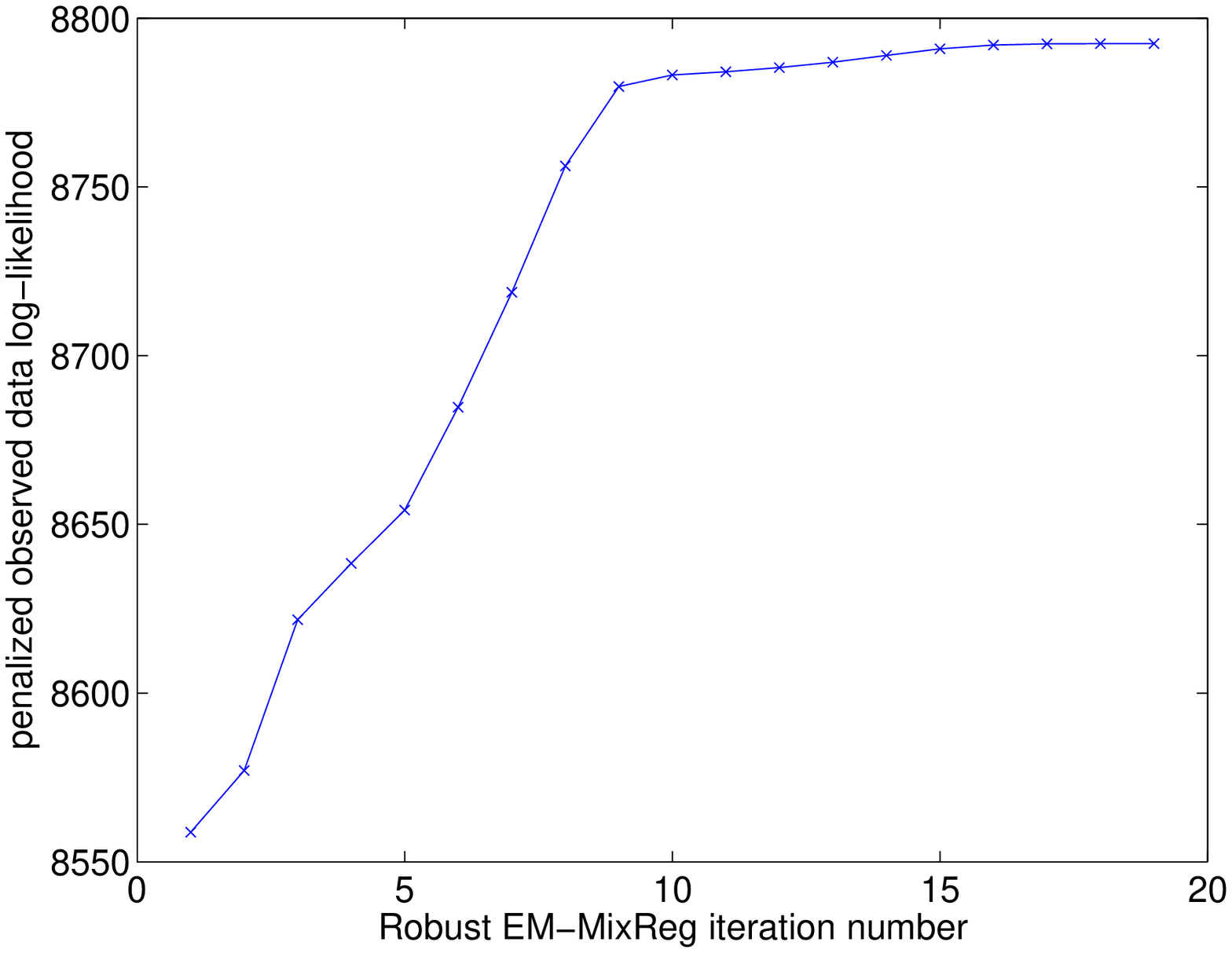}
   \caption{\label{fig: EM-MixReg stored-K pen-loglik toy nonlin}Variation of the number of clusters and the value of the objective function during the iterations of the algorithm for the PRM (left), PSRM (middle) and PbSRM (right) for the simulated data.}\end{figure}

\begin{figure*}[htbp]
 \centering 
 \includegraphics[width=5.2cm]{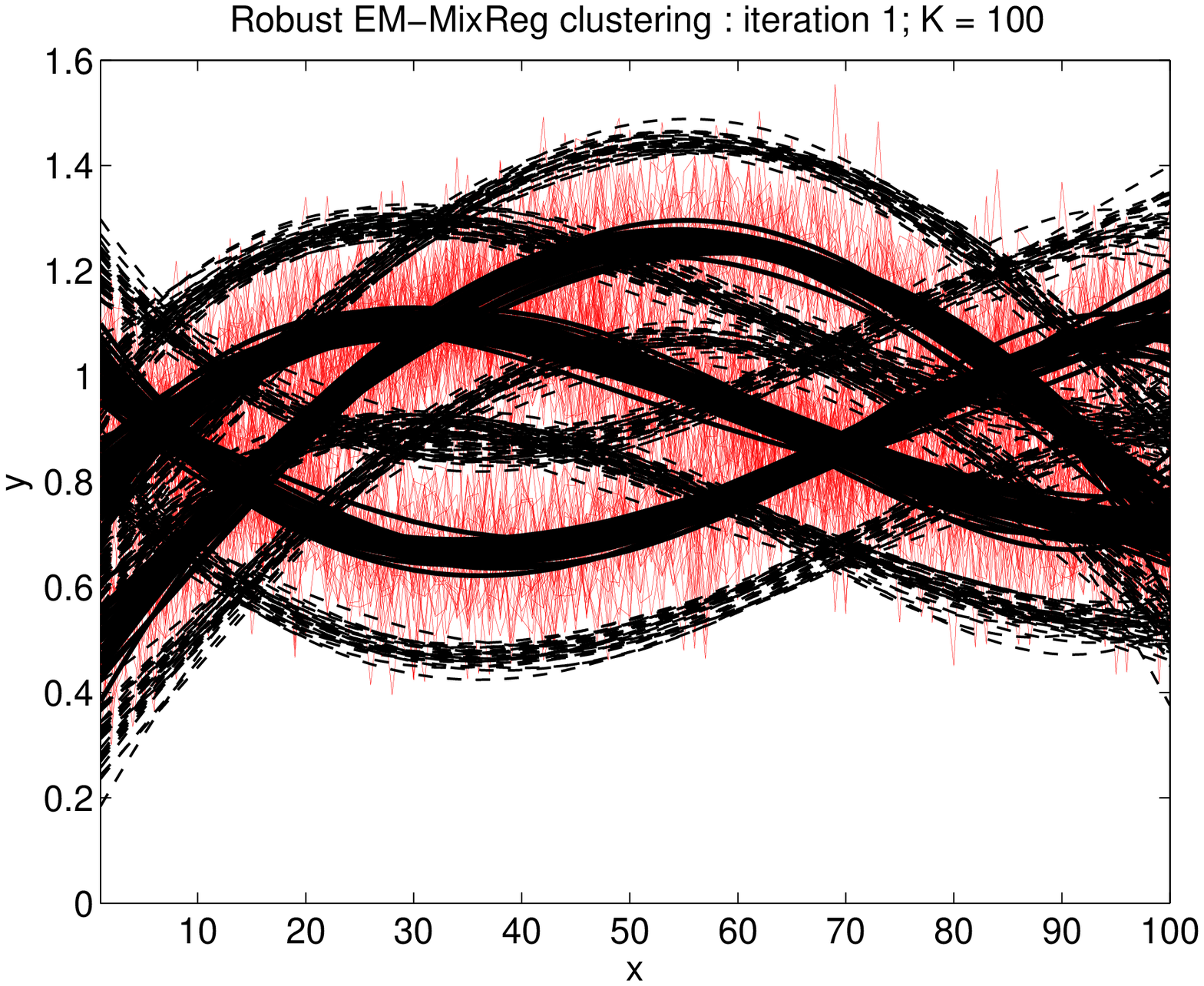} 
 \includegraphics[width=5.2cm]{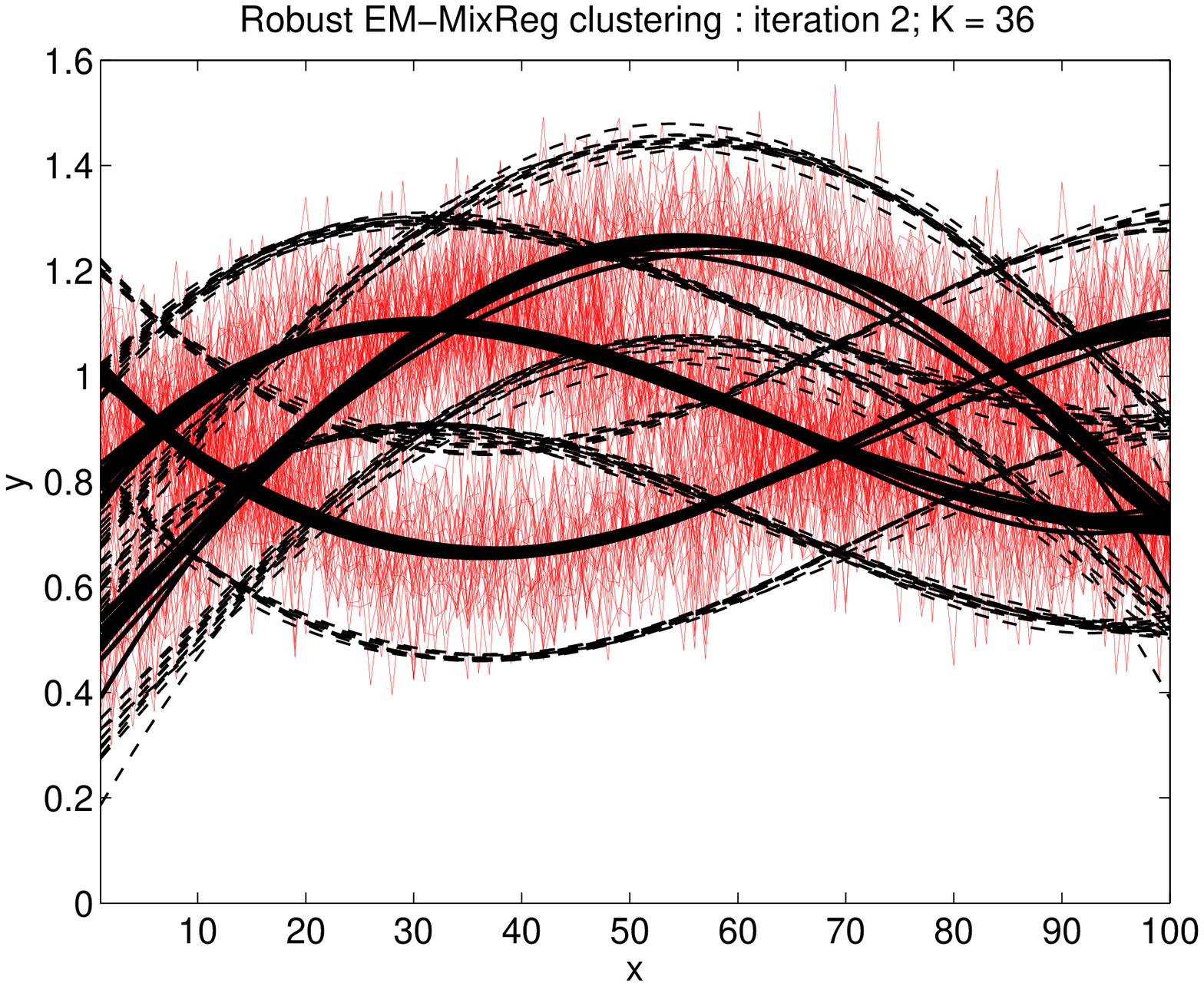} 
 \includegraphics[width=5.2cm]{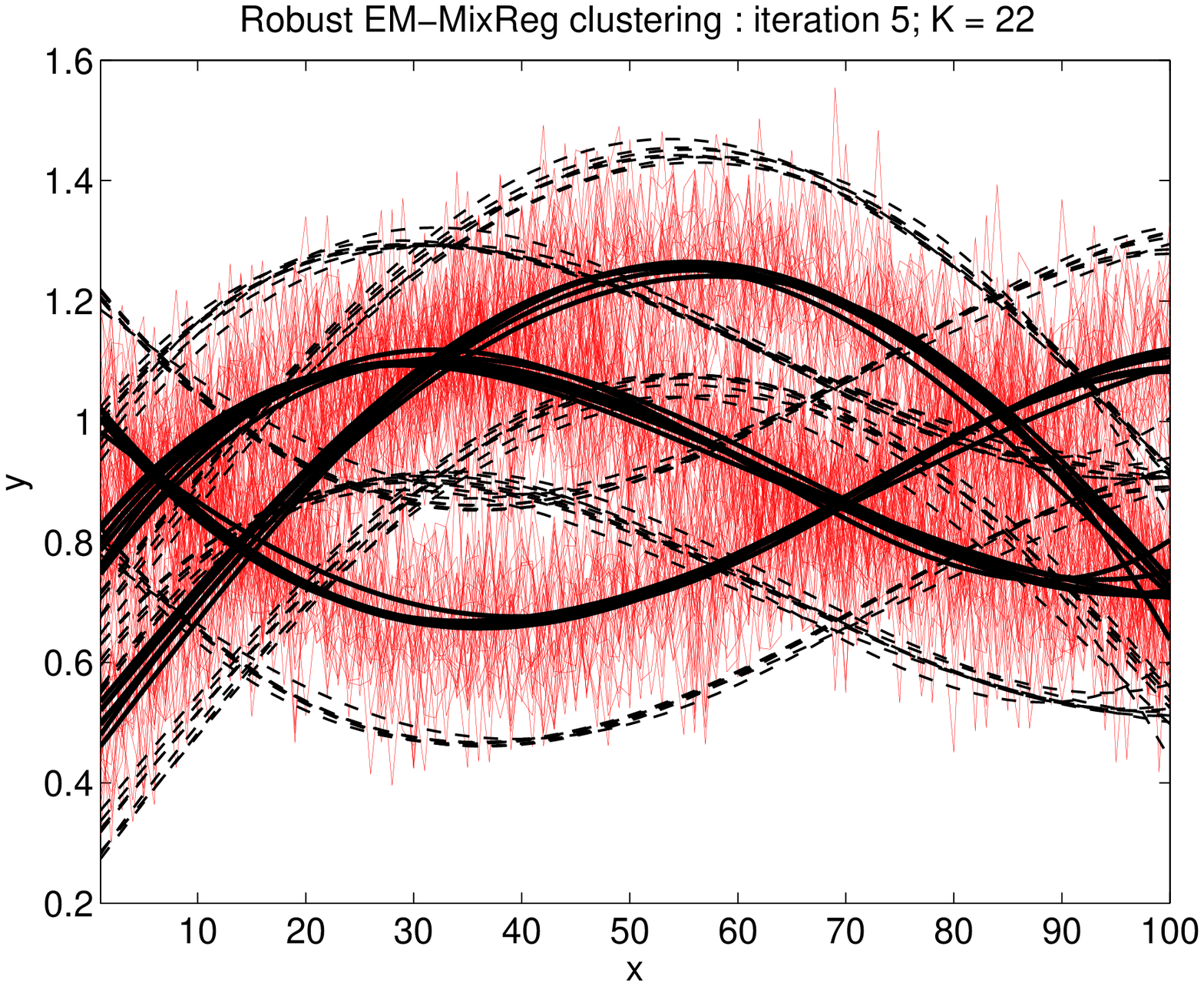} 
 \includegraphics[width=5.2cm]{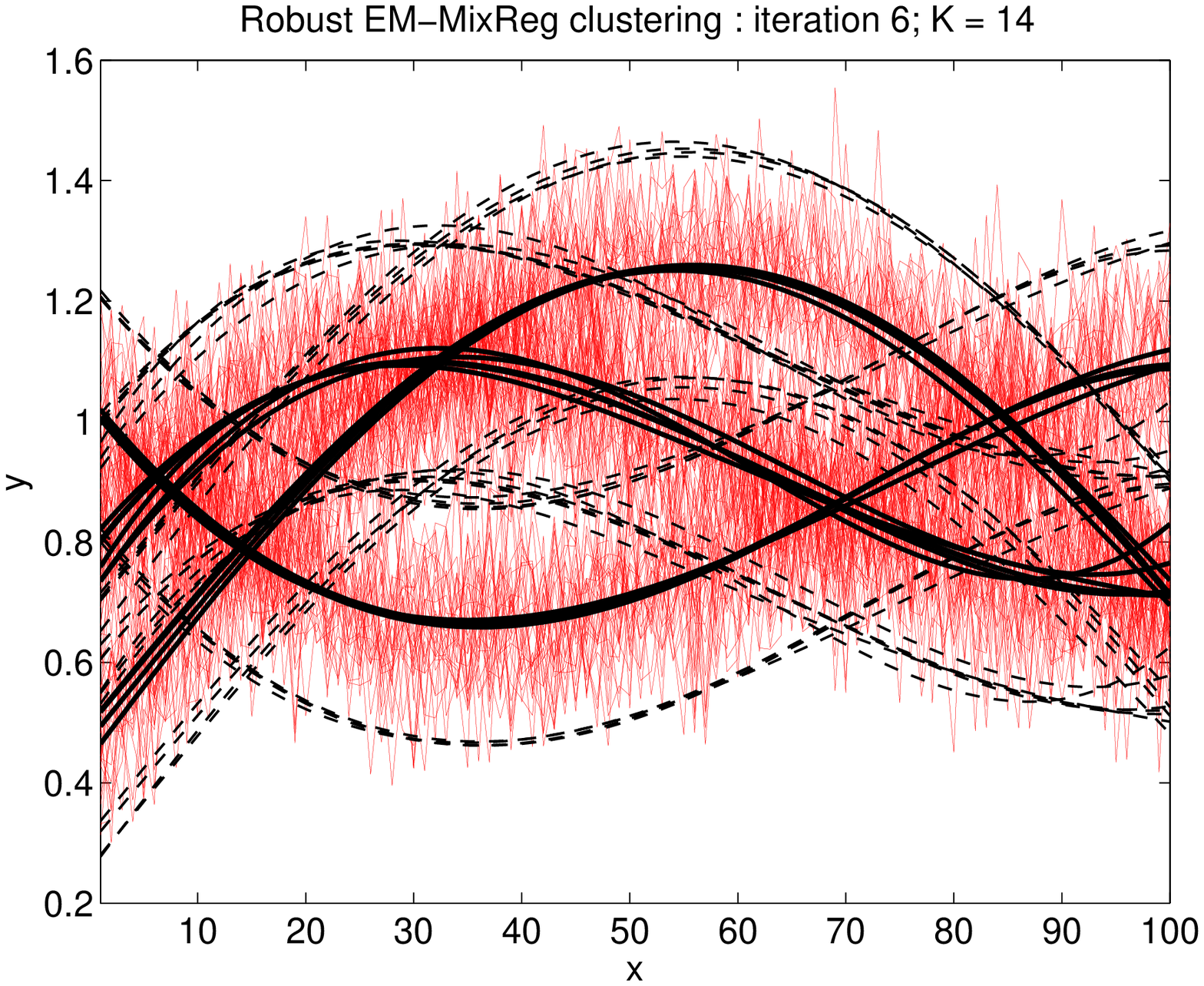} 
 \includegraphics[width=5.2cm]{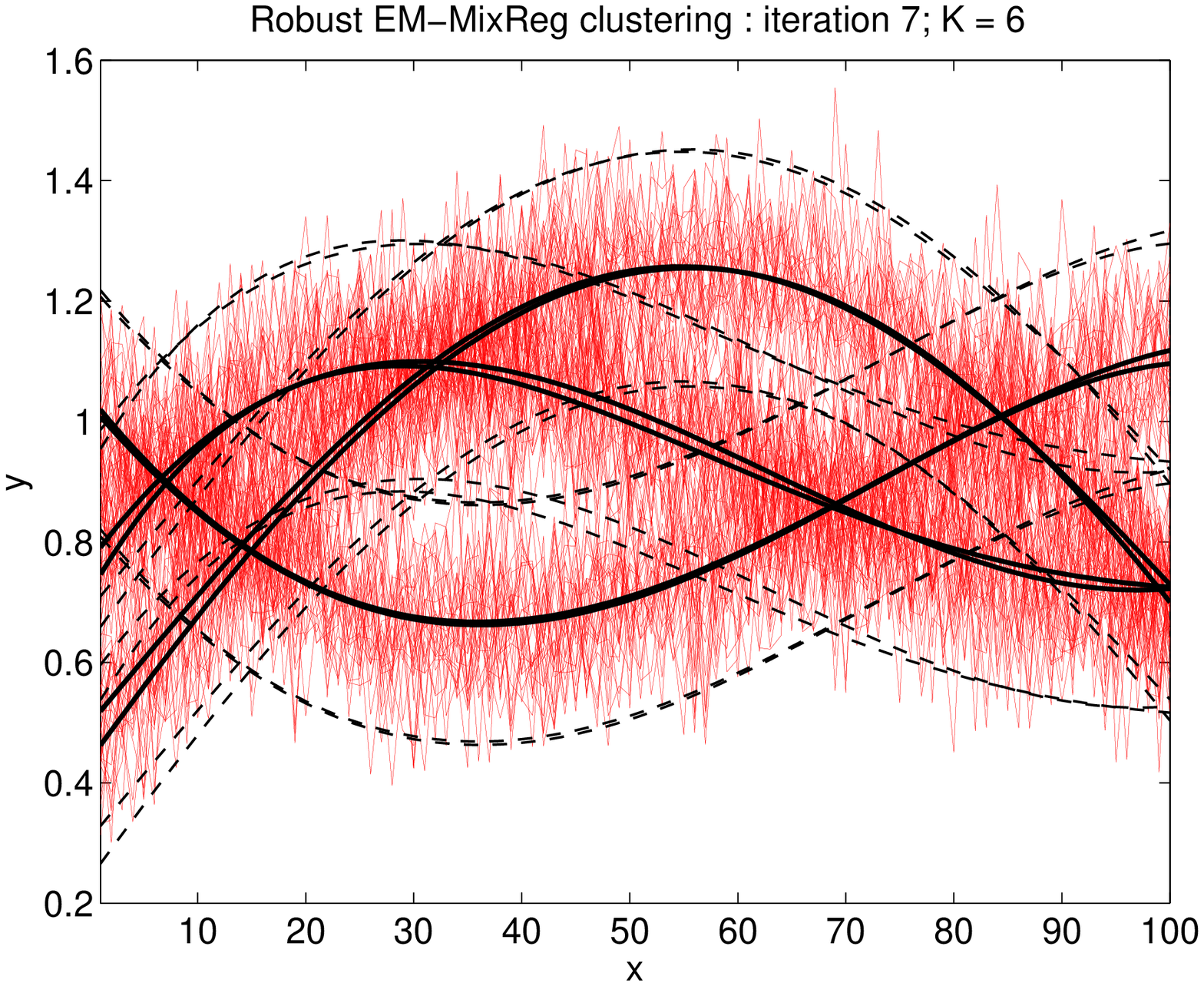} 
 \includegraphics[width=5.2cm]{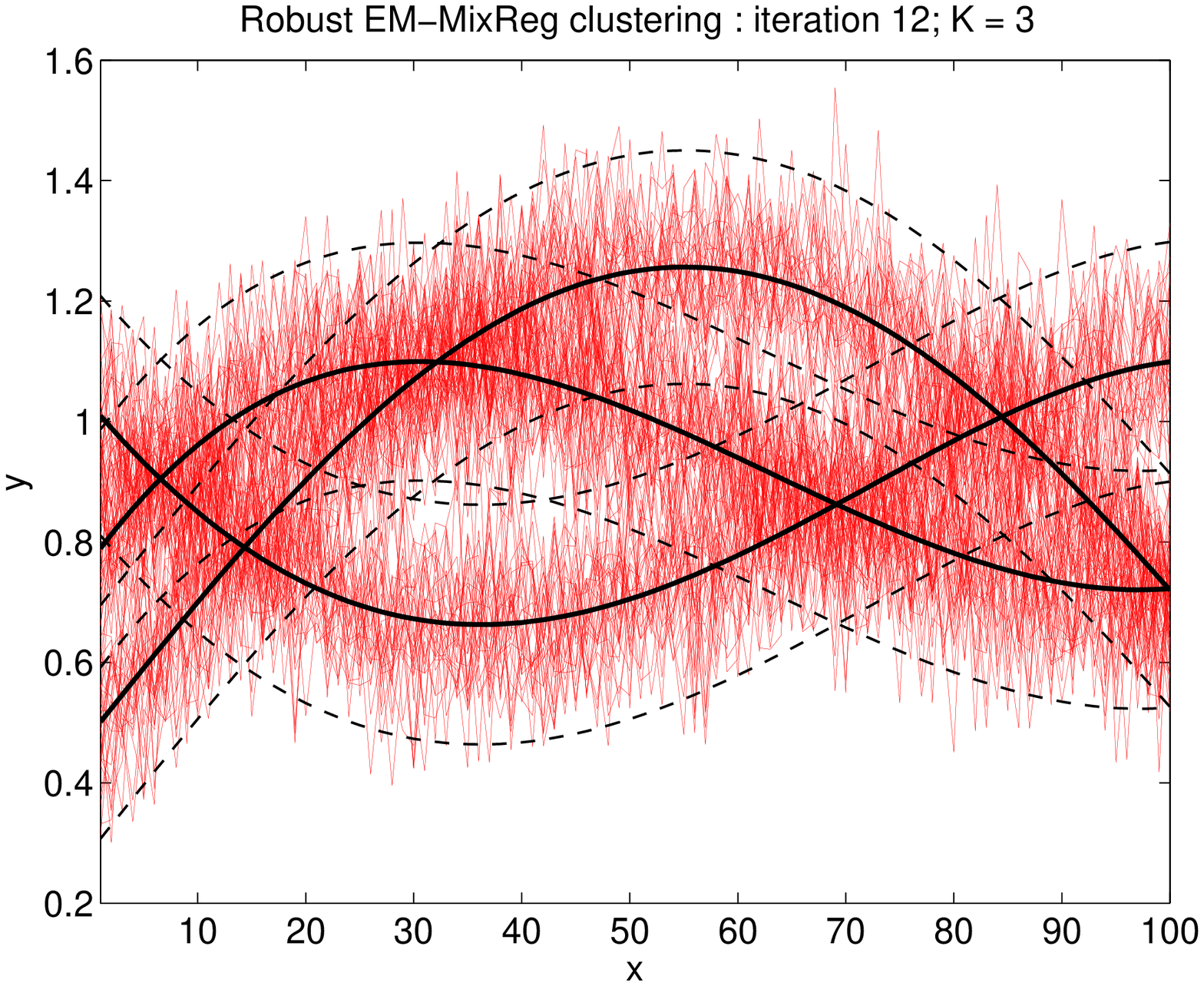} 
 \caption{\label{fig: robust EM-PRM iterations toy nonlin}Clustering results for the simulated data in Fig. \ref{fig: toy-nonlin-data} obtained by the proposed algorithm for polynomial regressions mixtures at different iterations of the algorithm.} 
\end{figure*}

Figure \ref{fig: EM-MixReg stored-K pen-loglik toy nonlin} shows, for each model, the variation of the number of clusters, and the objective function, during the iterations of the proposed robust EM algorithm. 
We see that  the number of clusters starting at $K=100$ equal to the number of curves, rapidly decreases after the two first iterations and the majority of illegitimate clusters is discarded (more than 50\%). Then, after few iteration, the algorithms converge to the actual number of clusters. We also see that in this case the penalized log-likelihood is increasing over the EM iterations and converges.
To highlight the variation of the clustering over the iterations, in Figure \ref{fig: robust EM-PRM iterations toy nonlin}, we show the clustering results obtained for the polynomial regression mixture model, at different iterations of proposed algorithm. We can see that after only one iteration, 64 illegitimate clusters are discarded. Then both the number of clusters and the clusters approximation are precisely updated until convergence. The convergence is achieved after only 12 iterations.
It can be seen that for the second data set of non-linear curves, the proposed algorithm also provides accurate results. After starting with a number of clusters $K=100$, the number of clusters decreases rapidly form 100 to 27 after only four iterations. Then the algorithm converges after 22 iterations and provides the actual number of clusters with precise clustering results.

\subsubsection{Further simulation: Waveform curves of Brieman}
\label{sssec: expermients using waveform data}

\begin{figure}[H]
 \centering
 \includegraphics[width=5cm]{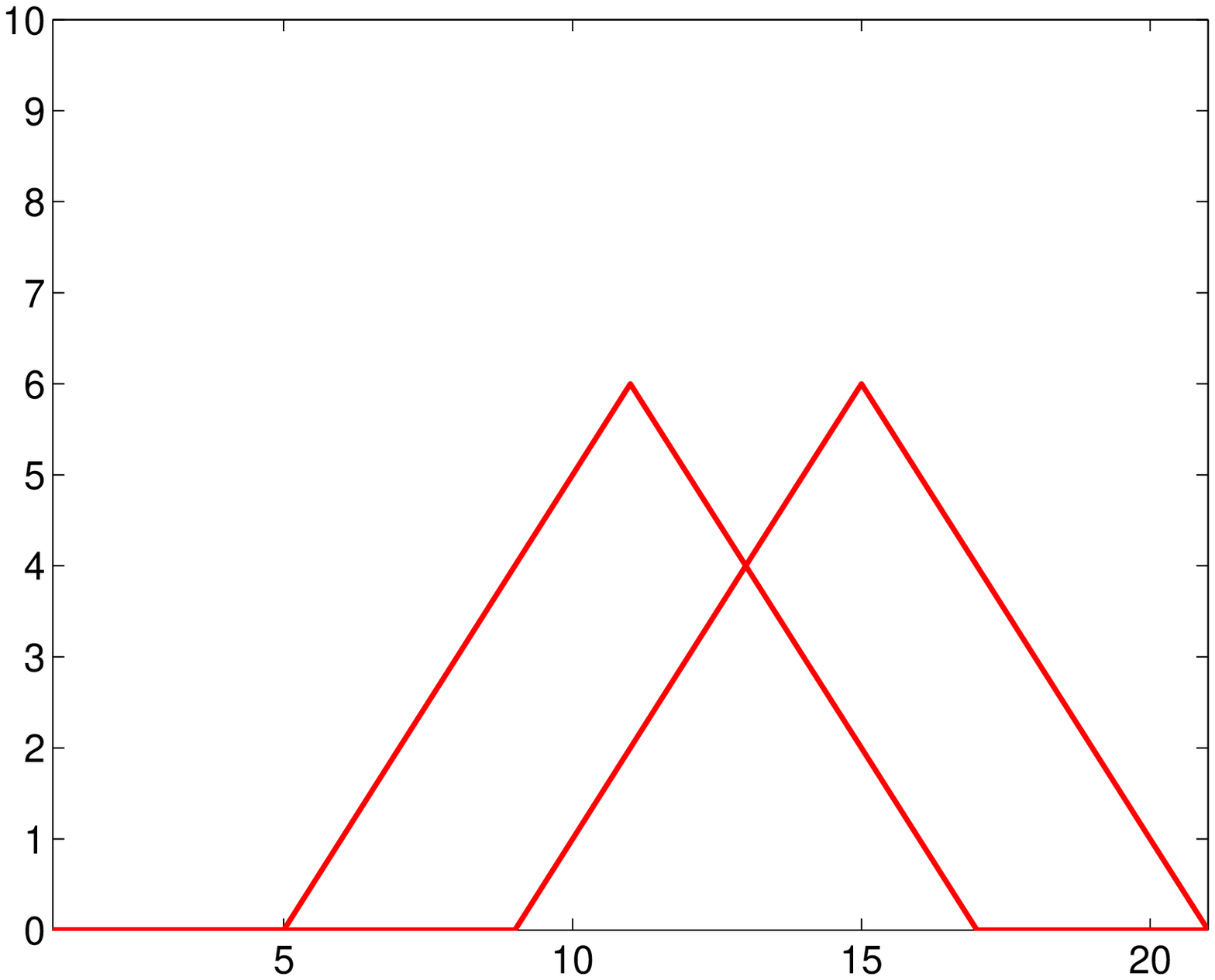}
  \includegraphics[width=5cm]{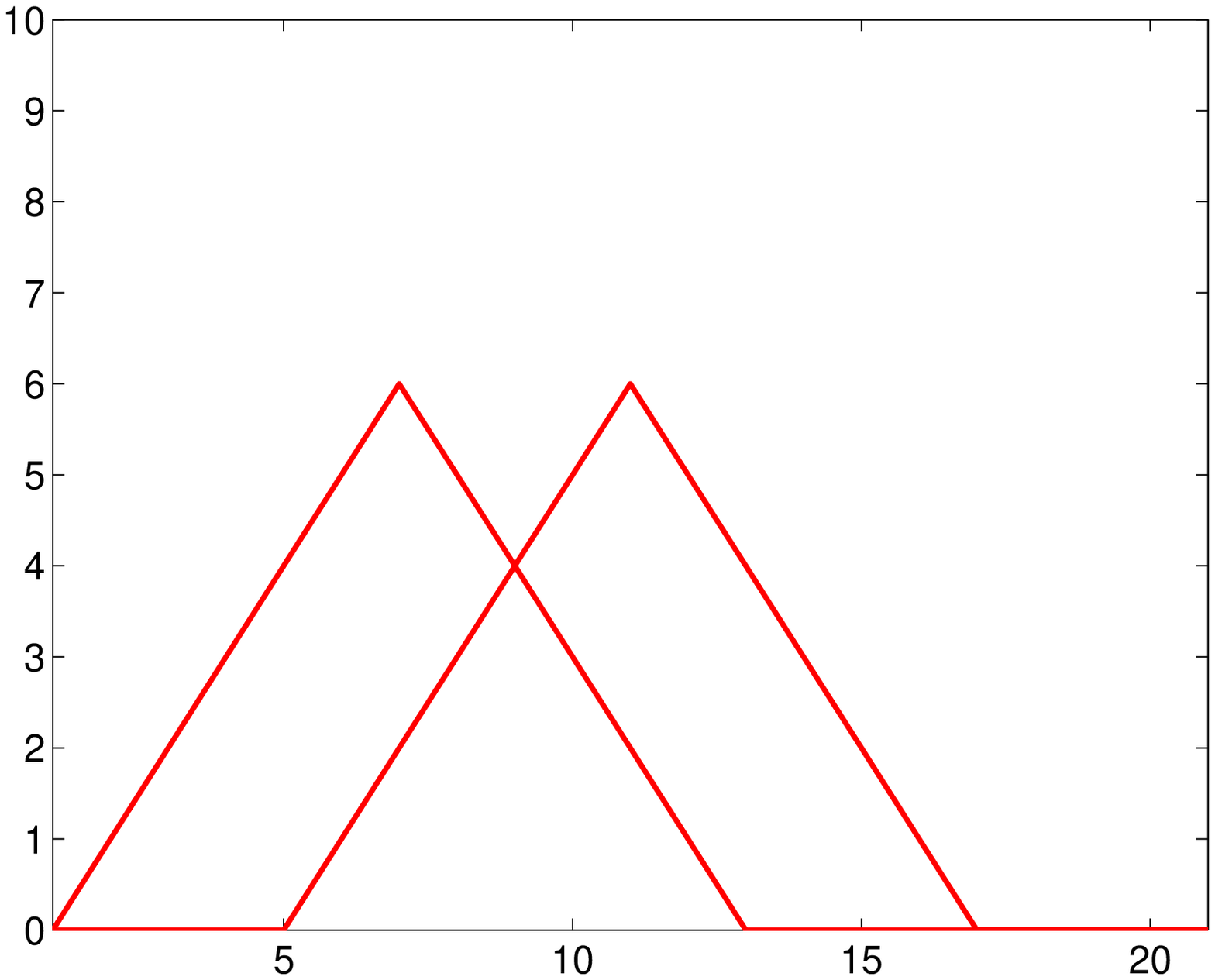}
   \includegraphics[width=5cm]{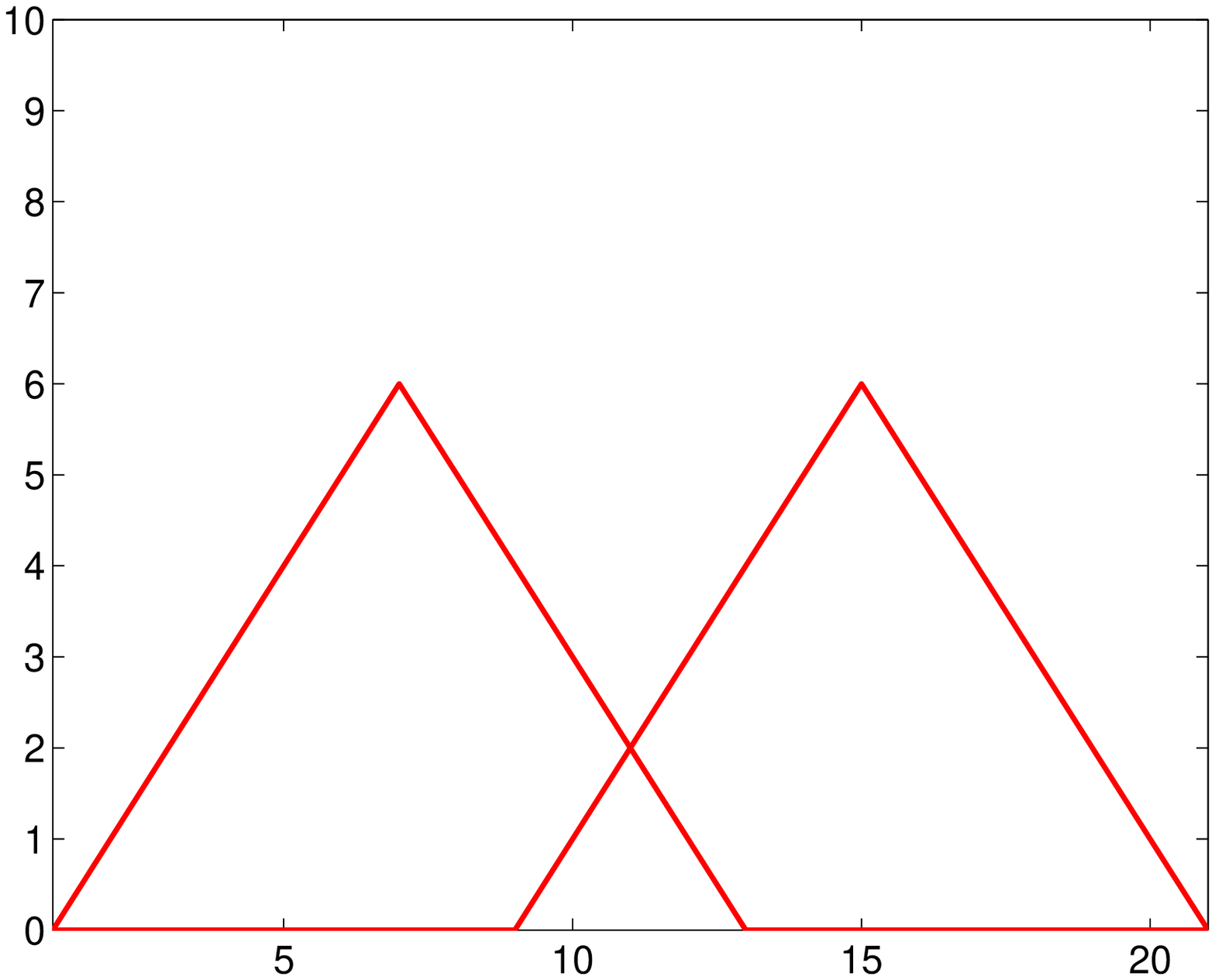}
  \caption{ \label{fig. waveform mean functions}The waveform mean functions from the generative model before the Gaussian noise is added.}
\end{figure}

\begin{figure}[H]
 \centering
 \includegraphics[width=6cm]{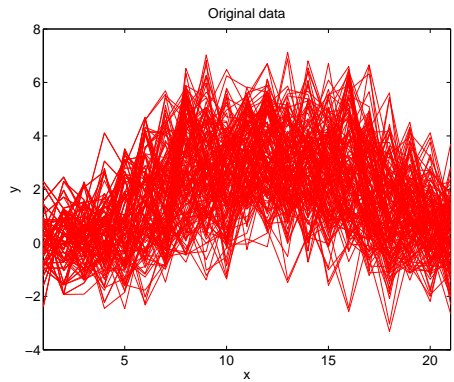}
  \caption{Waveform data.}
 \label{fig. waveform data}
\end{figure}

\begin{figure}[H]
   \centering 
   \includegraphics[width=5cm]{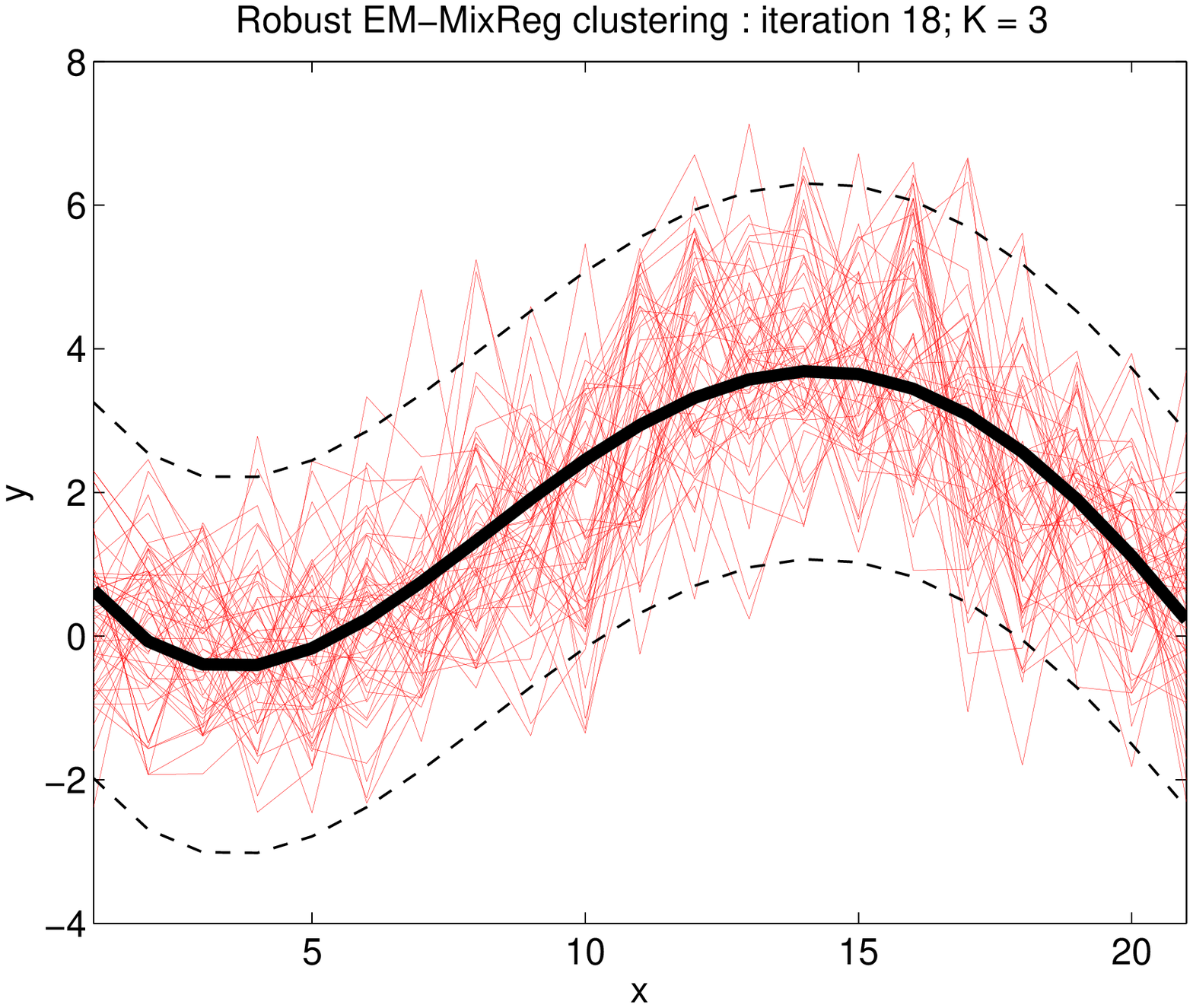}\includegraphics[width=5cm]{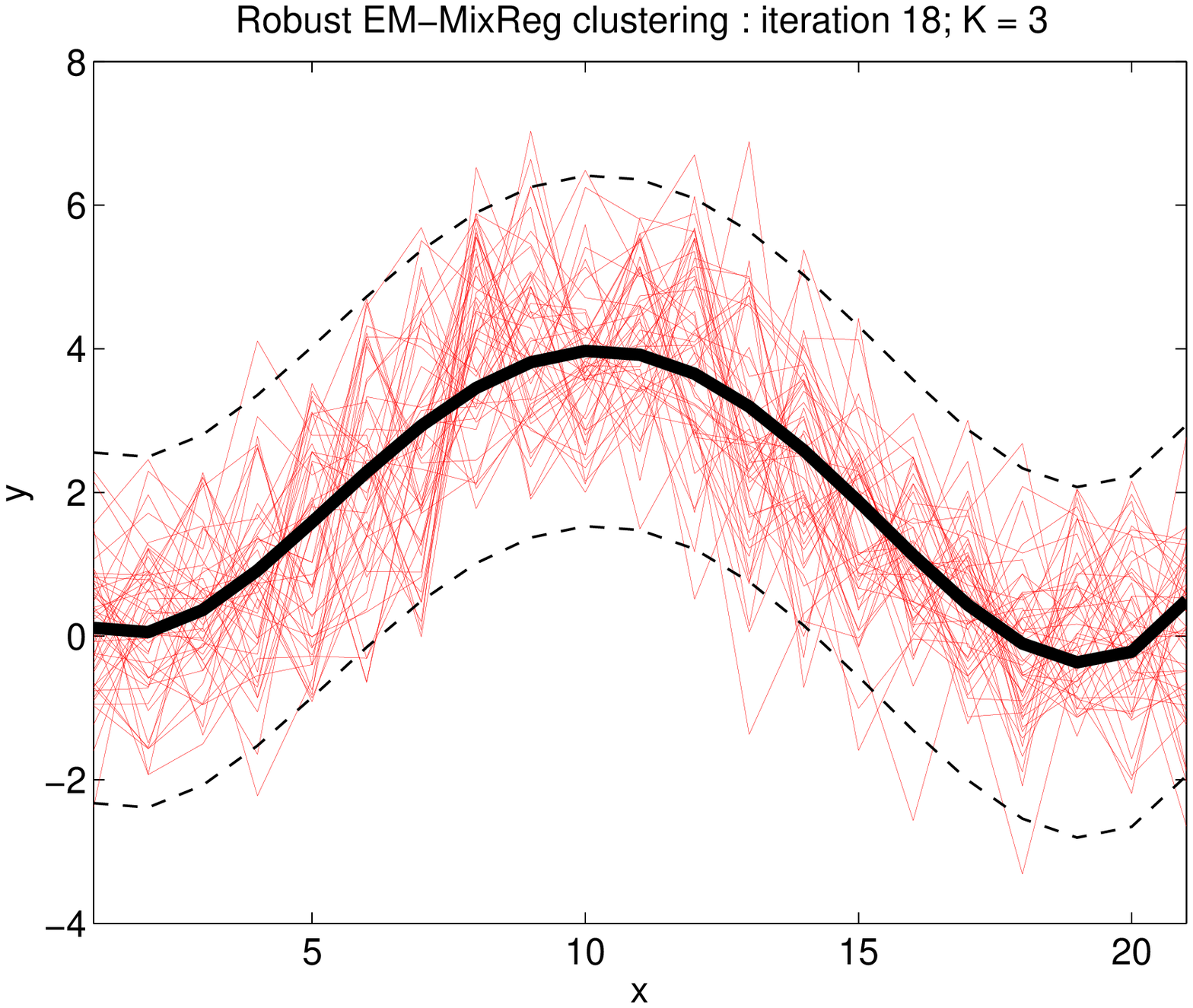}
    \includegraphics[width=5cm]{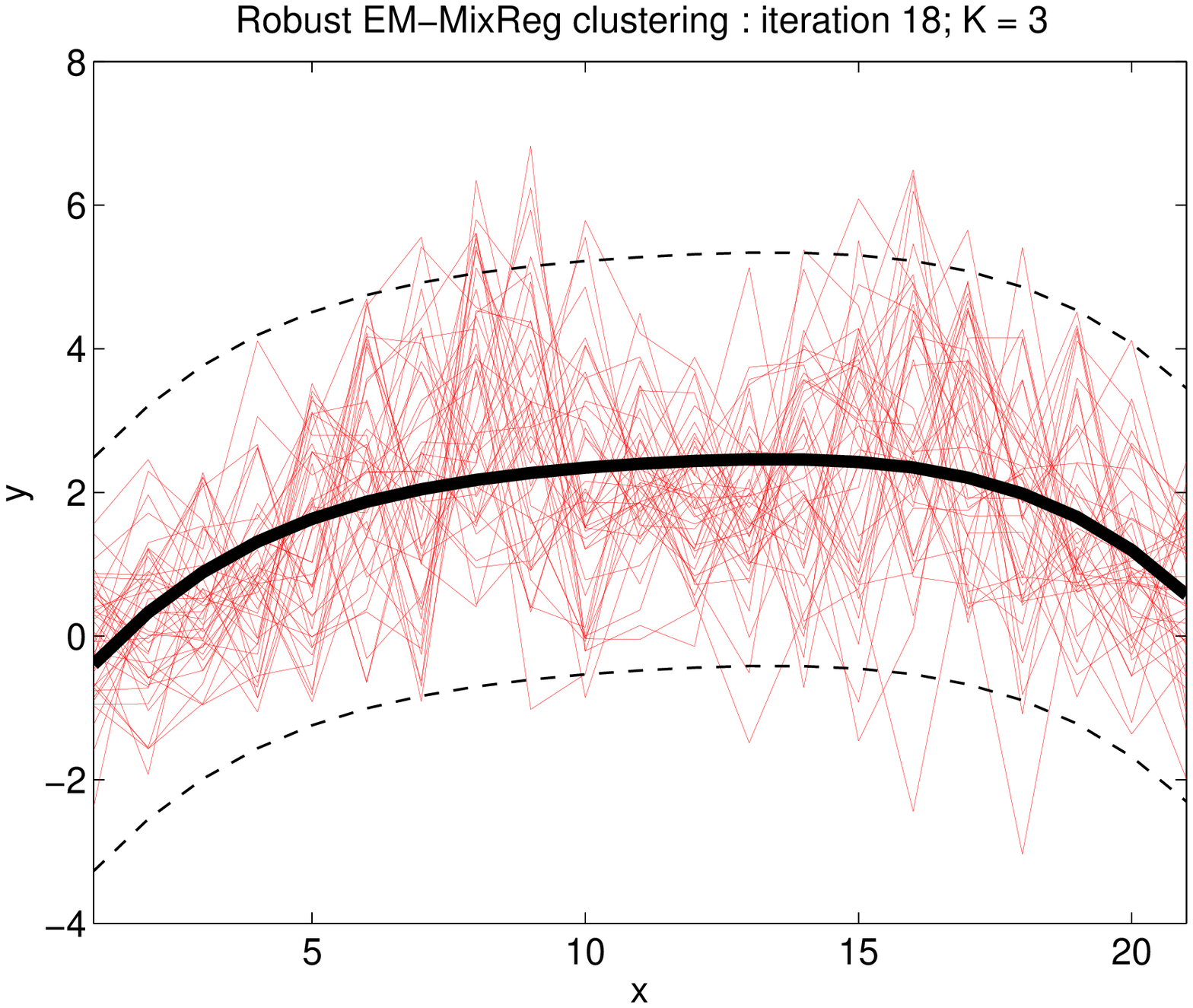} 
   \caption{\label{fig. robust EM-PRM waveform results}Clustering results obtained by the proposed robust EM algorithm and the PRM (plynomial degree $p=4$) model for the waveform data.}
\end{figure}

\begin{figure}[H]
   \centering 
   \includegraphics[width=5cm]{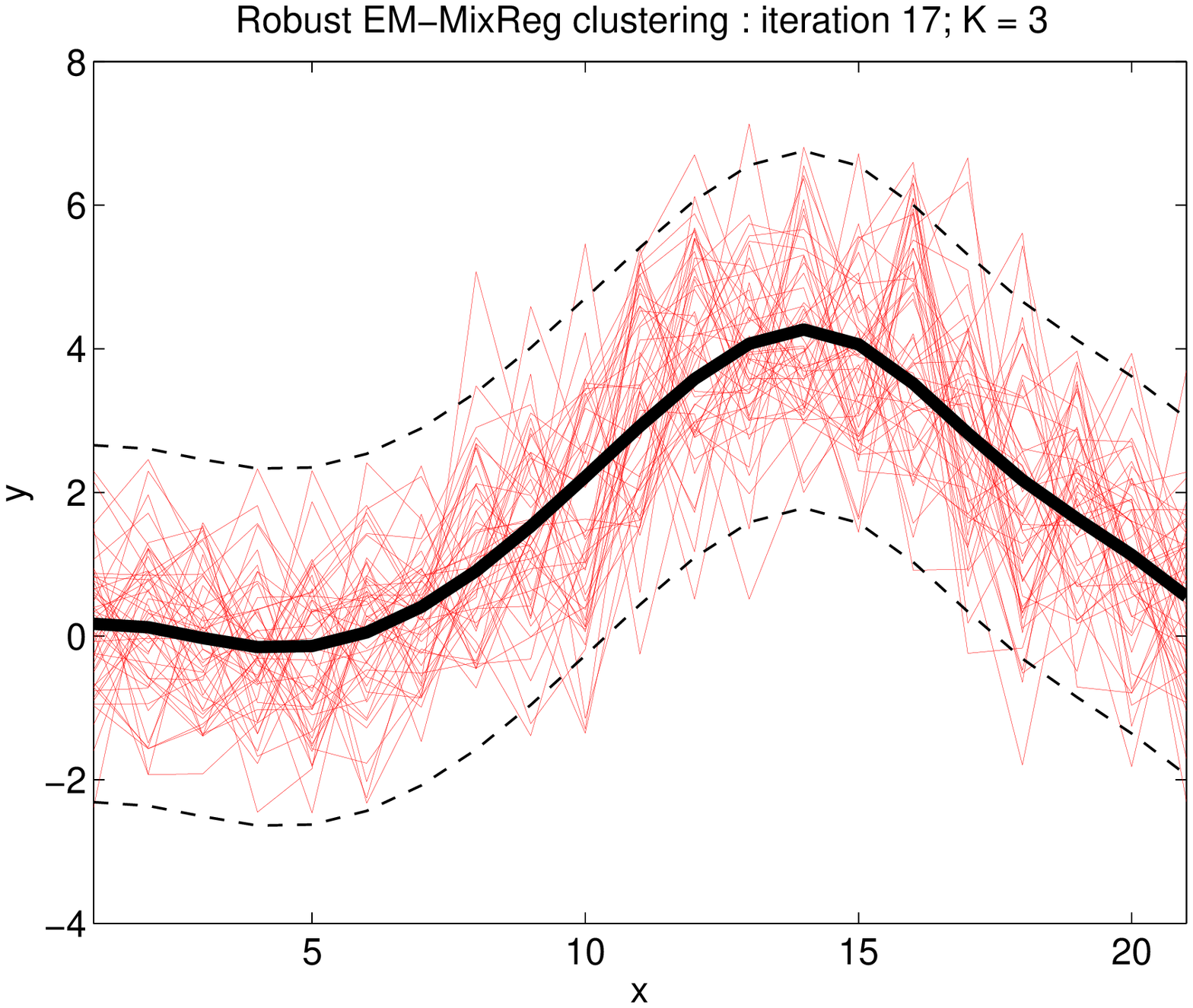}\includegraphics[width=5cm]{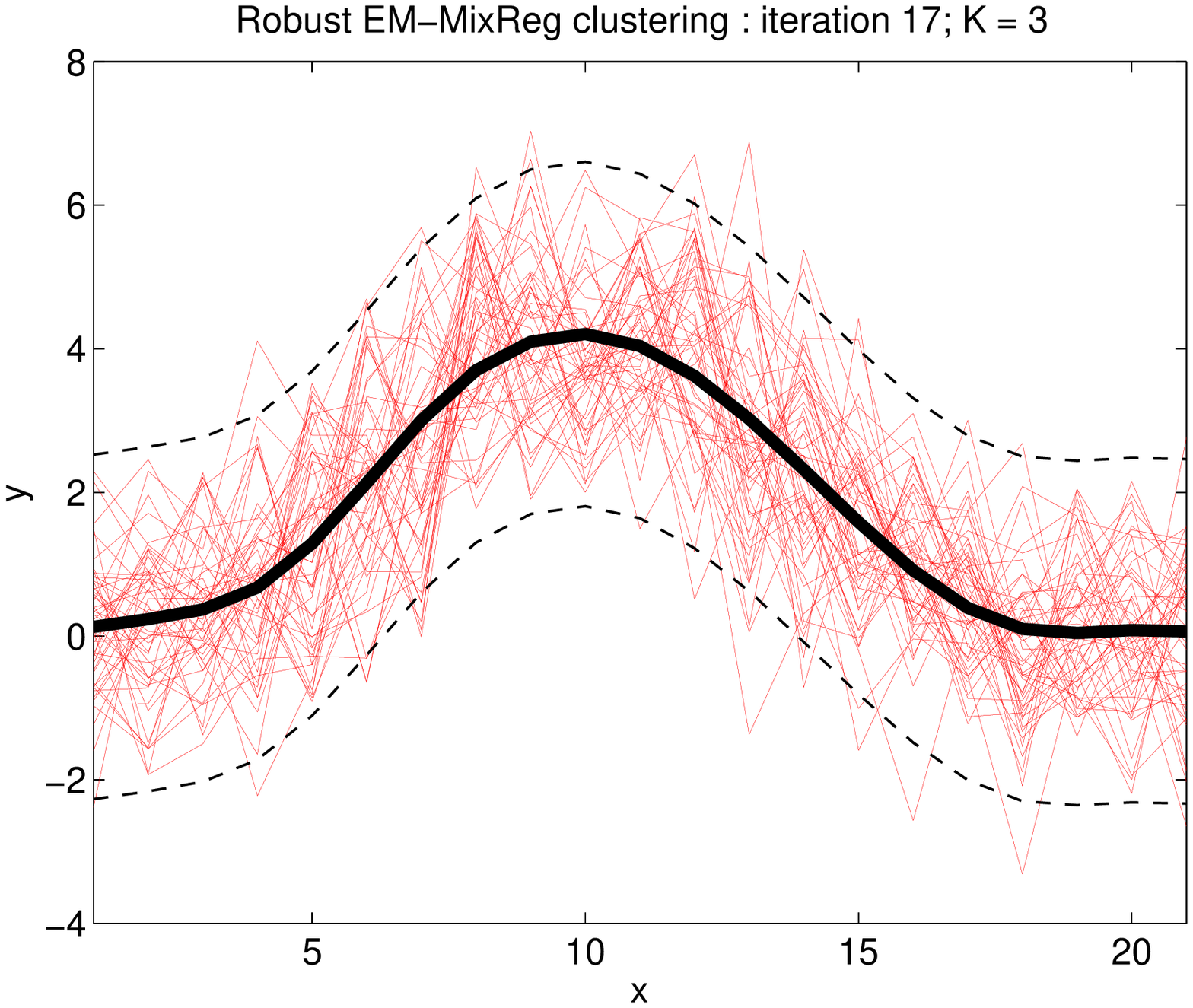}
    \includegraphics[width=5cm]{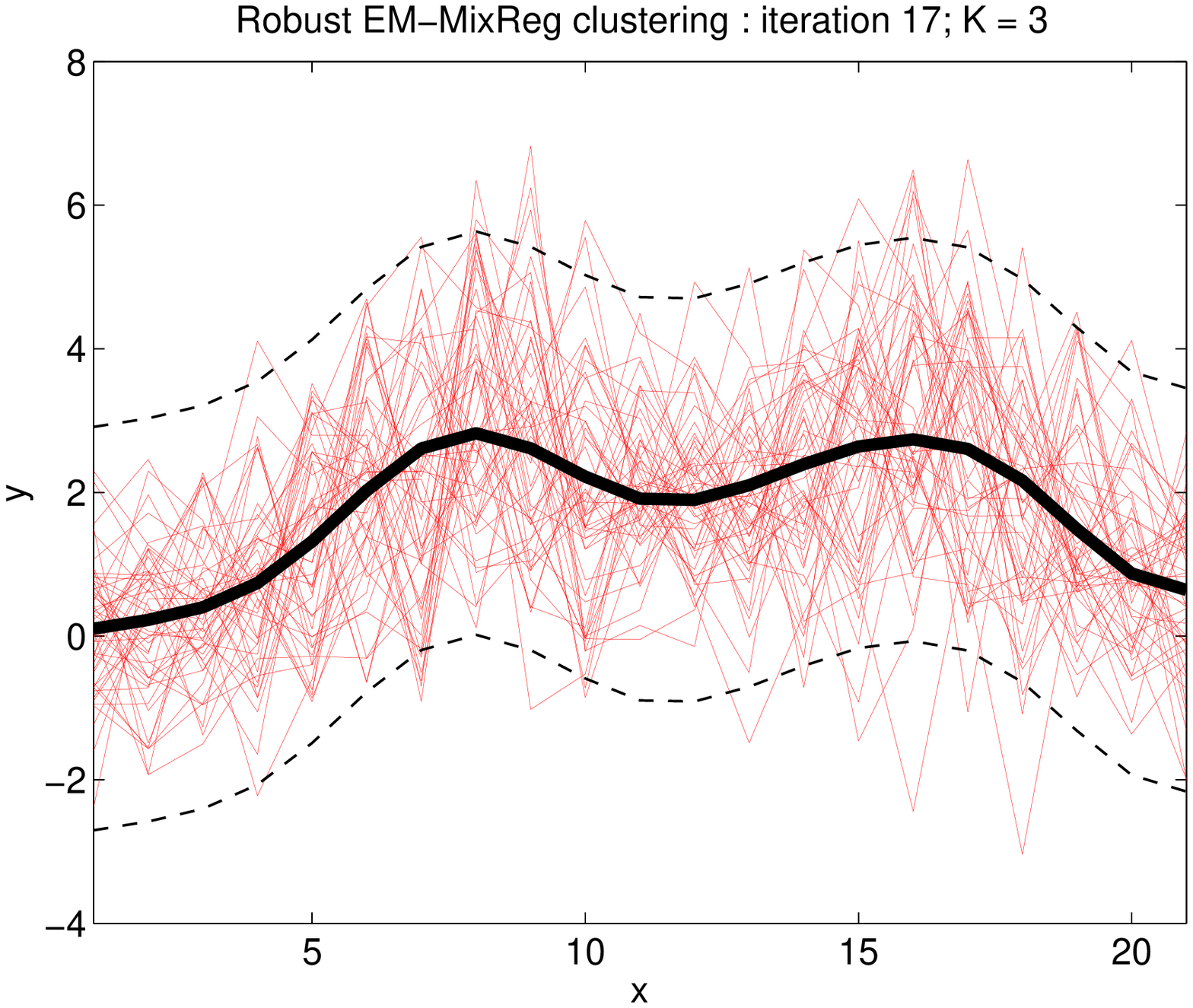}
   \caption{Clustering results obtained by the proposed robust EM algorithm and the SRM  with a cubic-spline of three knots for the waveform data.}
    \label{fig. robust EM-SRM waveform results}
\end{figure}

\begin{figure}[H]
   \centering 
   \includegraphics[width=5cm]{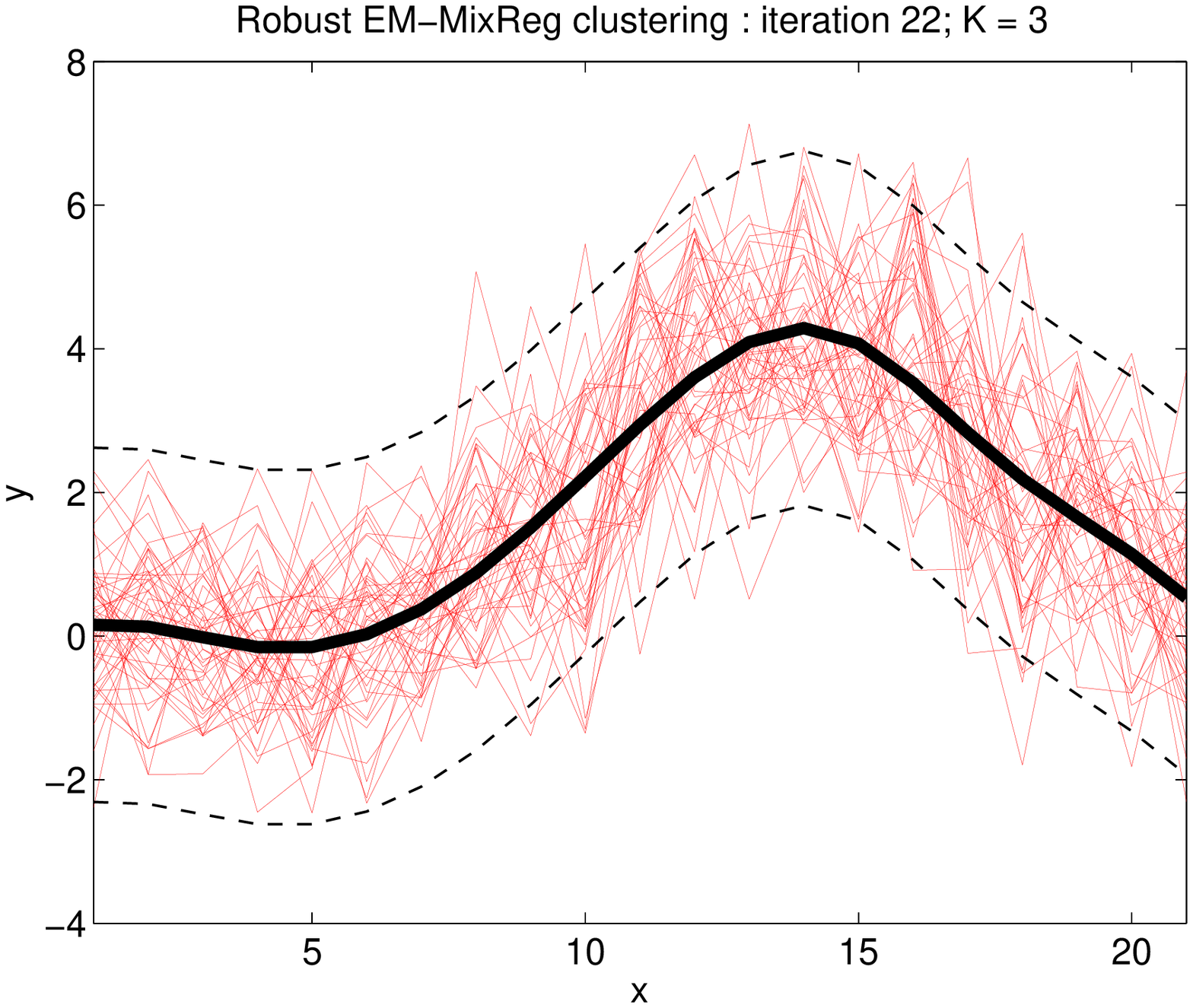}
   \includegraphics[width=5cm]{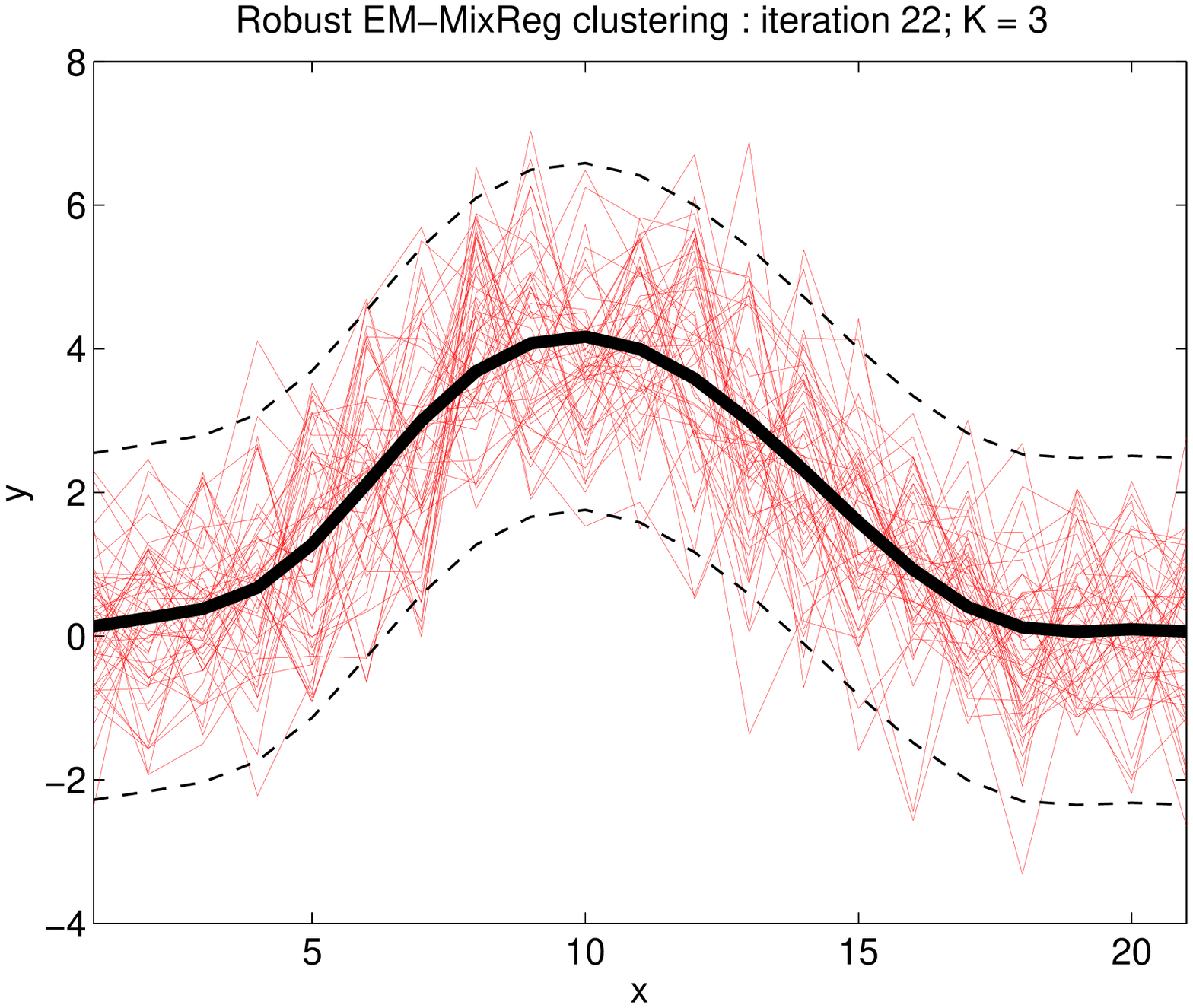}
   \includegraphics[width=5cm]{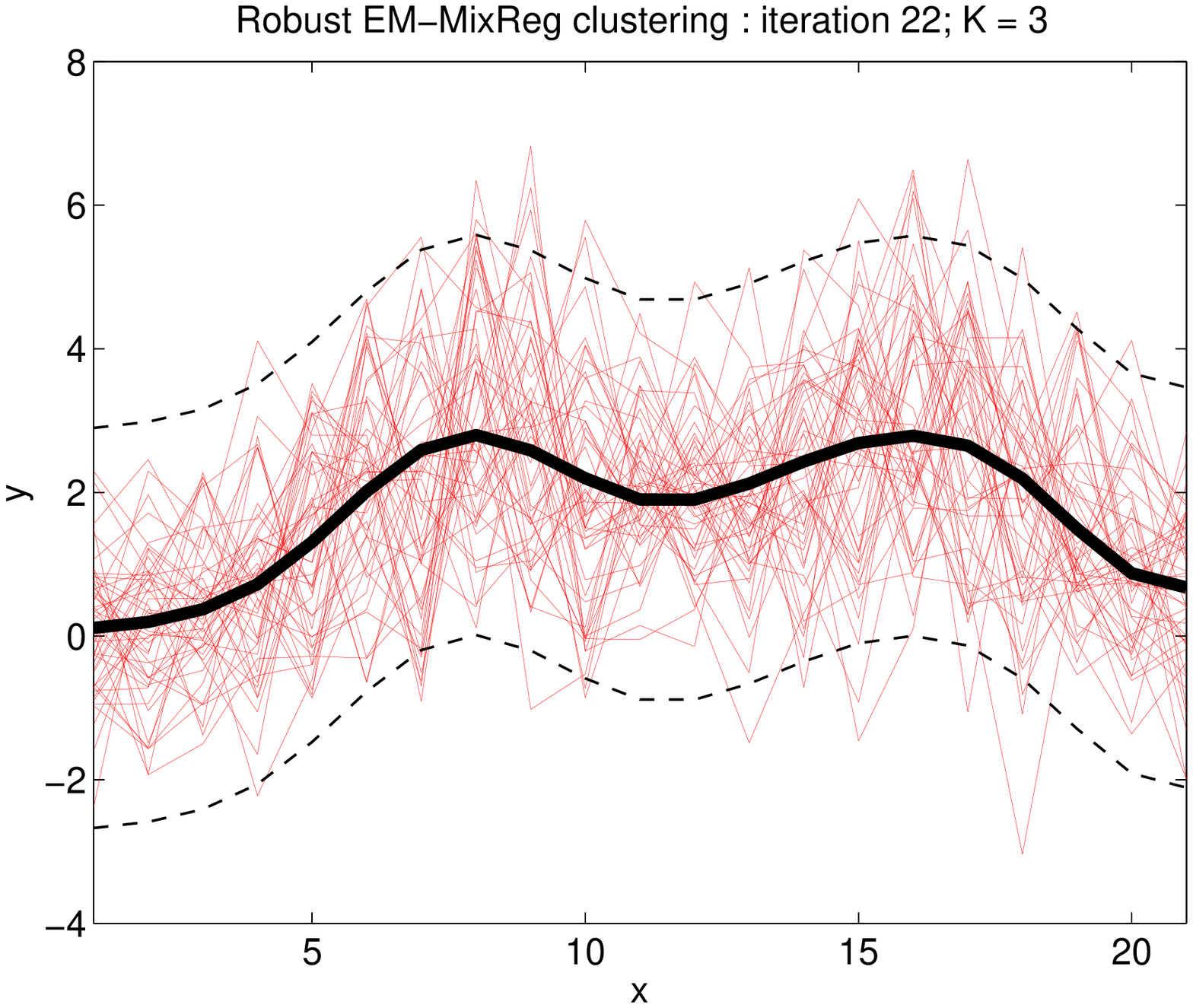}
   \caption{\label{fig. robust EM-bSRM waveform results}Clustering results obtained by the proposed robust EM algorithm and the bSRM with a cubic b-spline of three knots for the waveform data.}
\end{figure}

\begin{figure}[H]
   \centering  
   \includegraphics[width=5cm]{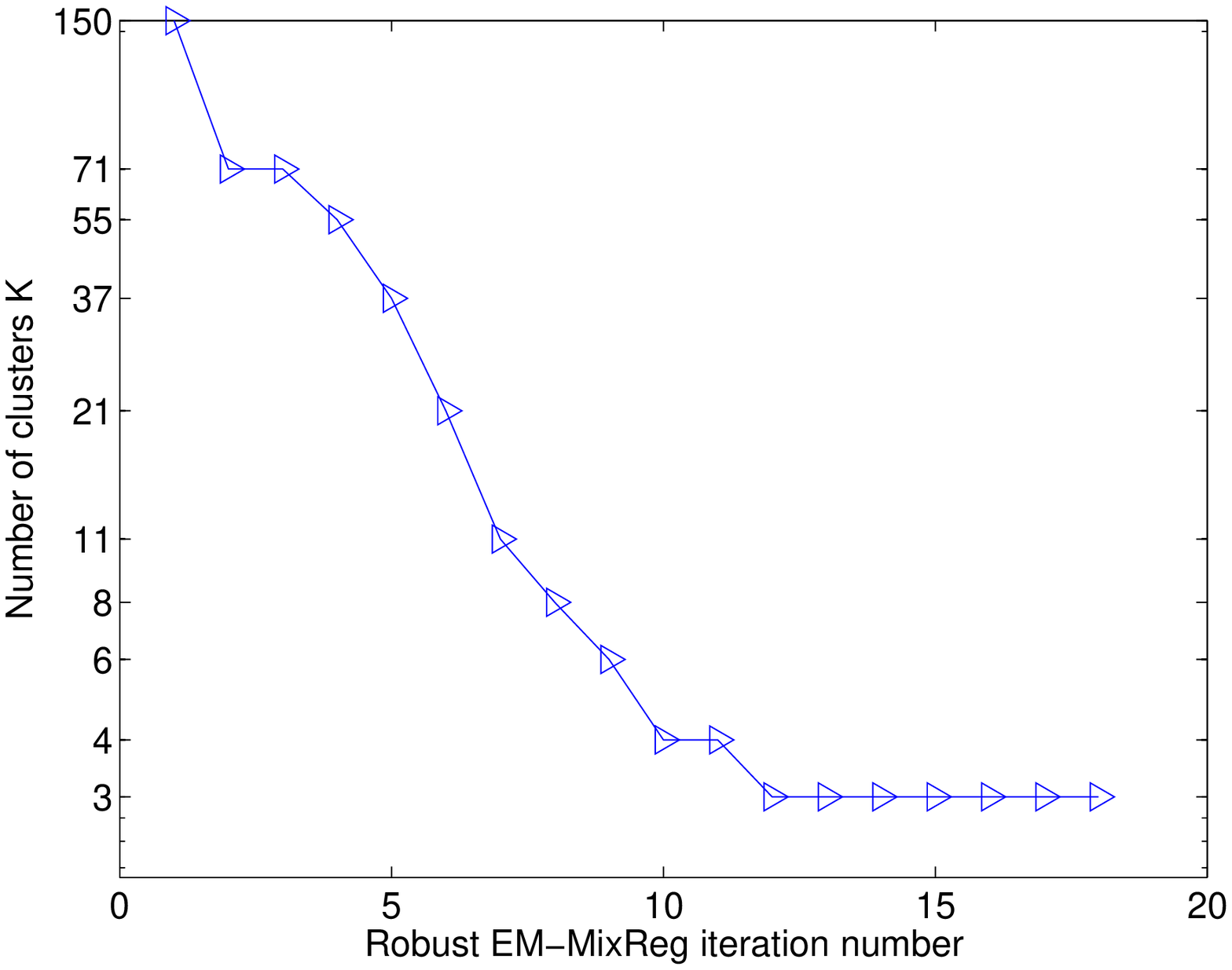}  
   \includegraphics[width=5cm]{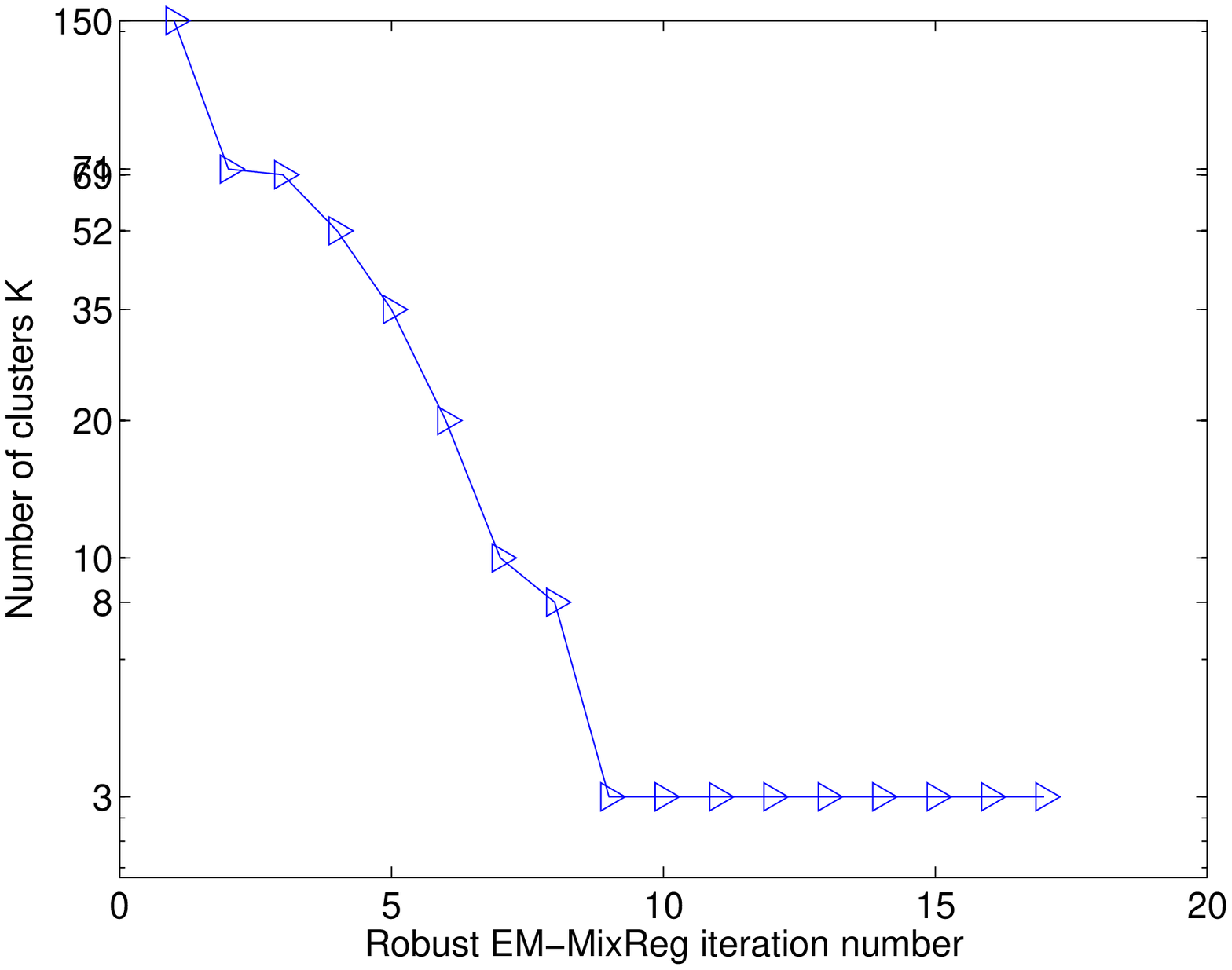}
   \includegraphics[width=5cm]{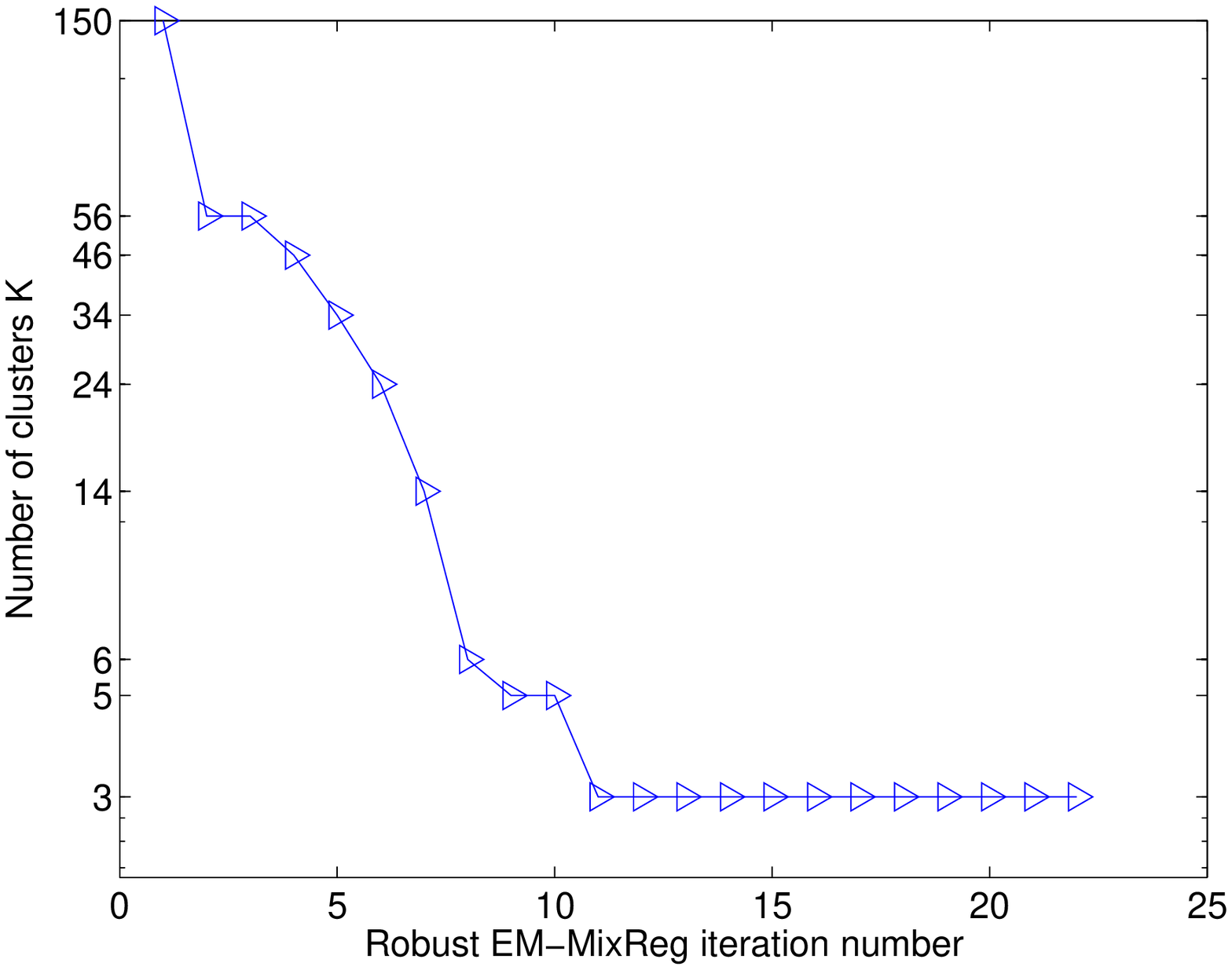}\\
   \includegraphics[width=5cm]{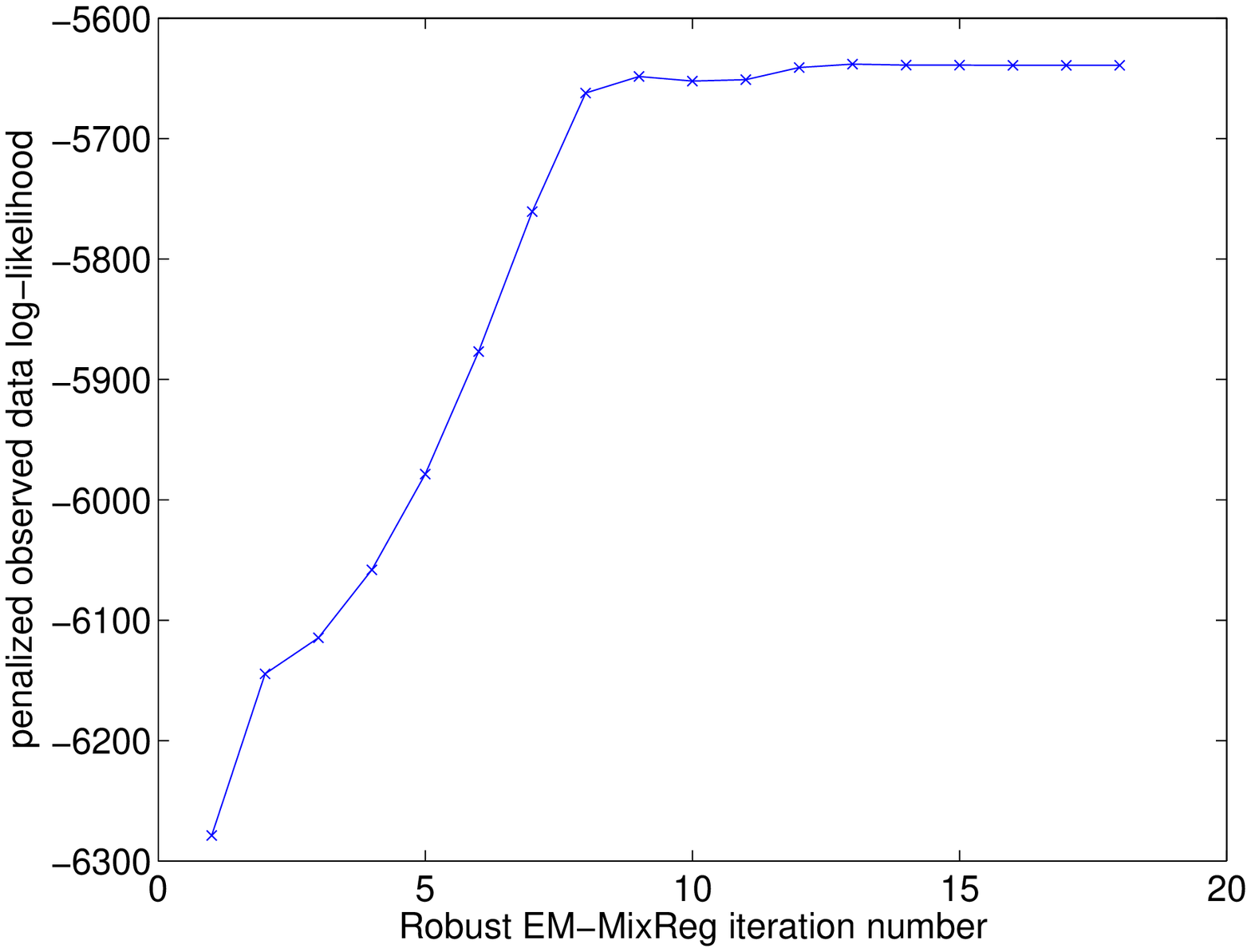}
   \includegraphics[width=5cm]{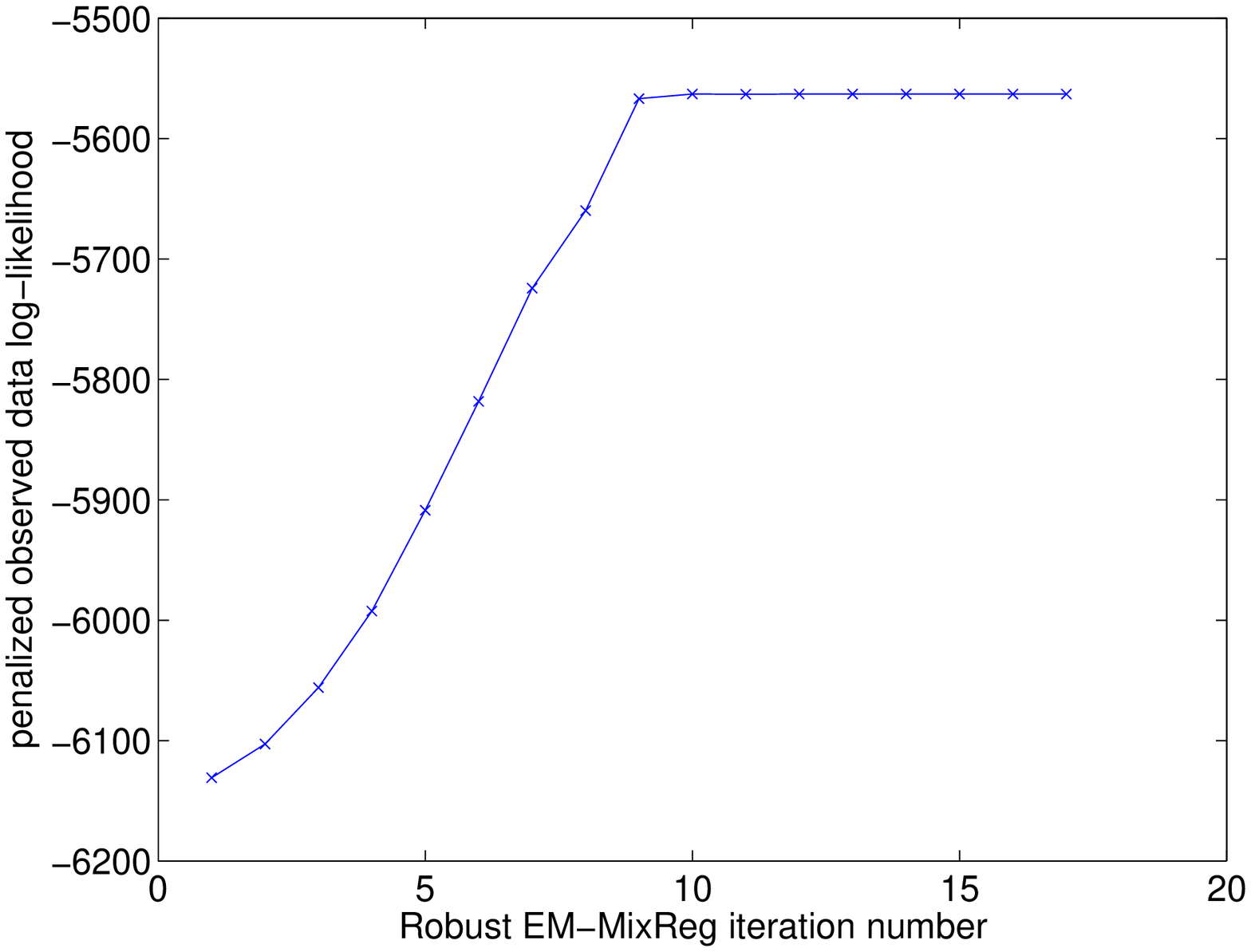}
   \includegraphics[width=5cm]{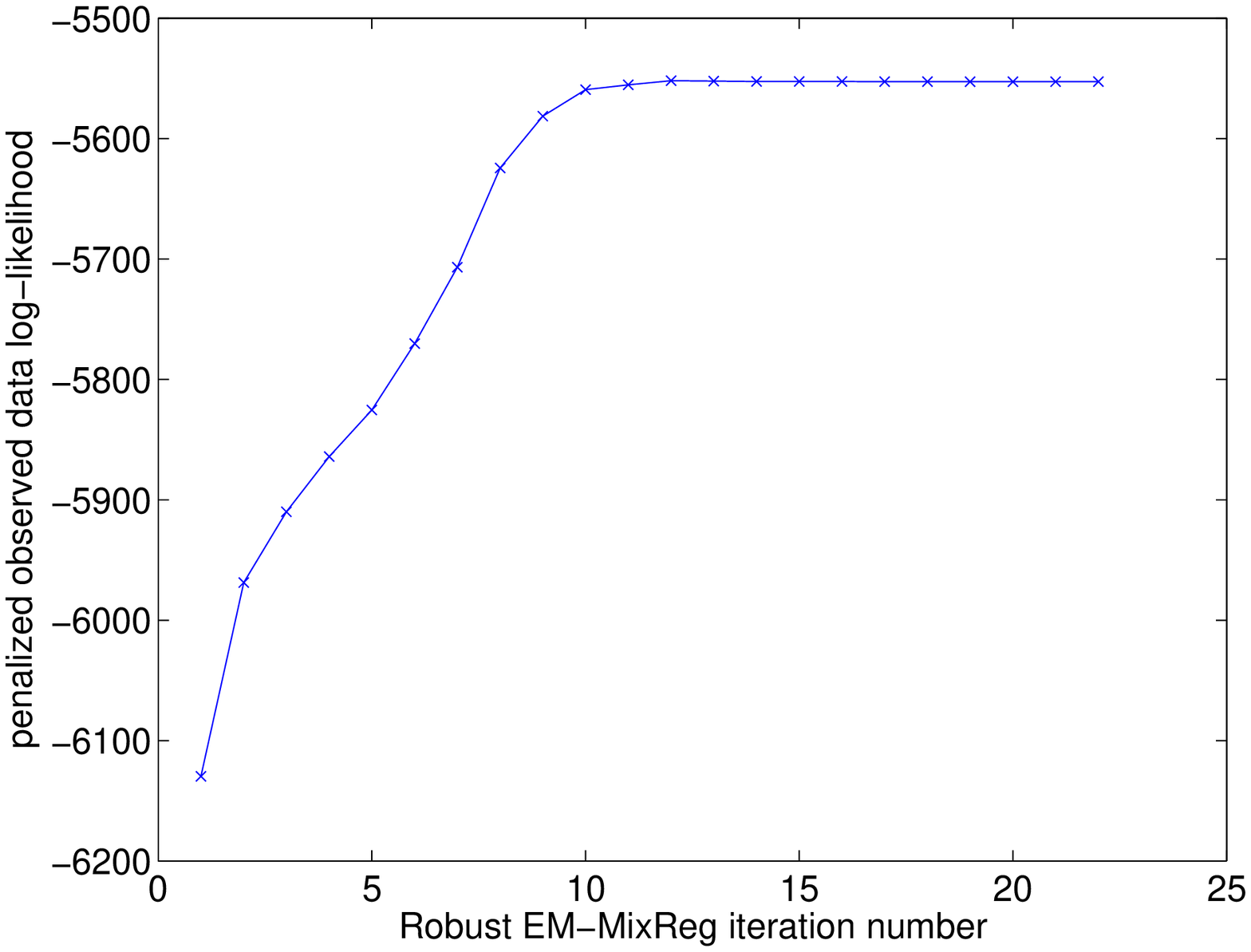}
   \caption{\label{fig: robust EM-MixReg stored-K pen-loglik waveform}Variation of the number of clusters and the value of the objective function during the iterations of the algorithm for the PRM (left), PSRM (middle) and PbSRM (right) for the waveform data.}
\end{figure}

In this experiment we consider the waveform data introduced in \cite{breiman} and studied in \cite{hastieANDtibshiraniMDA} and elsewhere. The  waveform data consist in a three-class problem where each curve is generated as follows:
\begin{itemize}
\item $\by_1(t)=uh_1(t) + (1-u)h_2(t) + \epsilon_t$ for the class 1;
\item $\by_2(t)=uh_2(t) + (1-u)h_3(t) + \epsilon_t$ for the class 2;
\item $\by_3(t)=uh_1(t) + (1-u)h_3(t) + \epsilon_t$ for the class 3.
\end{itemize}
where $u$ is a uniform random variable on $(0,1)$,
\begin{itemize}
\item $h_1(t)=\max (6-|t-11|,0)$;
\item $h_2(t)=h_1(t-4)$;
\item $h_3(t)=h_1(t+4)$.
\end{itemize}
and $\epsilon_t$ is a zero-mean Gaussian noise with unit standard deviation.
The temporal interval considered for each curve is $[1;21]$ with a constant period of sampling of 1 second. 
%
 Figure \ref{fig. waveform mean functions} shows the waveform data mean functions from the generative model before the Gaussian noise is added and Fig. \ref{fig. waveform data} shows a sample of 150 waveforms. 
Figures \ref{fig. robust EM-PRM waveform results}, \ref{fig. robust EM-SRM waveform results} and \ref{fig. robust EM-bSRM waveform results}  respectively show the obtained cluster results for the waveform data for the polynomial, spline and B-spline regression mixture. The number of clusters is correctly estimated by the three models. For this data, the spline regression models provide slightly better results in terms of clusters approximation than the polynomial regression mixture (here $p=4$). This can be seen for the third cluster. 
Table \ref{tab. classification results waveform data} presents the clustering results averaged over 20 different sample of 500 curves. It includes the estimated number of clusters, the misclassification error rate and the absolute error between the true clusters proportions and variances and the estimated ones.
One can observe that for all the models, the actual number of clusters is correctly retrieved. The misclassification error rate as well as the parameters estimation errors are slightly better for the spline regression models, in particular the B-spline regression mixture.
\begin{table}[H]
\centering
{\small
\begin{tabular}{|c c c c c|}
\hline
		   			& actual 				& 			EM-PRM 		     	& 			EM-SRM 			  &	 EM-bSRM\\
\hline
	$K$ 			& 	3		 				&	3								 	&			3						  &  3 \\
  misc. error	& 	-						&	4.31 $\pm$ (0.42)\%  	&	2.94 $\pm$ (0.88)\%   & 2.53 $\pm$ (0.70)\% \\
  $\sigma_1$	& 	1						&	0.128 $\pm$ (0.015)		& 0.108 $\pm$ (0.015)    & 0.103 $\pm$ (0.012) \\
  $\sigma_2$	& 	1						& 	0.102 $\pm$ (0.015)		& 0.090  $\pm$ (0.011) 	  & 0.079  $\pm$ (0.010) \\
  $\sigma_3$	& 	1						&	0.223 $\pm$ (0.021) 		& 0.180 $\pm$ (0.014)    & 0.141 $\pm$ (0.013)\\
  $\pi_1$		&	$\frac{1}{3}$		& 	0.0037 $\pm$ (0.0018)  & 0.0035 $\pm$ (0.0015)  & 0.0034 $\pm$ (0.0015) \\
  $\pi_2$		&	$\frac{1}{3}$		&	0.0029 $\pm$ (0.0023)  & 0.0018  $\pm$ (0.0015) & 0.0012  $\pm$ (0.0011) \\
  $\pi_3$ 		&	$\frac{1}{3}$		&	0.0040 $\pm$ (0.0062)	 & 0.0037 $\pm$ (0.0015)  & 0.0035 $\pm$ (0.0014)\\ 
 \hline
\end{tabular}}
\caption{\label{tab. classification results waveform data}Clustering results for the waveform data.}
\end{table}In Figure \ref{fig: robust EM-MixReg stored-K pen-loglik waveform}, one can see the variation of the estimated number of clusters as well as the value of the objective function from one iteration to another for the three models. These results highlight the capability of the proposed algorithm to provide an accurate partition with a well adapted number of clusters.

\subsection{Experiments on real data}

In this section, we consider real data sets to evaluate the proposed approach. The considered data are curves issued from three different application domains: the phonemes data, the Yeast cell cycle data and the Topex/Poseidon satellite data. 

\subsubsection{Phonemes data}

\begin{figure}[H]
   \centering  
   \includegraphics[width=6cm]{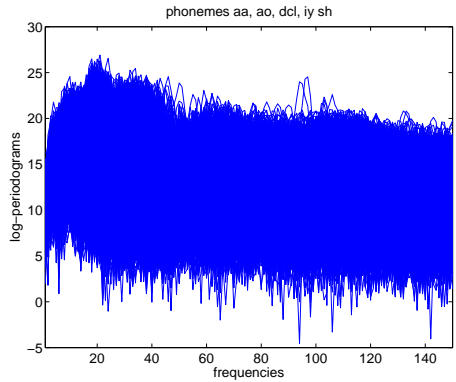}
   \caption{\label{fig. phonemes data}Phonemes data "ao", "aa", "yi", "dcl", "sh".}
\end{figure}

\begin{figure}[H]
   \centering 
   \begin{tabular}{cc}
 \includegraphics[width=6cm]{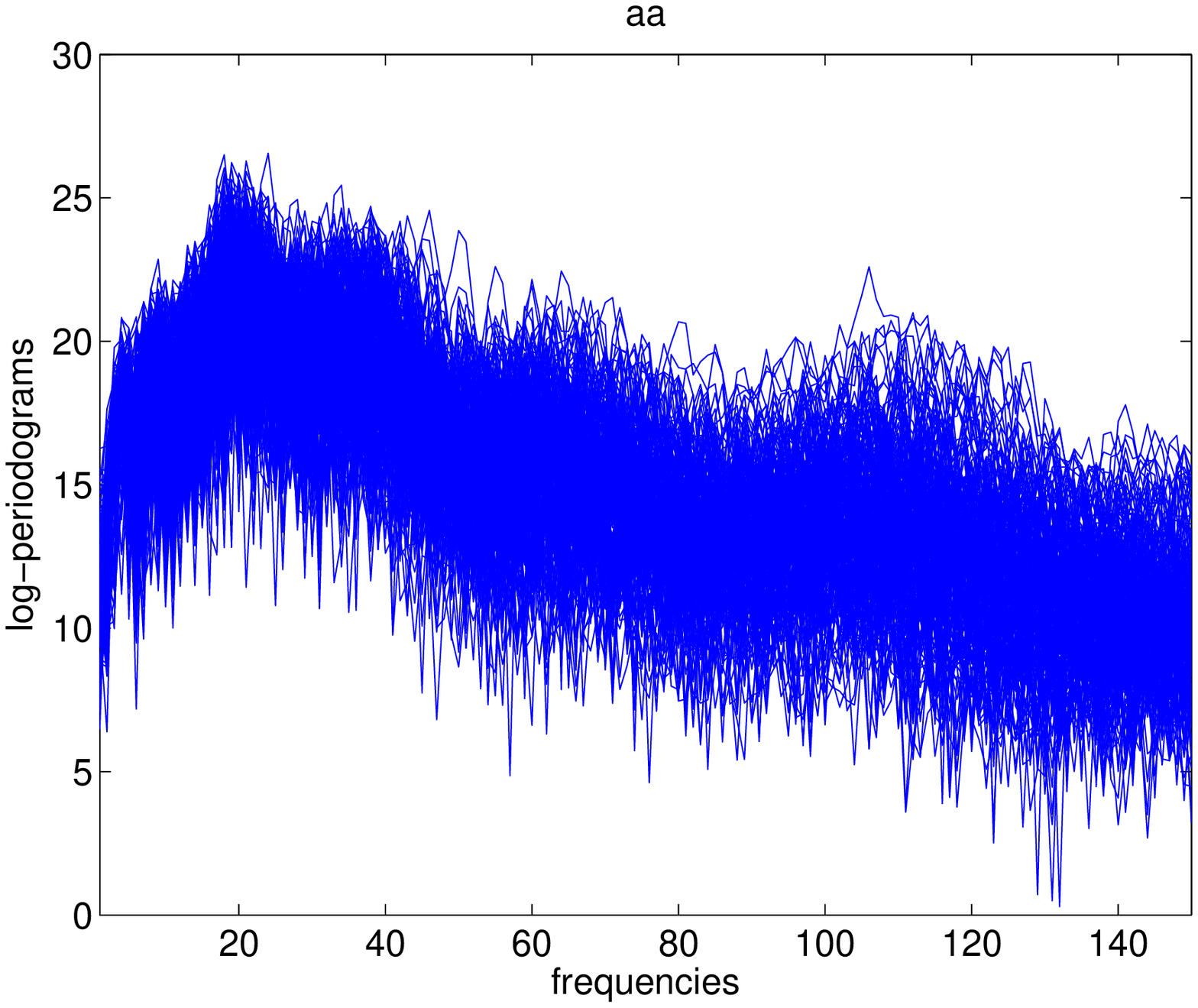}& \includegraphics[width=6cm]{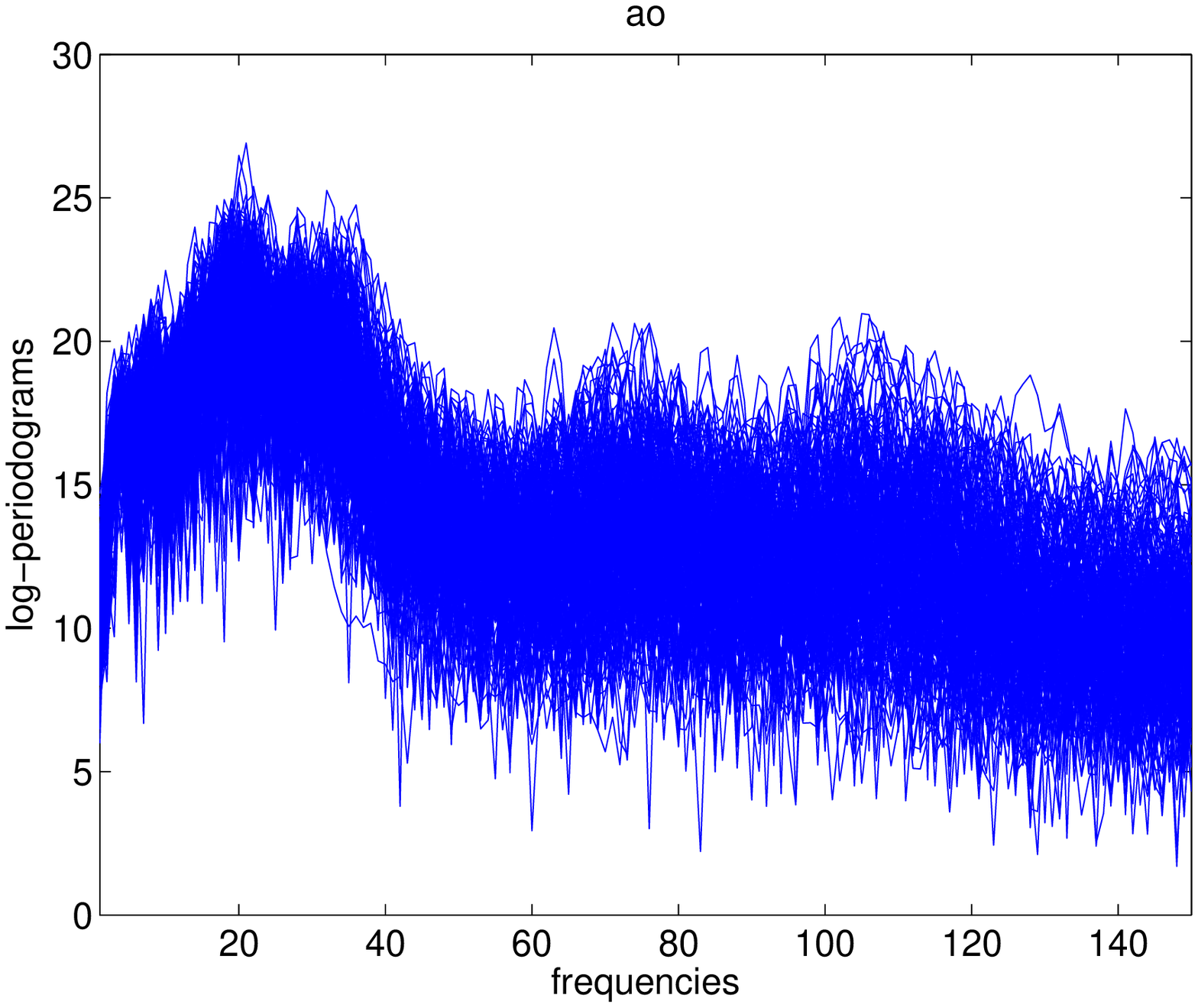} \\
   \includegraphics[width=6cm]{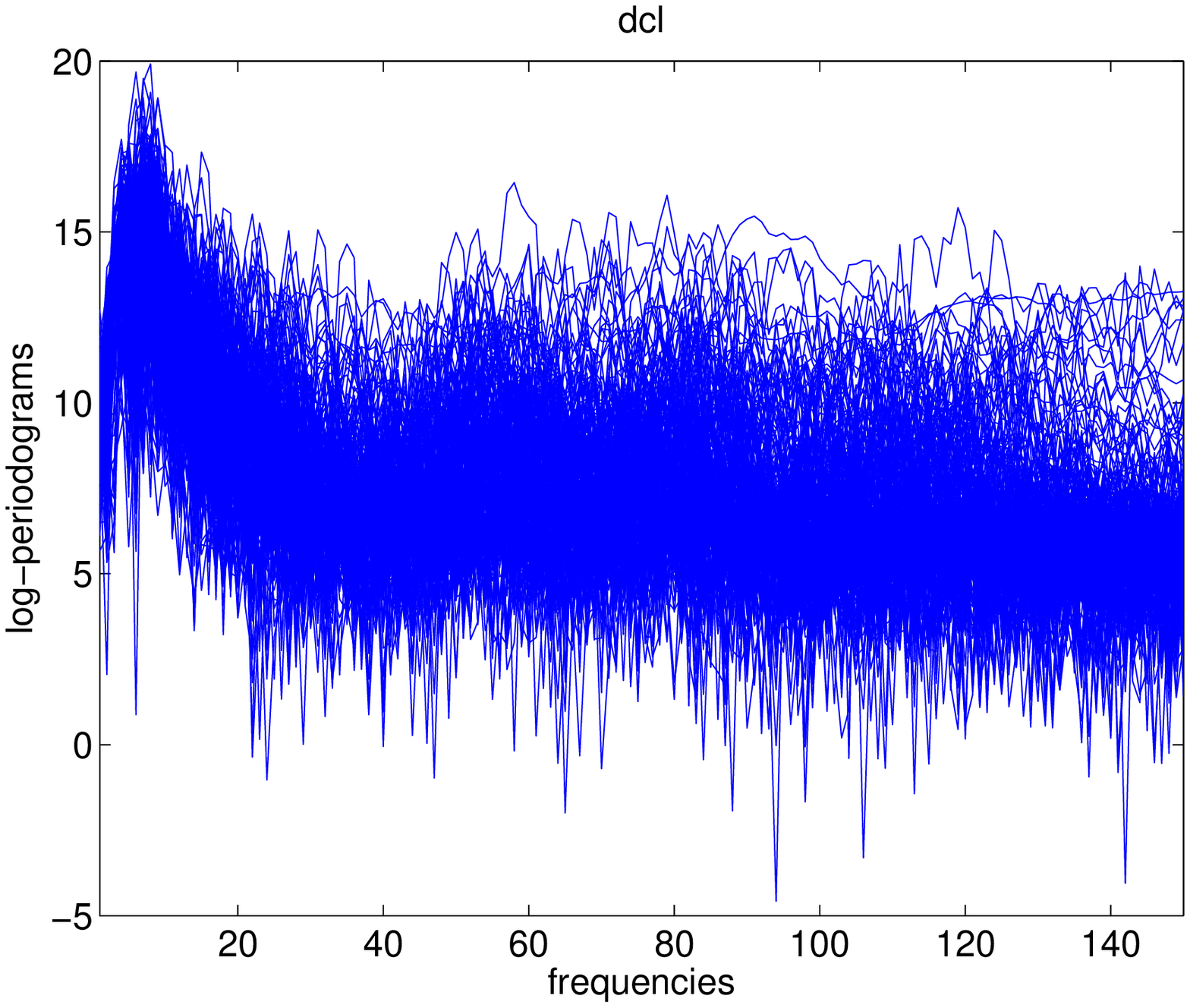} & \includegraphics[width=6cm]{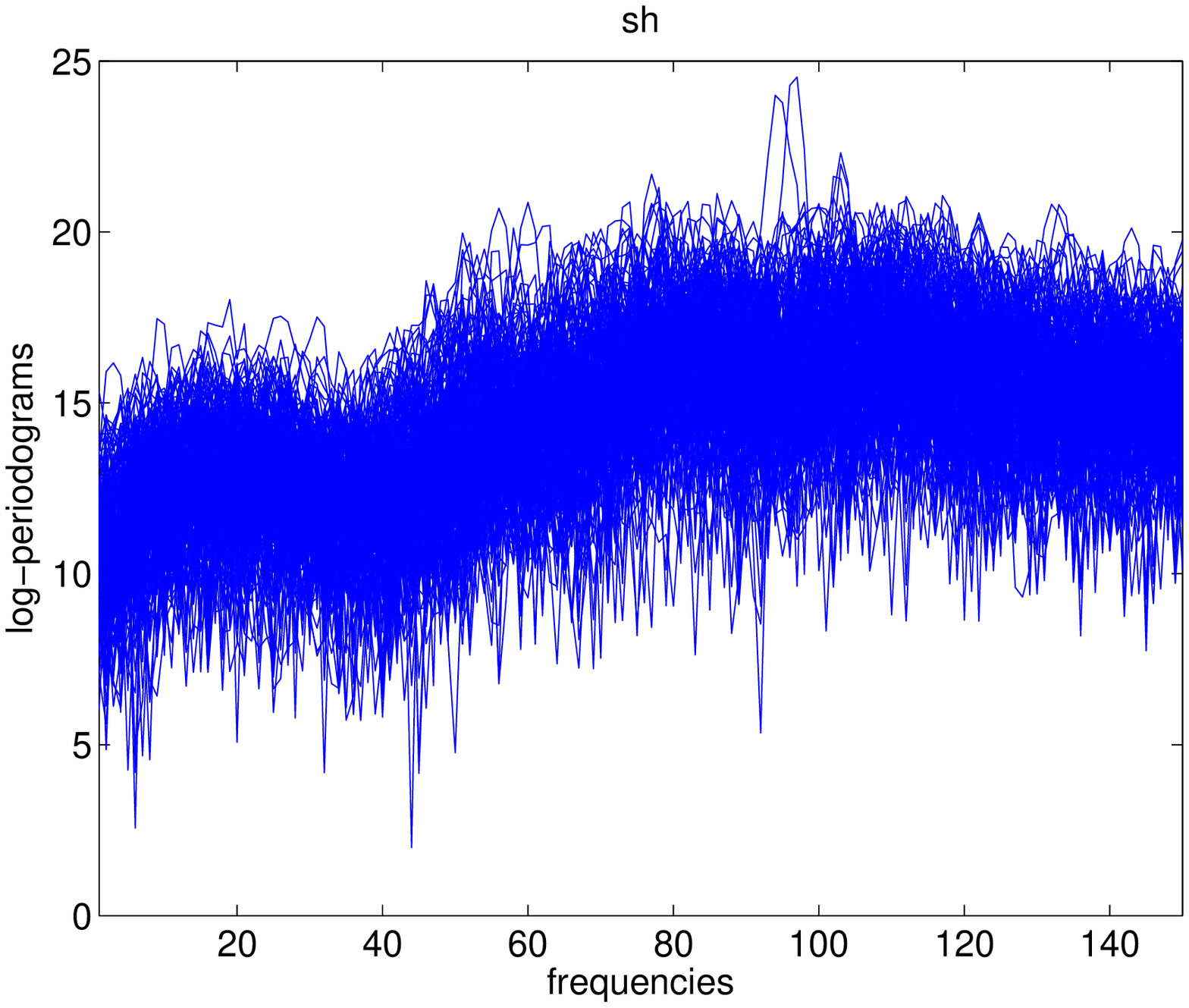} \\ \includegraphics[width=6cm]{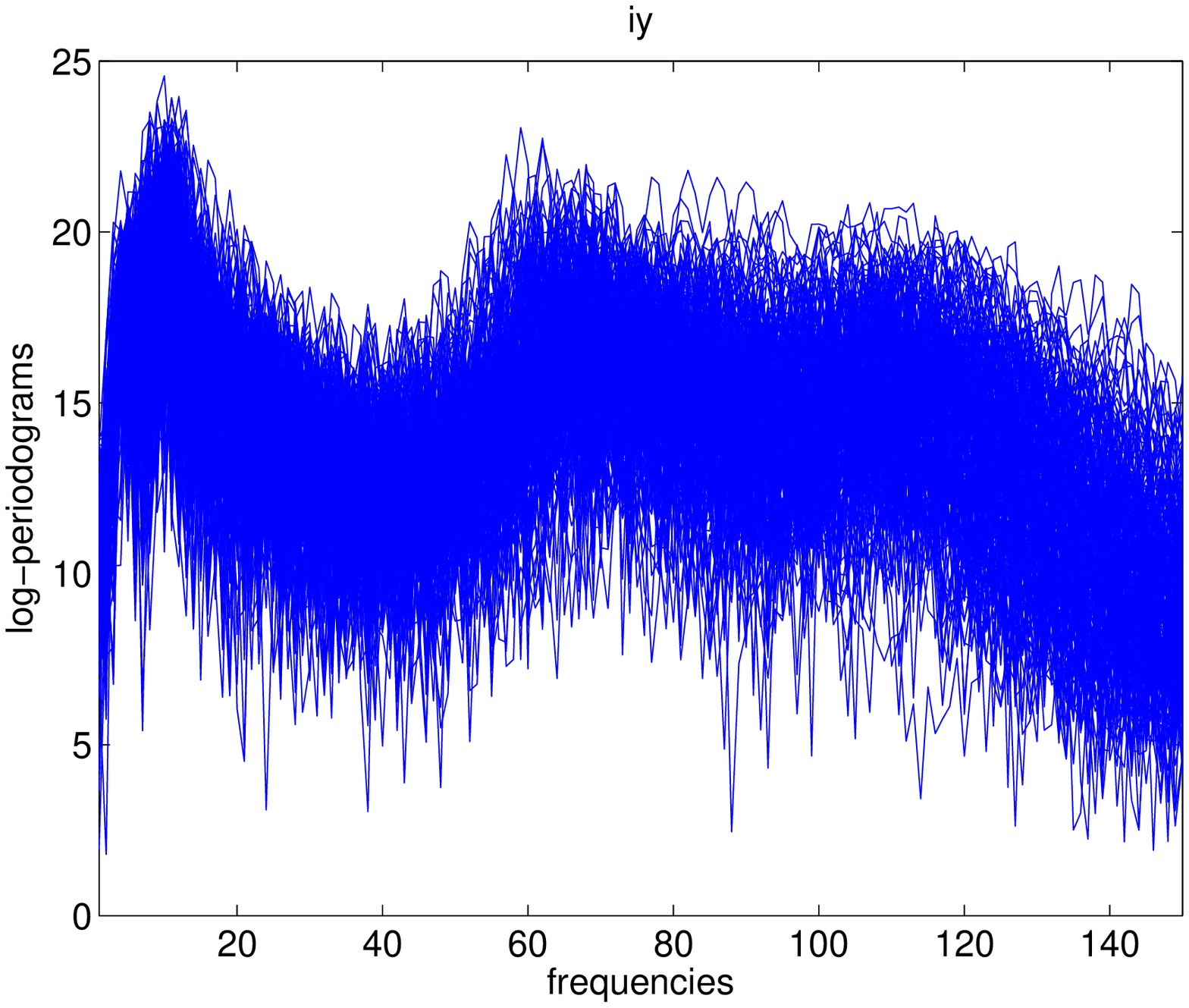}
   \end{tabular}
   \caption{\label{fig. phonemes data and actual clusters}Phonemes data classes: "ao", "aa", "yi", "dcl", "sh".}
\end{figure}

In this section, we use the phonemes dataset used in \cite{Ferraty2003}\footnote{Data from \url{http://www.math.univ-toulouse.fr/staph/npfda/}} which is a part of the original one available at \url{http://www-stat.stanford.edu/ElemStatLearn} and was described and used namely in \citep{Hastie95penalizeddiscriminant}. 
The application context related to this dataset is a phoneme classification problem. The phonemes dataset correspond to log-periodograms  $y$ constructed from recordings available at different equispaced frequencies $x$ for different phonemes. It contains five classes corresponding to the following five phonemes: "sh" as in "she", "dcl" as in "dark", "iy" as in "she", "aa" as in "dark" and "ao" as in "water". For each phoneme we have 400 log-periodograms at a 16-kHz sampling rate. We only retain  the first 150 frequencies from each subject as in \citep{Ferraty2003}. This dataset has been considered in a phoneme discrimination problem as in \cite{Hastie95penalizeddiscriminant, Ferraty2003} where the aim is to predict the phoneme class for a new log-periodogram.
Here we  reformulate the problem into a clustering problem where the aim is to automatically group the phonemes data  into classes. We assume that the cluster labels are missing. We also assume that the number of clusters is unknown. Thus, the proposed algorithm will be assessed in terms of estimating both the actual partition and the optimal  number of clusters from the data.
%
%
Fig. \ref{fig. phonemes data} 
shows the used 1000 log-periodograms (200 per cluster) and Fig. \ref{fig. phonemes data and actual clusters} shows the curves of the actual five phoneme classes, class by class.

\begin{figure}[H]
   \centering 
   \begin{tabular}{cc}
   \includegraphics[width=6cm]{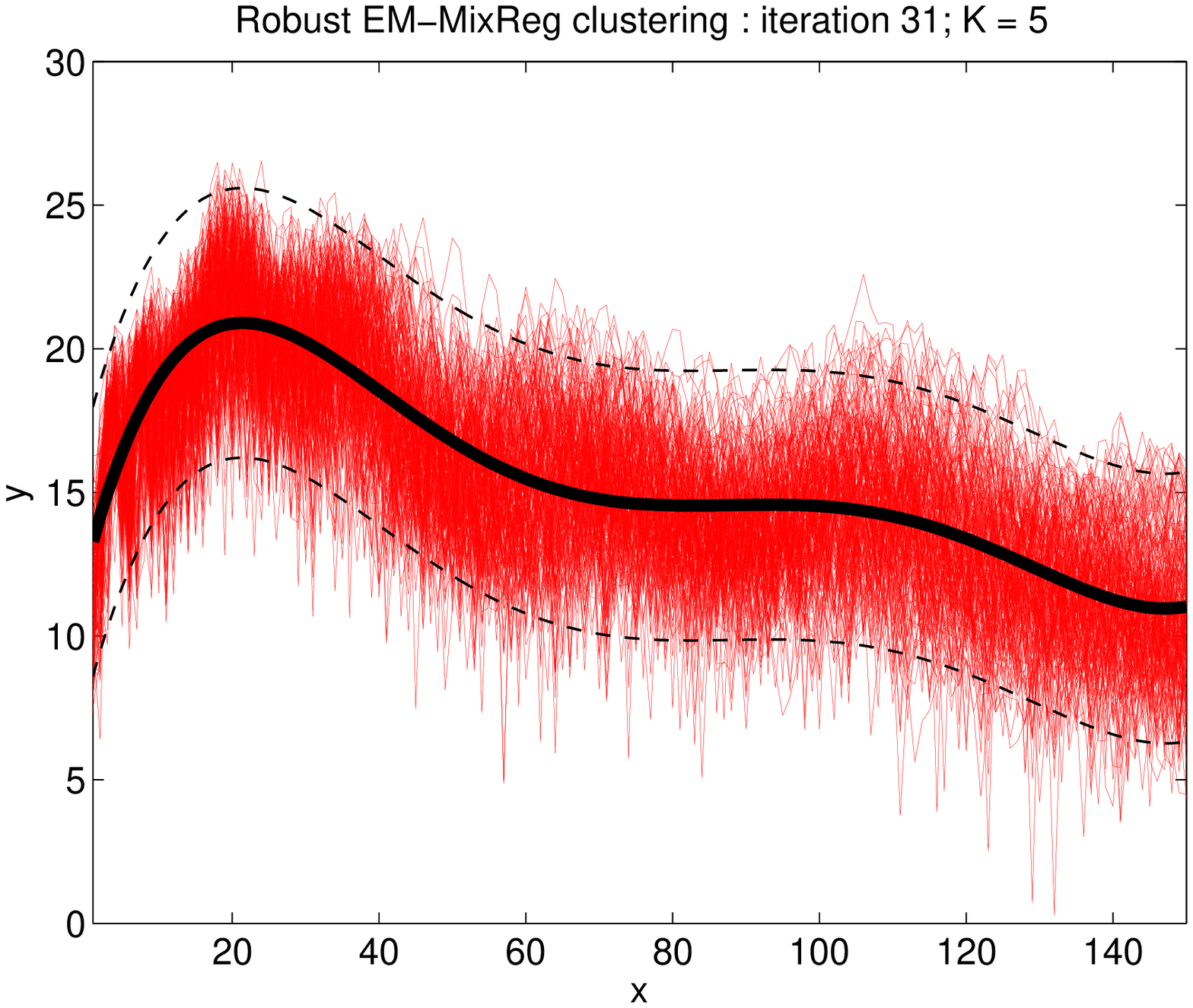}&
    \includegraphics[width=6cm]{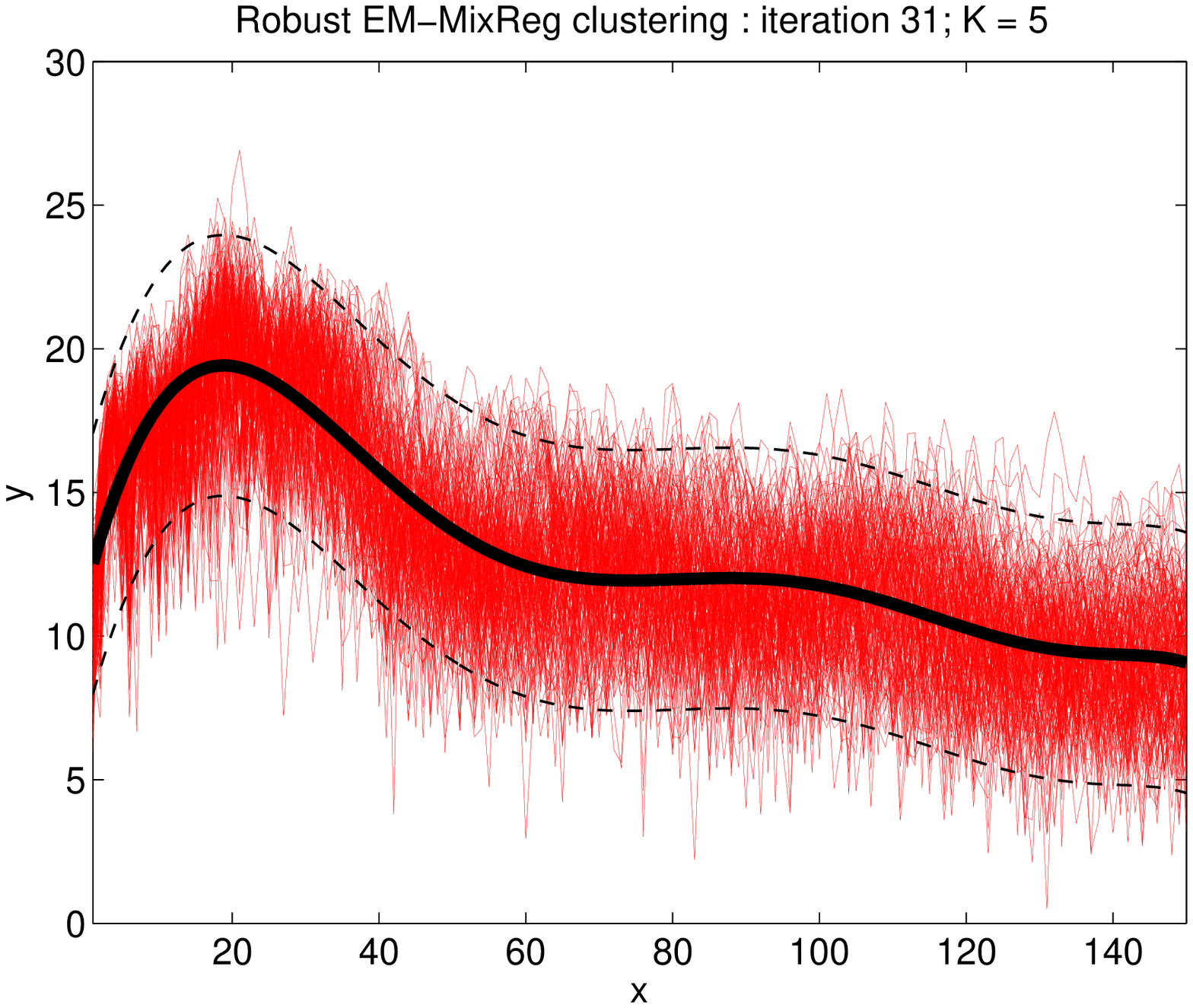}\\
    \includegraphics[width=6cm]{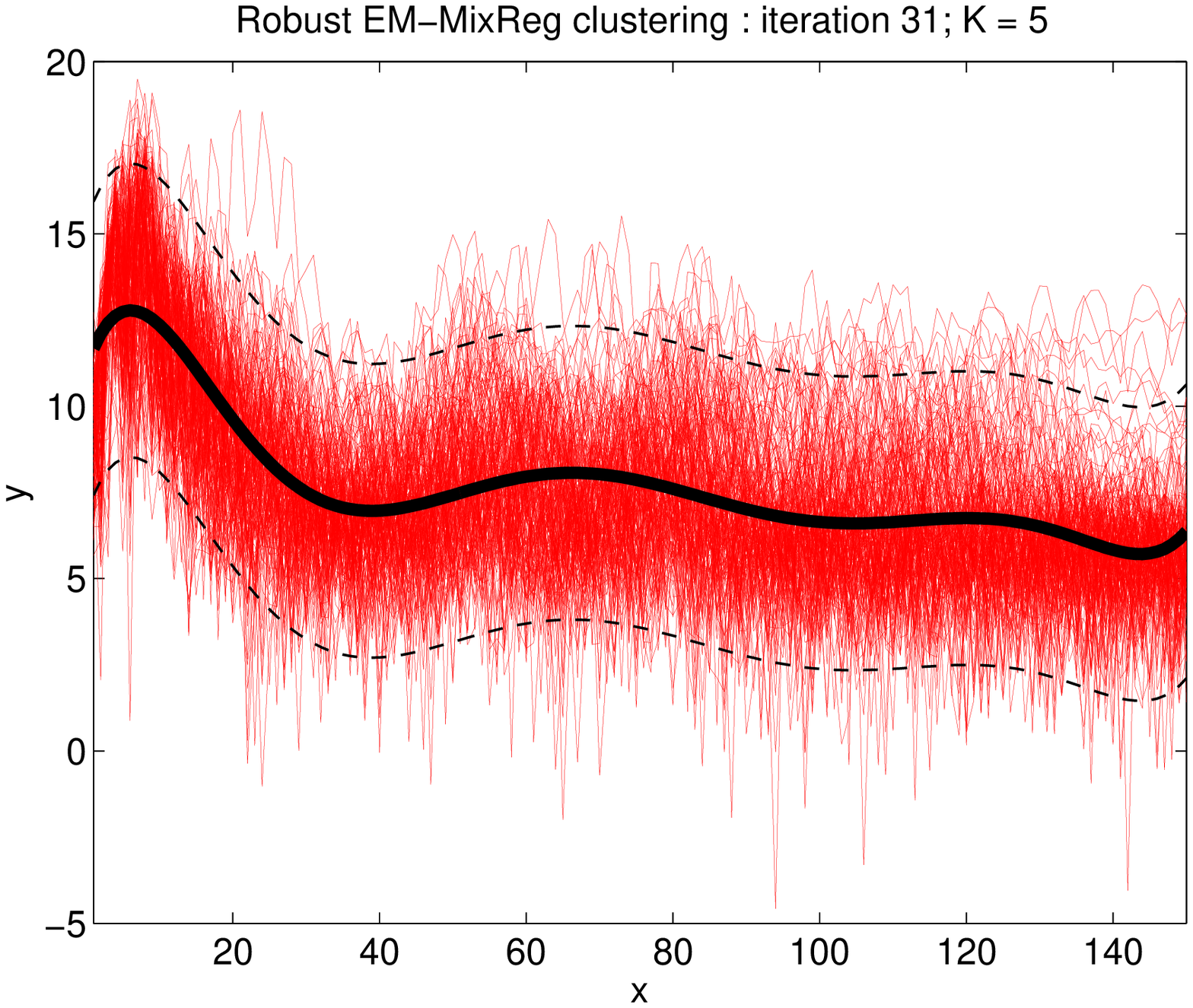}&
    \includegraphics[width=6cm]{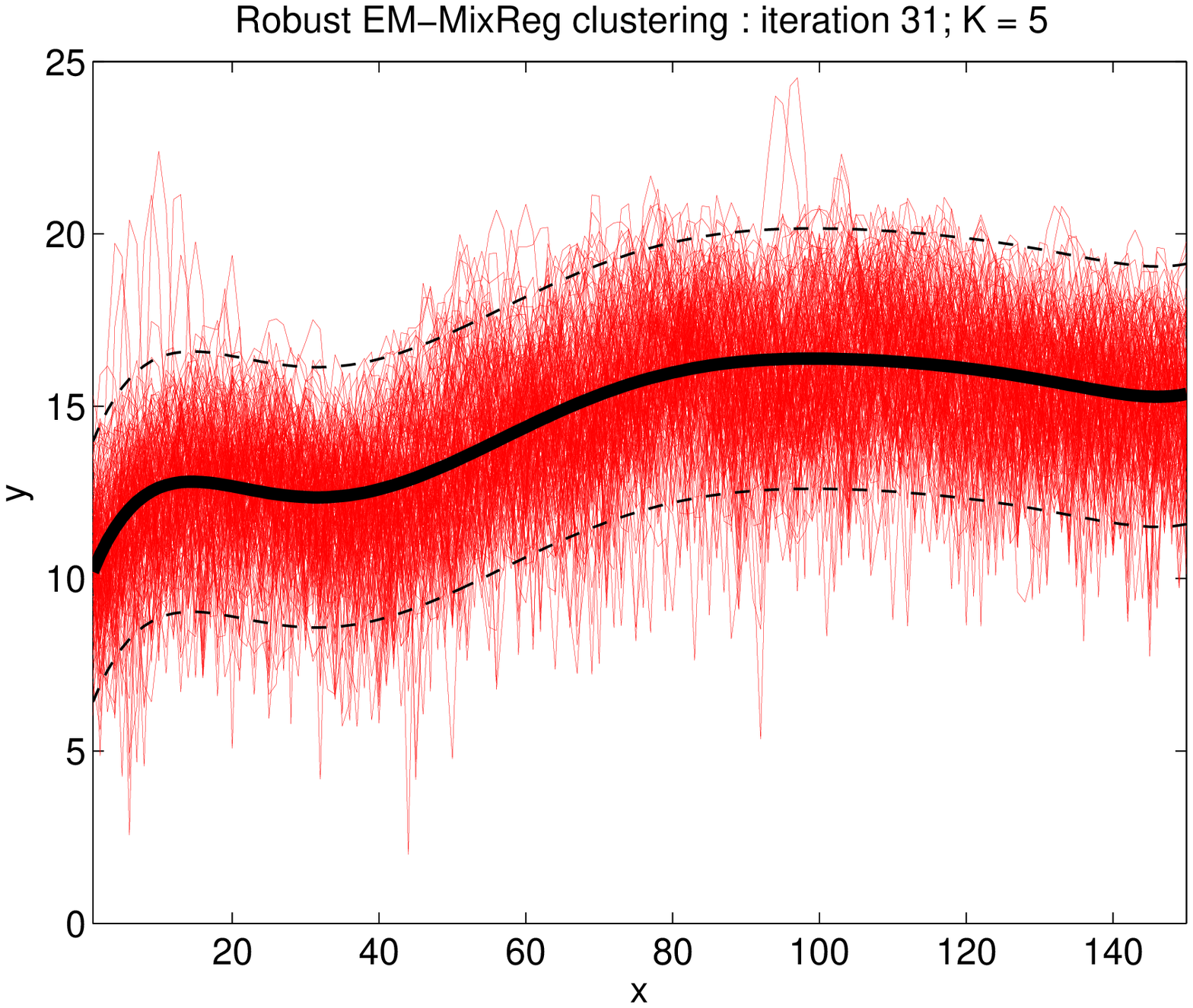}\\
    \includegraphics[width=6cm]{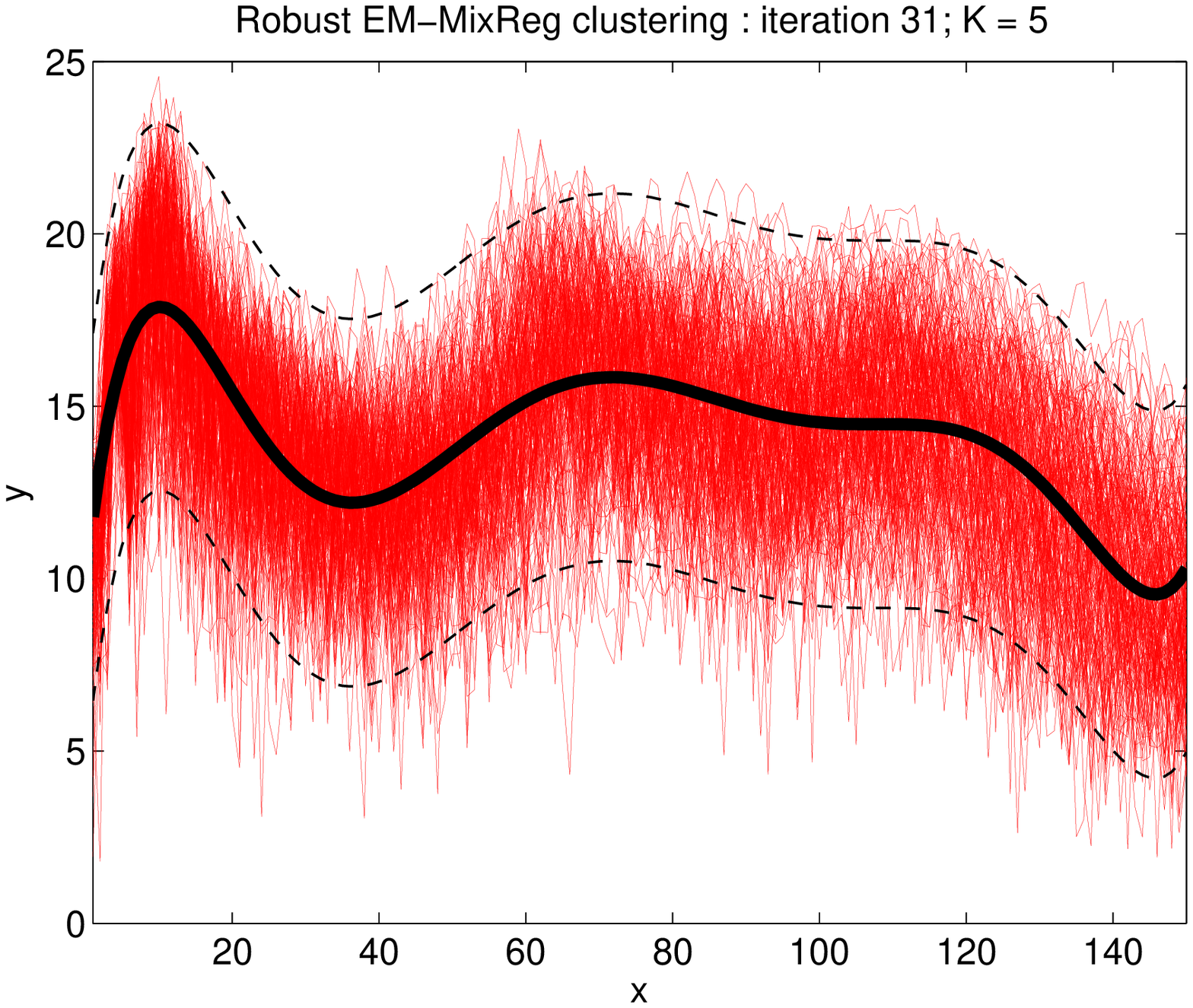} &
   \end{tabular}
   \caption{\label{fig. robust EM-PRM phonemes results}Clustering results obtained by the proposed robust EM algorithm and the PRM model (polynomial degree $p=7$) for the phonemes data.}
\end{figure}

\begin{figure}[H]
   \centering 
   \begin{tabular}{cc}
   \includegraphics[width=6cm]{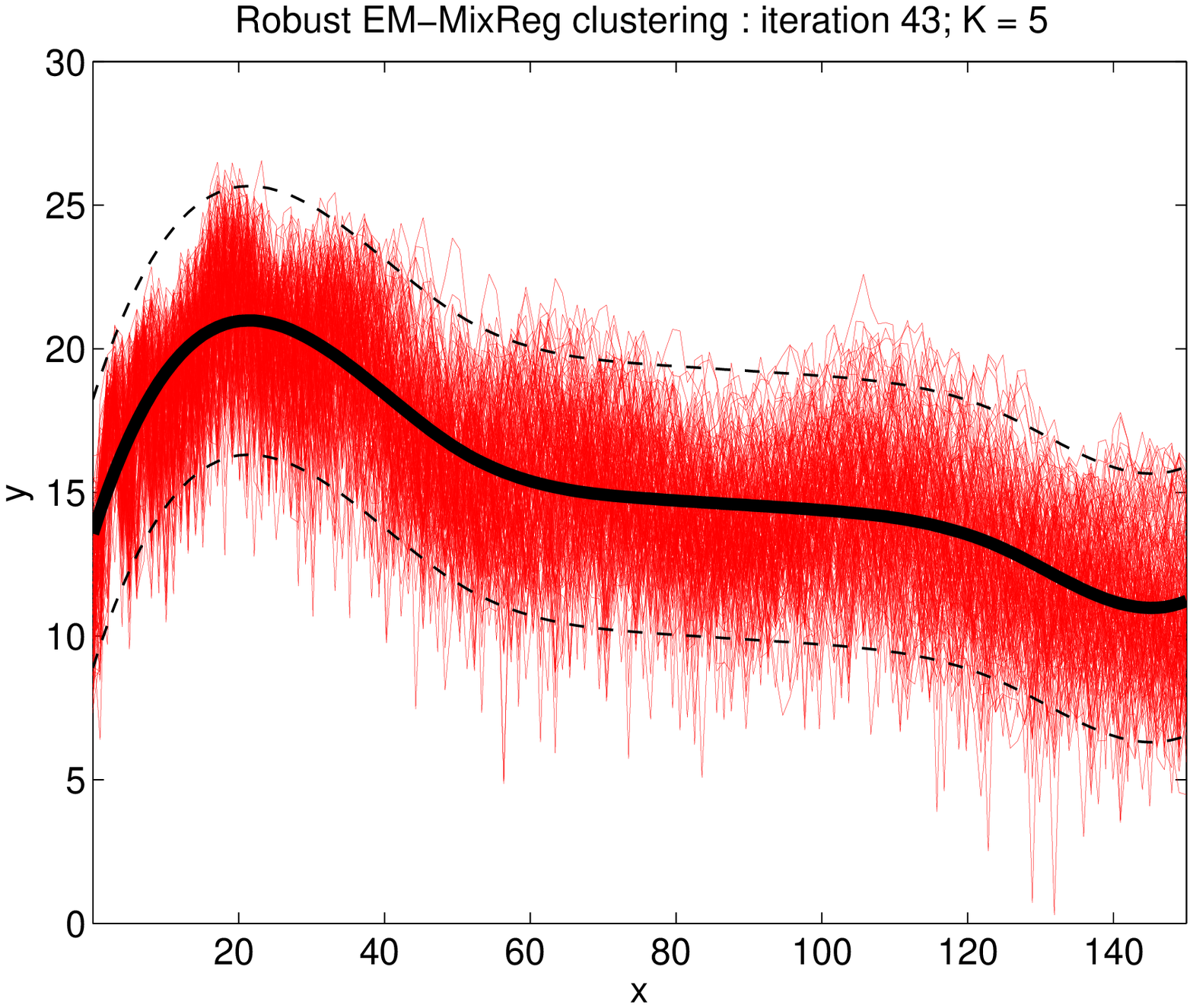}&  
   \includegraphics[width=6cm]{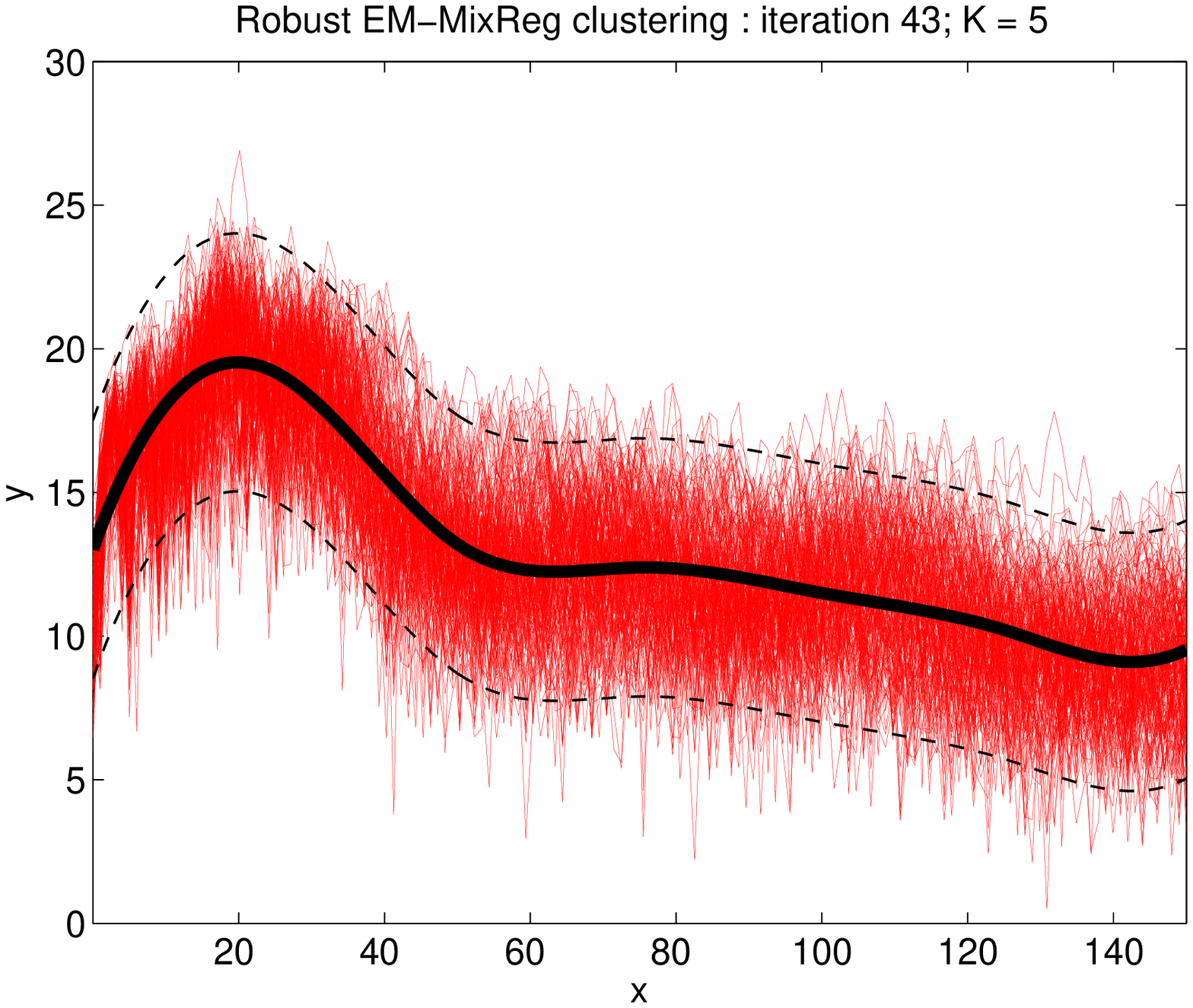}\\
   \includegraphics[width=6cm]{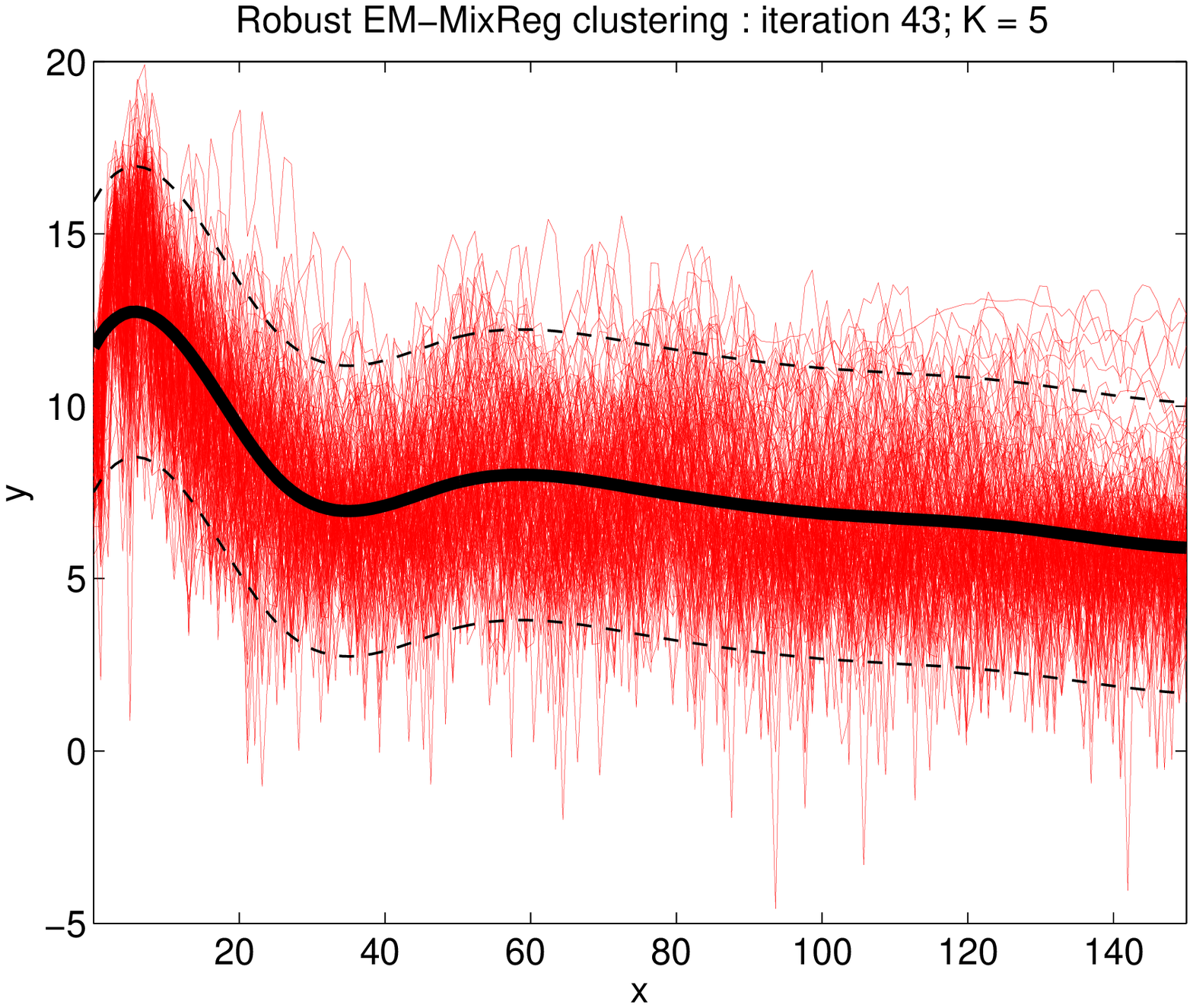}&
   \includegraphics[width=6cm]{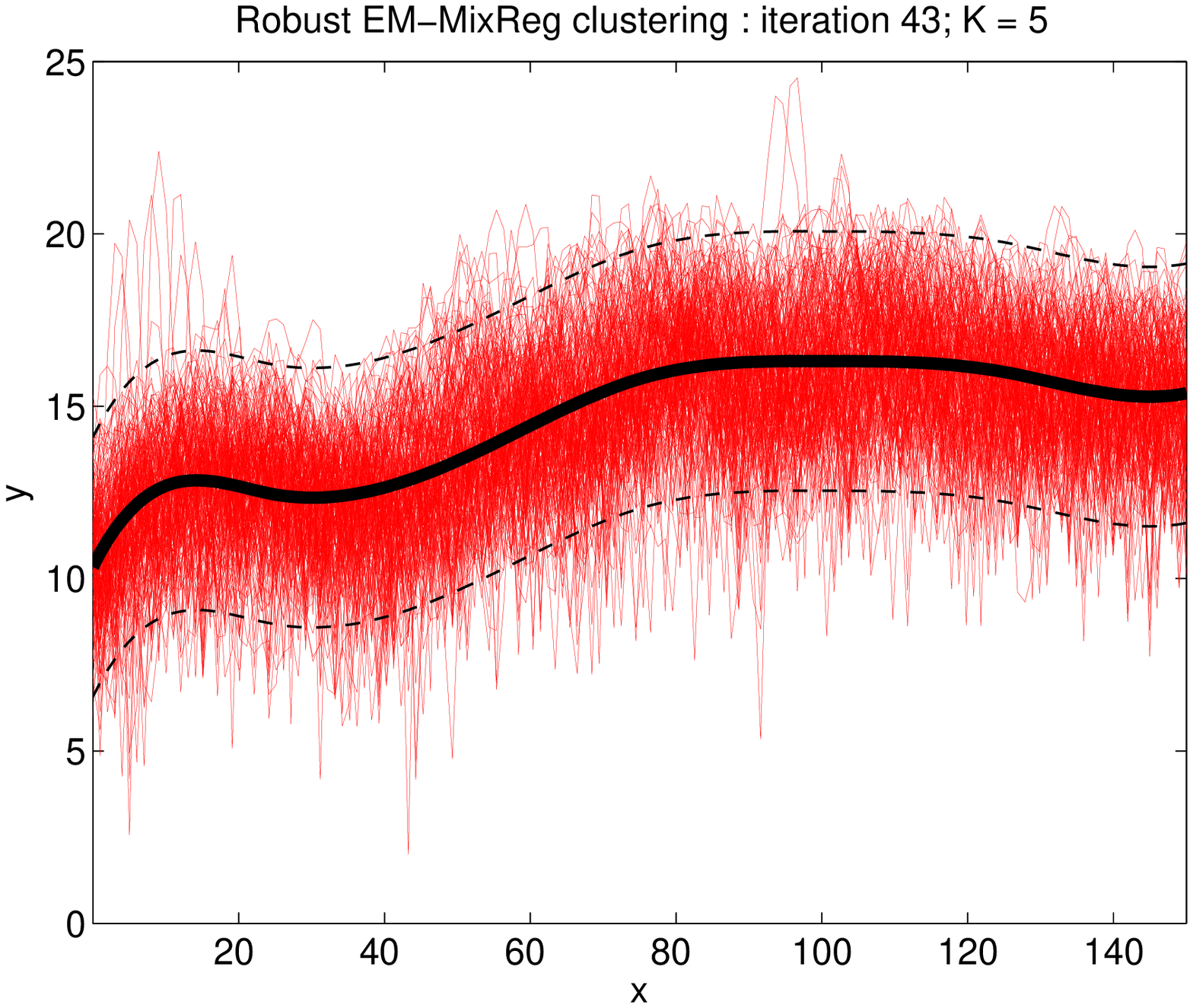}\\
   \includegraphics[width=6cm]{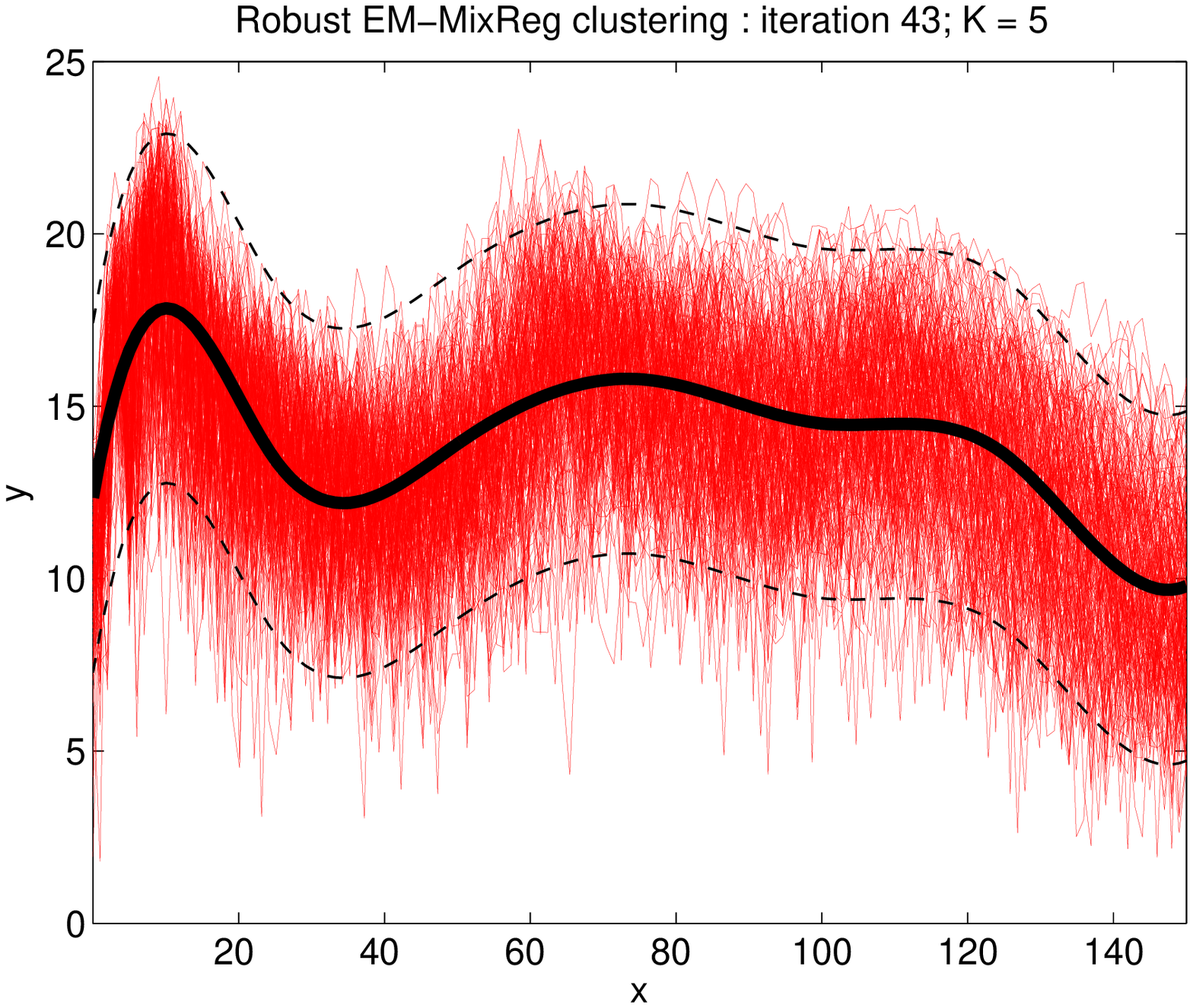} &
   \end{tabular}
      \caption{\label{fig. robust EM-SRM phonemes results}Clustering results obtained by the proposed robust EM algorithm and the SRM model with a cubic B-spline of seven knots for the phonemes data.}
\end{figure}

\begin{figure}[H]
   \centering 
   \begin{tabular}{cc}
   \includegraphics[width=6cm]{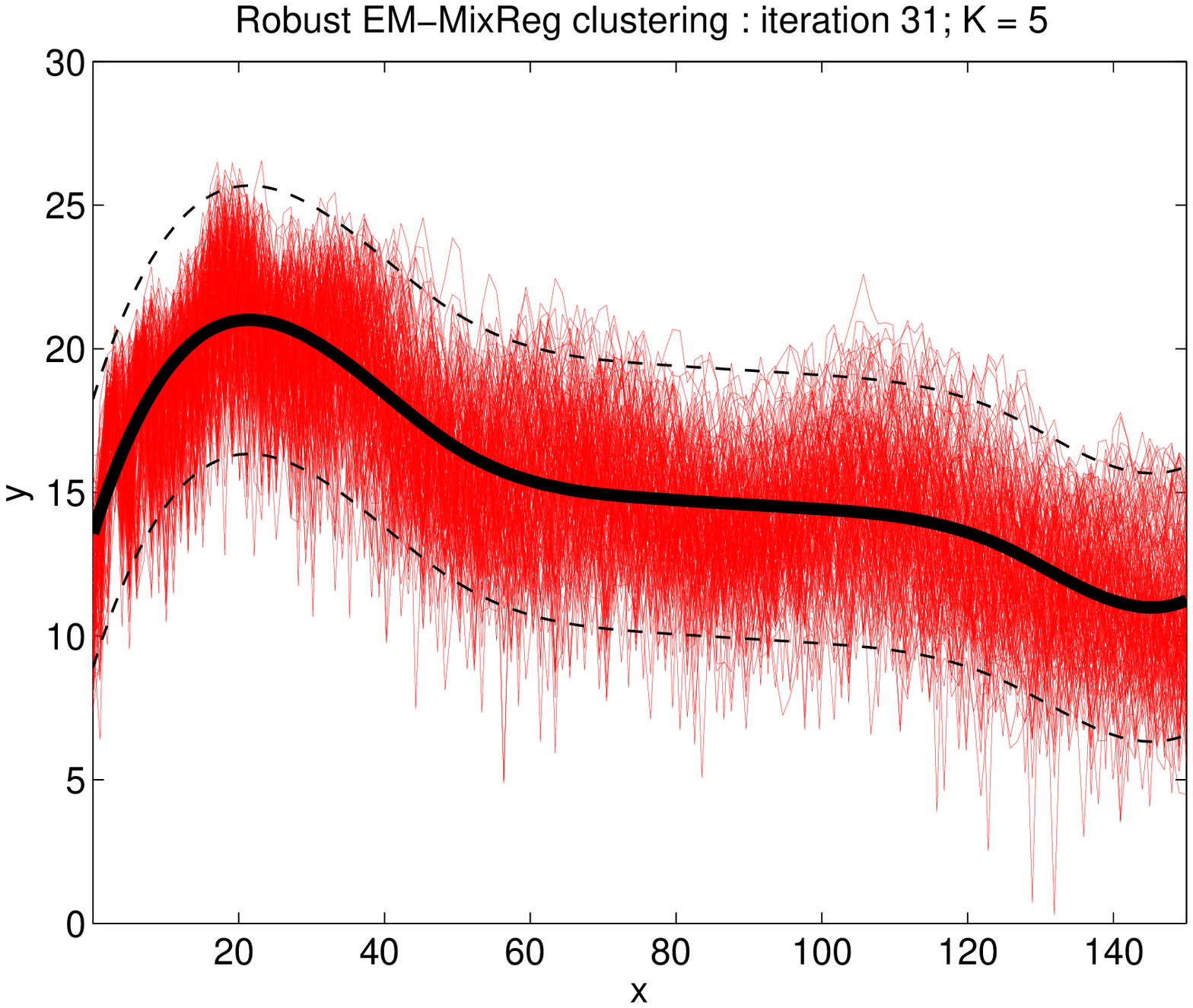}&
   \includegraphics[width=6cm]{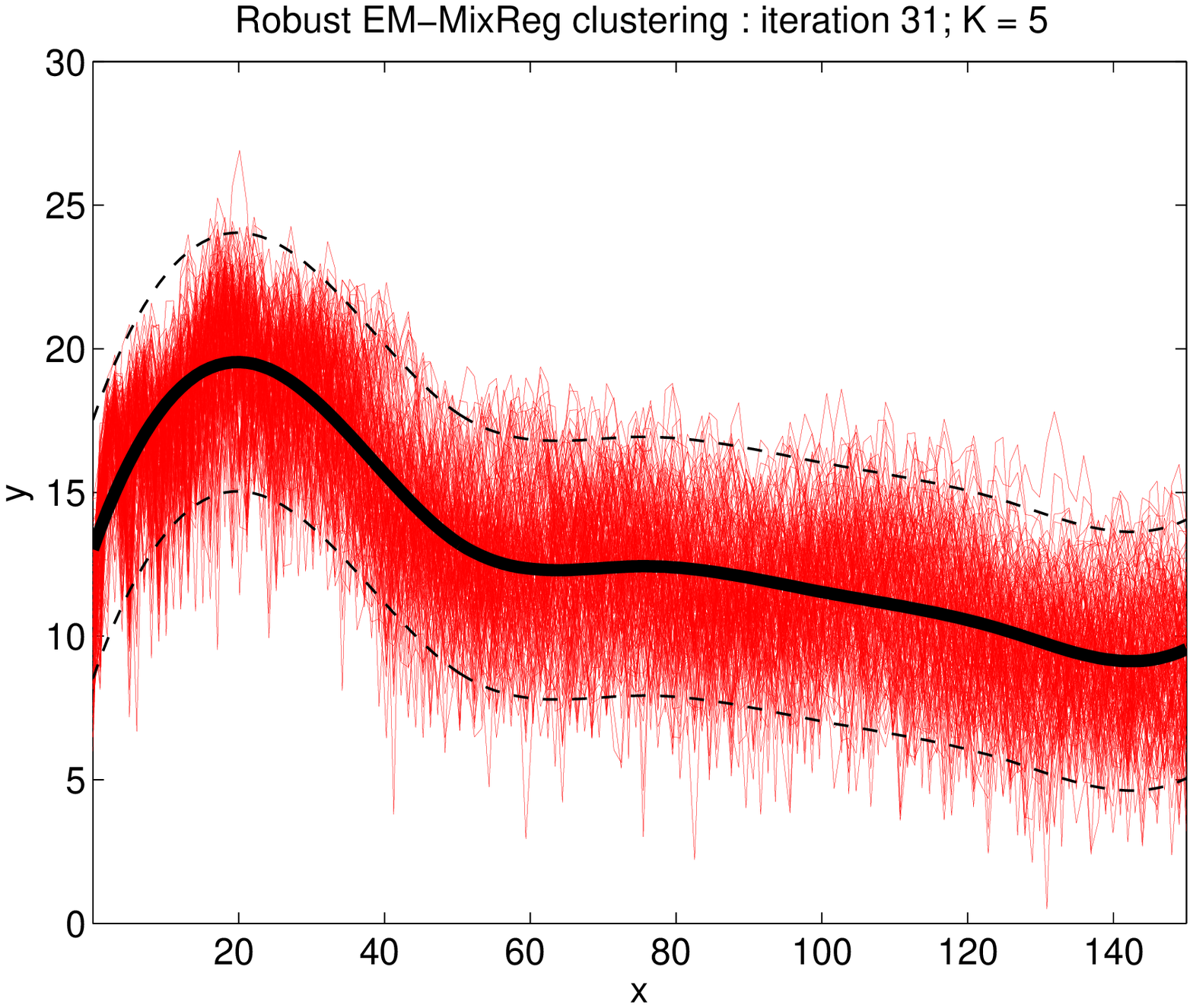}\\
   \includegraphics[width=6cm]{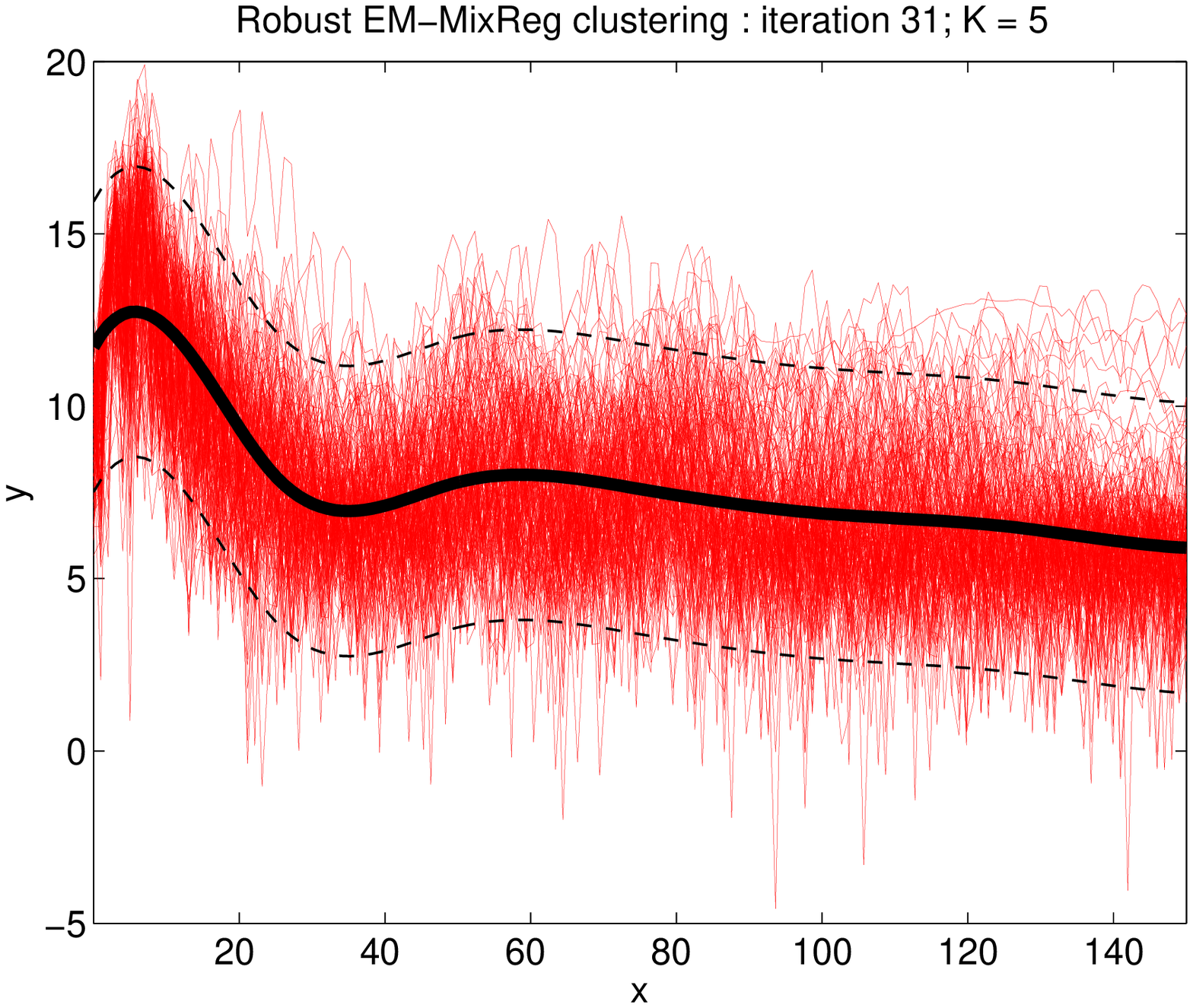}&
   \includegraphics[width=6cm]{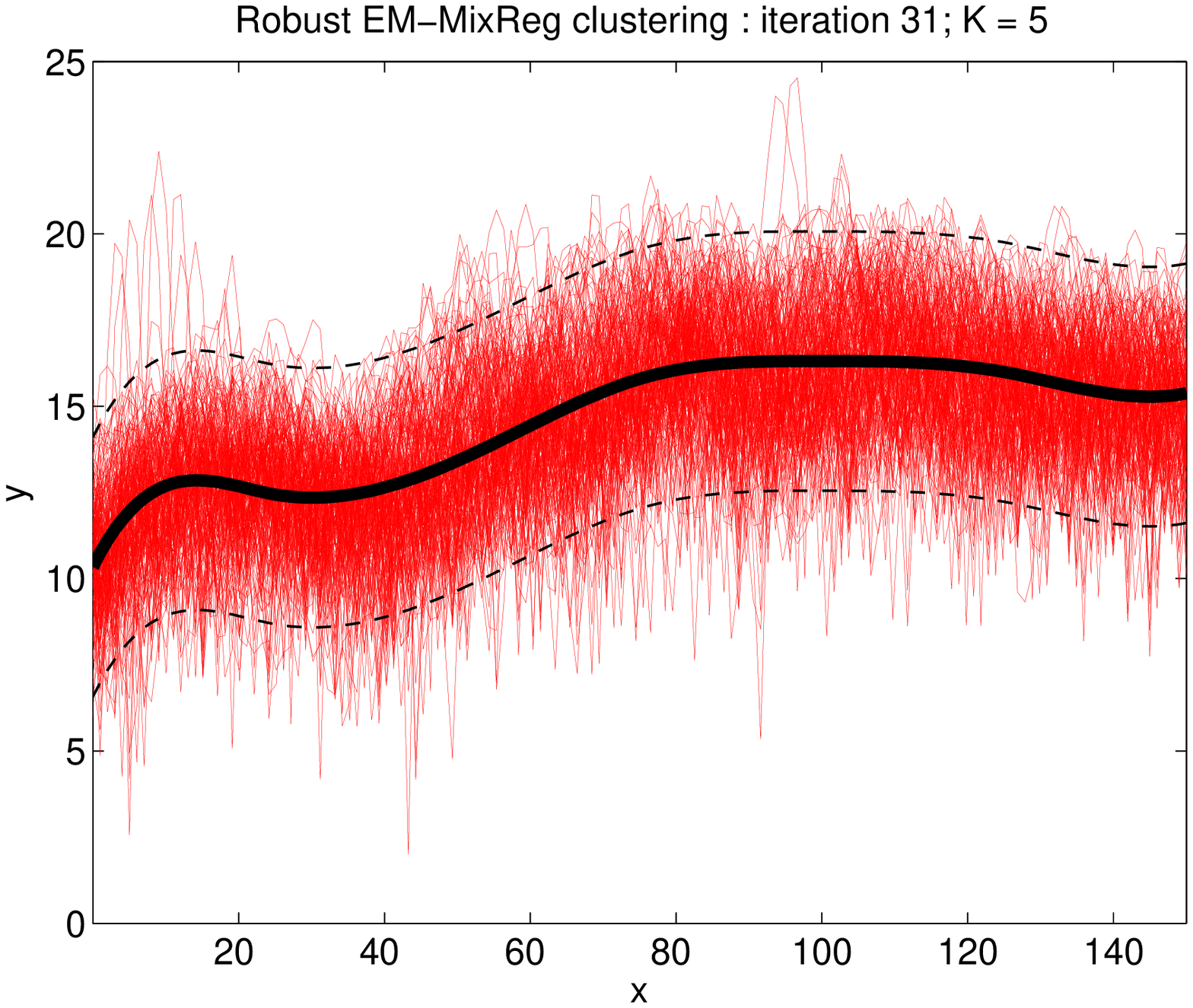}\\
   \includegraphics[width=6cm]{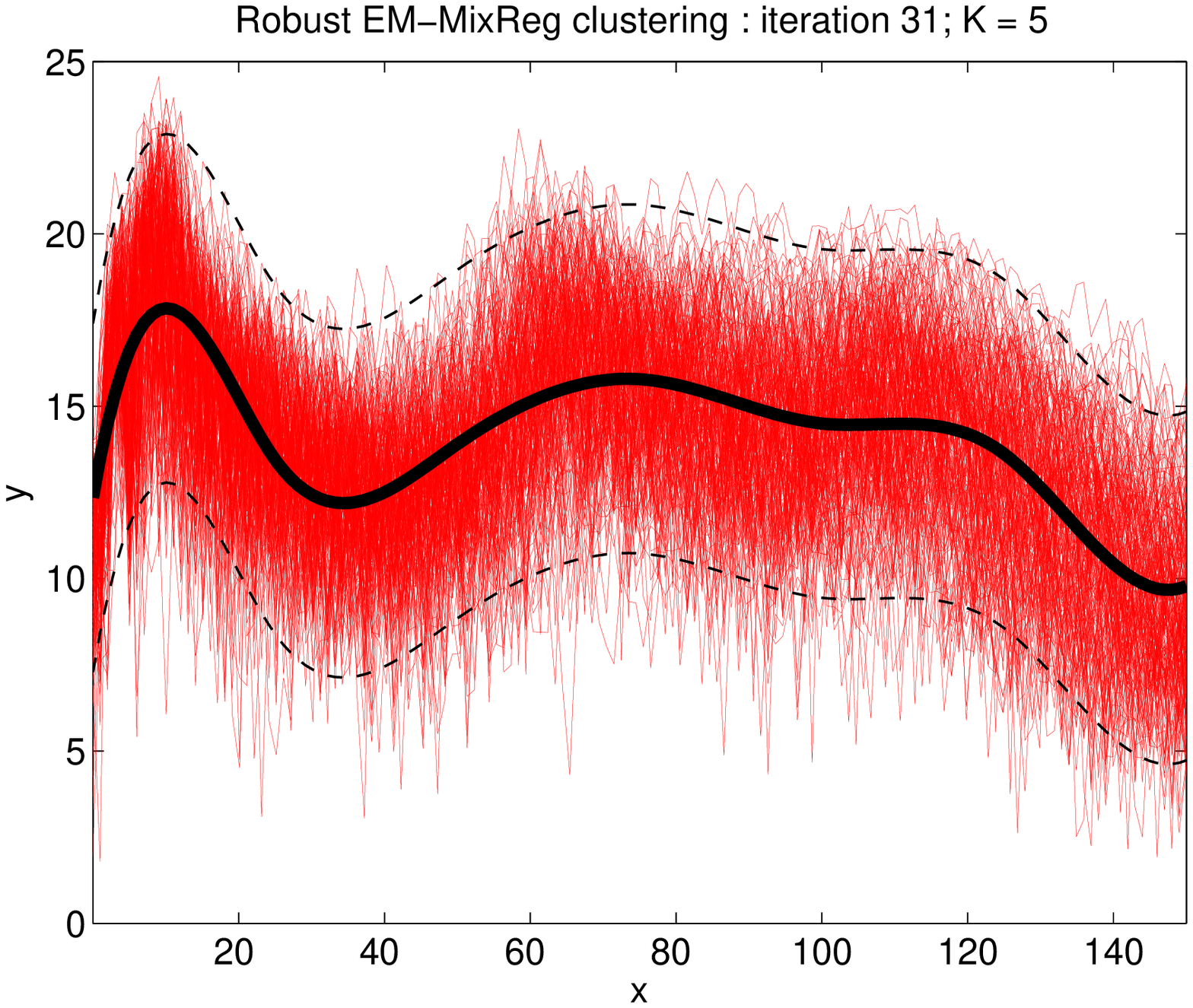} 
   \end{tabular}
   \caption{\label{fig. robust EM-bSRM phonemes results}Clustering results obtained by the proposed robust EM algorithm and the bSRM model with a cubic B-spline of seven knots for the phonemes data.}
\end{figure}

\begin{figure}[H]
   \centering  
\includegraphics[width=5cm]{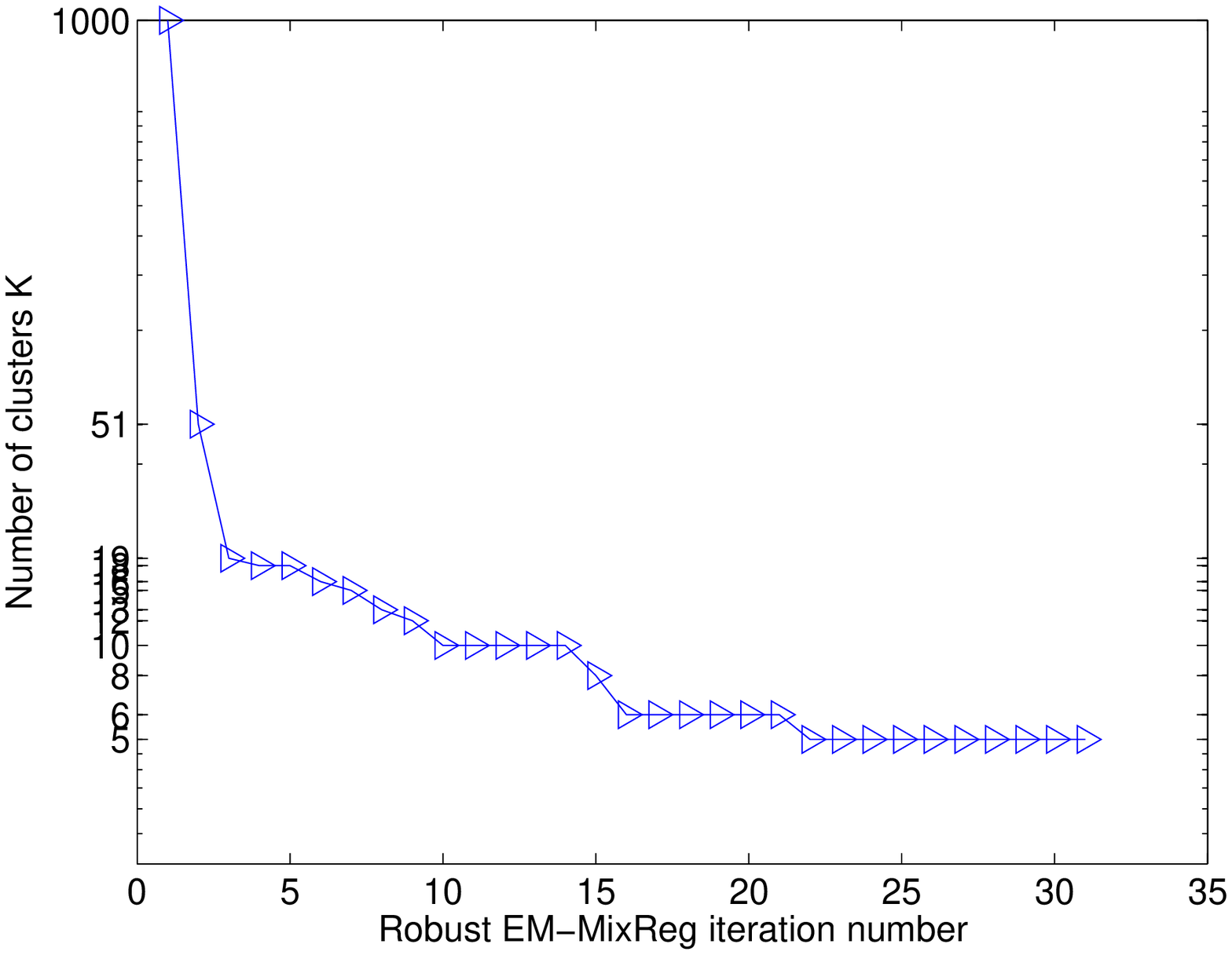}
\includegraphics[width=5cm]{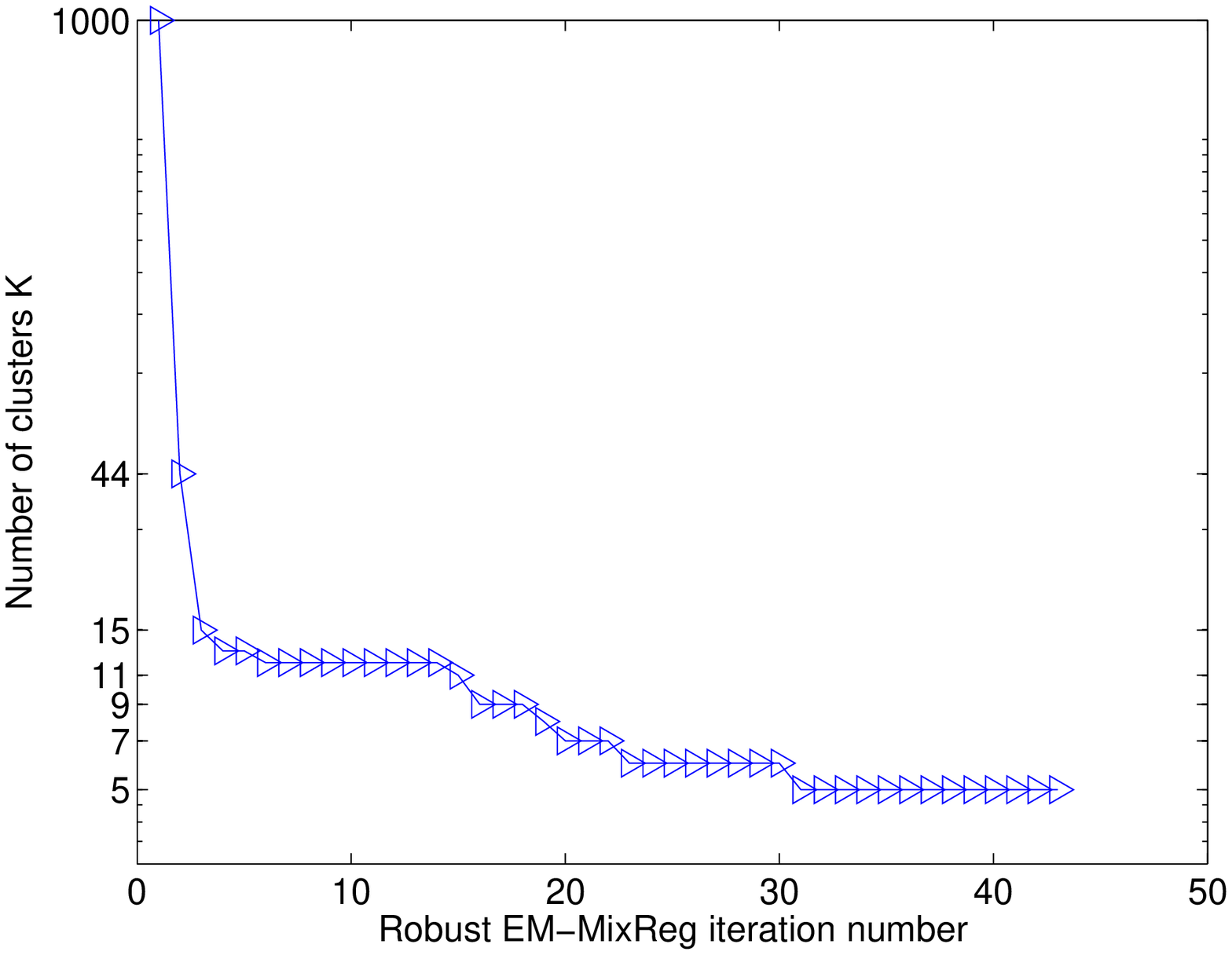}
\includegraphics[width=5cm]{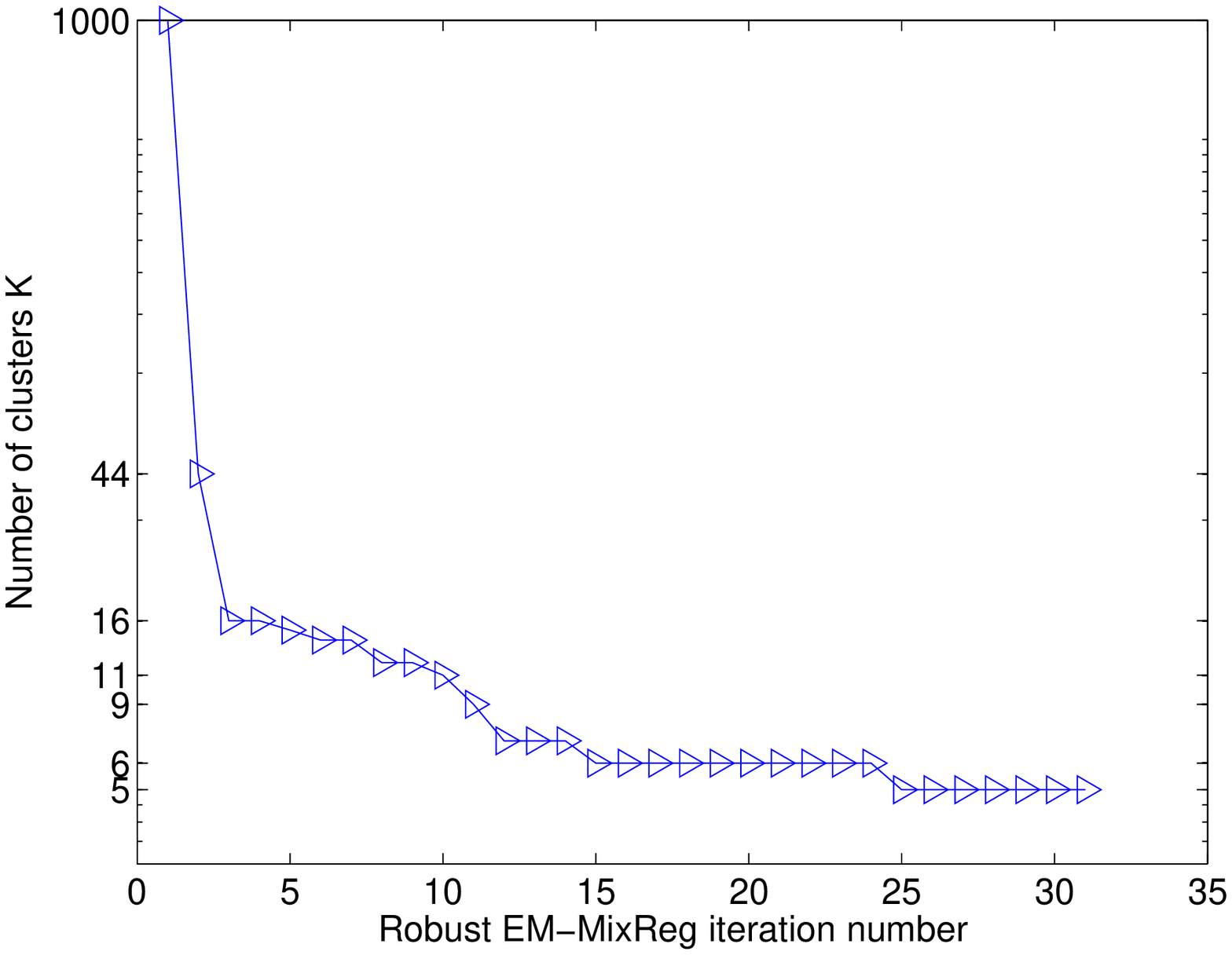}\\
\includegraphics[width=5cm]{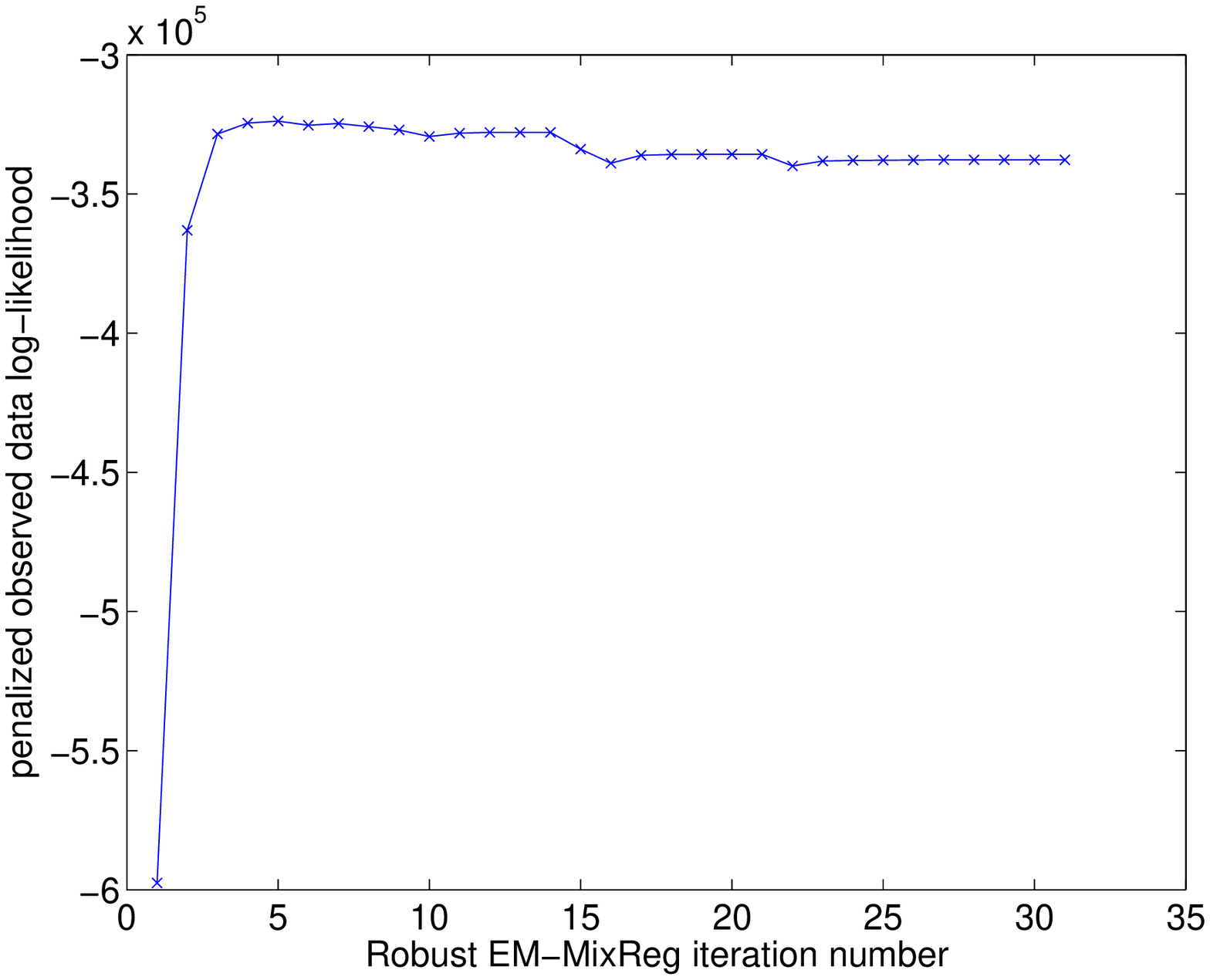}
\includegraphics[width=5cm]{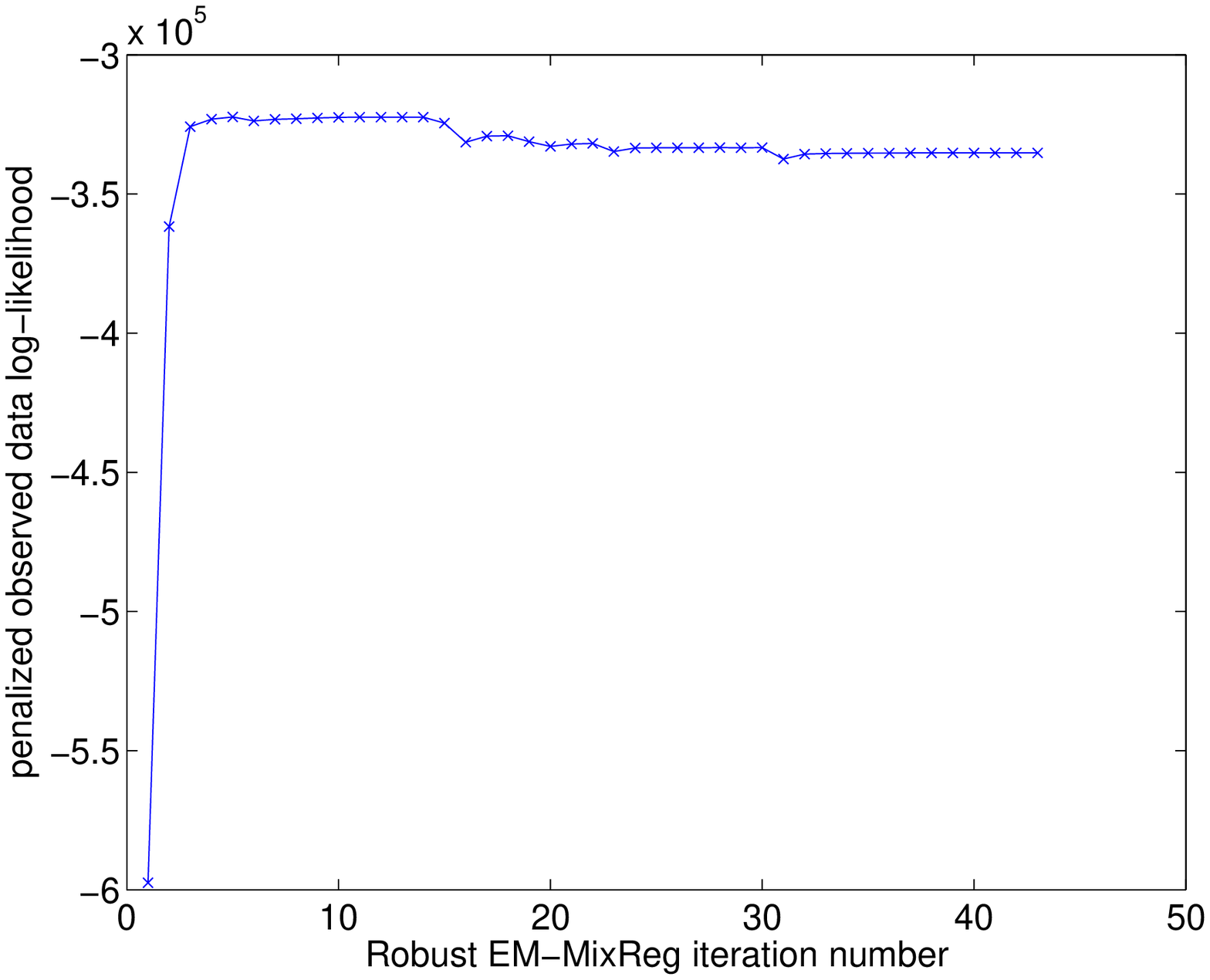}
\includegraphics[width=5cm]{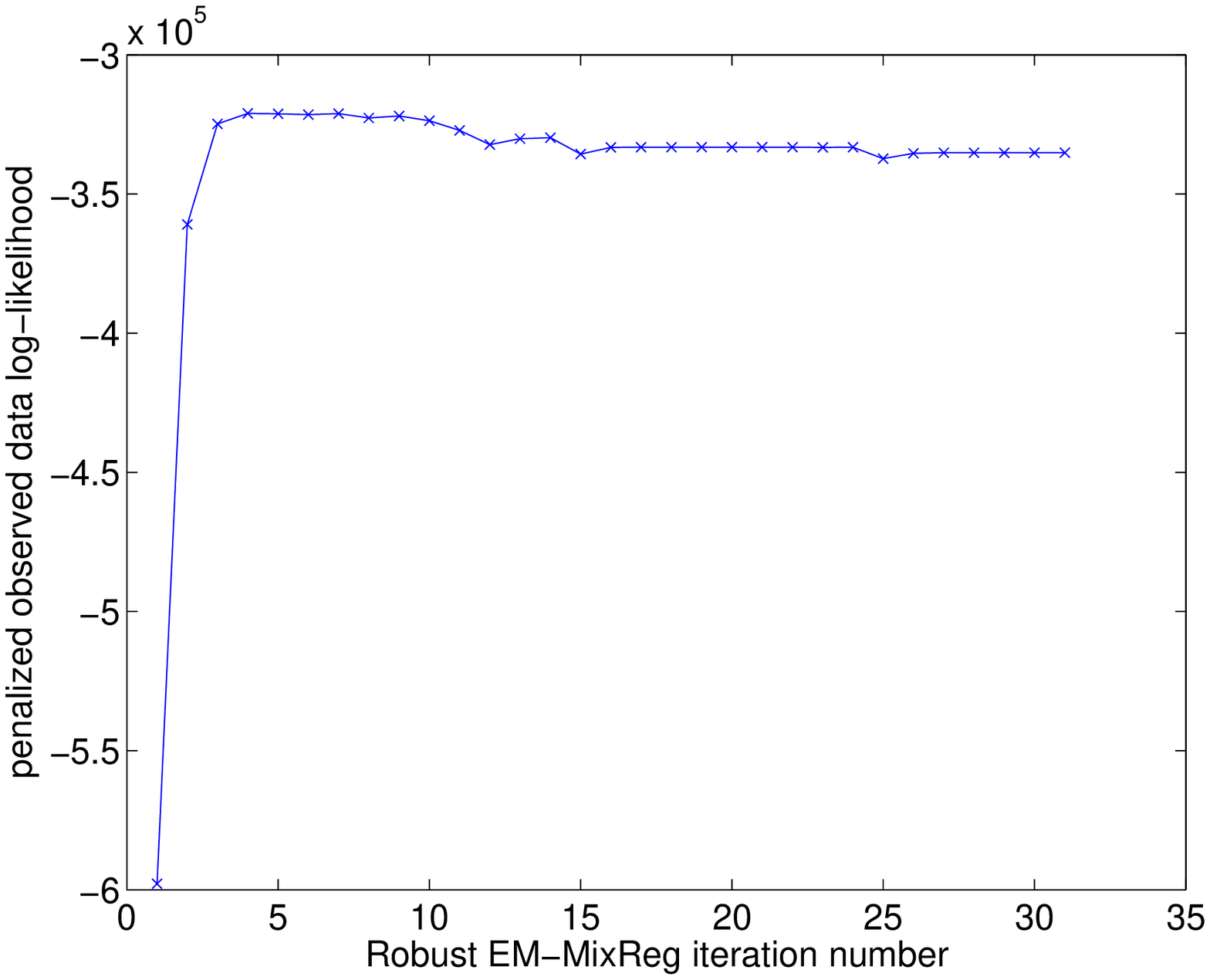}
   \caption{\label{fig: robust EM-MixReg stored-K pen-loglik phonemes}Variation of the number of clusters and the value of the objective function during the iterations of the algorithm for the PRM (left) PSRM (middle) and PbSRM (right) for the phonemes data.}
\end{figure}

Figures \ref{fig. robust EM-PRM phonemes results}, \ref{fig. robust EM-SRM phonemes results} and \ref{fig. robust EM-bSRM phonemes results}  respectively show the obtained cluster results for the phonemes log-periodograms for the polynomial, spline and B-spline regression mixtures. 
The number of phoneme classes (five) is correctly estimated by the three models. We can also see that the spline regression models provide better results in terms of clusters approximation than the polynomial regression mixture (here $p=7$). 
Notice that the value of $p=7$ correspond to the polynomial regression mixture model with the best error rate for $p$ varying from 4 to 8. The misclassification error rate in this case is 14.29 \%.
%
The values of the estimated number of clusters and the misclassification error rate corresponding to each of the three models are given in Table \ref{tab. results phonemes}. One can see that the spline regressions mixture  perform better than the polynomial regression mixture. In a general use of functional data modeling, the spline are indeed more adapted than simple polynomial modeling.
\begin{table}[H]
\centering 	
\begin{tabular}{cccc}
\hline
 & EM-PRM & EM-SRM &  EM-bSRM\\ 
$\hat K$ &  5 & 5 & 5 \\
Misc. error rate & 14.29 \% & 14.09 \% & 14.2 \% \\
\hline 
\hline
\end{tabular}
\caption{\label{tab. results phonemes}Clustering results for the phonemes data.}
\end{table}
In a similar way as previously, on Figure \ref{fig: robust EM-MixReg stored-K pen-loglik phonemes}, one can see the variation of the estimated number of clusters as well as the value of the objective function as the learning proceeds. 
It can be observed that the number of clusters decreases very rapidly from 1000 to 51 for the polynomial regression mixture model, and to 44 for the spline and B-spline regression mixture models one iteration to another for the three models. The grand majority of illegitimate clusters is discarded at the beginning the learning process.
Then, the number of clusters gradually decreases and the algorithm converge towards a partition with the actual number of clusters for the three models after at most 43 iterations.
 We can also see from the curve of the number of clusters and the objectives functions that the algorithm for the spline and B-spline regression mixture models behaves in a very similar way. We can also notice that the objective function becomes horizontal once the number of clusters is stabilized.

\subsubsection{Yeast cell cycle data}
\label{ssec: Yeast cell cycle data}
\begin{figure}[H]
   \centering 
   \includegraphics[width=6cm]{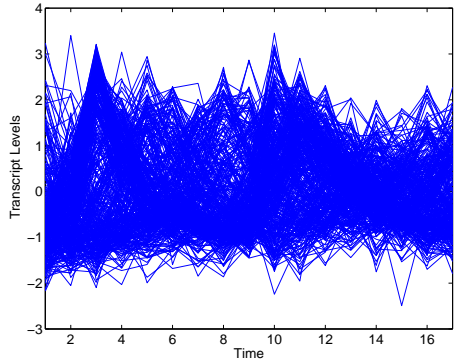}
   \caption{ \label{fig. yeast-cellcycle data}Yeast cell cycle data.}
\end{figure}

\begin{figure}[H]
   \centering 
   \begin{tabular}{cc}
   \includegraphics[width=5.2cm]{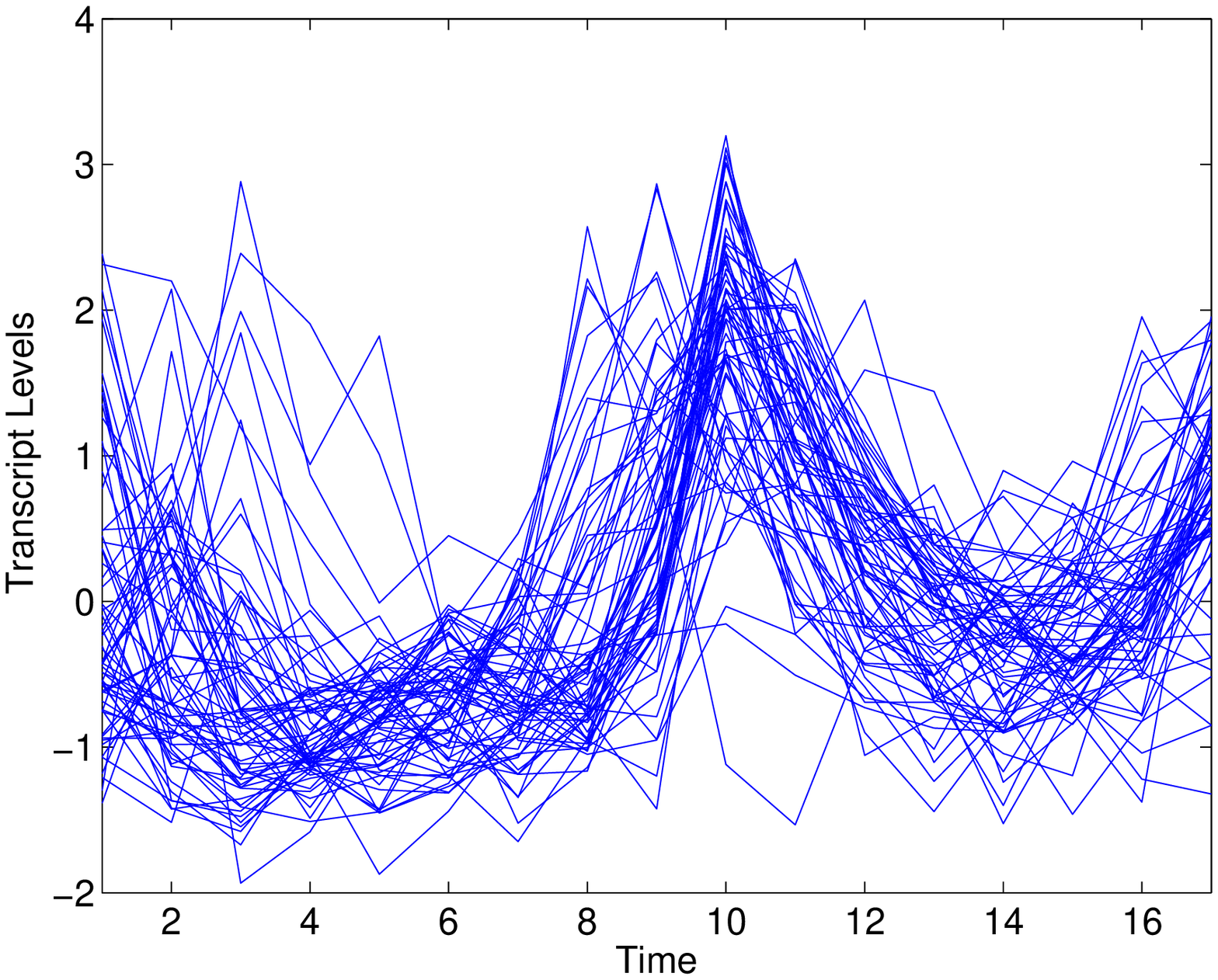}&
   \includegraphics[width=5.2cm]{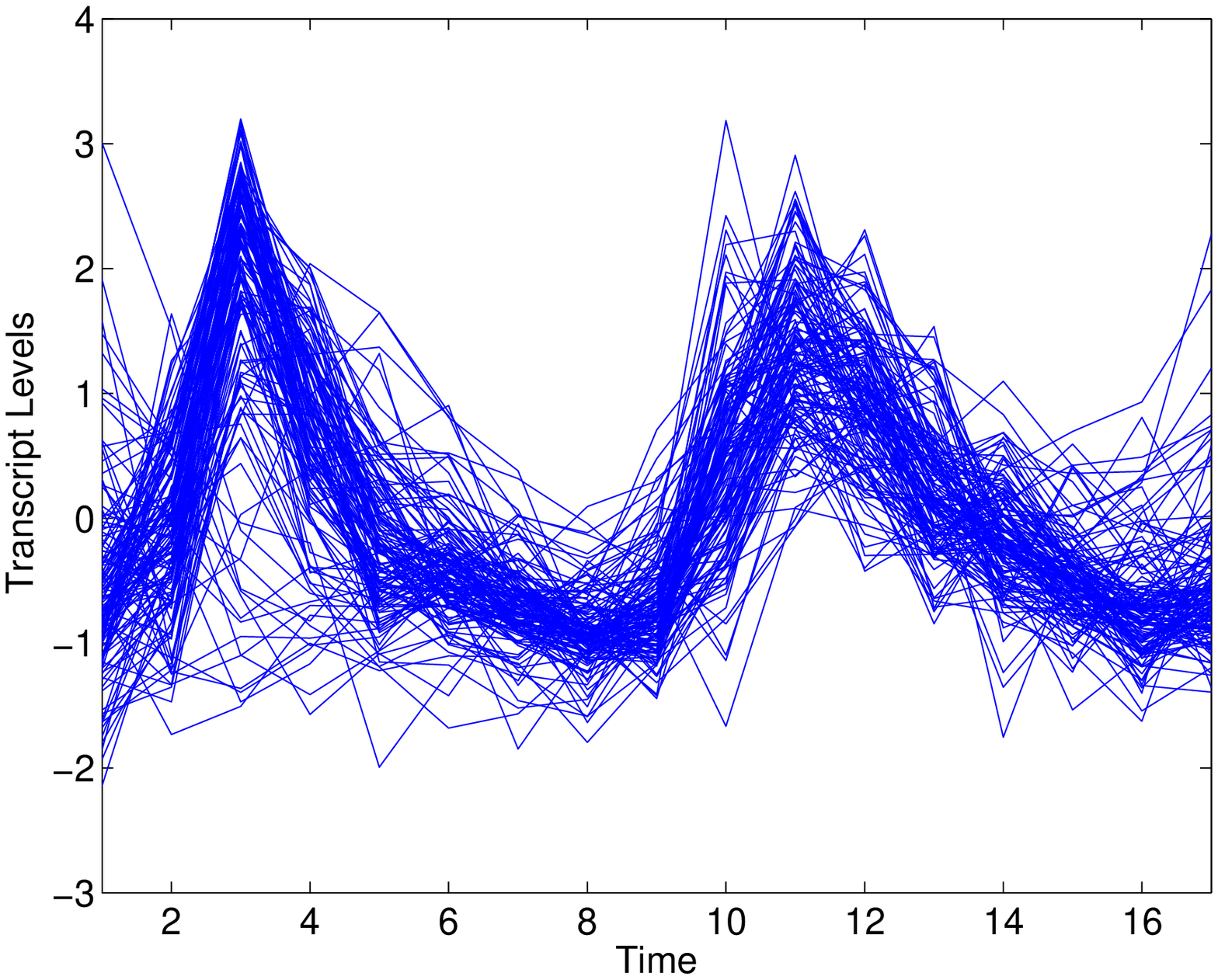} \\
    \includegraphics[width=5.2cm]{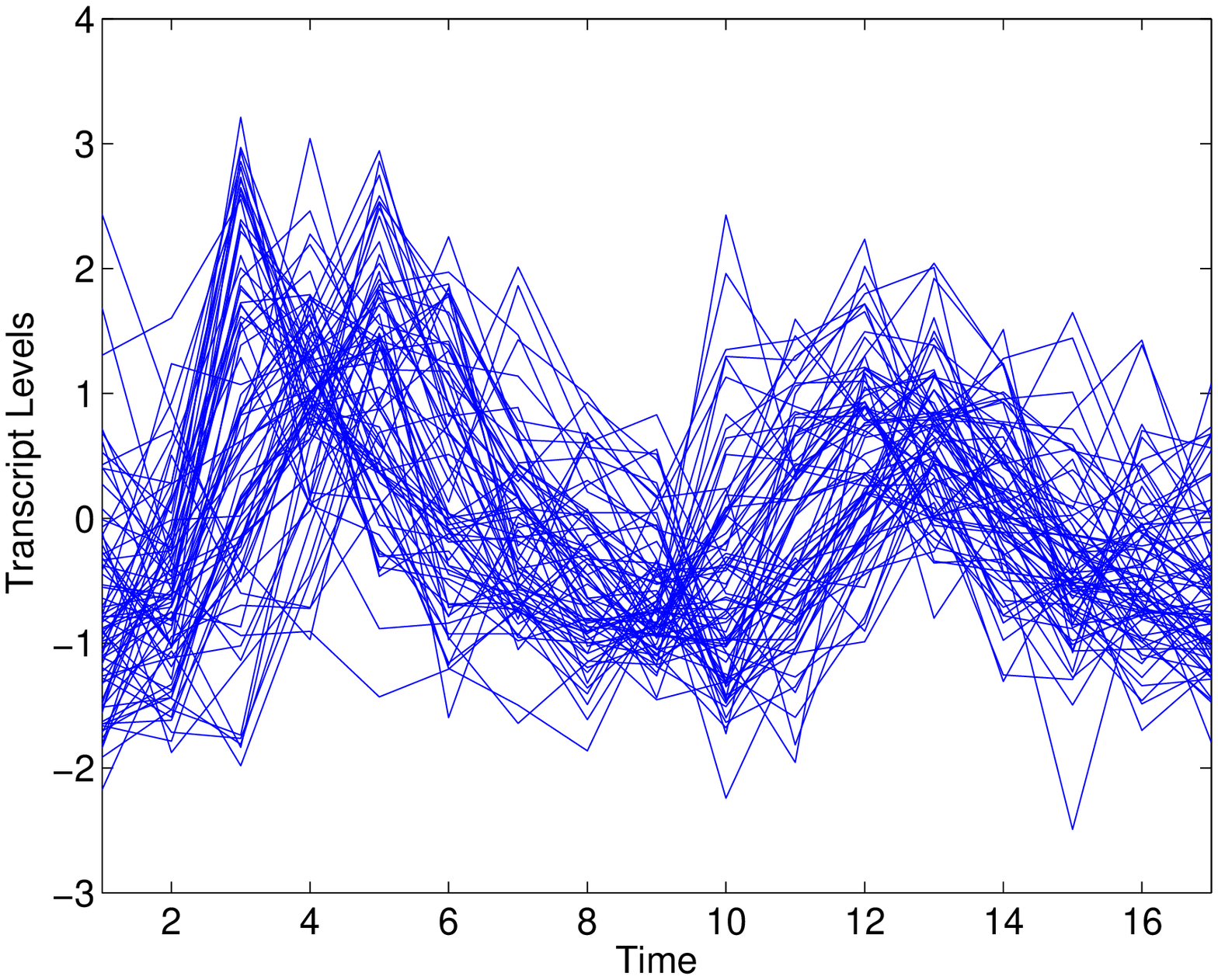}&  
    \includegraphics[width=5.2cm]{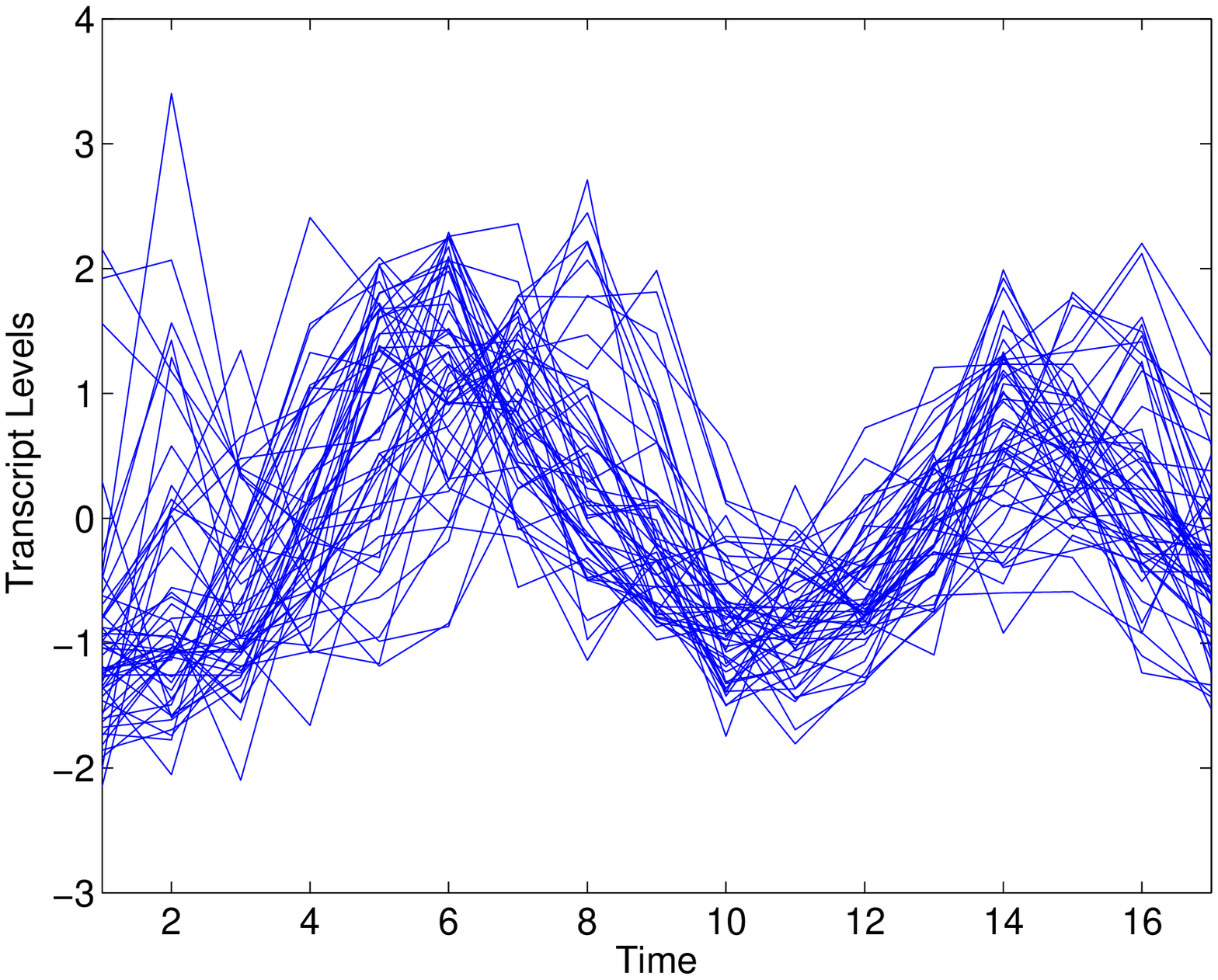} \\
     \includegraphics[width=5.2cm]{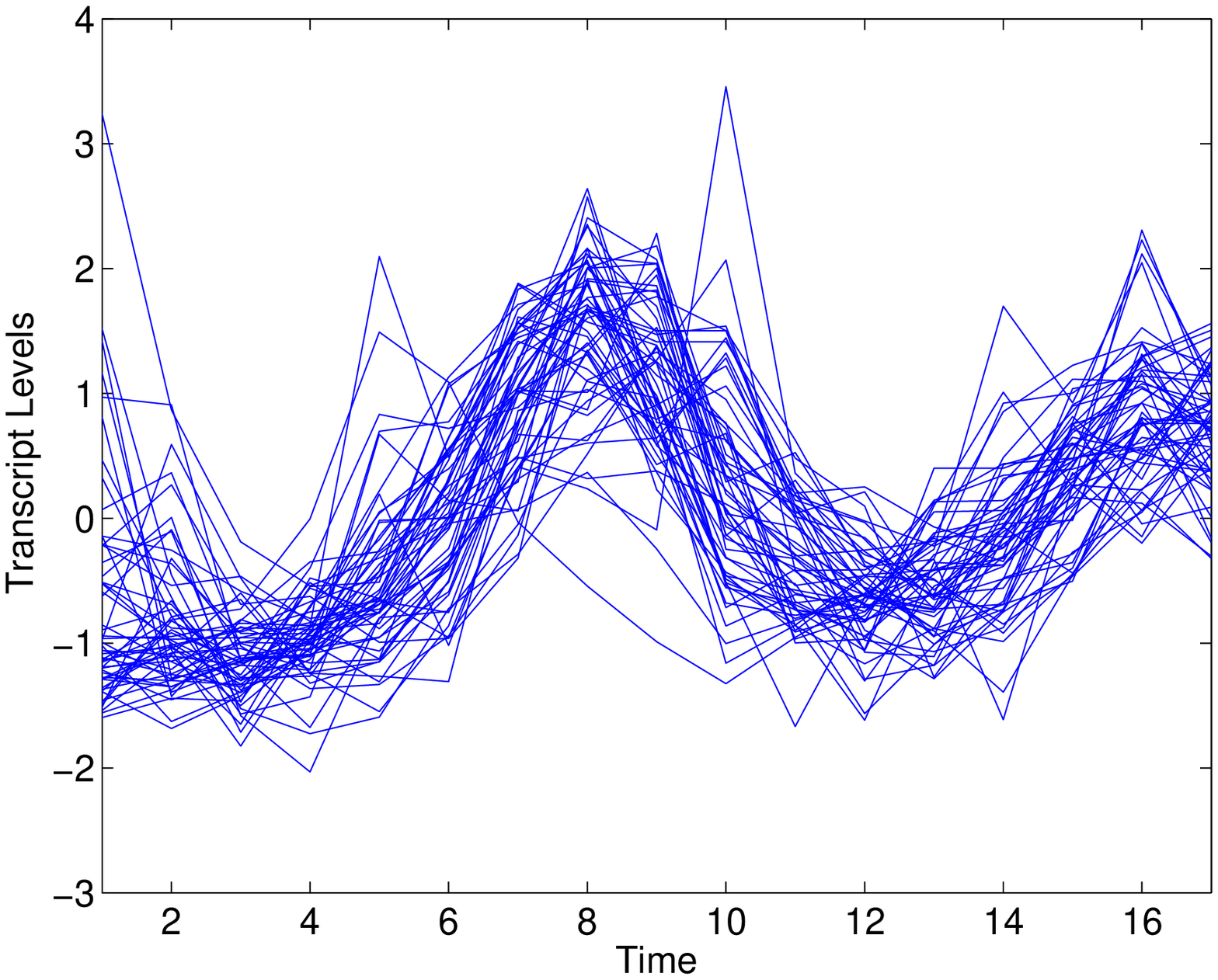}
   \end{tabular}
   \caption{\label{fig. yeast cell cycle data clusters}The five actual clusters of the  used yeast cell cycle data.}
\end{figure}

\begin{figure}[H]
   \centering 
   \begin{tabular}{cc}
   \includegraphics[width=5.2cm]{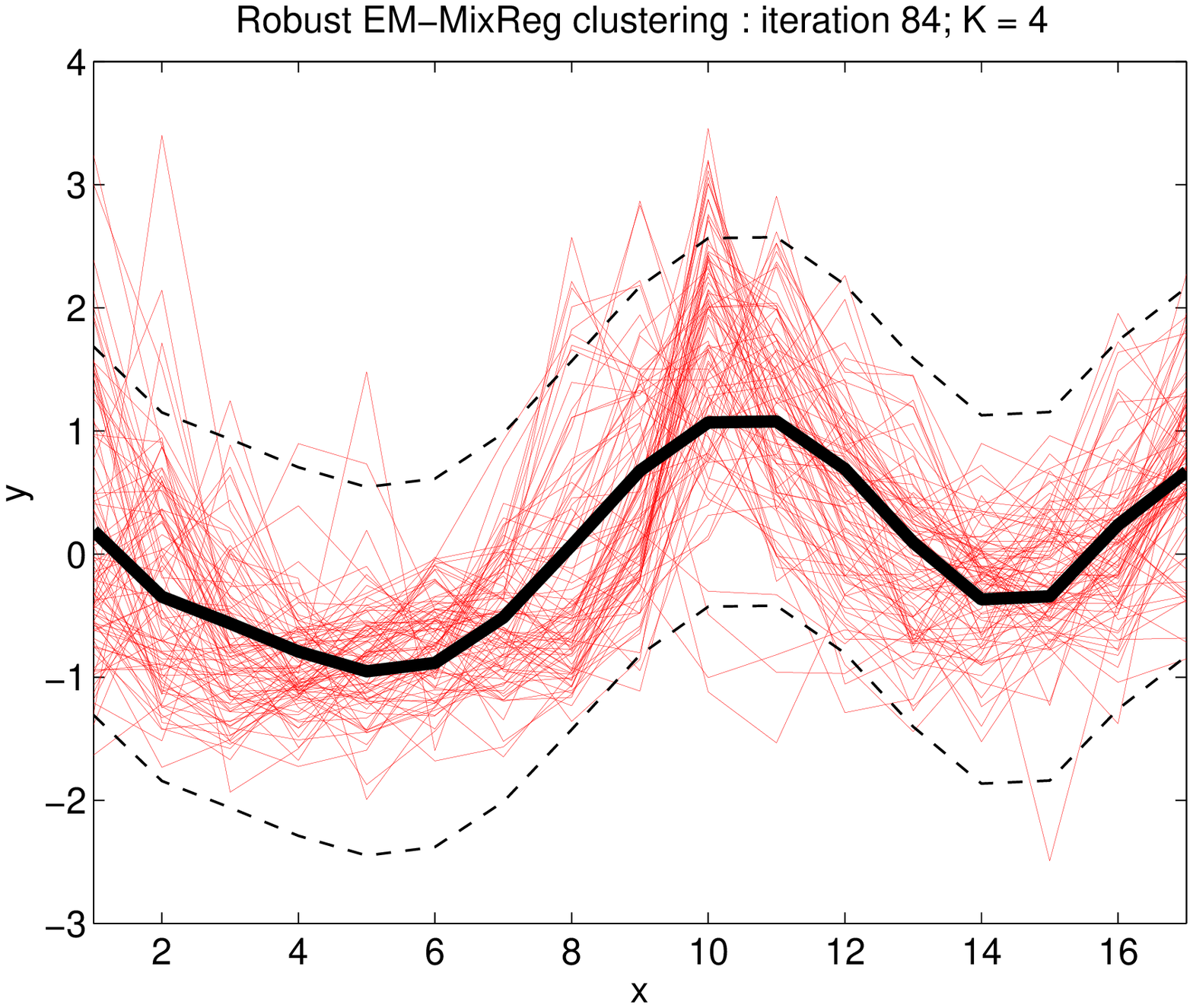}&
   \includegraphics[width=5.2cm]{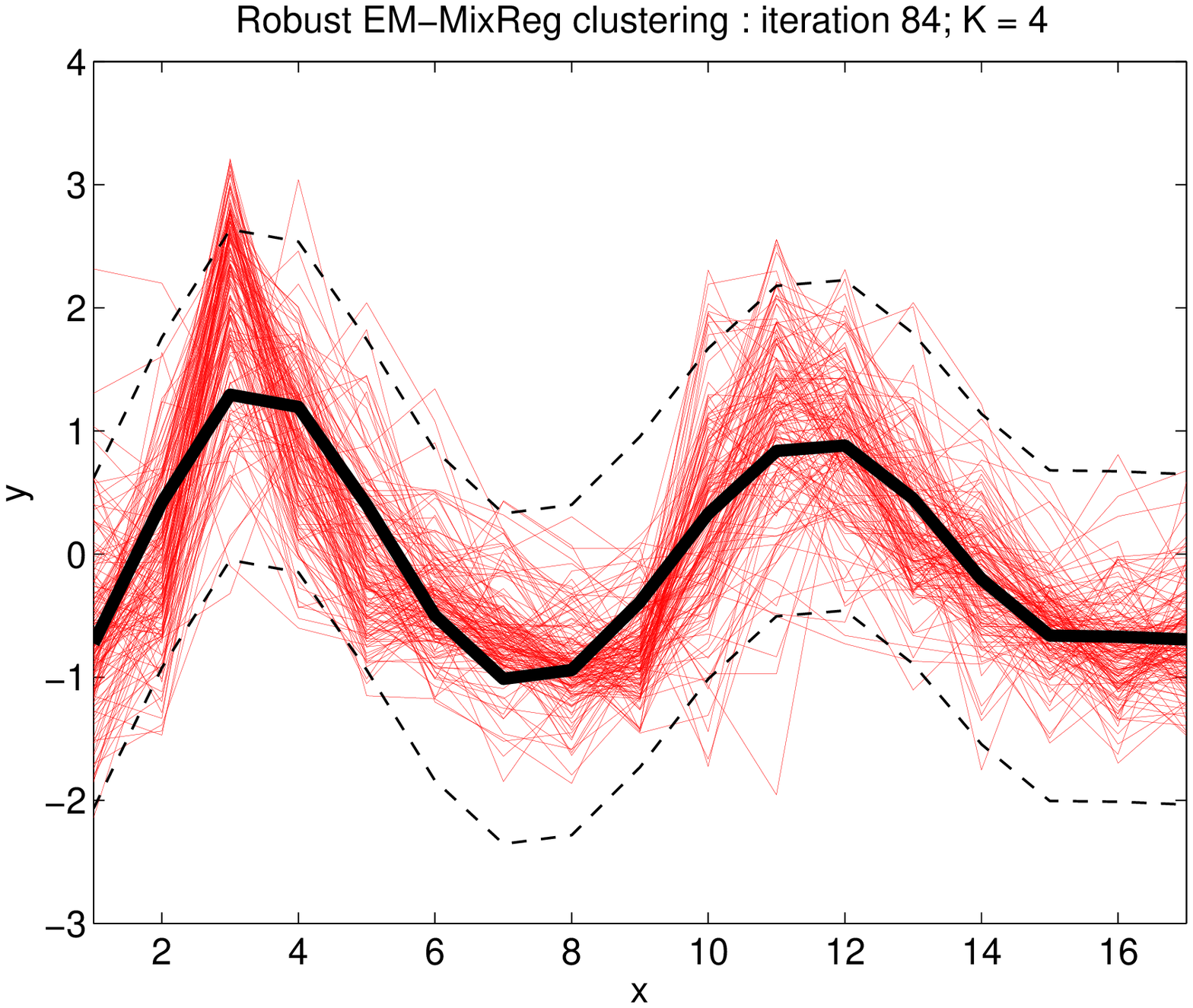}\\
    \includegraphics[width=5.2cm]{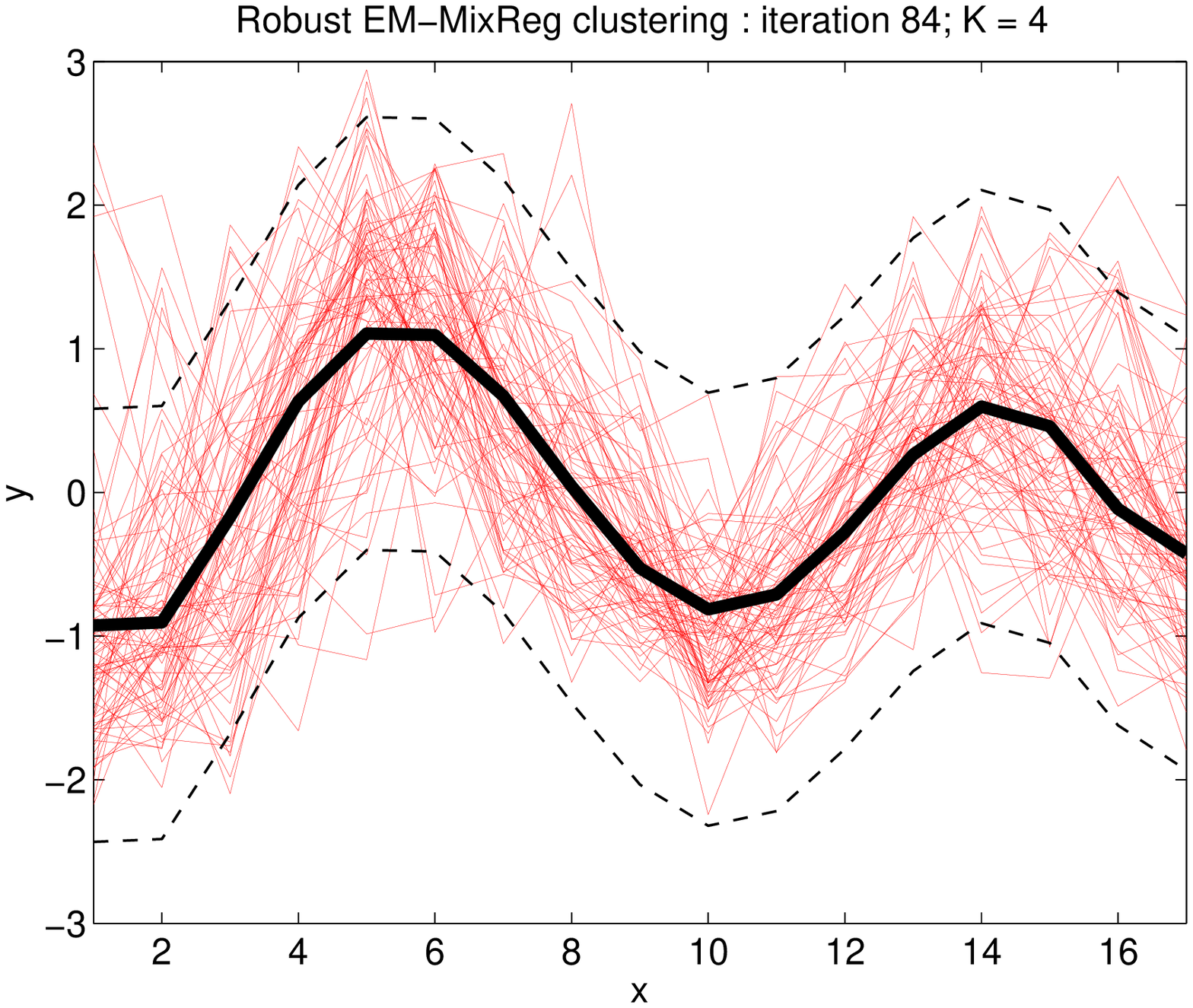}&
    \includegraphics[width=5.2cm]{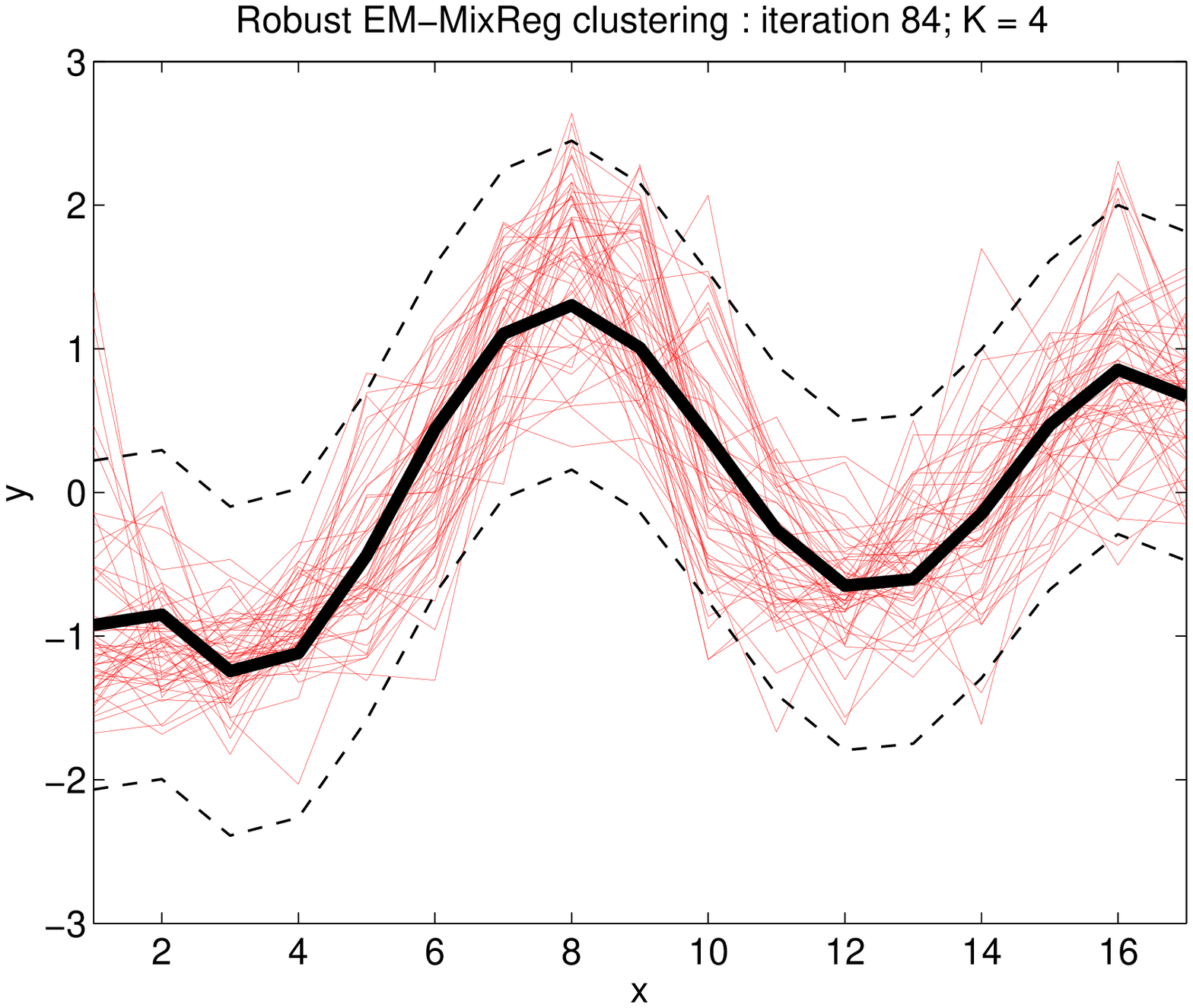}
   \end{tabular}
         \caption{\label{fig. robust EM-PRM yeast-cellcycle results}Clustering results obtained by the proposed robust EM algorithm and the PRM model (polynomial degree $p=6$) for the yeast cell cycle data.}
\end{figure}

\begin{figure}[H]
   \centering 
   \begin{tabular}{cc}
   \includegraphics[width=5.2cm]{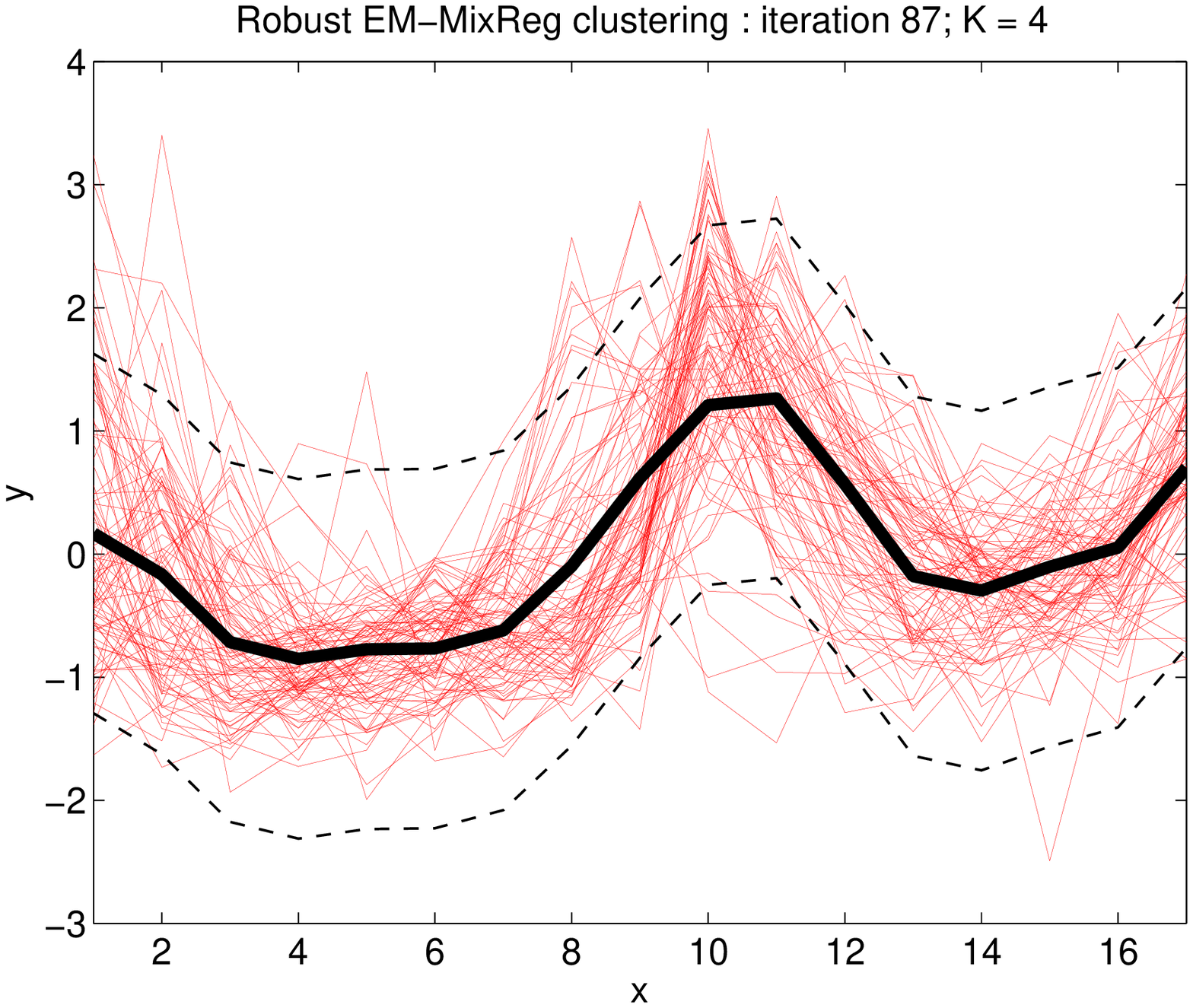}&
   \includegraphics[width=5.2cm]{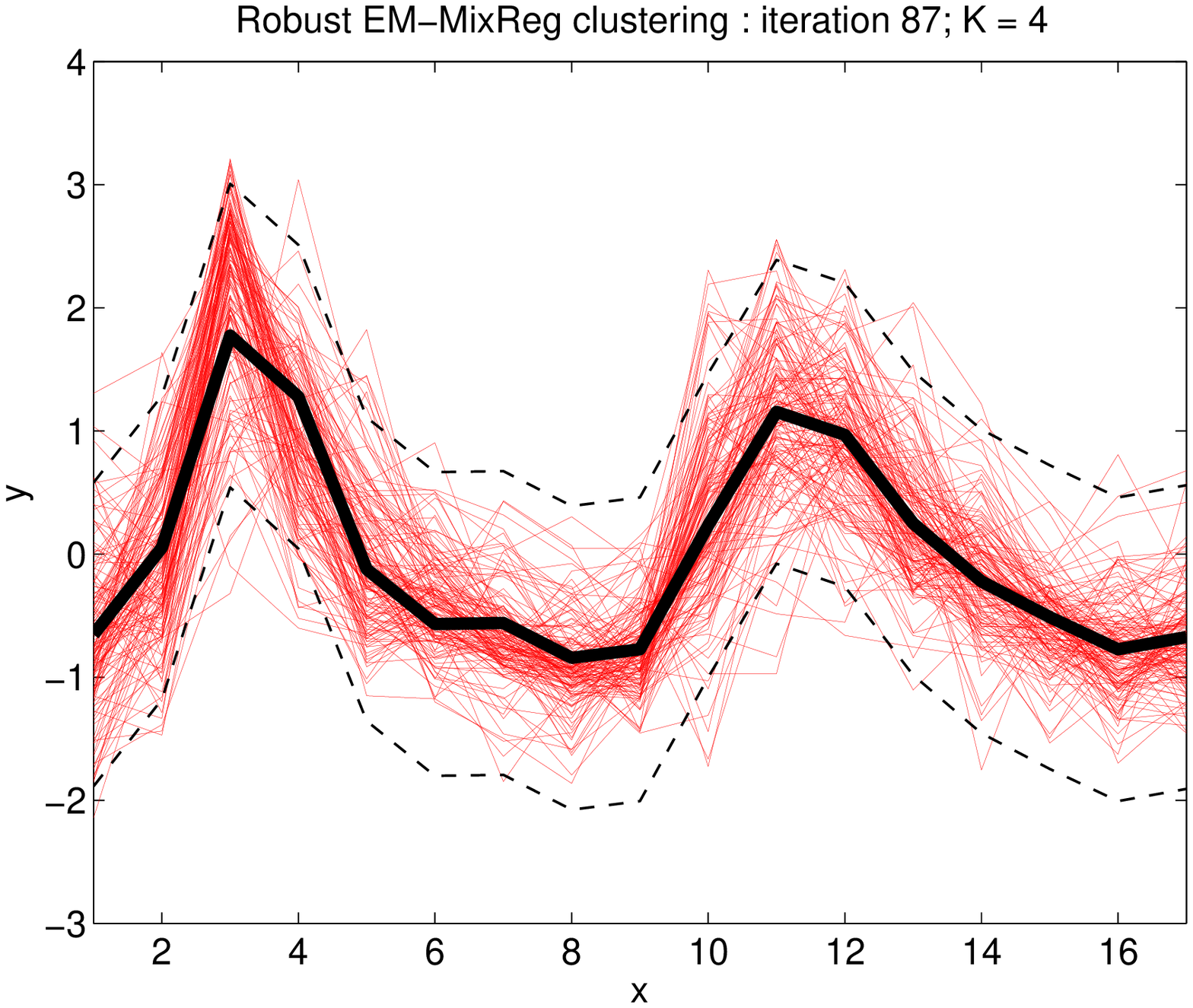}\\
    \includegraphics[width=5.2cm]{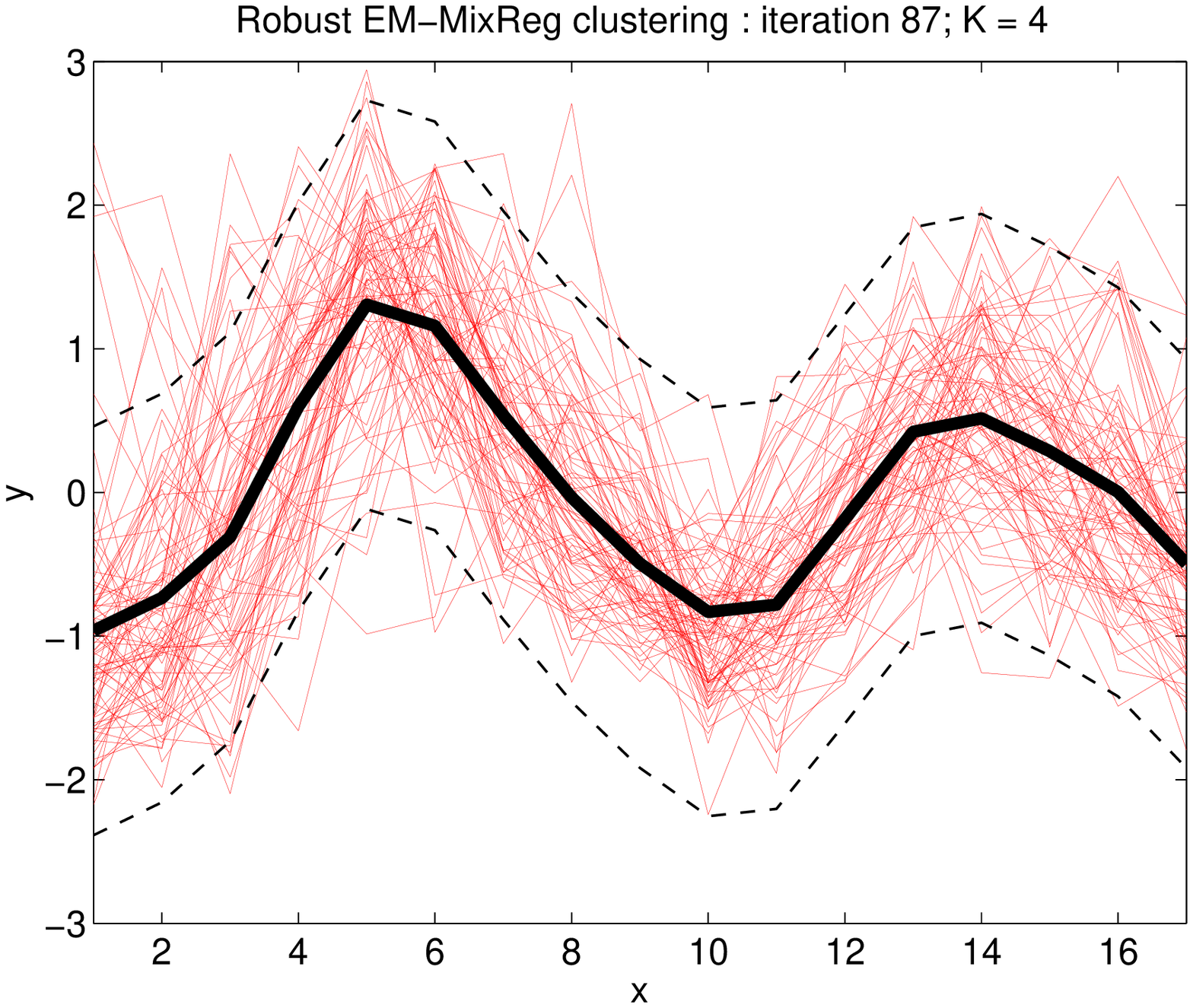}&
    \includegraphics[width=5.2cm]{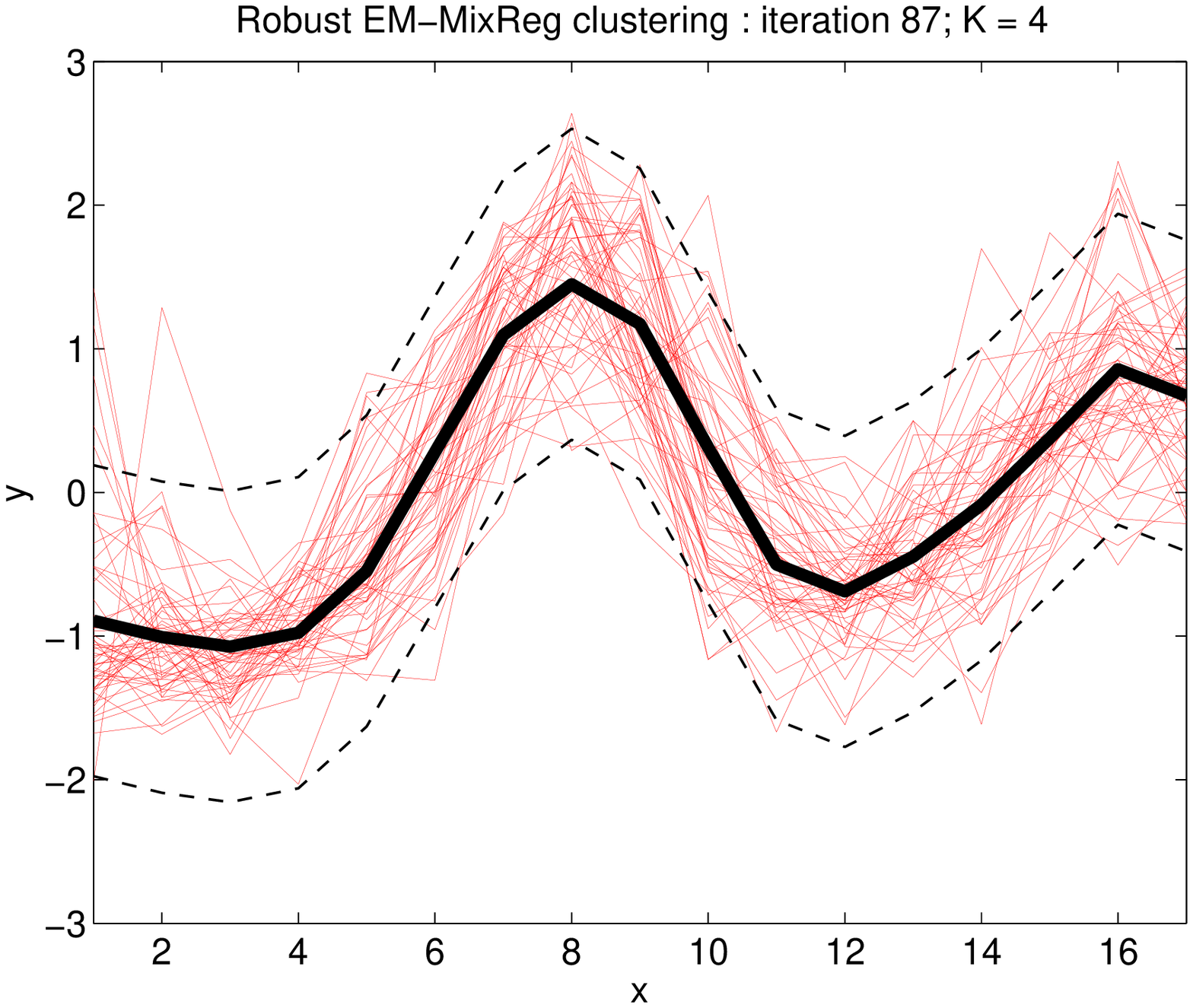}
   \end{tabular}
         \caption{\label{fig. robust EM-SRM yeast-cellcycle results}Clustering results obtained by the proposed robust EM algorithm and the SRM model with a cubic spline of 7 knots for the yeast cell cycle data.}
\end{figure}

\begin{figure}[H]
   \centering 
   \begin{tabular}{cc}
   \includegraphics[width=5.2cm]{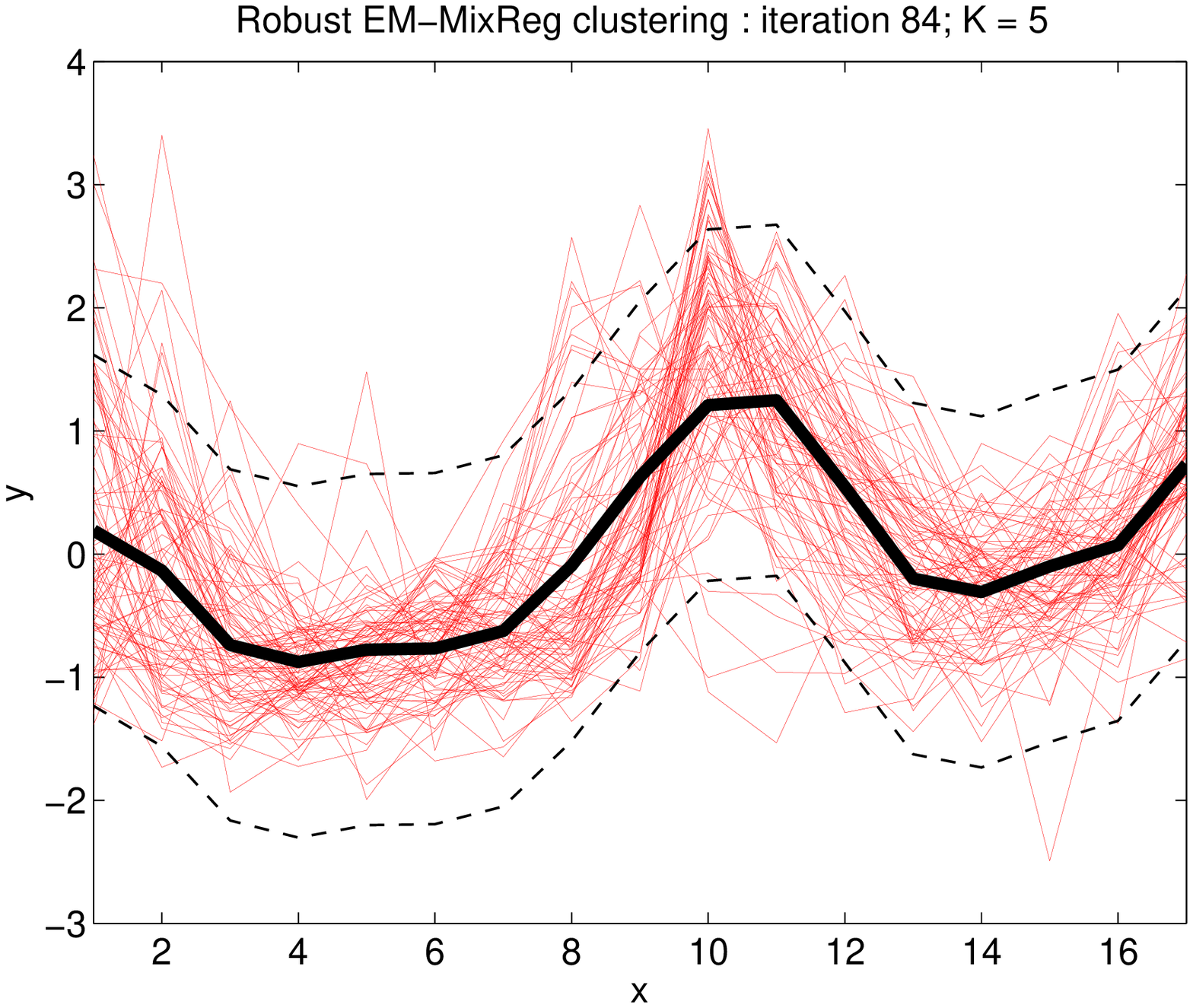}&
   \includegraphics[width=5.2cm]{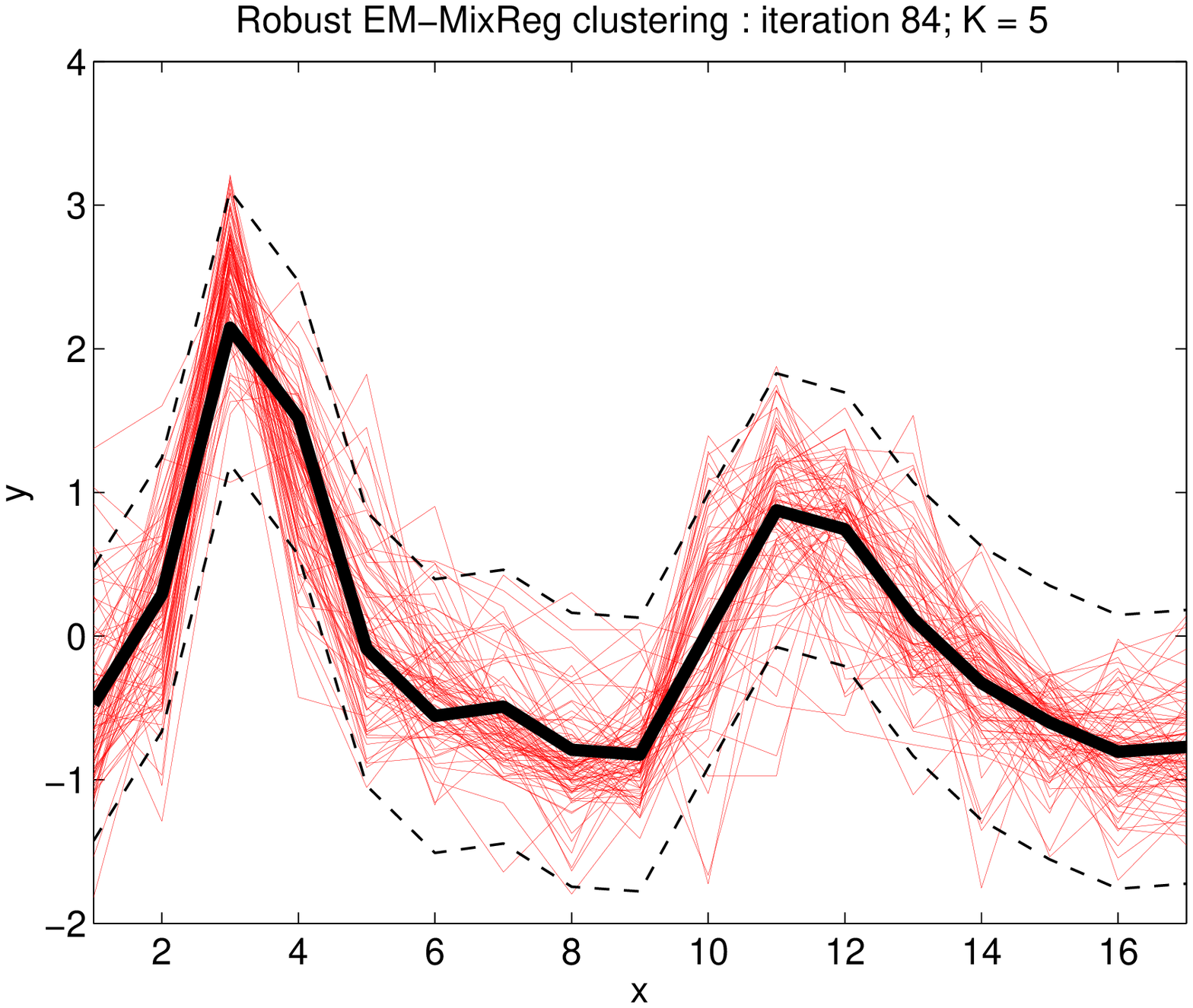}\\
   \includegraphics[width=5.2cm]{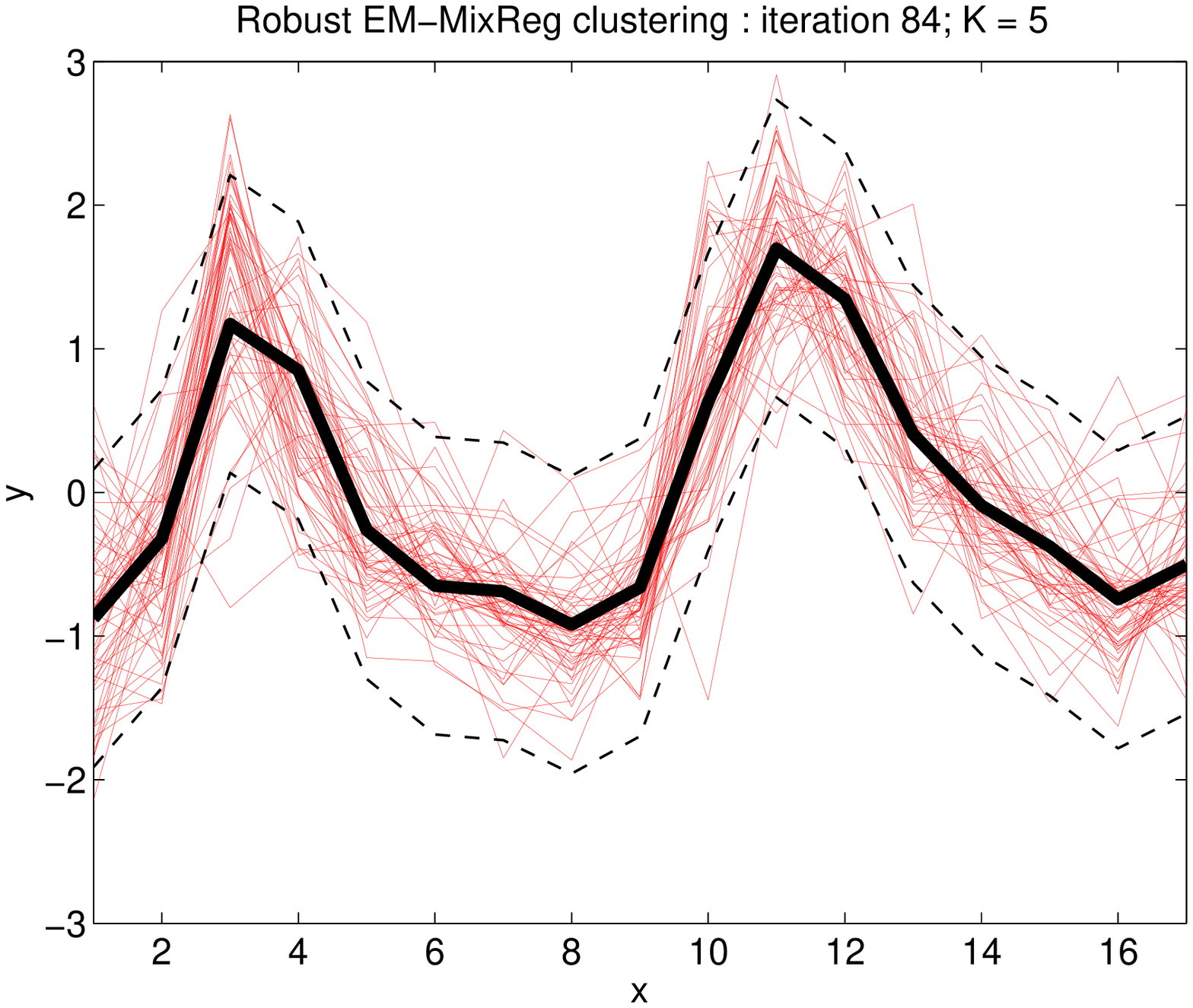}&
   \includegraphics[width=5.2cm]{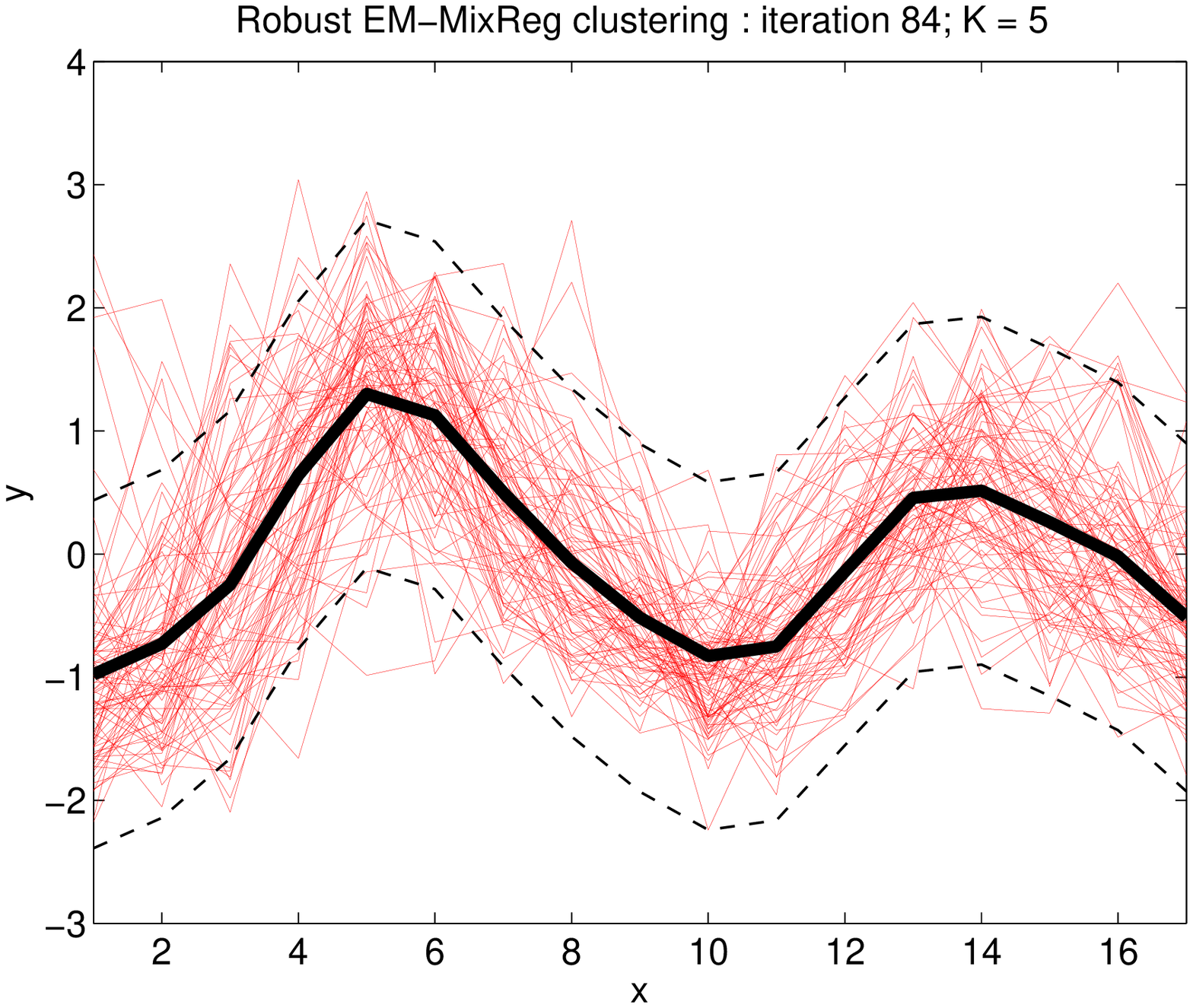}\\
      \includegraphics[width=5.2cm]{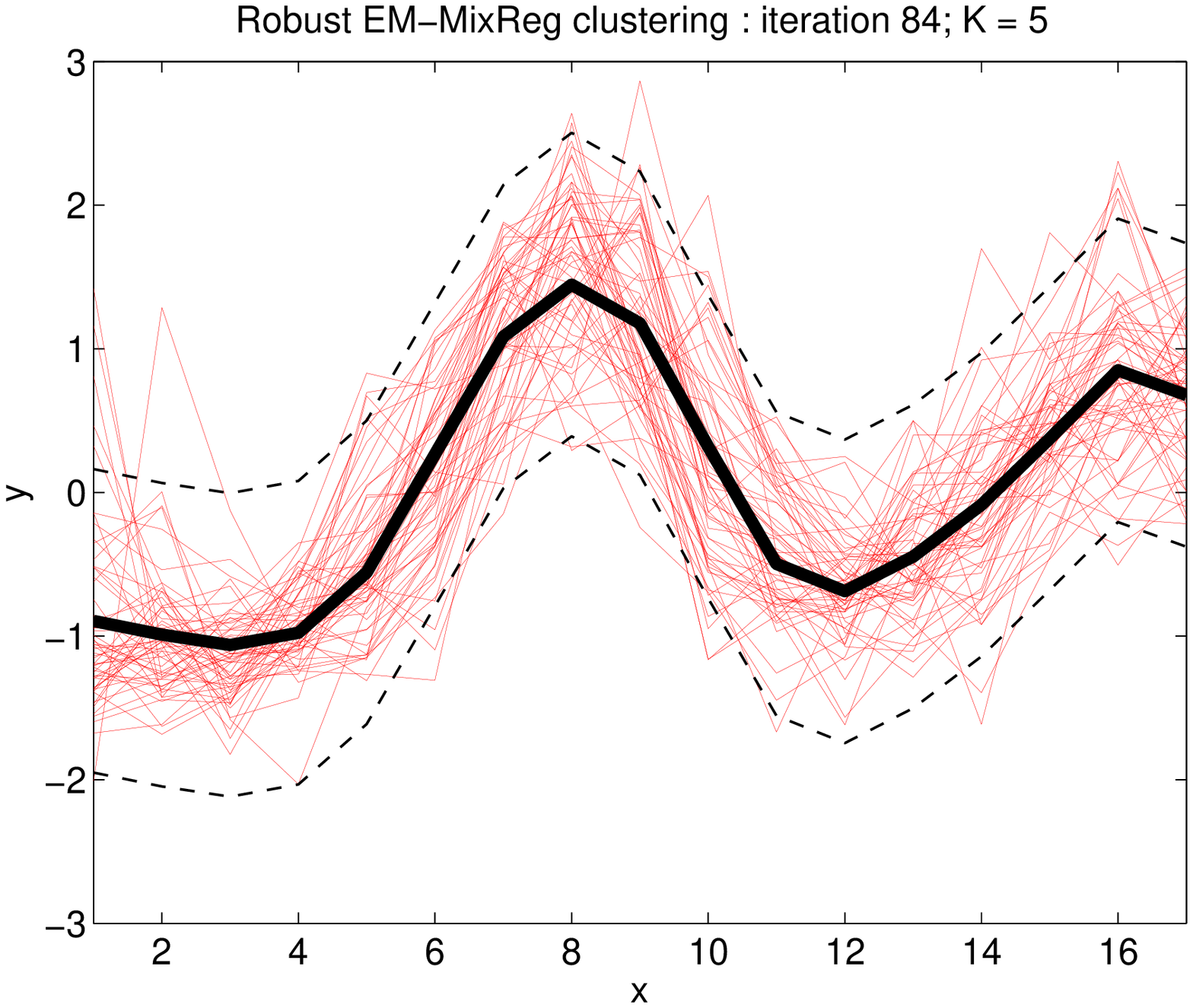}& 
   \end{tabular}
         \caption{\label{fig. robust EM-bSRM yeast-cellcycle results}Clustering results obtained by the proposed robust EM algorithm and the bSRM model with a cubic B-spline of 7 knots for the yeast cell cycle data.}
\end{figure}

In this experiment, we consider the  yeast cell cycle dataset \citep{Cho1998}. The original yeast cell cycle data represent the fluctuation of expression levels of approximately 6000 genes over 17 time points corresponding to two cell cycles \citep{Cho1998}. 
This data set has been used to demonstrate effectiveness of clustering techniques for for time course Gene expression data in bio-informatics such as model-based clustering as in\cite{YeungMBC2001}.  
We used the standardized subset constructed by \cite{YeungMBC2001}  available in \url{http://faculty.washington.edu/kayee/model/}\footnote{The complete data set was downloaded from \url{http://genome-www.stanford.edu/cellcycle/}.}. This dataset referred to as the subset of the 5-phase criterion in \cite{YeungMBC2001} 
contains 384 genes expression levels over 17 time points. 
The utility of the cluster analysis in this case is therefore to reconstruct this five class partition. 
%
 Fig. \ref{fig. yeast-cellcycle data} 
shows the 384  curves of the yeast cell cycle data and Fig. \ref{fig. yeast cell cycle data clusters} shows the curves of each of five clusters.
The clustering results are shown in Figures \ref{fig. robust EM-PRM yeast-cellcycle results}, \ref{fig. robust EM-SRM yeast-cellcycle results} and \ref{fig. robust EM-bSRM yeast-cellcycle results} respectively for the polynomial, spline and B-spline regression mixture models. Both the PRM model and the SRM provide similar partitions with four clusters.
The actual clusters in middle left and bottom in Figure \ref{fig. yeast cell cycle data clusters} look to be merged into the cluster in middle left in Figure \ref{fig. robust EM-PRM yeast-cellcycle results} (respectively in Figure \ref{fig. robust EM-SRM yeast-cellcycle results}). 
%
%
However, the bSRM model infers the actual number of clusters. The Rand index for the obtained partition is 0.7914.

\begin{figure}[H]
   \centering 
       \includegraphics[width=5cm]{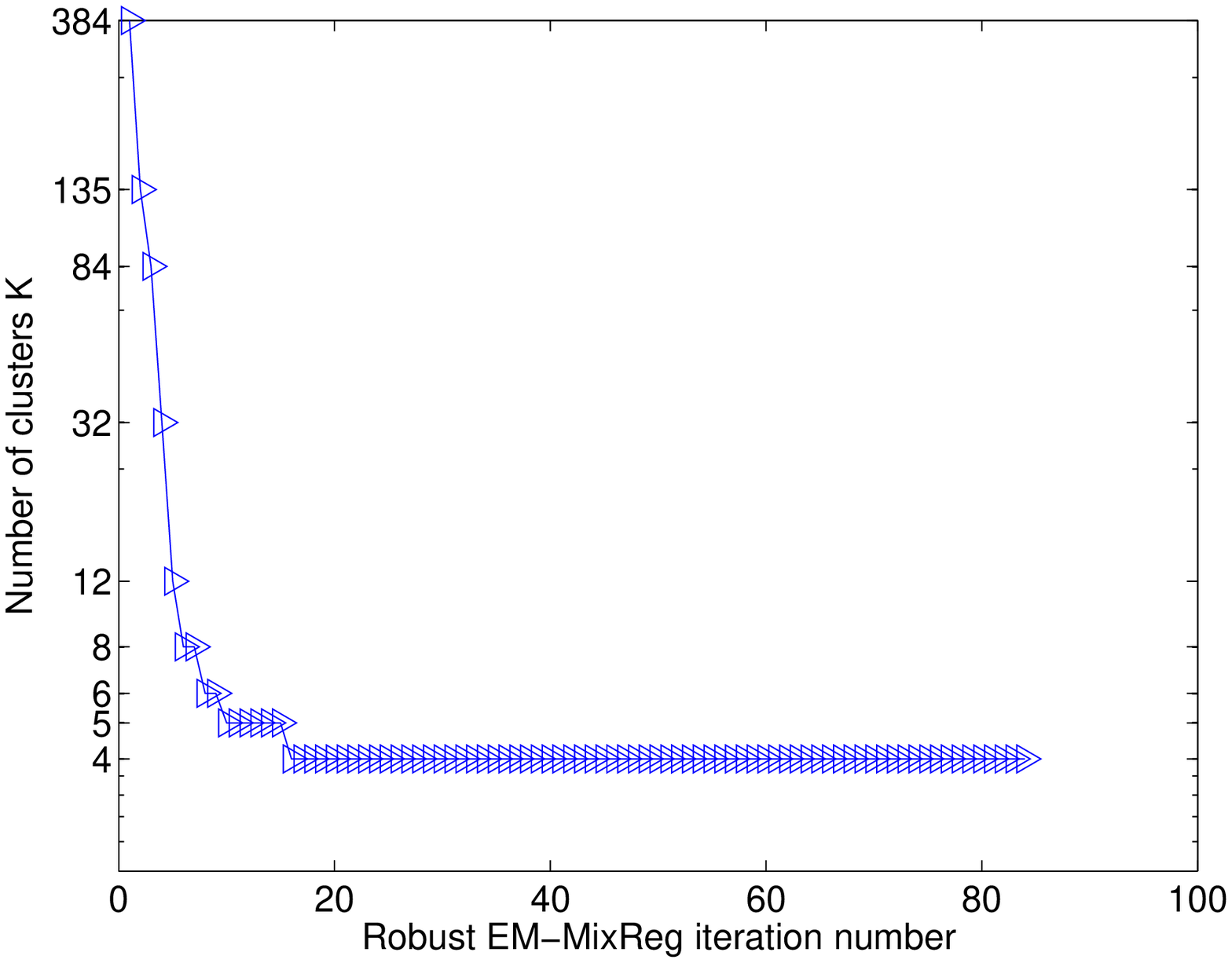}
   \includegraphics[width=5cm]{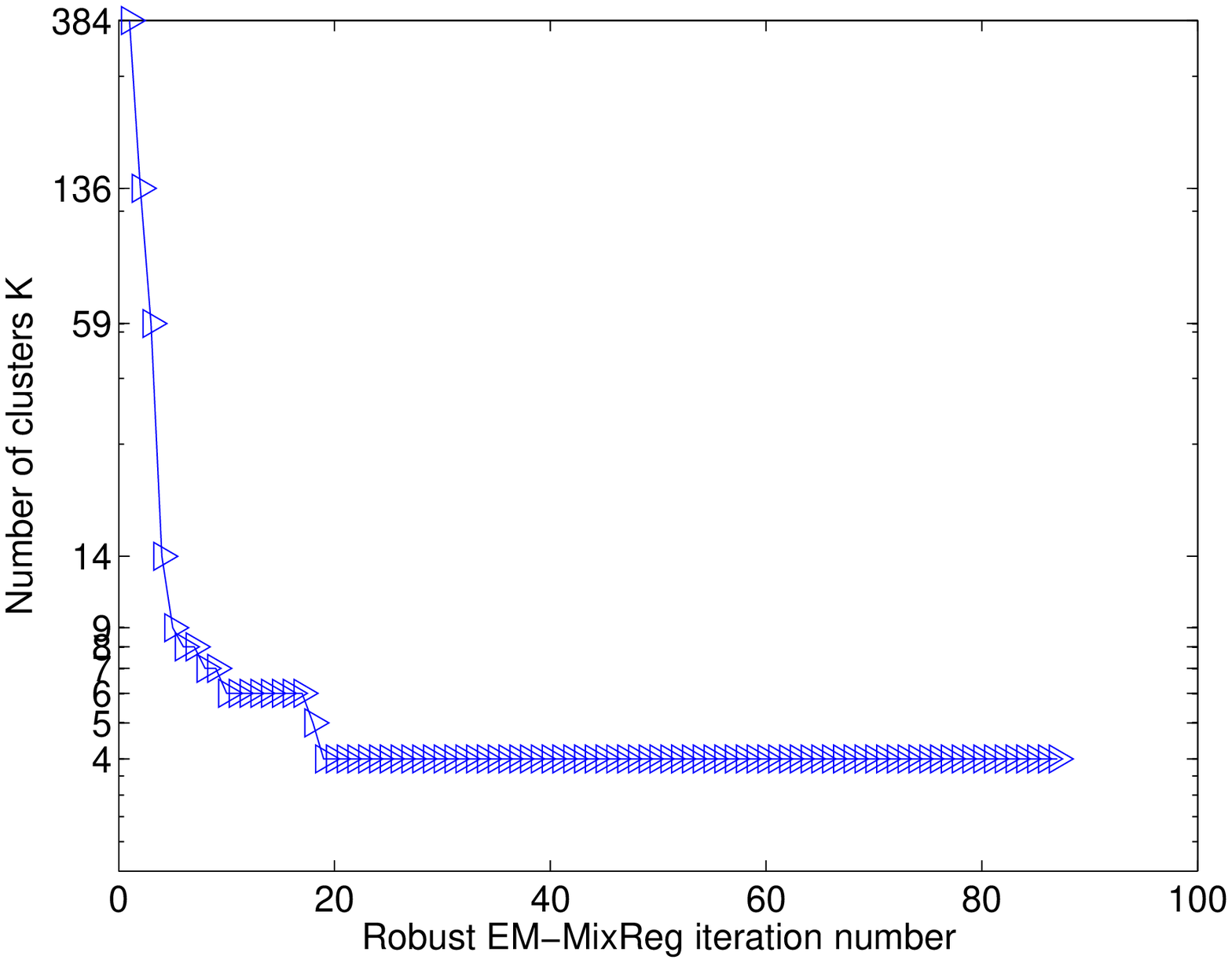}
   \includegraphics[width=5cm]{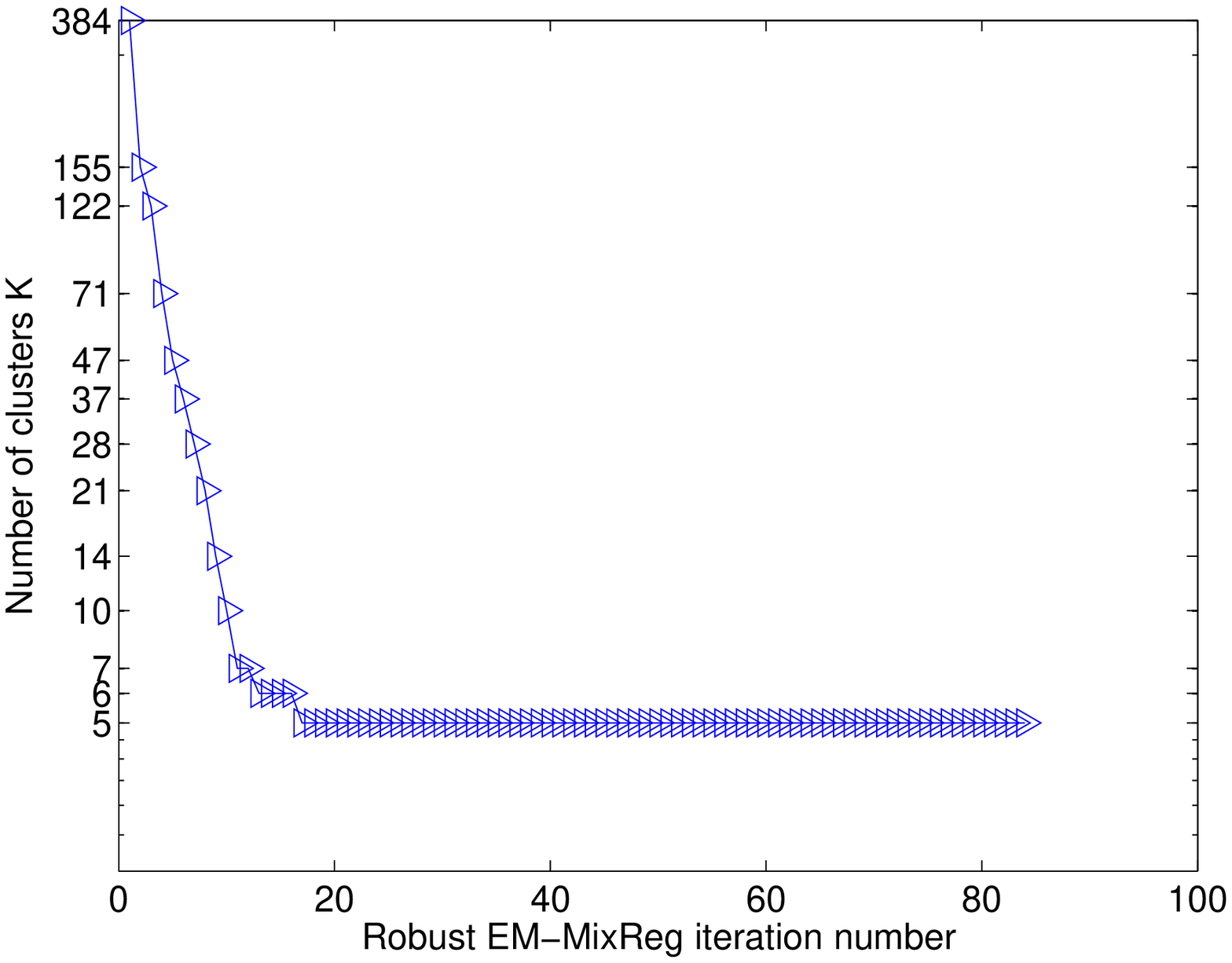}\\
   \includegraphics[width=5cm]{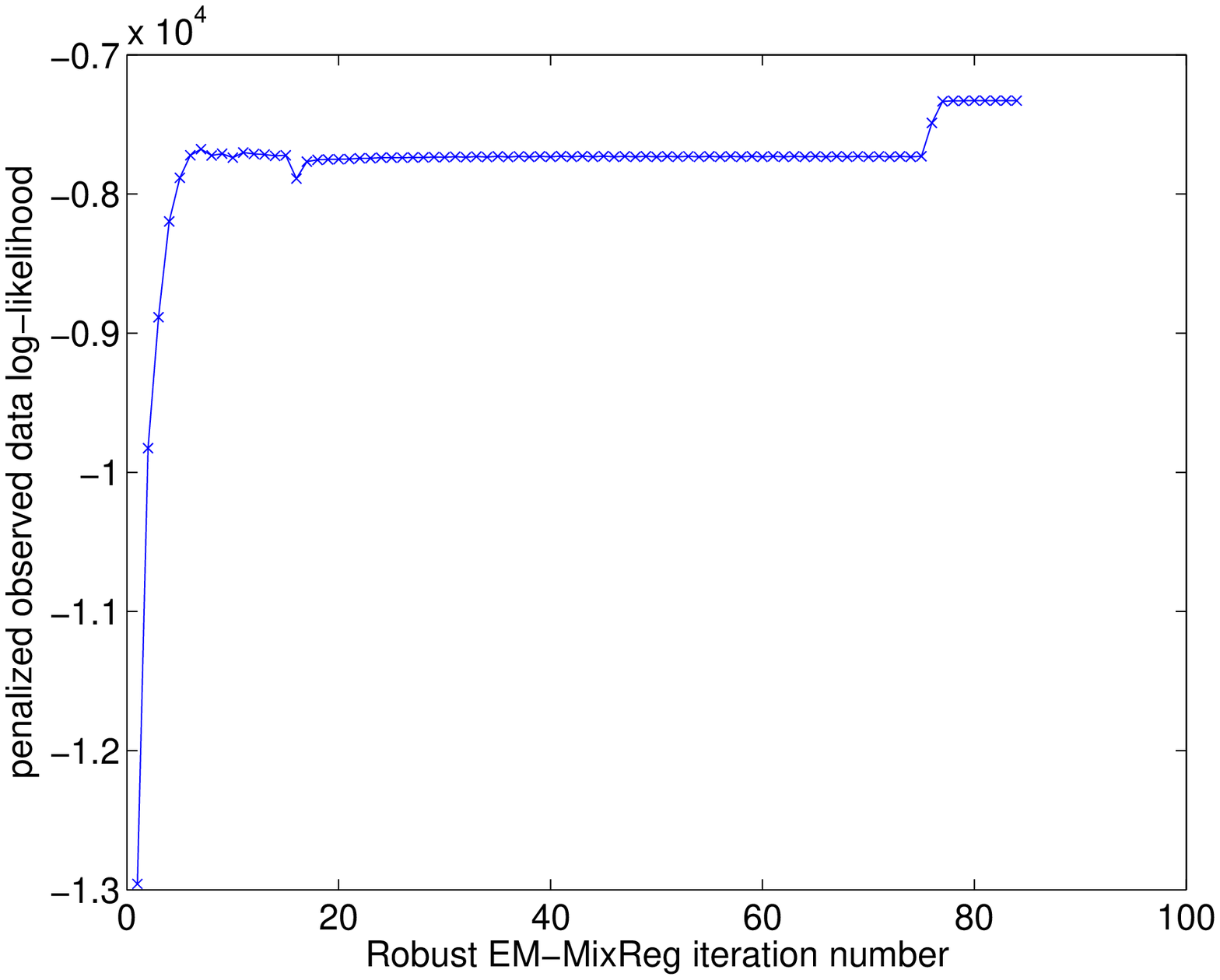}
   \includegraphics[width=5cm]{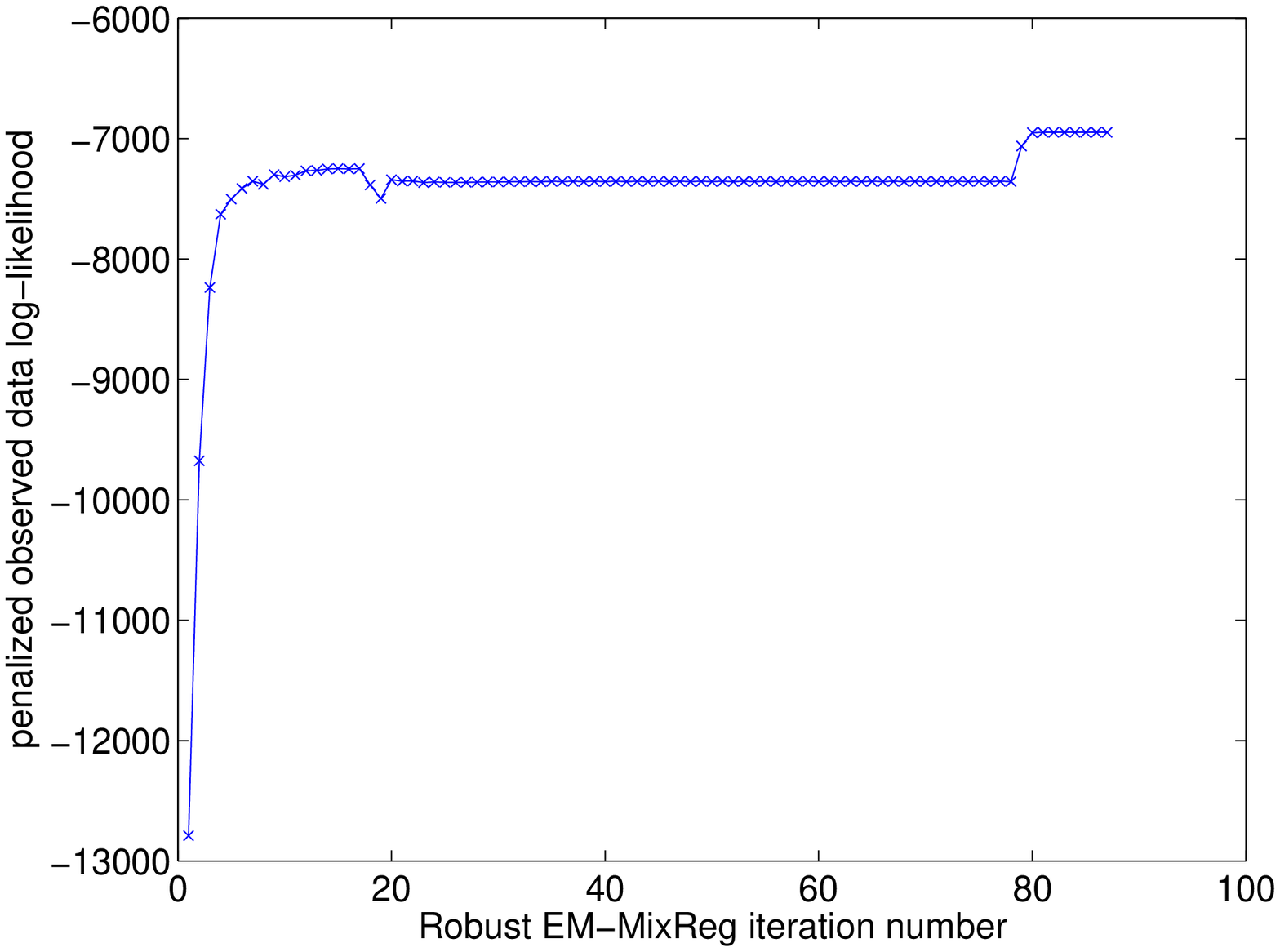}
   \includegraphics[width=5cm]{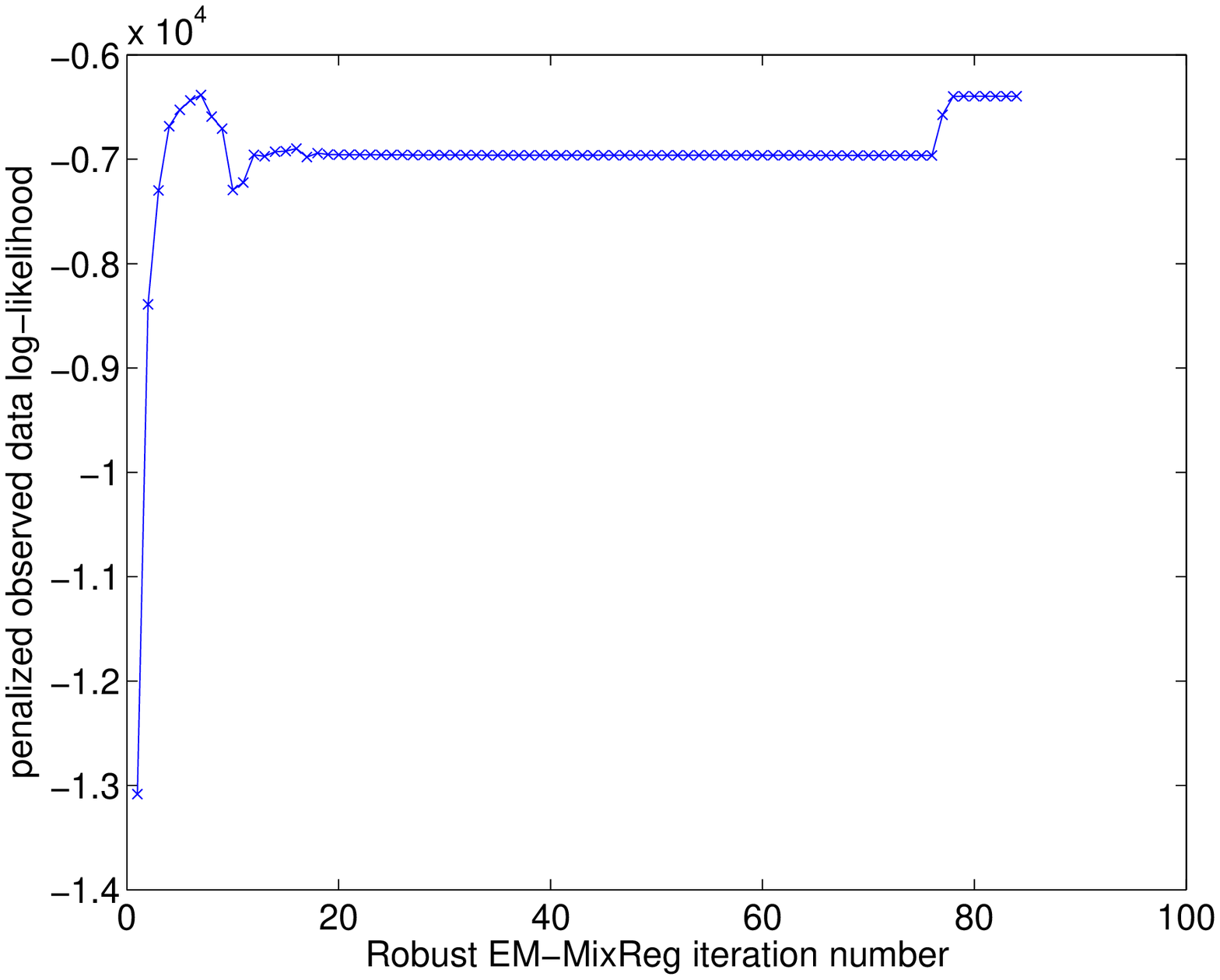}
   \caption{\label{fig: robust EM-MixReg stored-K pen-loglik yeast}Variation of the number of clusters and the value of the objective function during the iterations of the algorithm for the PRM (left) PSRM (middle) and PbSRM (right) for the yeast cell cycle data.}
\end{figure}

Figure \ref{fig: robust EM-MixReg stored-K pen-loglik yeast} shows the variation of the number of clusters and the value of the objective function during the iterations of the algorithm for three models. We can see that the number of clusters starts with $n=384$ clusters and more than half is discarded after one iteration. Then it gradually decreases and stabilized until convergence. The shape of the objective function also becomes horizontal when it is converged.

\subsubsection{Topex/Poseidon satellite data}

\begin{figure}[htbp] 
\centering 
\includegraphics[width=6cm]{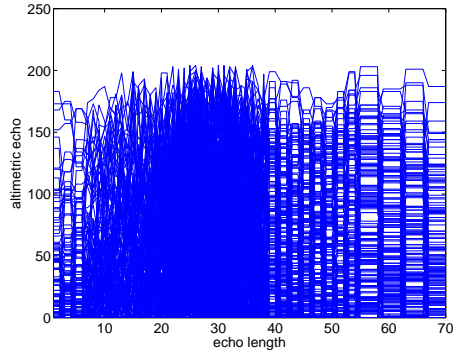}
\caption{\label{fig. satellite data}Topex/Poseidon satellite curves.}
\end{figure}

The last used real dataset is the Topex/Poseidon radar satellite data namely used in \cite{DaboNiang2007,hebrailEtal:2010} \footnote{Satellite data are available at \url{http://www.lsp.ups-tlse.fr/staph/npfda/npfda-datasets.html}.} 
This dataset were registered by the satellite Topex/Poseidon around an area of 25 kilometers upon the Amazon River. The data  contain $n=472$ waveforms of the measured echoes, sampled at $m=70$ (number of echoes). The actual number of clusters and the actual partition are unknown for this dataset. 
The curves of this data set are shown in 
Figure \ref{fig. satellite data}.

\begin{figure}[H]
   \centering  
   \includegraphics[width=5.2cm]{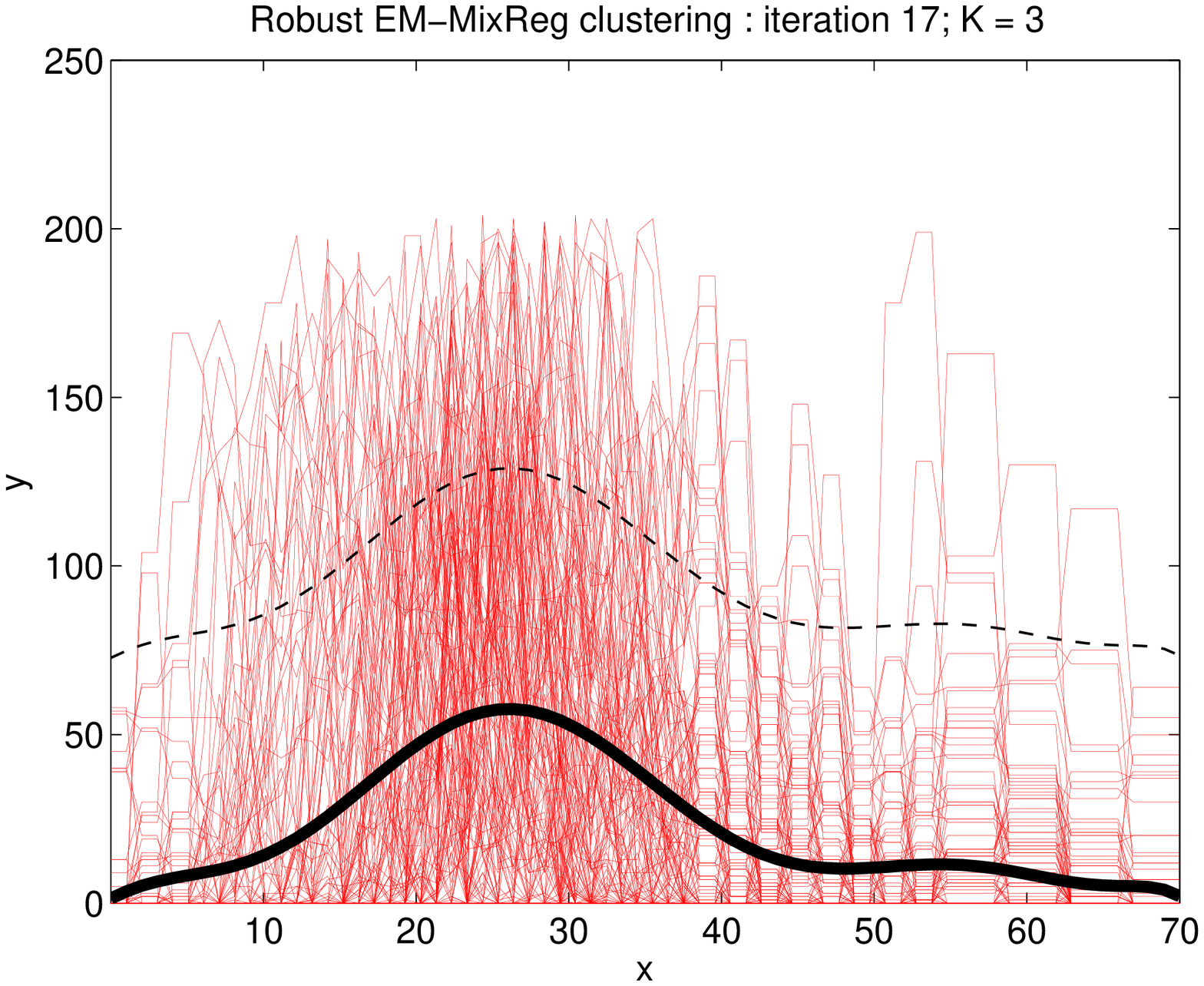}
   \includegraphics[width=5.2cm]{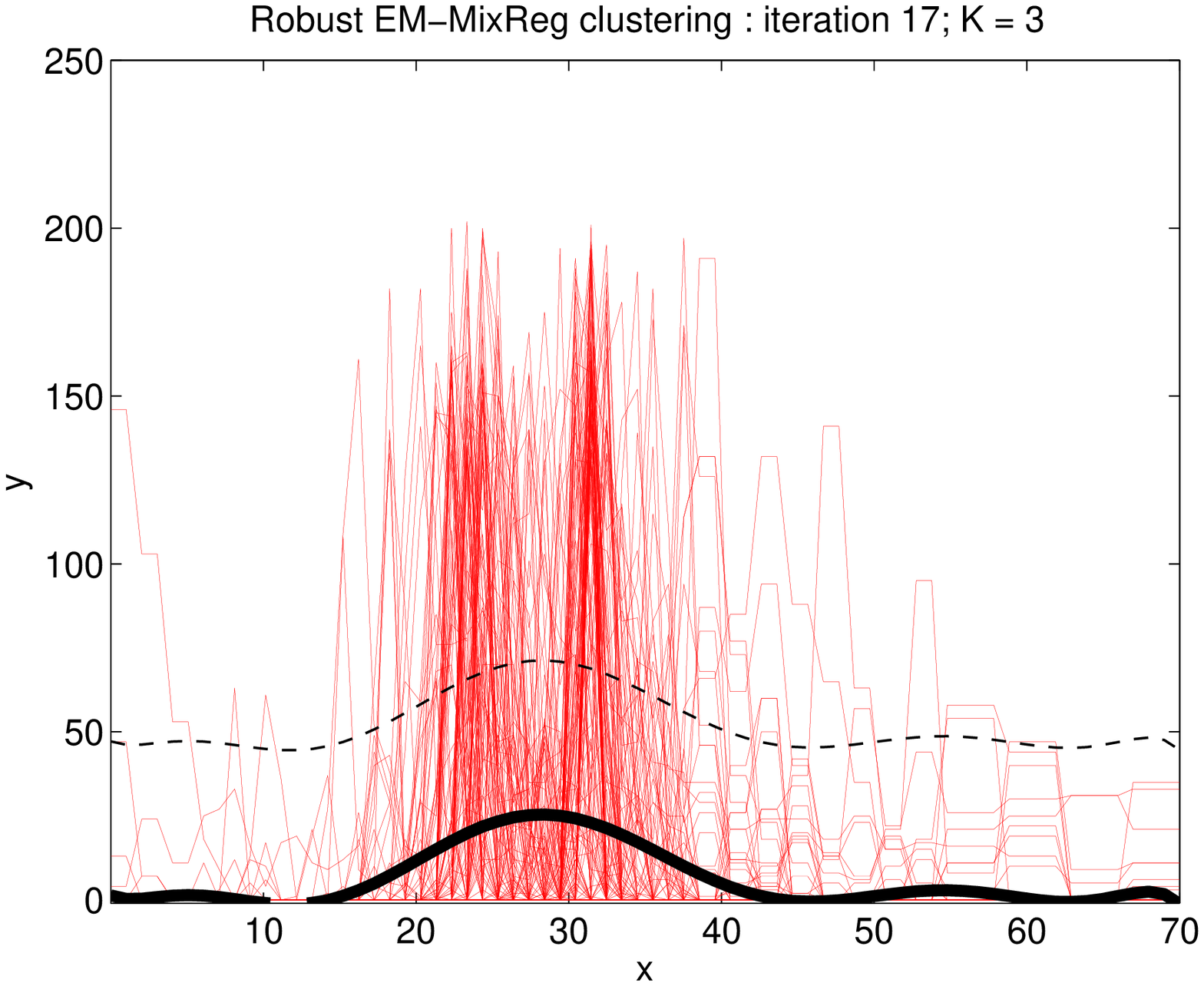}
    \includegraphics[width=5.2cm]{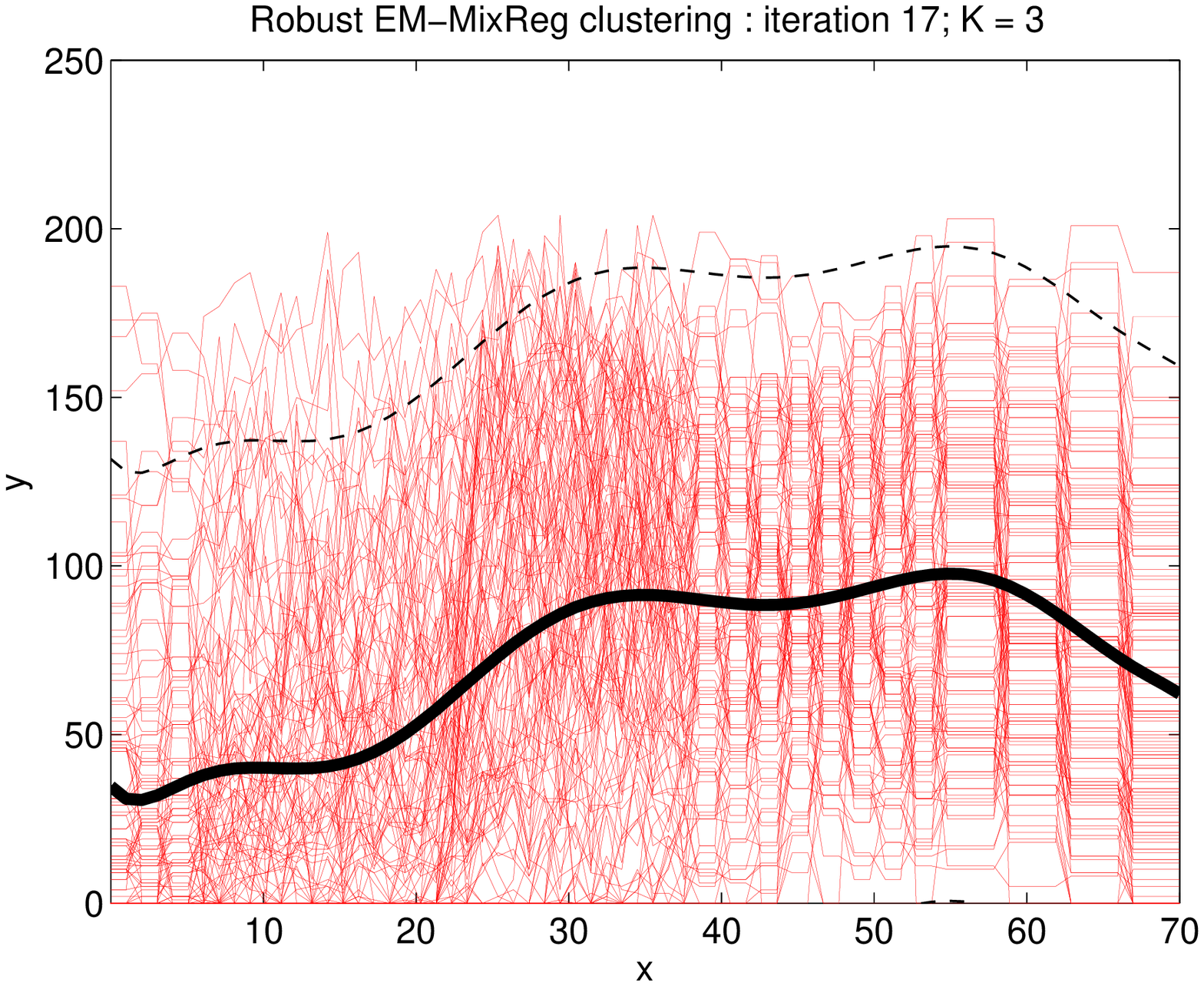} 
         \caption{\label{fig. robust EM-PRM satellite results}Clustering results obtained by the proposed robust EM algorithm and the PRM with $p=9$ for the satellite data.}
\end{figure}

\begin{figure}[H]
   \centering  
   \begin{tabular}{cc}
   \includegraphics[width=5.5cm]{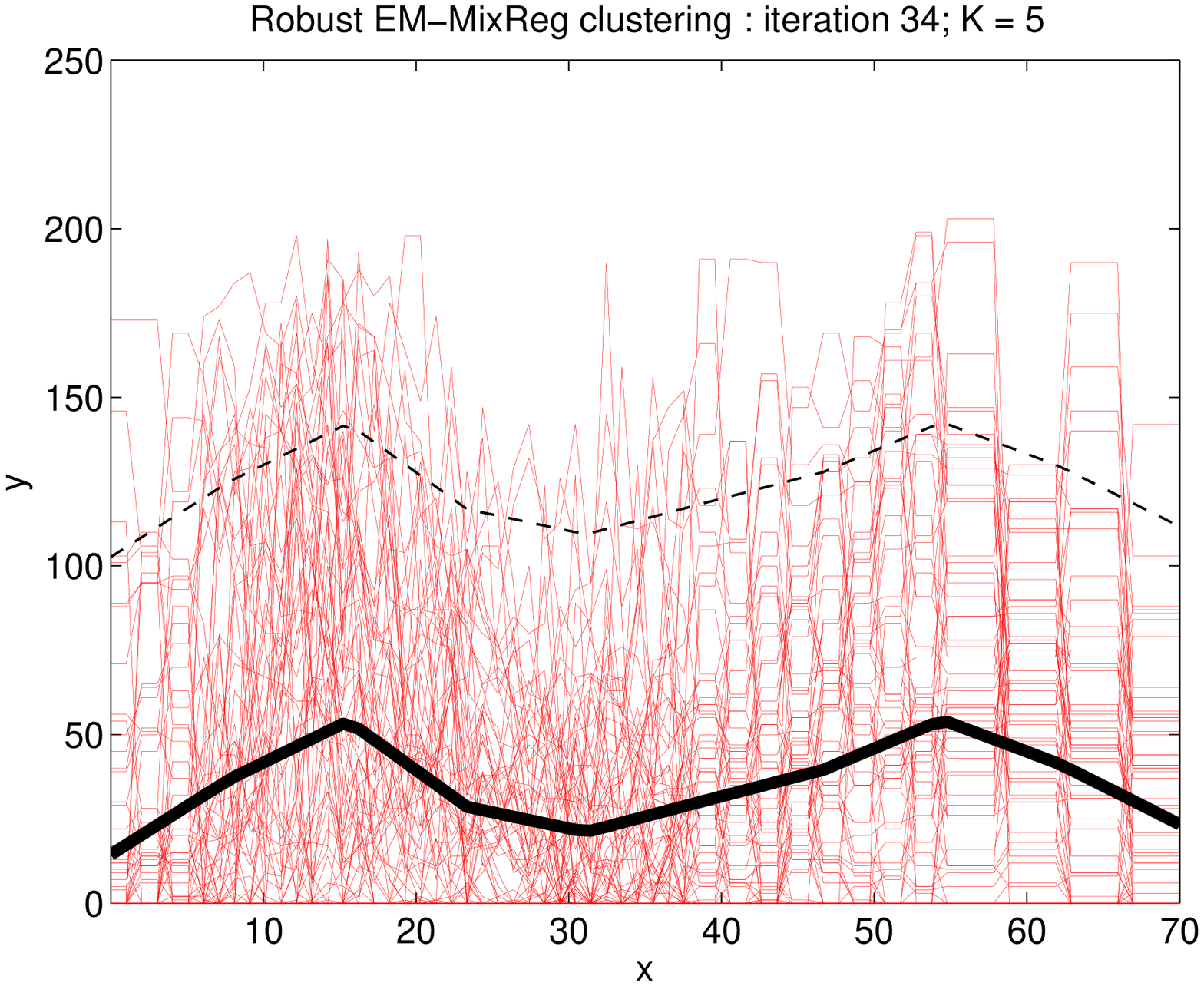}&
   \includegraphics[width=5.5cm]{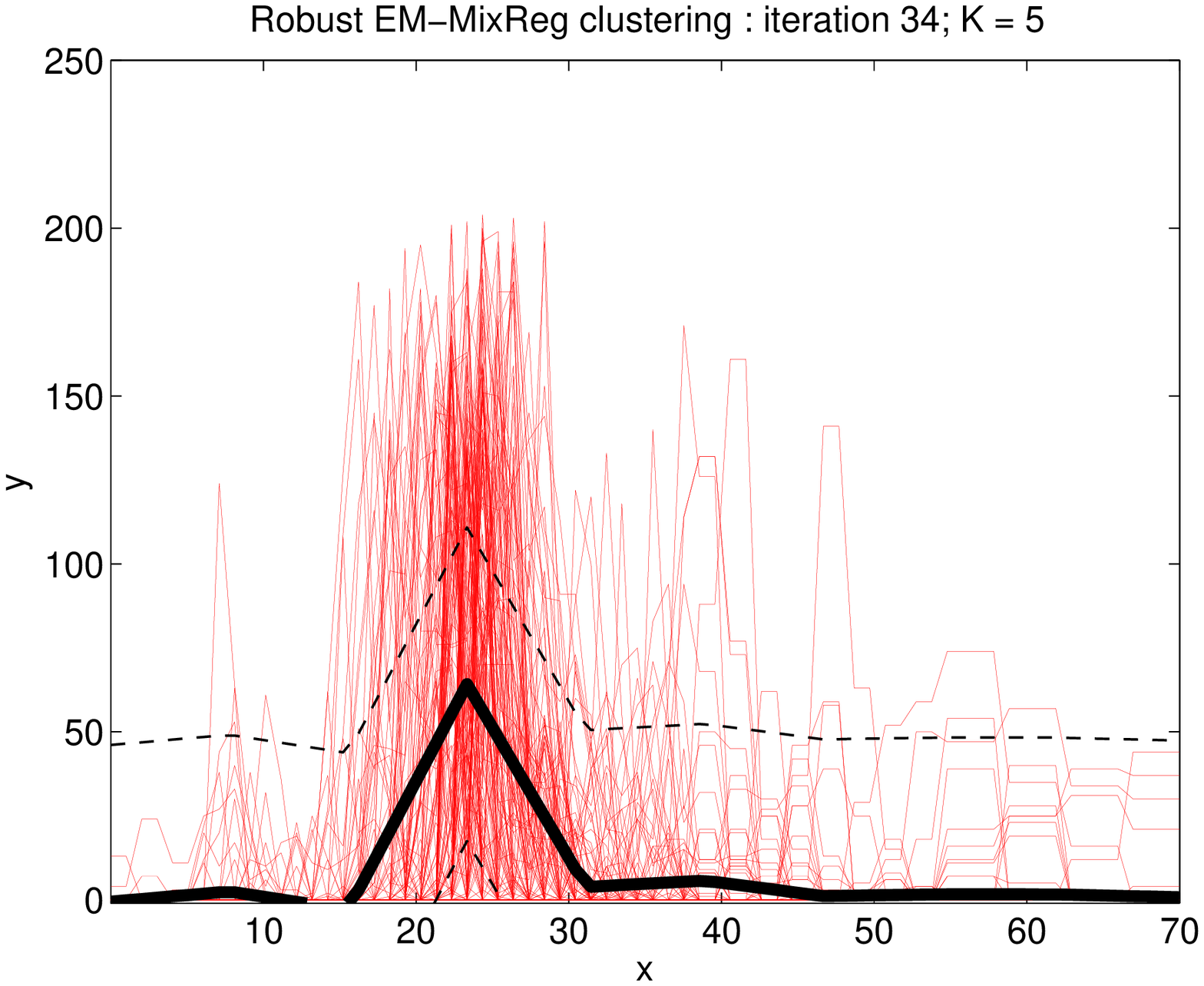}\\
    \includegraphics[width=5.5cm]{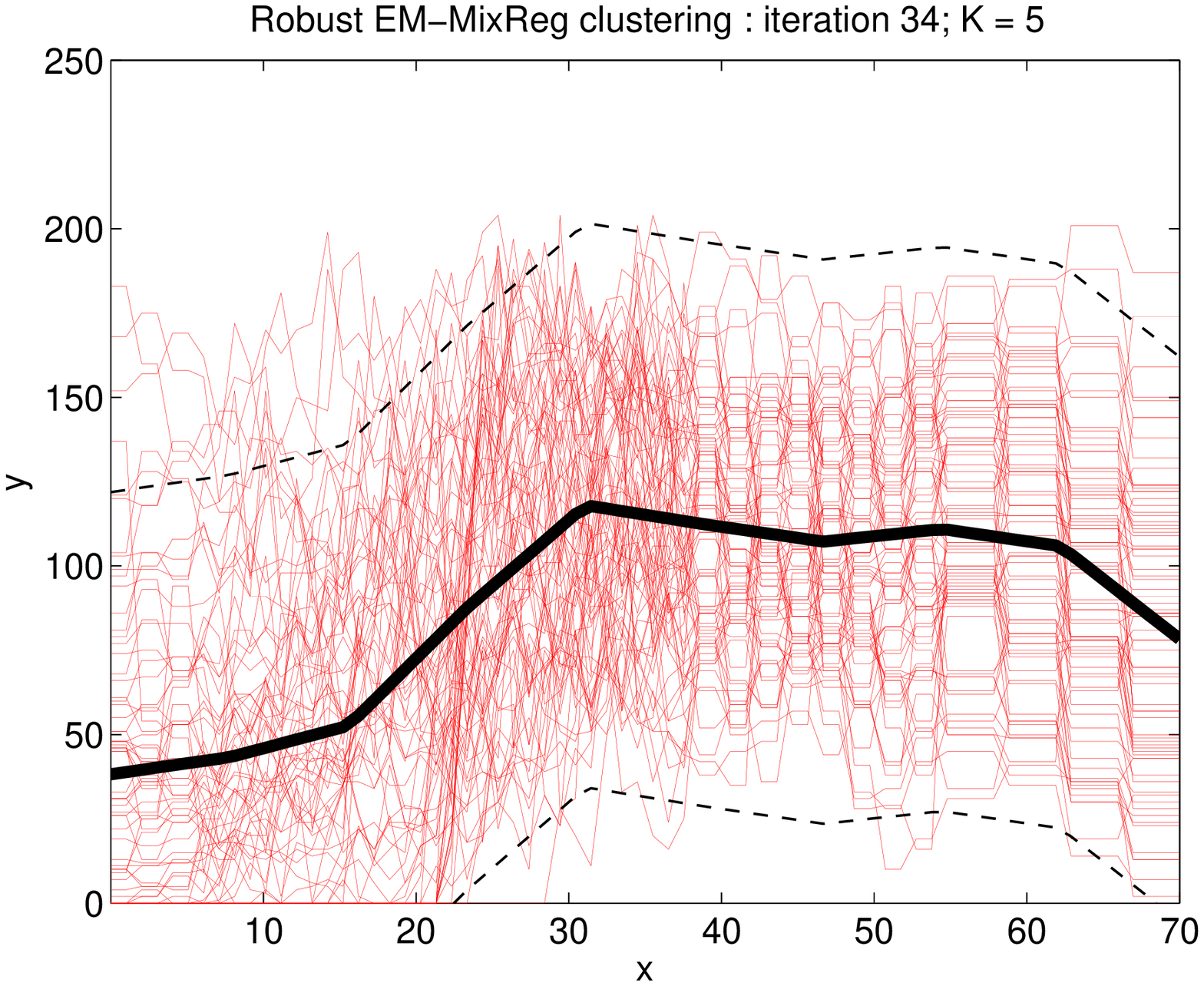}&
        \includegraphics[width=5.5cm]{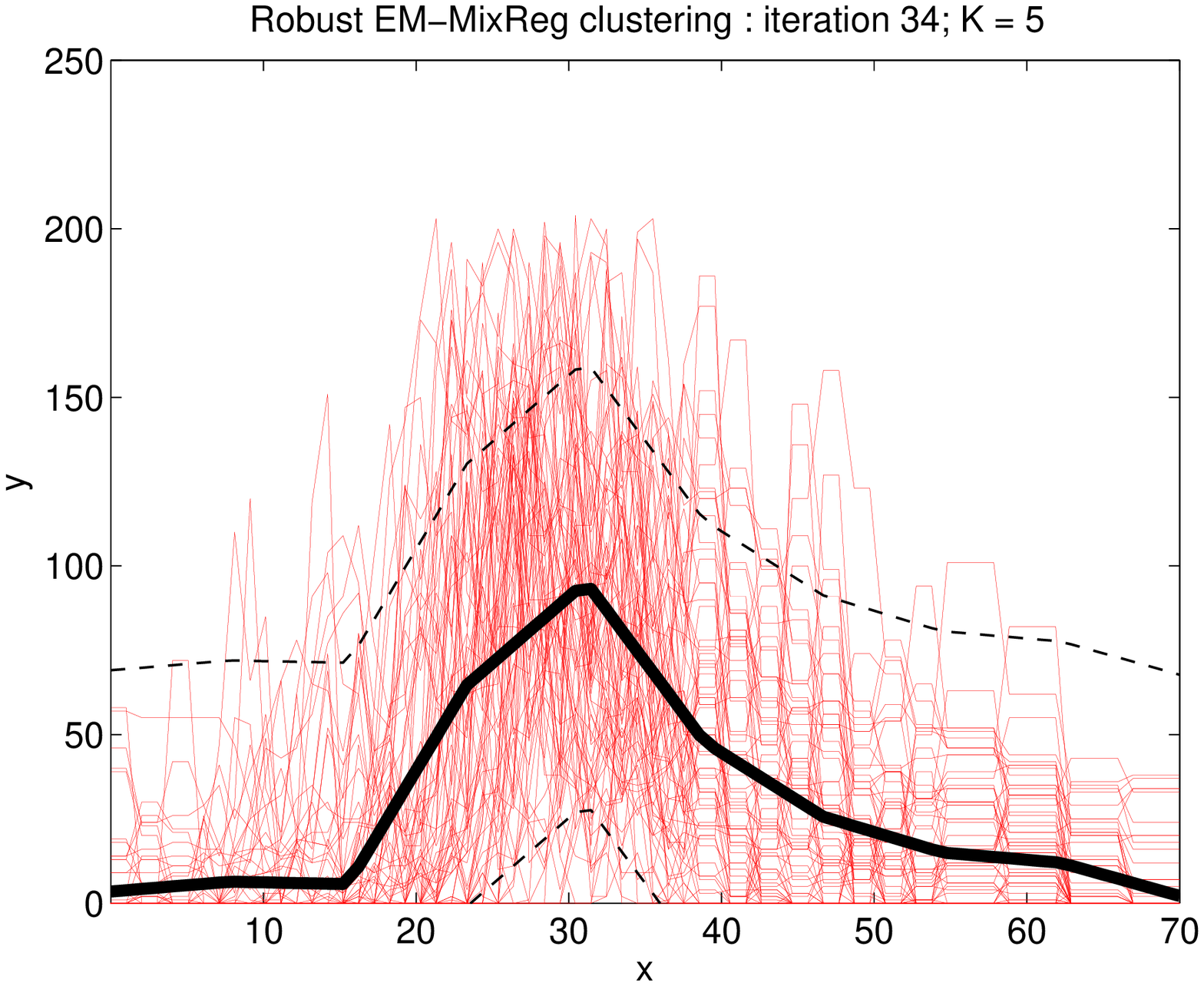}\\
            \includegraphics[width=5.5cm]{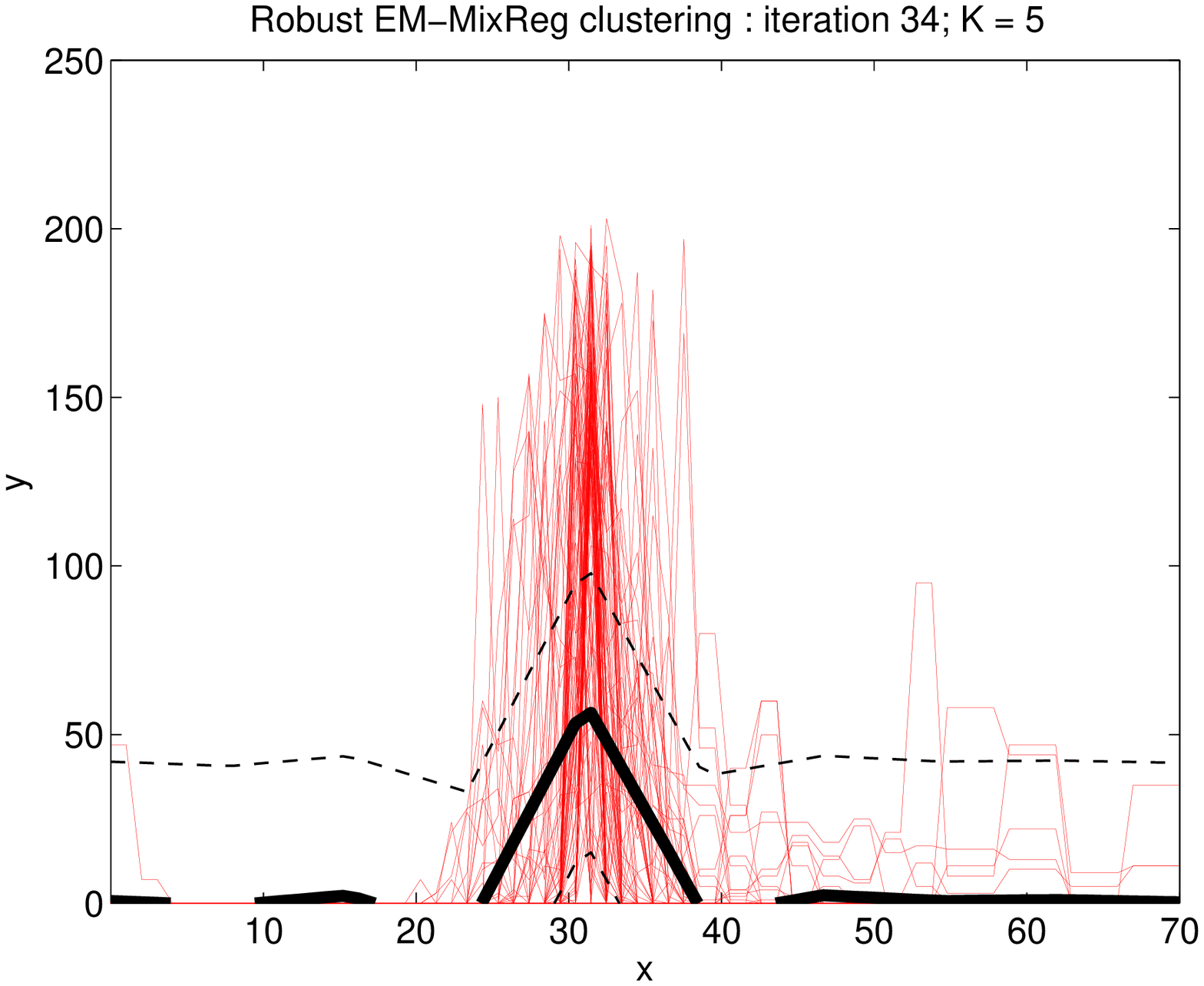} &
   \end{tabular} 
         \caption{\label{fig. robust EM-SRM satellite results}Clustering results obtained by the proposed robust EM algorithm and the SRM model with a linear spline of 8 knots for the satellite data.}
\end{figure}

\begin{figure}[H]
   \centering  
   \begin{tabular}{cc}
   \includegraphics[width=5.5cm]{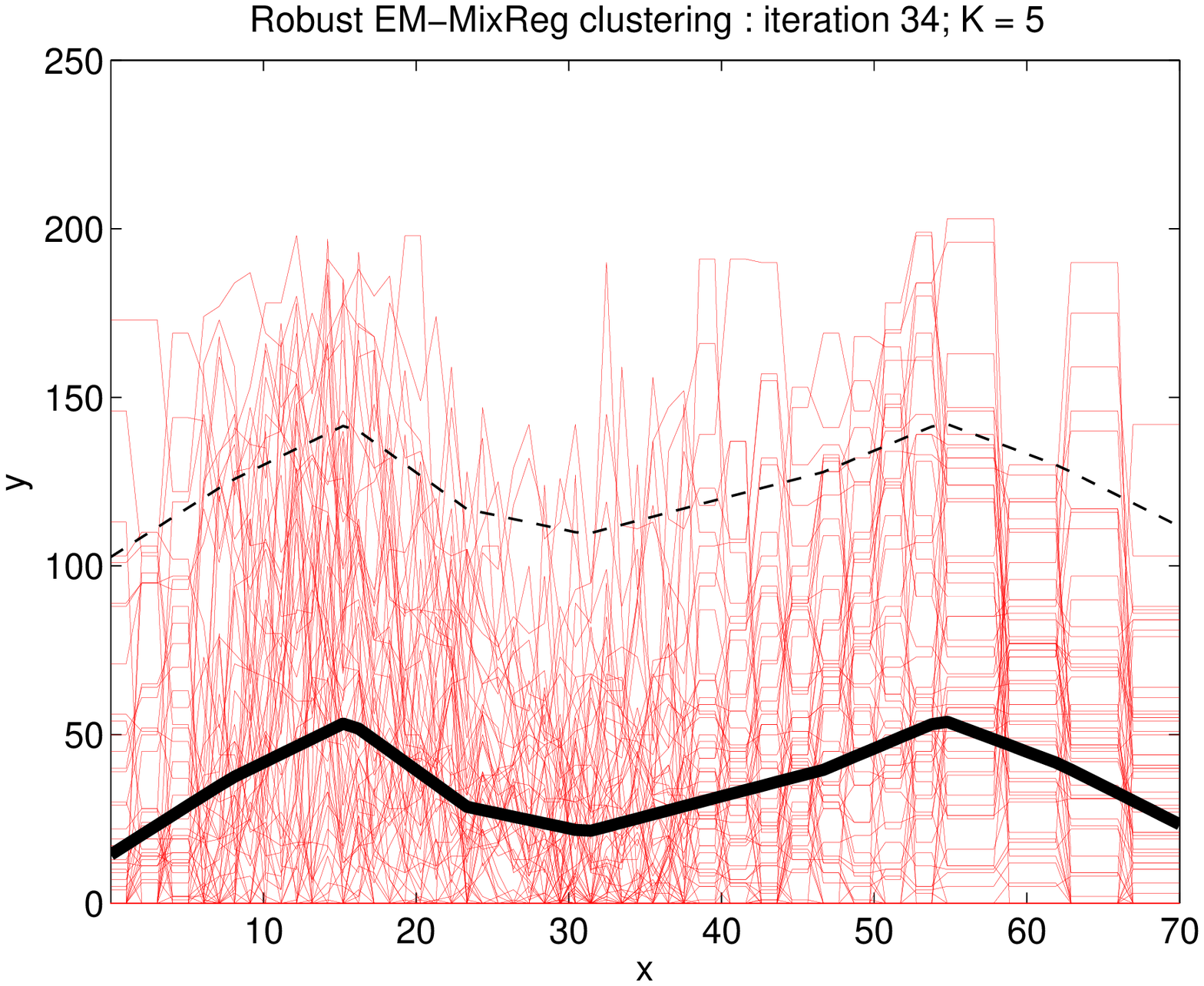}&
   \includegraphics[width=5.5cm]{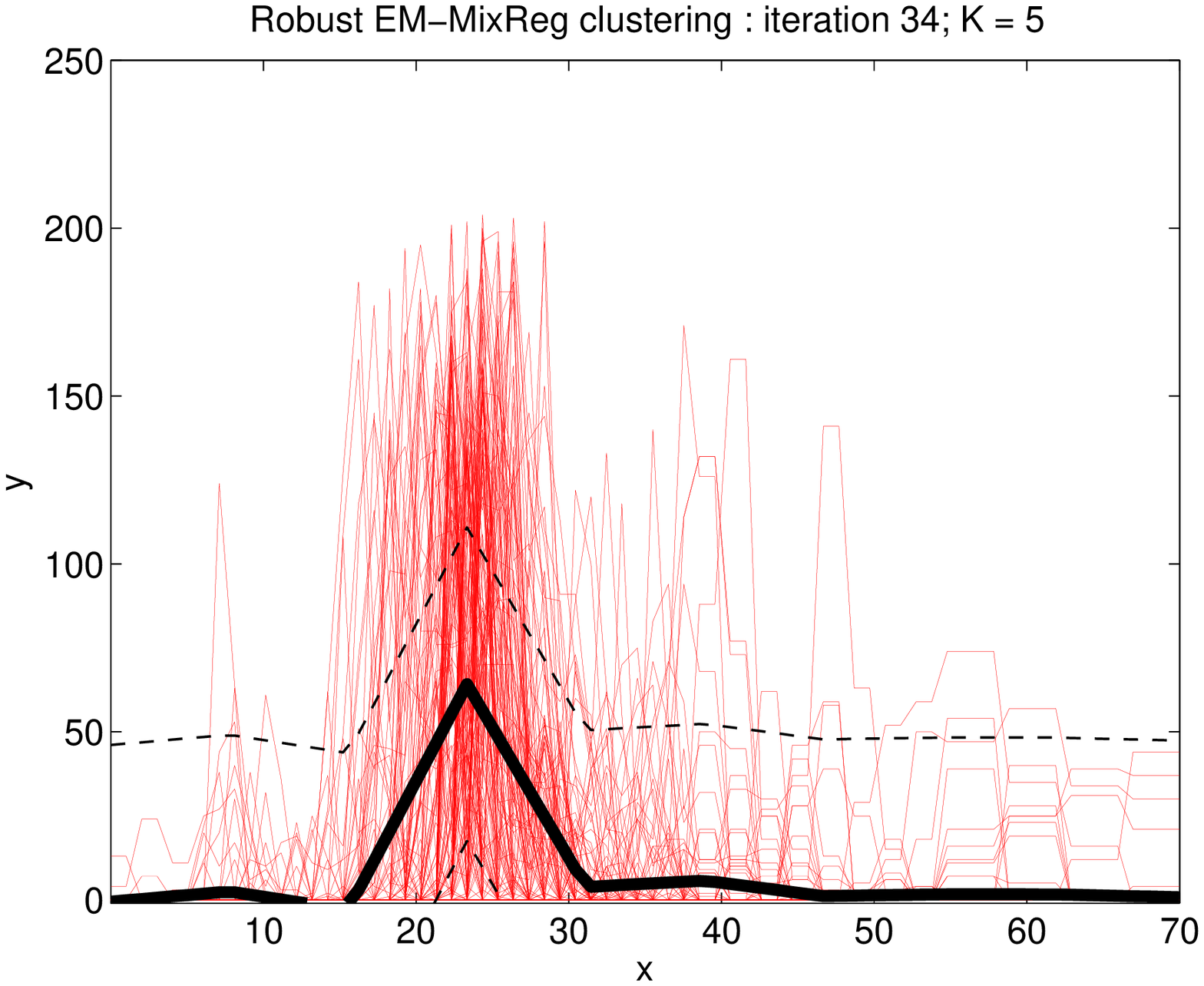}\\
    \includegraphics[width=5.5cm]{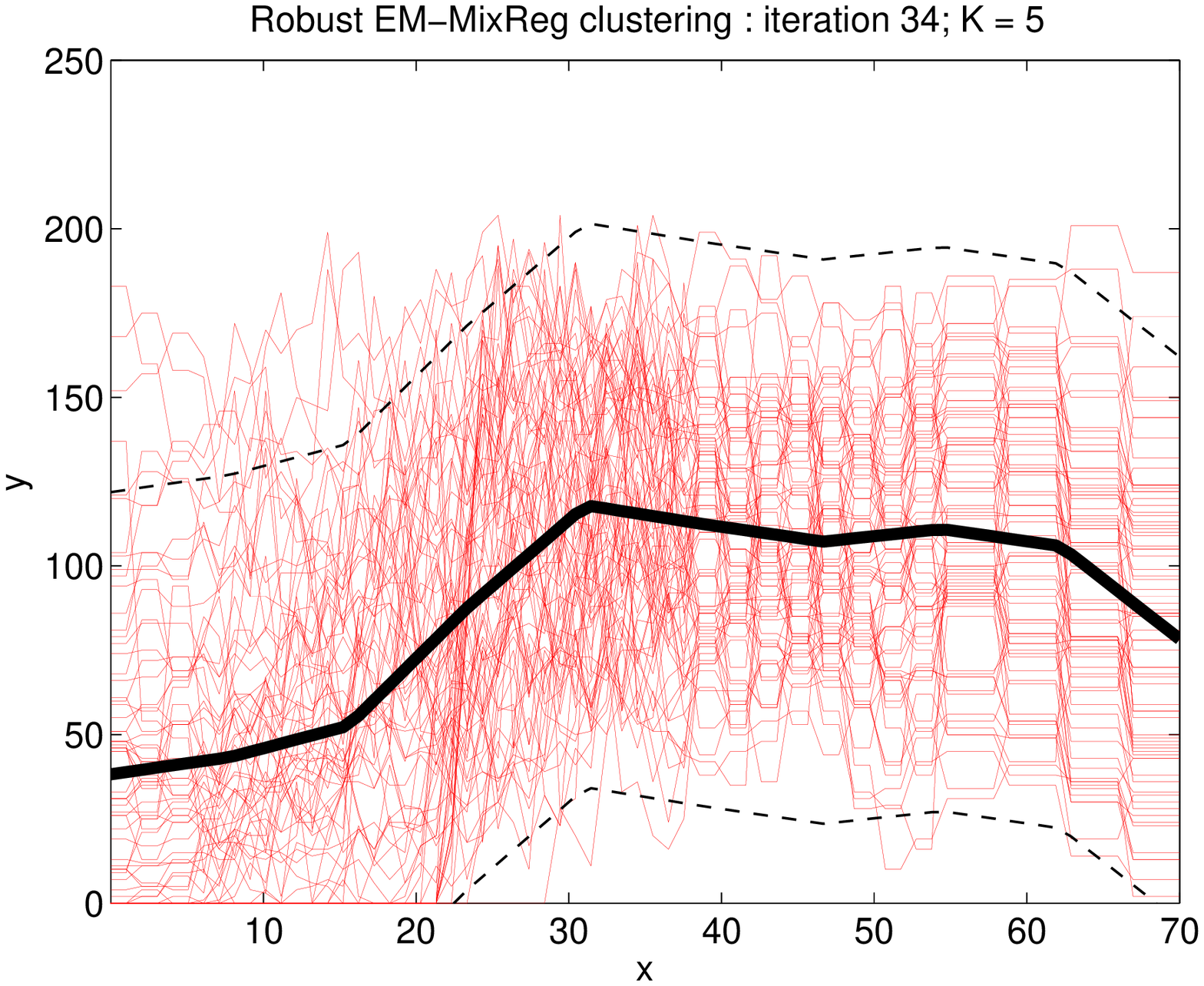}&
        \includegraphics[width=5.5cm]{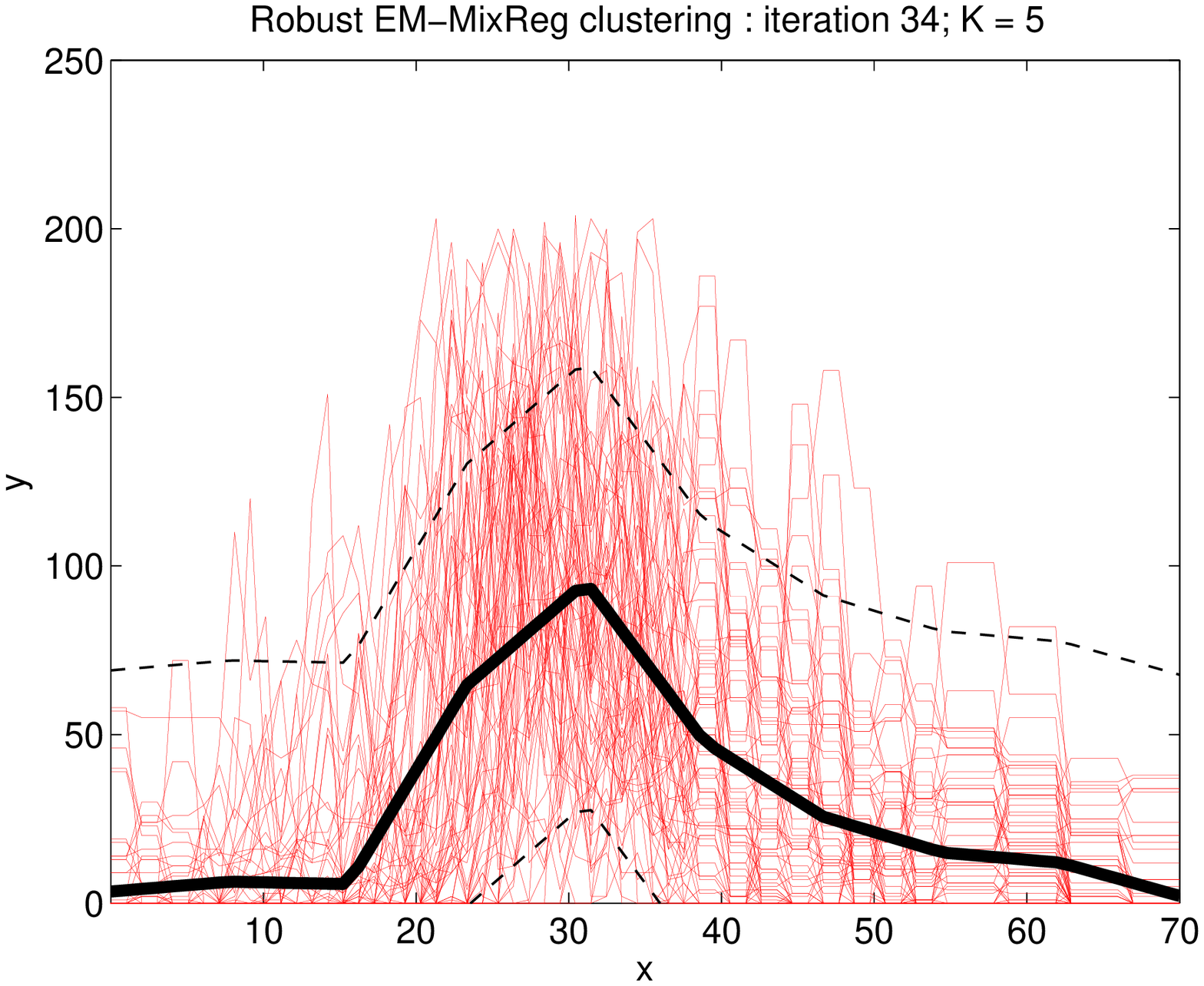}\\
            \includegraphics[width=5.5cm]{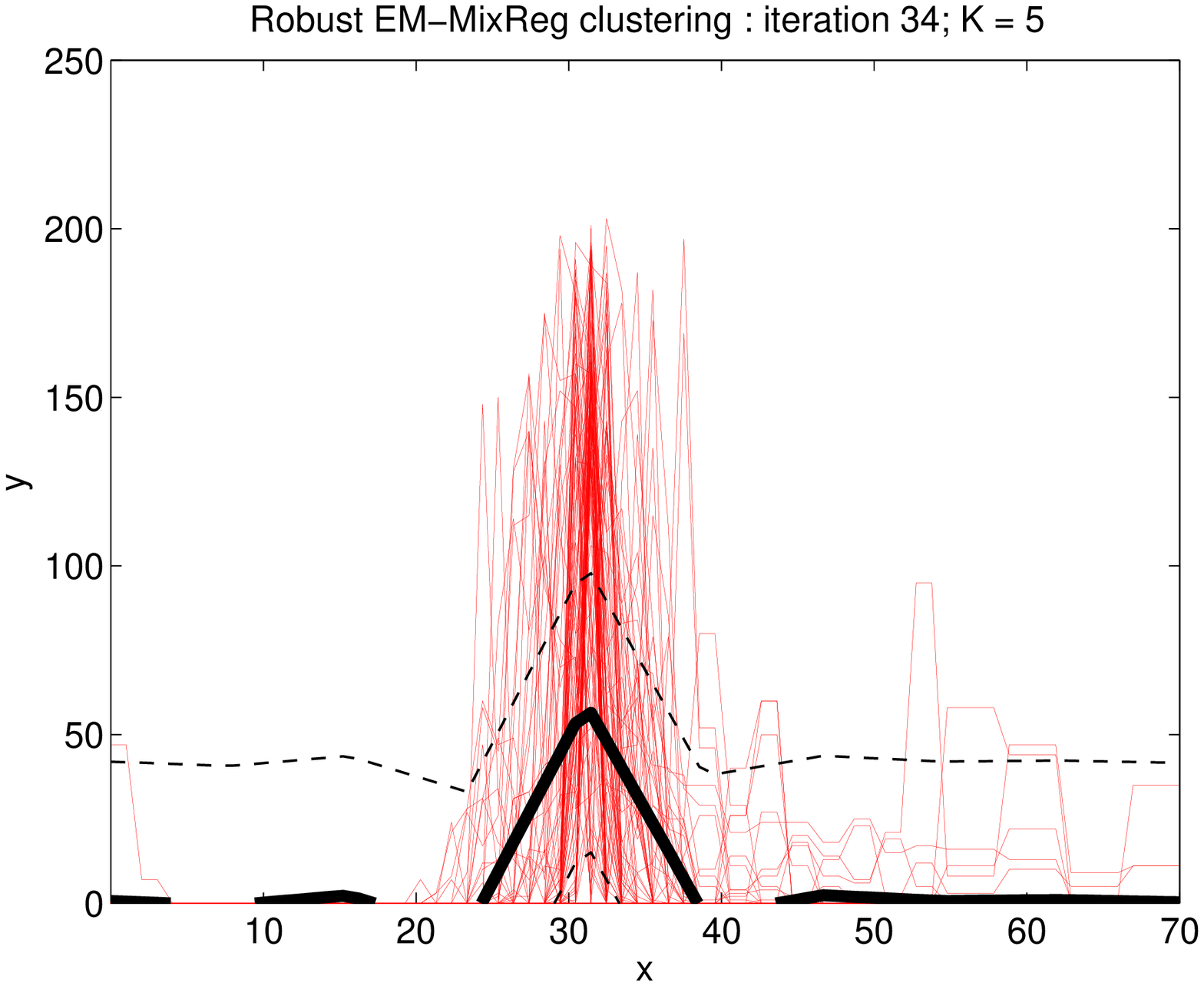} &
   \end{tabular} 
         \caption{\label{fig. robust EM-bSRM satellite results}Clustering results obtained by the proposed robust EM algorithm and the bSRM model with a linear B-spline of 8 knots for the satellite data.}
\end{figure}

\begin{figure}[H]
   \centering  
   \includegraphics[width=5cm]{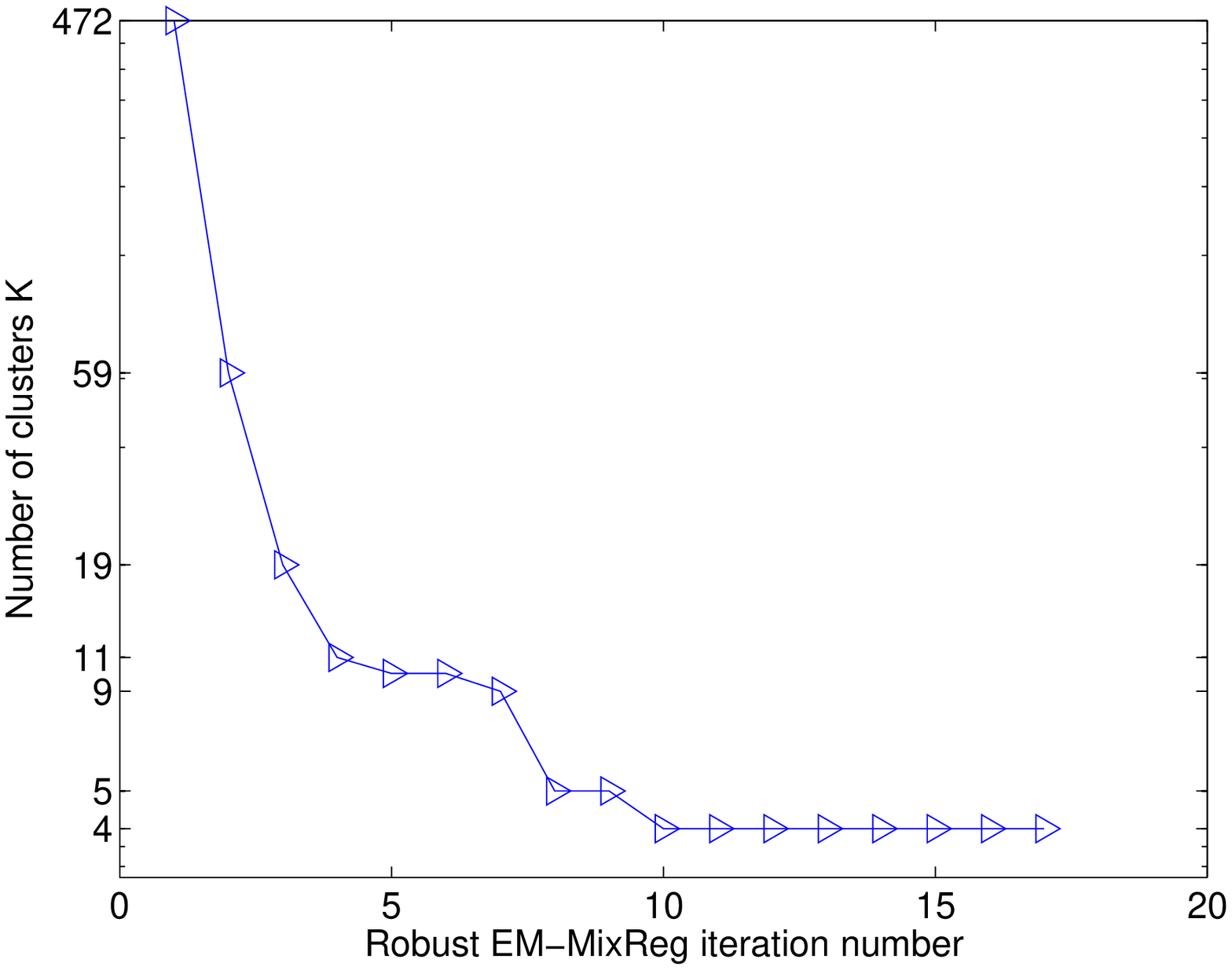}
   \includegraphics[width=5cm]{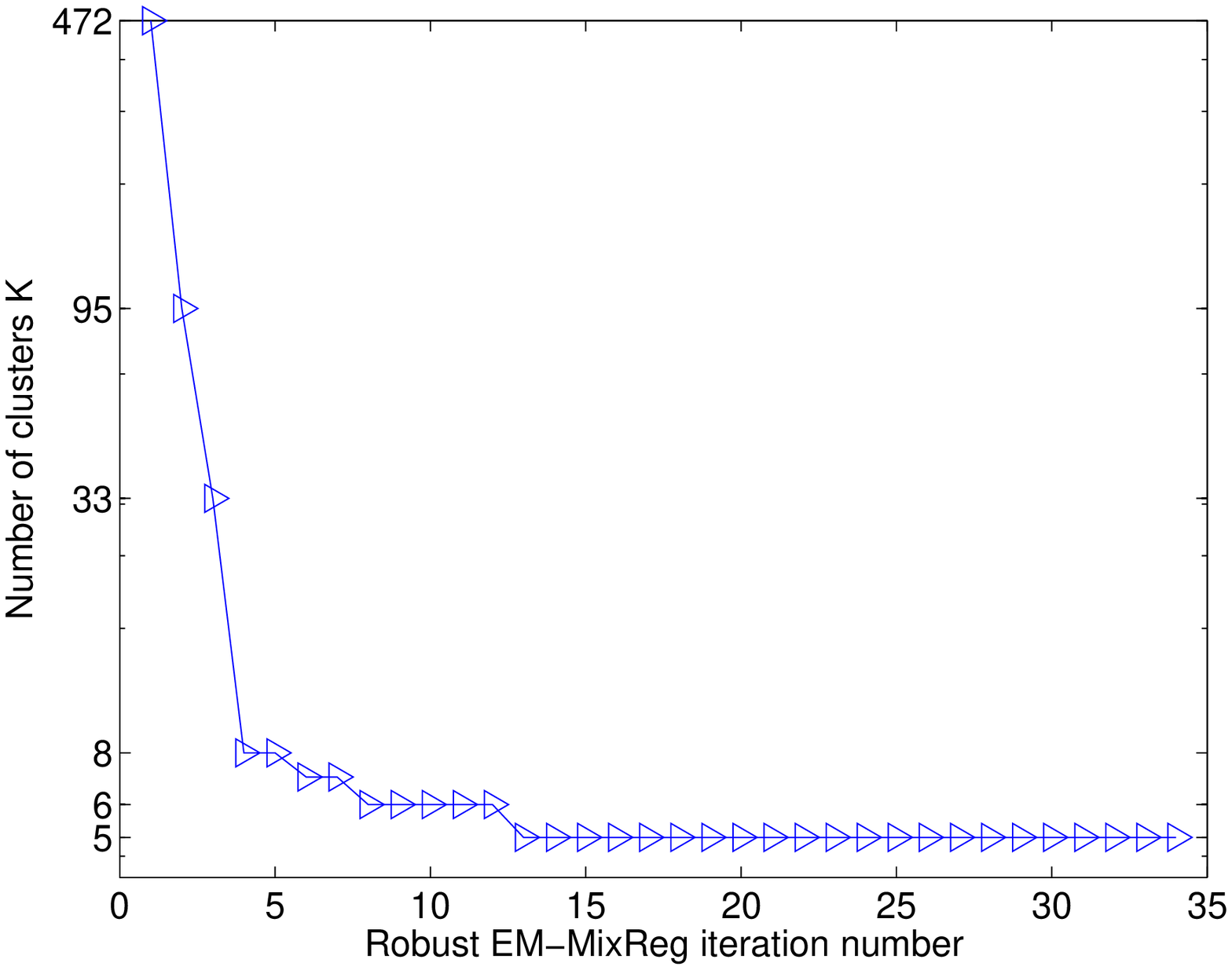}
   \includegraphics[width=5cm]{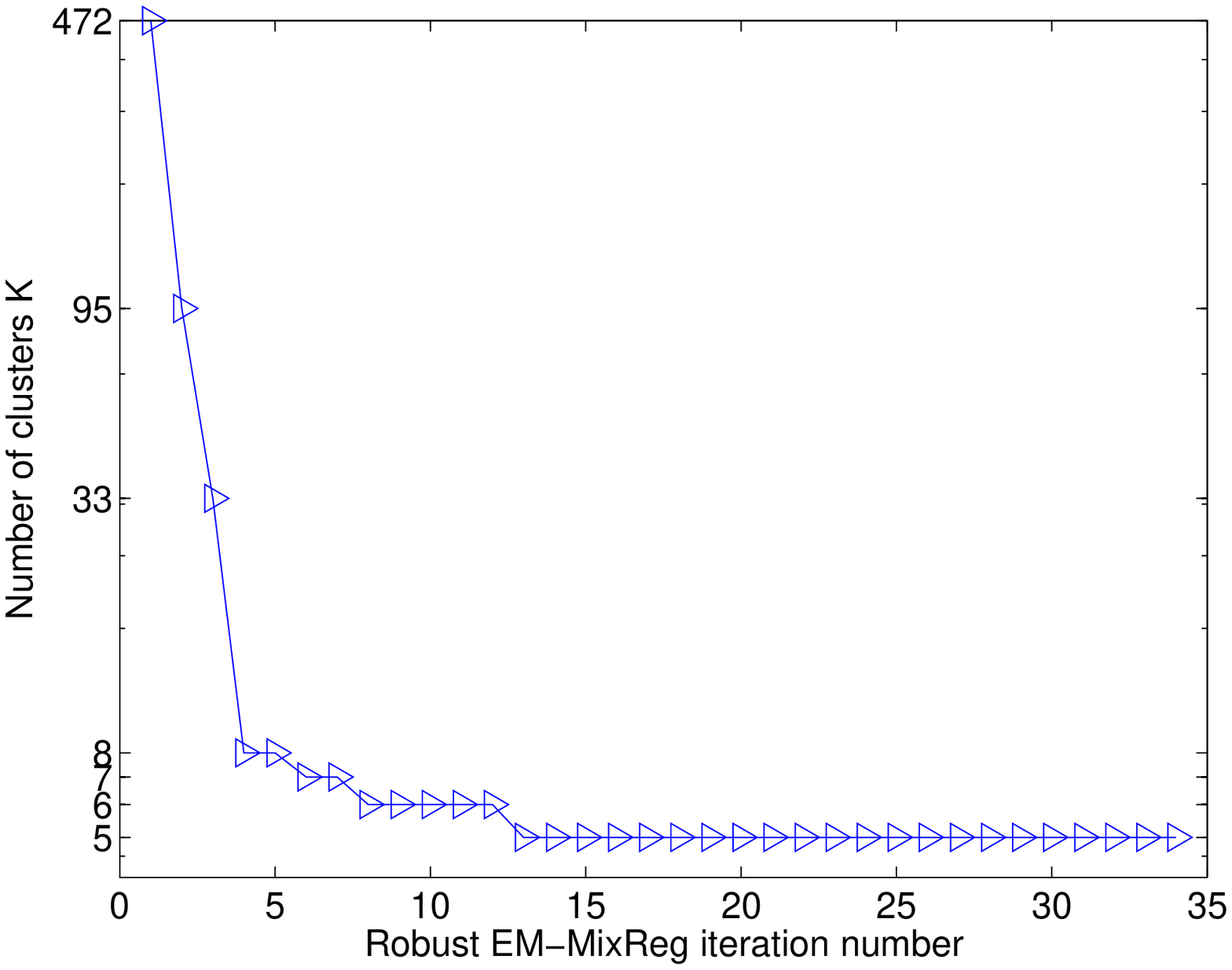}\\
   \includegraphics[width=5cm]{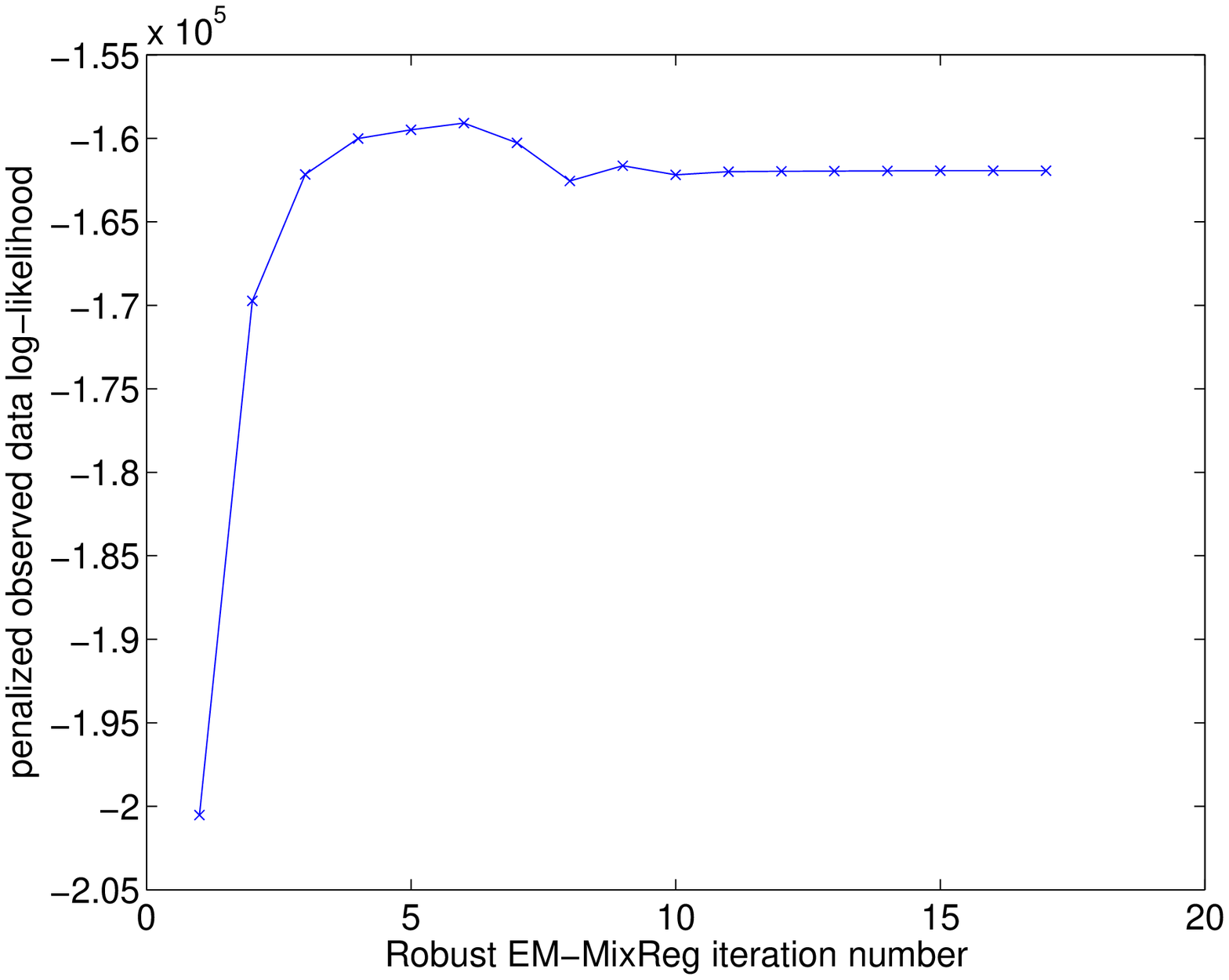}	
   \includegraphics[width=5cm]{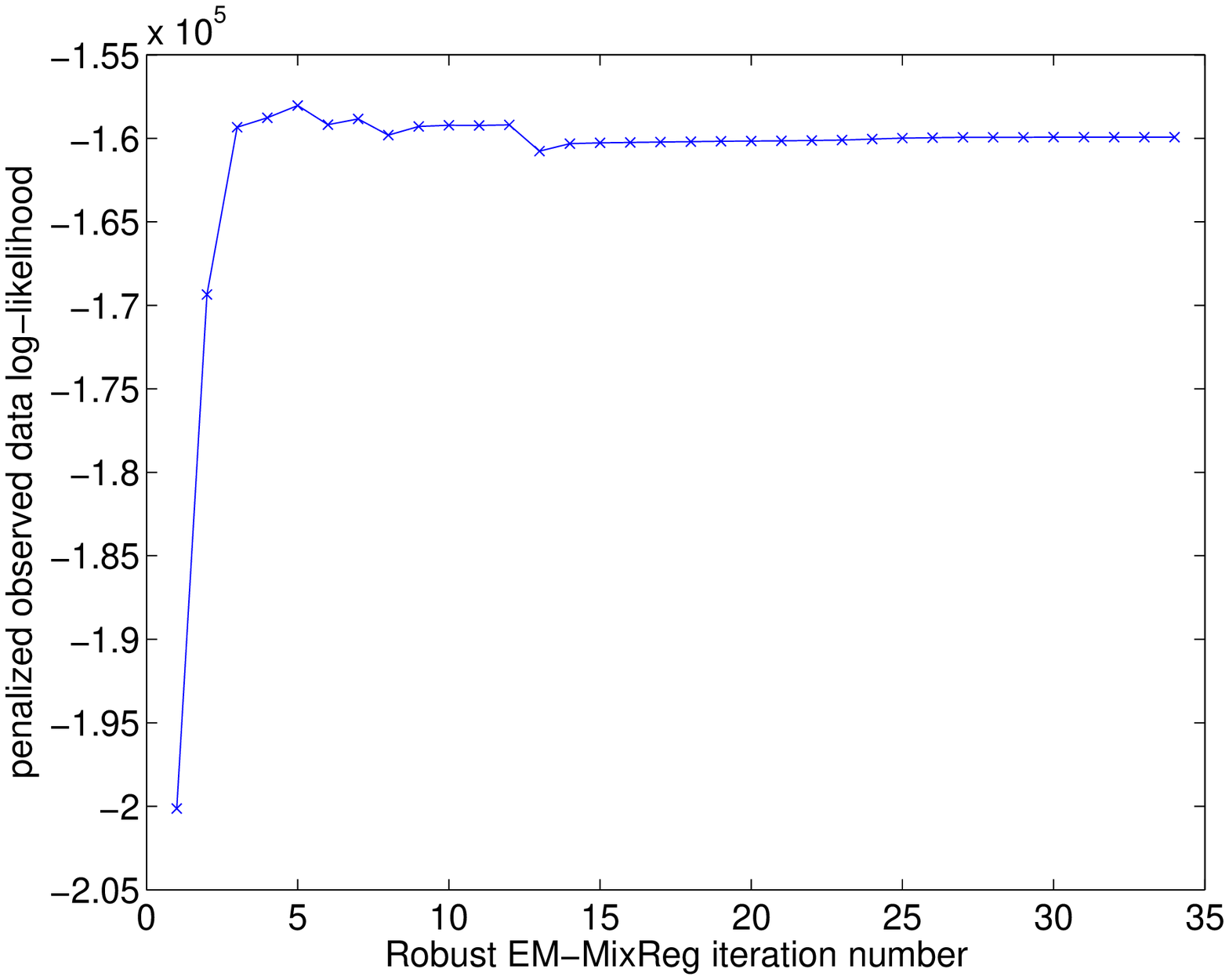}
   \includegraphics[width=5cm]{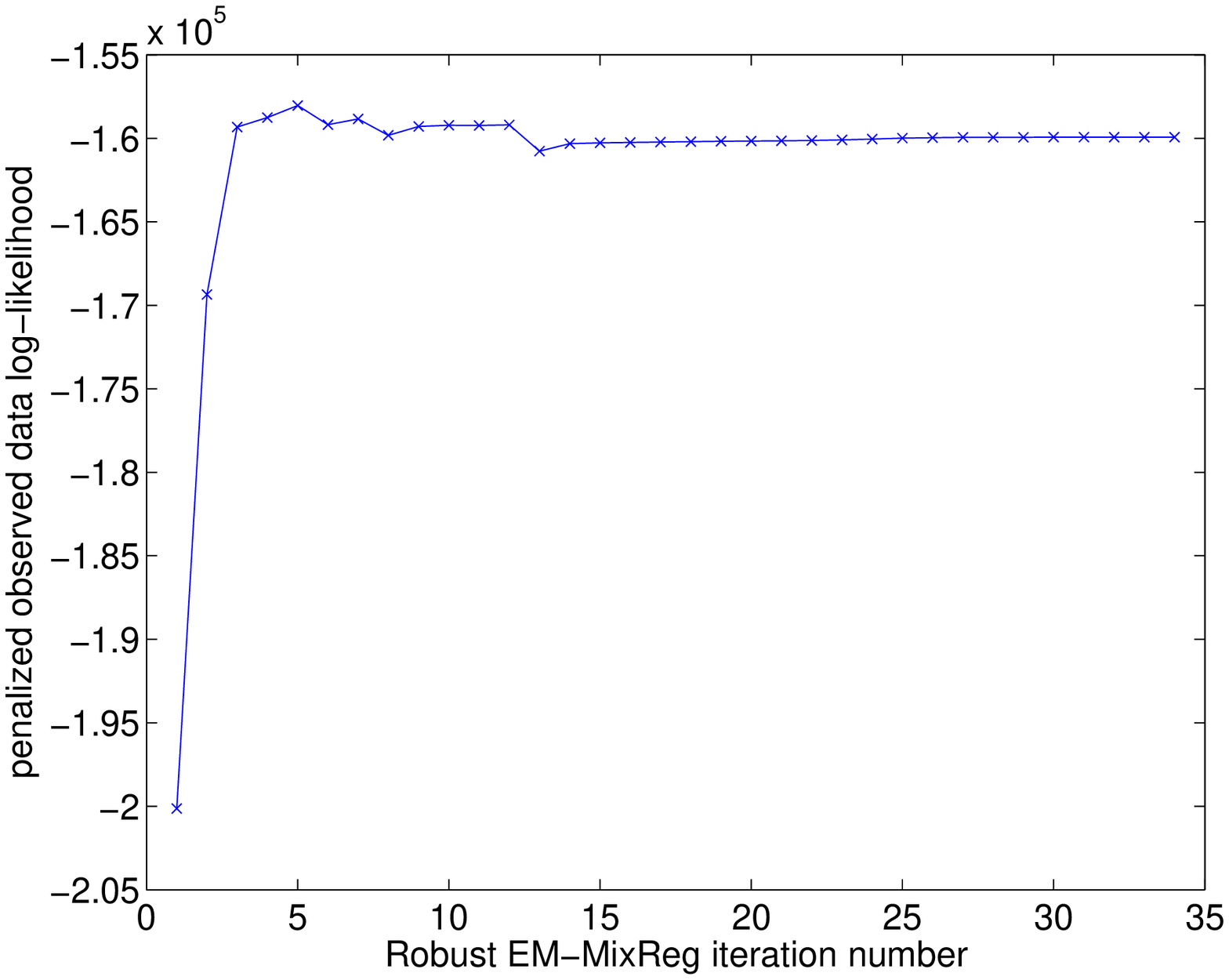}
   \caption{\label{fig: robust EM-MixReg stored-K pen-loglik satellite}Variation of the number of clusters and the value of the objective function during the iterations of the algorithm for the PRM (left) PSRM (middle) and PbSRM (right) for the satellite data.}
\end{figure}

Figures \ref{fig. robust EM-PRM satellite results}, \ref{fig. robust EM-SRM satellite results} and \ref{fig. robust EM-bSRM satellite results}  show the obtained clustering results. 
First, one can see that the provided solution for the polynomial regression mixture fitting is rather an overall rough approximation that provides three clusters. Even they don't look similar, the polynomial fitting for this type of curves is not adapted. This is because the curves namely present peaks and transitions. The solutions provided by the spline and B-spline regression mixture is more informative about the underlying structure of this dataset. We indeed used a linear (B)-spline for this dataset in order to allow piecewise linear function approximation and thus to better recover the possible peaks and transitions in the curves. As a result, both the SRM and the bSRM provide a five class partition. The partitions are quasi-identical and contain clearly informative clusters. We can indeed see different forms of waves that summarize the general structure underlying of this data. 
We can see two clusters containing curves presenting one narrow peak (the cluster in top left and the one in bottom). The two clusters differ with  the peak location in $x$. The cluster in middle left contains curves with one less narrow peak. The first cluster contains curves that look to have two large peaks. Finally, the third cluster in middle left looks to contain curves without peaks and with a part rather flat.
Furthermore, we can see that the structure more clear with the cluster mean (prototypes) than with the raw curves. The spline regression mixture models thus helps to better understand the underlying structure of the data as well as to recover a plausible number of clusters from the data.
%
%
In addition, the found number of cluster (five) also equals the one found by \cite{DaboNiang2007} by using another hierarchical nonparametric kernel-based unsupervised classification technique. The mean curves for the five terminal groups reflecting the hidden structure provided by the proposed approach for both the PRM and the bSRM are similar to those in \cite{DaboNiang2007}.
On the other hand, one can also see that this result is similar to the one found in \cite{hebrailEtal:2010}, most of the profiles are present in the two results. The slight difference can be attributed to the fact that the result in \cite{hebrailEtal:2010} is provided from a two-stage scheme which includes and additional pre-clustering step using the Self Organizing Map (SOM), rather by directly applying the piecewise regression model to the raw data.
We also notice that, in \cite{hebrailEtal:2010}, the number of clusters was set to twenty and the clustering procedure was two-fold. The authors indeed used a two-fold scheme and first performed a topographic clustering step using the SOM, and then apply a $K$-means-like approach to the results of the SOM. However, in our approach, we directly apply the proposed algorithm to the raw satellite data without a preprocessing step.  In addition, the number of clusters is automatically inferred from the data. 
We also can observe that, the found five clusters here do summarize the general behavior of the twenty clusters in \cite{hebrailEtal:2010} which can be summarized in clusters with one narrow shifted peak, less narrow peak, two large peaks and finally a cluster containing curves with sharp increase followed by a slow decrease.
%
Figure \ref{fig: robust EM-MixReg stored-K pen-loglik satellite} shows that the algorithms converge after at most 35 iterations. 
The variation of the number of clusters during
the iterations of the algorithm shows that after starting with 472 ($n$) clusters, the number of clusters rapidly decreases to 59 for the PRM and to 95 for both the SRM and the bSRM models. Then it gradually decreases until the number of clusters is stabilized.
The variation of the value of the objective function during the iterations of the algorithm also shows that it becomes horizontal at converge which correspond to the stabilization of the partition.

\section{Conclusions and discussion}
\label{sec: conclusion and discussions}

In this paper, we presented a new robust EM algorithm for model-based clustering using regression mixtures. It optimizes a penalized observed-data log-likelihood and overcomes both the problem of sensitivity to initialization and determining the optimal number of clusters for standard EM for regression mixtures.
The experimental results on simulated data and real-world data demonstrate the benefit of the proposed approach for applications in curve clustering.

We note that, for the polynomial regression mixture, the choice of the polynomial degree can be performed in such a way to obtain the best partition. In practice, we varied the polynomial degree from 3 to 7 for the simulated data and from 4 to 7 for the waveform data. The obtained clustering results was closely similar and the number of clusters was always correctly selected. For the phonemes data and the yeast cell cycle data, the polynomial degree with the best solution was retained. However, for a more general use in functional data clustering and approximation, the splines are clearly more adapted.
In practice, for the spline and B-spline regression mixtures, we used cubic (B)-splines because cubic splines, which correspond to a spline of order 4 which are are sufficient to approximate smooth functions. However, when the data present  irregularity, such as a kind of piecewise non continuous functions, which is the case of the the Topex/poseidon sateliite data, we use a linear (B)-spline approximation. 
We also note that the algorithm is fast for the three models and converges after a few number of iterations, and took at most less than one 45 seconds for the phonemes data. For the other data, it took only few seconds.  This makes is useful for real practical situations.

In this paper, we considered the problem of unsupervised fitting of regression mixtures with unknown number of components. %
The regression mixture models are similar to the mixture of experts (ME) model \cite{jacobsME}. Although similar, mixture of experts differ from curve clustering models in many respects. One of the main differences is that the ME model consists in a fully conditional mixture while in the regression mixture, only the component densities are conditional. Indeed, the mixing proportions are constant for the regression mixture, while in the mixture of experts, they mixing proportions (known as the gating functions of the network) are modeled as a function of the inputs, generally as a logistic or a softmax function. 
One interesting future direction is to extend the proposed approach to the problem of fitting mixture of experts \citep{jacobsME} and hierarchical mixture of experts \citep{Jordan1994} with unknown number of experts. 
%

\appendix
\section{Construction of B-splines basis functions}
\label{apxeq: construction of B-spline basis functions}

Given the sequence of knots $\xi_0<\xi_1<\ldots<\xi_{L+1} $ ($\xi_0$ and $\xi_{L+1}$ are the two bounds of $x$), let us define the augmented knot sequence $\zeta$  such that
\begin{itemize}
\item $\zeta_1 \leq \zeta_2 \ldots \leq\zeta_M \leq \xi_0$;
\item $\zeta_{M+\ell} = \xi_\ell , \; \ell = 1,\ldots, L;$
\item $\xi_{L+1} \leq \zeta_{L+M+1} \leq \zeta_{KL+M+2} \ldots \leq \zeta_{L + 2M}.$
\end{itemize}
The actual values of these additional knots beyond the boundary are arbitrary, and a common choice is to make them all the same and equal to $\xi_0$ and $\xi_{L+1}$ respectively.
Let us denote by $B_{\ell,M} (t)$ the $\ell$th B-spline basis function of order $M$ for the knot-sequence $$\zeta_1\leq \zeta_2 \ldots \leq \zeta_M \leq \xi_0 <\xi_1< \ldots <\xi_L <\xi_{L+1} \leq \zeta_{L+M+1} \leq \zeta_{L+M+2} \ldots \leq \zeta_{L + 2M}.$$
These basis functions are defined recursively as follows:
\begin{itemize}
\item $B_{\ell,1}(x_{ij})=\Indicatrice_{[\zeta_j,\zeta_{j+1}]}, \; \forall \ell = 1,\ldots, L + 2M - 1$;
\item $B_{\ell,M}(x_{ij})= \frac{x_{ij} - \zeta_\ell}{\xi_{\ell+M-1}-\zeta_\ell} B_{\ell,M-1}(x_{ij}) + \frac{\zeta_{\ell+M}-x_{ij}}{\zeta_{\ell+M}-\zeta_{\ell+1} } B_{\ell+1,M-1}(x_{ij}), \; \forall \ell = 1,\ldots, L + M.$
\end{itemize}
For the B-spline regression model, the $j$th  row $\bb_j$ ($j=1,\ldots,m_i$) of the $m_i \times (L+M)$ B-spline regression matrix $\bB_i$ for the $i$th curve  is then constructed as follows:
$$\bb_j=[B_{1,M}(x_{ij}), \; B_{2,M}(x_{ij}),\ldots,\;B_{L+M,M}(x_{ij})].$$

\section{Estimation of the mixing proportion}
\label{apx. Estimation of the mixing proportion}
 Consider the problem of finding the maximum of the function  (\ref{eq: J(pik) pinalized log-lik PRM})
{\small \begin{equation}
Q_{\pi}(\lambda,\pi_1,\ldots,\pi_K; \bstheta^{(q)}) = \sum_{i=1}^{n}\sum_{k=1}^{K}\tau_{ik}^{(q)} \log \pi_k + \lambda  n \sum_{k=1}^K  \pi_k \log \pi_k
\end{equation}}
w.r.t the mixing proportions $(\pi_{1},\ldots,\pi_{K})$ subject to the constraint $\sum_{k=1}^{K} \pi_{k} = 1$. 
To perform this constrained maximization, we introduce the Lagrange multiplier $\alpha$ such that the resulting Lagrangian function is given by:
{\small \begin{equation}
L(\pi_1,\ldots,\pi_K) = \sum_{i=1}^{n}\sum_{k=1}^{K}\tau_{ik}^{(q)} \log \pi_k + \lambda  n \sum_{k=1}^K  \pi_k \log \pi_k + \alpha (1-\sum_{k=1}^K \pi_k).
\end{equation}}
Taking the derivatives of the Lagrangian with respect to $\pi_k$ for $k=1,\ldots,K$ we obtain:
{\small \begin{equation}
\frac{\partial L(\pi_1,\ldots,\pi_K)}{\partial \pi_k} = \frac{\sum_{i=1}^{n} \tau^{(q)}_{ik}}{\pi_{k}} + \left(\lambda \sum_{i=1}^{n} (\log \pi_k + 1)\right)  - \alpha
\end{equation}}
Then, setting these derivatives to zero yields:
\begin{equation}
\frac{\sum_{i=1}^{n} \tau^{(q)}_{ik}}{\pi_{k}} + n \lambda \log \pi_k + n \lambda =  \alpha
\label{apx: Lagrange value for the mixing proportions}
\end{equation} 
By multiplying each hand side of (\ref{apx: Lagrange value for the mixing proportions}) by $\pi_k$ and summing over $k$ we get 
\begin{equation}
\sum_{k=1}^{K}  \pi_k \times \left(\frac{\sum_{i=1}^{n} \tau^{(q)}_{ik}}{\pi_{k}} + n \lambda \log \pi_k + n \lambda\right) = \sum_{k=1}^{K} \alpha \times \pi_k
\end{equation}
which implies that 
\begin{equation}
n + n \lambda \sum_{k=1}^{K} \pi_k \log \pi_k + n \lambda = \alpha\cdot
\end{equation}
Then, from (\ref{apx: Lagrange value for the mixing proportions}) we can write
\begin{equation}
 \sum_{i=1}^{n} \tau^{(q)}_{ik} + n \lambda \pi_{k} \log \pi_k +  n \lambda \pi_{k} =  n \pi_{k} + n \lambda \pi_{k} \sum_{h=1}^{K} \pi_h \log \pi_h + n \lambda  \pi_k
\end{equation} 
\begin{equation}
 n \pi_{k} = \sum_{i=1}^{n} \tau^{(q)}_{ik} + n \lambda \pi_k \log \pi_k  - n \lambda \sum_{h=1}^{K} \pi_h \log \pi_k 
\end{equation}
Finally we therefore get the updating formula for the mixing proportions $\pi_k$'s, that is
\begin{equation}
\pi_{k}^{(q+1)} = \frac{\sum_{i=1}^{n} \tau^{(q)}_{ik}}{n} + \lambda \pi^{(q)}_k \left(\log \pi^{(q)}_k  - \sum_{h=1}^{K} \pi^{(q)}_h \log \pi^{(q)}_h\right)
\end{equation}$\quad \forall k \in \{1,\ldots,K\}$.


\bibliographystyle{plainnat}

\bibliography{references}  

\end{document}